\RequirePackage{etex}
\documentclass[12pt,a4paper]{article} 
\pdfoutput=1
\usepackage{datetime} 
\usepackage{graphicx}
\usepackage{etex}
\usepackage{jheppub}
\usepackage{epstopdf}
\usepackage{epsfig}
\usepackage{dcolumn}  
\usepackage{bm}    
\usepackage{amssymb} 
\usepackage{amsmath,bm}
\usepackage{amsfonts}    
\usepackage{slashed}  
\usepackage{youngtab}
\usepackage[mathscr]{euscript}
\usepackage{epsfig}
\usepackage[utf8]{inputenc}
\usepackage{tikz}
\usetikzlibrary{tikzmark,calc,,arrows,shapes,decorations.pathreplacing}
\usepackage{xcolor}
\usepackage{mathrsfs}
\usepackage{verbatim}
\usepackage{cases}
\usepackage{bbding}
\usepackage{pifont}
\usepackage{ulem}
\usepackage{dcolumn}
\usepackage{pgf}
\usepackage{pstricks}
\usepackage{pst-node}
\usepackage[all]{xy}
\usepackage[figuresright]{rotating}
\usepackage{mathtools}
\usepackage{braket}
\usepackage{titlesec}
\usepackage[most]{tcolorbox}
\usepackage{stmaryrd}
\usepackage{cancel}
\usepackage{textcomp}
\usepackage{extarrows}
\usepackage{accents}
\usepackage{ulem}
\usepackage{color}

\usepackage{tikz}
\tikzstyle{vertex}=[circle, draw, minimum size=0pt]
\newcommand{\vertex}{\node[vertex]}

\usetikzlibrary{backgrounds,calc,positioning}

\usetikzlibrary{arrows.meta}

\def\lm{\limits}

\makeatletter
\DeclareFontFamily{OMX}{MnSymbolE}{}
\DeclareSymbolFont{MnLargeSymbols}{OMX}{MnSymbolE}{m}{n}
\SetSymbolFont{MnLargeSymbols}{bold}{OMX}{MnSymbolE}{b}{n}
\DeclareFontShape{OMX}{MnSymbolE}{m}{n}{
    <-6>  MnSymbolE5
   <6-7>  MnSymbolE6
   <7-8>  MnSymbolE7
   <8-9>  MnSymbolE8
   <9-10> MnSymbolE9
  <10-12> MnSymbolE10
  <12->   MnSymbolE12
}{}
\DeclareFontShape{OMX}{MnSymbolE}{b}{n}{
    <-6>  MnSymbolE-Bold5
   <6-7>  MnSymbolE-Bold6
   <7-8>  MnSymbolE-Bold7
   <8-9>  MnSymbolE-Bold8
   <9-10> MnSymbolE-Bold9
  <10-12> MnSymbolE-Bold10
  <12->   MnSymbolE-Bold12
}{}

\let\llangle\@undefined
\let\rrangle\@undefined
\DeclareMathDelimiter{\llangle}{\mathopen}%
                     {MnLargeSymbols}{'164}{MnLargeSymbols}{'164}
\DeclareMathDelimiter{\rrangle}{\mathclose}%
                     {MnLargeSymbols}{'171}{MnLargeSymbols}{'171}
\makeatother

\makeatletter
\newsavebox{\mstrut}
\newcommand{\bbra}[1]{%
    \sbox{\mstrut}{\(#1\)}%
    \mathinner{\left\langle\kern-0.5\ht\mstrut\left\langle{#1}\right|\mkern-2mu\right|}%
}
\newcommand{\kett}[1]{%
    \sbox{\mstrut}{\(#1\)}%
    \mathinner{\left|\mkern-2mu\left|{#1}\right\rangle\kern-0.5\ht\mstrut\right\rangle}%
}
\makeatother

\titleclass{\subsubsubsection}{straight}[\subsection]

\newcounter{subsubsubsection}[subsubsection]
\renewcommand\thesubsubsubsection{\thesubsubsection.\arabic{subsubsubsection}}
\titleformat{\subsubsubsection}
  {\normalfont\normalsize\bfseries}{\thesubsubsubsection}{1em}{}
\titlespacing*{\subsubsubsection}
{0pt}{3.25ex plus 1ex minus .2ex}{1.5ex plus .2ex}

\makeatletter
\renewcommand\paragraph{\@startsection{paragraph}{5}{\z@}%
  {3.25ex \@plus1ex \@minus.2ex}%
  {-1em}%
  {\normalfont\normalsize\bfseries}}
\renewcommand\subparagraph{\@startsection{subparagraph}{6}{\parindent}%
  {3.25ex \@plus1ex \@minus .2ex}%
  {-1em}%
  {\normalfont\normalsize\bfseries}}
\def\toclevel@subsubsubsection{4}
\def\toclevel@paragraph{5}
\def\toclevel@paragraph{6}
\def\l@subsubsubsection{\@dottedtocline{4}{7em}{4em}}
\def\l@paragraph{\@dottedtocline{5}{10em}{5em}}
\def\l@subparagraph{\@dottedtocline{6}{14em}{6em}}
\makeatother

\usepackage{young}
\usepackage{graphicx,rotating,booktabs}

\usepackage{varwidth}

\newdimen\tableauside\tableauside=1.0ex
\newdimen\tableaurule\tableaurule=0.4pt
\newdimen\tableaustep
\def\phantomhrule#1{\hbox{\vbox to0pt{\hrule height\tableaurule width#1\vss}}}
\def\phantomvrule#1{\vbox{\hbox to0pt{\vrule width\tableaurule height#1\hss}}}
\def\sqr{\vbox{%
		\phantomhrule\tableaustep
		\hbox{\phantomvrule\tableaustep\kern\tableaustep\phantomvrule\tableaustep}%
		\hbox{\vbox{\phantomhrule\tableauside}\kern-\tableaurule}}}
\def\squares#1{\hbox{\count0=#1\noindent\loop\sqr
		\advance\count0 by-1 \ifnum\count0>0\repeat}}
\def\tableau#1{\vcenter{\offinterlineskip
		\tableaustep=\tableauside\advance\tableaustep by-\tableaurule
		\kern\normallineskip\hbox
		{\kern\normallineskip\vbox
			{\gettableau#1 0 }%
			\kern\normallineskip\kern\tableaurule}%
		\kern\normallineskip\kern\tableaurule}}
\def\gettableau#1 {\ifnum#1=0\let\next=\null\else
	{{\tiny\yng(1)}}s{#1}\let\next=\gettableau\fi\next}

\renewcommand{\(}{\left(}
\renewcommand{\)}{\right)}

\newcommand{\includeCroppedPdf}[2][]{%
\IfFileExists{./#2-crop.pdf}{}{%
\immediate\write18{pdfcrop #2 #2-crop.pdf}}%
\includegraphics[#1]{#2-crop.pdf}}        
               
\newcommand{\be}{ \begin{equation}}
\newcommand{\ee}{\end{equation}}
\newcommand{\bea}[1]{\begin{eqnarray}\label{#1} }
\newcommand{\eea}{\end{eqnarray}}
\newcommand{\eq}[1]{(\ref{#1})}
\def\ZZZ{{\hskip-3pt\hbox{ Z\kern-1.6mm Z}}}
\def\zzz{{\hskip-3pt\hbox{ z\kern-1mm z}}}

\newcommand{\IZ}{{\mathbb{Z}}}

\newcommand{\K}{{\rm K_1}}

\def\bal#1\eal{\begin{align}#1\end{align}}

\newcommand{\W}{{\cal W}}

\renewcommand{\(}{\left(}
\renewcommand{\)}{\right)}

\newcommand\Kappa{\mathrm{K}}

\def\mye{e}
\def\myf{f}
\def\myk{\psi}

\def\mys{\texttt{s}}

\def\myphi{\boldsymbol{\varphi}}
\def\myp{{\mathfrak{p}}}

\def\extra{{\ft}}
\def\ft{\mathfrak{t}}

\newcommand{\longsquiggly}{\xymatrix{{}\ar@{~>}[r]&{}}}

\def\mye{e}
\def\myf{f}
\def\myk{\psi}

\newcommand\sqbox[1]{{
	\setbox0=\hbox{\scalebox{1}{\mbox{$\Box$}}}
	\setbox1=\hbox{\mbox{\raisebox{0.35ex}{\scriptsize #1}}}
	\mbox{\raisebox{-0.2ex}{\rlap{\hbox to \wd0{\hss{\box1}\hss}}\box0}}
}}

\newcommand\sqboxs[1]{{
	\setbox0=\hbox{\scalebox{1.3}{\mbox{$\Box$}}}
	\setbox1=\hbox{\mbox{\raisebox{0.5ex}{\scriptsize #1}}}
	\mbox{\raisebox{-0.2ex}{\rlap{\hbox to \wd0{\hss{\box1}\hss}}\box0}}
}}

\newcommand\sqboxss[1]{{
	\setbox0=\hbox{\scalebox{2}{\mbox{$\Box$}}}
	\setbox1=\hbox{\mbox{\raisebox{0.6ex}{\scriptsize #1}}}
	\mbox{\raisebox{-0.2ex}{\rlap{\hbox to \wd0{\hss{\box1}\hss}}\box0}}
}}

\title{Quiver algebras and their representations \\
for arbitrary quivers}
\author{Wei Li}
\affiliation{Institute of Theoretical Physics, Chinese Academy of Sciences,\\
\hspace*{0.3cm}100190 Beijing, P.R.\ China} 

\emailAdd{weili@mail.itp.ac.cn}

\date{\today}

\abstract{
The quiver Yangians were originally defined for the quiver and superpotential from string theory on general toric Calabi-Yau threefolds, and serve as BPS algebras of these systems.
Their characters reproduce the unrefined BPS indices, which are related to classical Donaldson-Thomas (DT) invariants.
We generalize this construction in two directions.
First, we show that this definition extends to arbitrary quivers with potentials.
Second, we explore how one can refine the characters of the quiver Yangian to incorporate the refined BPS indices, which are related to motivic DT invariants.
We focus on two main classes of quivers: the BPS quivers of 4D $\mathcal{N}=2$ theories and the quivers from the knot-quiver correspondence. 
The entire construction allows for straightforward generalizations to trigonometric, elliptic, and generalized cohomologies. 
}
\begin{document}

\maketitle
	
\section{Introduction}

The BPS algebra, or the algebra \textit{of}
BPS states, captures essential information about the BPS sector of an 4D $\mathcal{N}=2$ supersymmetric gauge theory \cite{Harvey:1996gc}. 
For example, various different counting problems within a given theory can be described by different representations of the BPS algebra for that theory \cite{Galakhov:2021xum}. 
The BPS algebra also provides the underlying reason behind various correspondences involving BPS sectors, such as the AGT correspondence \cite{Alday:2009aq,Wyllard:2009hg}, the Gauge/Bethe correspondence \cite{Nekrasov:2009uh,Nekrasov:2009ui,Nekrasov:2009rc,Galakhov:2022uyu},\footnote{For recent developments on on Gauge/Bethe correspondence see e.g.\ \cite{Dedushenko:2021mds,Bullimore:2021rnr,Gu:2022dac}.} etc.

\medskip

For type IIA string theory compactified on arbitrary toric Calabi-Yau threefolds (CY$_3$'s), the BPS algebras are given by the quiver Yangians defined in \cite{Li:2020rij}.
The quiver Yangians correspond to the rational cohomology of the BPS moduli space and have been further generalized for the trigonometric,  elliptic, and generalized cohomologies, giving rise to toroidal quiver algebras, elliptic quiver algebras, and more \cite{Galakhov:2021omc}.

The quiver Yangians $\textsf{Y}(Q,W)$ in \cite{Li:2020rij} were defined for the quiver and superpotential pair $(Q,W)$ that arises from compactifying type II string theory on toric CY$_3$'s.
These quiver-superpotential pairs satisfy certain properties such that the representations are given by certain 3D crystals \cite{Ooguri:2008yb}, which we call BPS crystals.
In \cite{Li:2020rij}, the quiver Yangians $\textsf{Y}(Q,W)$ were obtained by first fixing their actions on the set of BPS crystals, by demanding that they reproduce the fusion process of the BPS states, and then determining the algebraic relations based on these actions.
Namely, one can bootstrap the BPS algebra from its action on the set of BPS crystals. 
The quadratic part of the algebra is very easy to write down, and its dependence on the quiver and superpotential is manifest. 

\medskip

In this paper, we first generalize the construction of quiver Yangians in \cite{Li:2020rij} from quivers associated to toric CY$_3$'s to arbitrary quivers.\footnote{
Note that the quivers considered in this paper only have directed arrows, i.e.\ we do not include those with unoriented links, such as those appearing in string theory on Calabi-Yau four-folds. In addition, in this paper, the term ``quiver" often refers to a quiver with potential.
}
This allows us to apply the quiver Yangians to study the BPS quivers of all 4D $\mathcal{N}=2$ gauge theories, as well as the quivers from the knot-quiver correspondence \cite{Kucharski:2017poe,Kucharski:2017ogk}.
As we will see, for the quadratic part of the quiver Yangian, the definition applies to all quivers, not just those from toric CY$_3$'s.\footnote{
Already for the toric CY$_3$ quivers, the higher order relations need to be defined in a case-by-case manner. 
} 
The crucial difference is that for quivers not from toric CY$_3$'s, the representations were previously unknown. 
We will show how to construct similar representations for arbitrary quivers, for which the representations lose the 3D crystal structure.
Instead, they are given by certain posets (i.e.\ partially ordered sets) defined from the quiver.
These posets serve as representations of the quiver Yangian for the given quiver.
We will first construct the vacuum-like representations and then use the vacuum-like representations as building blocks to define non-vacuum representations.\footnote{
Similar to the toric CY$_3$ case, the quiver Yangian for an arbitrary quiver can be bootstrapped once its action on its representations is given. 
However in this paper, once the representation space is constructed, the detailed action of the algebra on the representation will be determined from the quadratic part of the algebraic relations. } 

\medskip

Once the representations are constructed, it is easy to compute their characters by simply enumerating their states (as finite-codimensional ideals of the corresponding Jacobian algebra), and these characters correspond to the unrefined framed BPS degeneracies (related to the numerical or classical Donaldson-Thomas invariants).
We show that they are consistent with results computed using other methods.
By matching the characters computed this way with the characters computed by counting independent states built from applying the modes of raising operators on the ground states, we can also determine all the higher order relations of quiver Yangian. 

We also explore how to refine the characters of quiver Yangians in order to reproduce the refined framed BPS degeneracies (which are related to the motivic DT invariants).
First of all, one can try to define the refinement of the quiver Yangian representations by utilizing the spin-grading of the raising operators. 
To use this method to compute the refined character of a representation, one first needs to enumerate all the states by applying the raising operators on the ground state and eliminate dependent states using the quadratic and higher-order relations. 
Although the quadratic relations of the quiver Yangian have a universal form for all quivers, the higher order relations need to be determined in a case-by-case or at least class-by-class manner.

Therefore, in this paper, besides discussing the standard refinement, we also explore an alternative way to refine the characters of the representations by directly assigning gradings to the arrows of the quiver.  
Such an assignment can be fixed by matching with results computed using other means at low orders, and can often offer closer-form expressions. 

\medskip

We focus on two main types of quivers.
The first type are the quivers that capture the BPS spectra of 4D $\mathcal{N}=2$ gauge theories (the so-called BPS quivers).
There are a few prominent subtypes:
\begin{enumerate}
\item For the 4D $\mathcal{N}=2$ theories \cite{Cecotti:2011rv,Alim:2011ae,Alim:2011kw}, the BPS quivers are two-acyclic (i.e.\ without loop of length one or two), and hence are all chiral quivers. 
	
\item For 4D $\mathcal{N}=2$ theories from compactifying type IIA string on toric Calabi-Yau threefold $X$, the 1/2-BPS sectors are giving by D-brane bound states that consist of one D6 brane wrapping the full CY$_3$ and multiple D2 and D0 branes wrapping holomorphic two and zero cycles.
	The worldvolume theory is captured by the so-called toric Calabi-Yau quivers, equipped with the associated superpotentials.
	%The quiver Yangians and their representations for this class of quivers have already been studied extensively before \cite{Li:2020rij,Galakhov:2020vyb,Galakhov:2021xum,Galakhov:2021vbo}. 
	%We review them and discuss their refinements. 
	
	\item A closely related class is when the Calabi-Yau three is  $(\mathbb{C}^2/G) \times \mathbb{C}$, with finite group $G\subset SL(2,\mathbb{C})\subset SL(3,\mathbb{C})$. 
	The corresponding quivers are the ``triple quivers" of the original extended ADE quiver. 
	
\end{enumerate}
For the first case, we can refine the characters using the physical spin-grading of the $e^{(a)}_n$ generators; for simple examples, these results can also be reproduced by computing in terms of ideal poset, using rather simple prescriptions.
For the last two cases, the definition of refined BPS invariants is often not unique and depends, e.g.\ on the choice of a  certain subtorus of the CY$_3$, which  can be traced back to the non-compactness of the moduli space.
In this paper, we focus on one particularly simple choice, and compute the corresponding refined vacuum characters. 
We show that our results agree with those obtained using the same choice, and compare them with results obtained using other choices.

As a bonus, we demonstrate that it is easy to see from these refined vacuum characters that the affine Yangian of $\mathfrak{g}$, where $\mathfrak{g}$ is a Lie superalgebra $\mathfrak{gl}_{m|n}$ or $D(2,1|\alpha)$, or $\mathfrak{g}=\mathfrak{h}\oplus \mathfrak{u}(1)$ with $\mathfrak{h}$ being a  Lie algebra of ADE type, is isomorphic to the universal enveloping algebra (UEA) of the $\mathfrak{g}$-extended $\mathcal{W}_{1+\infty}$ algebra.\footnote{
	For earlier studies on the $\mathfrak{gl}_{n}$-extended $\mathcal{W}_{1+\infty}$ algebras see e.g.\ \cite{Creutzig:2018pts,Eberhardt:2019xmf}, and for the $\mathfrak{gl}_{m|n}$-extended $\mathcal{W}_{1+\infty}$ algebras see e.g.\ \cite{Creutzig:2019qos,Rapcak:2019wzw}.
} 
This result provides a $\mathcal{W}$ algebra explanation for the plethystic exponential form of the BPS partition function. 

\medskip

The second class of examples consists of the quivers that come from the knot-quiver correspondence, which  are all symmetric and without potential. 
There is no ambiguity in the definition of the motivic DT invariants, which are to be translated to the corresponding knot invariants.

We first study $m$-loop quivers, which can be viewed as the building blocks for all the quivers in this class, and then the quiver from the trefoil knot.
We show that the refined vacuum characters of the quiver Yangians reproduce the motivic DT invariants of these quivers and thus the LMOV invariants of the corresponding knots.
This is strong evidence that the quiver Yangians are also the BPS algebras for the 3D $\mathcal{N}=2$ gauge theories that underlie the knot-quiver correspondence.

\medskip

The quiver Yangians lie at the bottom of the rational/trigonometric/elliptic hierarchy.  
Once the quiver Yangian is known, it is straightforward to write down the corresponding trigonometric and elliptic versions, and the resulting algebras are called toroidal quiver algebra and elliptic quiver algebras, respectively \cite{Galakhov:2021omc}.\footnote{For earlier studies on quantum toroidal algebras see e.g.\ \cite{MR1324698,Ding:1996mq,Miki2007,MR2793271,Feigin:2013fga,MR2566895,Bezerra:2019dmp}, and for the elliptic versions see e.g.\ \cite{MR3262444,Nieri:2015dts}.} 
The quiver Yangian/toroidal quiver algebra/elliptic quiver algebra are associated to the  (equivariant) cohomology/K-theory/elliptic cohomology of the moduli space of the quiver gauge theory.
As shown in \cite{Galakhov:2021omc}, this can even be extended to a generalized cohomology theory, with the corresponding algebra defined in terms of the inverse logarithm of the formal group law of the generalized cohomology theory.  
As we will see, all these remain true for arbitrary quivers: we can define, for arbitrary quivers, the trigonometric, elliptic, and generalized cohomology versions of the quiver algebras.

\medskip

The plan of the paper is as follows. 
In Sec.\ \ref{sec:QuiverYangians}, we give the definition of the quiver Yangian for arbitrary quivers.
In Sec.\ \ref{sec:Representation}, we explain how to construct the vacuum-like representations of a quiver Yangian for an arbitrary quiver, and determine the action of the quiver Yangian on this representation. 
In Sec.\ \ref{sec:BPSInvariants}, we study the BPS quivers of 4D $\mathcal{N}=2$ theories and discucc how to refine the vacuum characters of these quiver Yangians.
We also present new examples of quiver Yangians, such as for the $m$-Kronecker quiver and the affine Yangian of ADE type.
In Sec.\ \ref{sec:KnotQuiver}, we change gears and study the quivers from the knot-quiver correspondence. 
We give a refinement prescription that reproduces the motivic DT invariants, which in turn gives the knot invariants. 
Sec.\ \ref{sec:Summary} contains a summary and discussion.

In addition, in Appx.\ \ref{appsec:NonVacuumReps} we explain in detail how to construct non-vacuum representations, using the vacuum-like representations constructed in Sec.\ \ref{sec:Representation} as building blocks.
Appx.\ \ref{appsec:AddingRulen=23} contains an illustration of the proofs of the Adding Rule and the Rule of Simple Pole for the cases of $n=2$ and $n=3$.
Appx.\ \ref{appsec:ExamplePosetConstruction} contains  detailed examples of the construction of representations for some quiver Yangians discussed in the main text; we first confirm in Appx. \ref{appssec:toricCY3ex} that for toric CY$_3$ quivers, the poset-generating procedure reproduces the 3D colored crystals, and then we give detailed examples for the affine Yangians of $A_n$ and $E_{6,7,8}$ type in Appx.\ \ref{appssec:PosetAn} and \ref{appssec:PosetE678}, respectively.
In Appx.\ \ref{appsec:TFquiver}, we show in detail how to determine the refinement prescription for the  trefoil knot quiver.
Finally, in Appx.\ \ref{appsec:TrigElliptic}, we summarize how to generalize the quiver Yangians to the trigonometric, elliptic, and generalized cohomology cases. 

\section{Quiver Yangians for arbitrary quivers}
\label{sec:QuiverYangians}

In this section, we define the quiver Yangians for arbitrary quivers. 
For their generalizations to the trigonometric, elliptic, and generalized cohomology cases 
see Appx.\ \ref{appsec:TrigElliptic}.

\subsection{Quiver and potential}
\label{ssec:QuiverPotential}

\subsubsection{Quiver and weights of the arrows}

A quiver consists of four pieces of data 
\begin{equation}
Q=\{Q_0,Q_1, s, t\}\,,
\end{equation}
where $Q_0$ is the set of vertices and $Q_1$ is the set of arrows between these vertices:
\begin{equation}
Q_0=\{\textrm{vertex } a\} \qquad \textrm{and} \qquad  Q_1=\{\textrm{arrow } I  \}\,,
\end{equation}
and $s$ (resp.\ $t$) is the source (resp.\ target) function, which maps an arrow to the vertex at the beginning (resp.\ end) of the arrow, namely, for an arrow $I$ from vertex $a$ to vertex $b$, denoted as $I^{a\rightarrow b}$, we have
\begin{equation}
s(I^{a\rightarrow b})=a \qquad \textrm{and}\qquad t(I^{a\rightarrow b})=b\,.
\end{equation}
We also define the notations 
\begin{equation}\label{eq:atobDef}
\begin{aligned}
\{a\rightarrow b\}&\equiv \textrm{the set of arrows from $a$ to $b$}\,,\\
|a\rightarrow b|&\equiv |\{a\rightarrow b\}|= \textrm{the number of arrows from $a$ to $b$}\,,
\end{aligned}
\end{equation} 
and the chirality for a pair of vertices $(a,b)$: 
\begin{equation}\label{eq:chiabDef}
\chi_{ab}\equiv |a\rightarrow b| - |b\rightarrow a|\,.
\end{equation}
Throughout the paper, we will often refer to the label of a vertex as its color. 

To define the quiver algebra associated to a quiver $Q$, we assign a weight (sometimes called charge) to each arrow $I\in Q_1$:
\begin{equation}
h_I \qquad  \textrm{for }\qquad I\in Q_1 \,,
\end{equation}
which parameterize the algebra.
There are $|Q_1|$ weights to start with.
As we will see shortly, the quiver can be endowed with a potential, which imposes further constraints on these weights, and thus reduces the number of parameters. 

\subsubsection{Potential}

The quiver $Q$ can be supplemented by a potential, which has the form:
\begin{equation}\label{eq:Potential}
W=\sum^{\textrm{F}}_{d=1} \pm M_{d}(\{I\})\,,
\end{equation}
where $M_{d}$ is a monomial of the arrows,  and the number of terms in the potential is denoted by $\textrm{F}$.
Each monomial consists of arrows that form a closed loop and is cyclically equivalent, $I_1\cdot I_2 \cdots I_n\simeq I_2 \cdots I_n \cdot I_1$. 

For the quiver algebra, the effect of the potential on the quiver is to impose constraints on the weights of the arrows: 
each monomial  $M_{d}$ of the potential $W$ is translated into a constraint on $\{h_I\}$:\footnote{
These constraints are called ``loop constraints" in \cite{Li:2020rij} because, for a toric CY$_3$ quiver, each term in $W$ corresponds to a loop in the periodic quiver, see below.
}
\begin{equation}\label{eq:LoopConstraints}
\sum_{I\in M_{d}} h_I=0 \qquad d=1,2,\dots, \textrm{F}\,.
\end{equation}

\subsection{Quiver Yangian algebra}
\label{ssec:QYalgebra}

\subsubsection{Generators}

The generators of the quiver Yangian are organized into $|Q_0|$ triplets, one for each vertex $a\in Q_0$:
\begin{equation}\label{eq:epsif}
\begin{aligned}
\textrm{raising:} \qquad & e^{(a)}_j \qquad j\in \mathbb{N}_{0} \,,\\
\textrm{Cartan:} \qquad & \psi^{(a)}_j \qquad j \in \mathbb{Z} \,,\\
\textrm{lowering}: \qquad & f^{(a)}_j \qquad j \in \mathbb{N}_0\,.
\end{aligned}
\end{equation}
Note that the ranges of the modes are for generic quivers. 
For special quivers, the range can be smaller: for example, for the quiver associated to a toric CY$_3$ without compact $4$-cycles, the ranges for the Cartan generators $\psi^{(a)}_j$ are $j\in\mathbb{N}_0$, as will be shown below.

For later convenience, we organize these modes into ``fields" using a spectral parameter $z$:
\begin{equation}\label{eq:ModeExpansions}
\begin{aligned}
\textrm{raising:} \qquad &  e^{(a)}(z)=\sum^{\infty}_{j=0} \frac{e^{(a)}_j}{z^{j+1}} \,, \\
\textrm{Cartan:} \qquad & \psi^{(a)}(z)=\sum_{j\in\mathbb{Z}} \frac{\psi^{(a)}_j}{z^{j+1+\mys^{(a)}}}\,,\\
\textrm{lowering}: \qquad & f^{(a)}(z)=\sum^{\infty}_{j=0} \frac{f^{(a)}_j}{z^{j+1}}\,.
\end{aligned}
\end{equation}
Note that in the definition of $\psi^{(a)}$, the summation range can have a finite lower bound for special quivers, as will be shown below. 

\subsubsection{Bonding factor}

For each (ordered) pair of vertices $(a,b)$ in $Q_0$, we define a bonding factor\footnote{\label{fn:tildevarphi}This can also be rewritten as  $\varphi^{a\Leftarrow b} (u) 
=
e^{\pi i \tilde{t}_{ab}}\frac{\prod_{I\in \{a\rightarrow b\}} (-u-h_I)}{\prod_{J\in \{b\rightarrow a\}} (u-h_J)}$, with $\tilde{t}_{ab}=t_{ab}+|a\rightarrow b|$.}
\begin{equation}\label{eq:BondingFactorDef}
\varphi^{a\Leftarrow b} (u) 
\equiv 
 e^{\pi i t_{ab}}\frac{\prod_{I\in \{a\rightarrow b\}} (u+h_I)}{\prod_{J\in \{b\rightarrow a\}} (u-h_J)}\,,
\end{equation}
where $\{a\rightarrow b\}$ was defined in \eqref{eq:atobDef}, and the statistical factor $t_{ab}$ will be constrained later.

\subsubsection{Quadratic relations}

The quadratic relations of the quiver Yangian, in terms of the fields $(e^{(a)}(z),\psi^{(a)}(z),$\\
$f^{(a)}(z))$, take a universal form for generic quivers \cite{Li:2020rij}: 
\begin{tcolorbox}[ams align]\label{eq:QuadraticFields}
\begin{aligned}
\psi^{(a)}(z)\, \psi^{(b)}(w)&=   \psi^{(b)}(w)\, \psi^{(a)}(z)  \;,\\
\psi^{(a)}(z)\, e^{(b)}(w)&\simeq  \varphi^{a\Leftarrow b}(\Delta)\, e^{(b)}(w)\, \psi^{(a)}(z)  \;, \\
e^{(a)}(z)\, e^{(b)}(w)&\sim  e^{\pi i s_{ab}}\, \varphi^{a\Leftarrow b}(\Delta) \, e^{(b)}(w)\, e^{(a)}(z)  \;, \\
\psi^{(a)}(z)\, f^{(b)}(w)&\simeq   \varphi^{a\Leftarrow b}(\Delta)^{-1} \, f^{(b)}(w)\,\psi^{(a)}(z) \;,\\
f^{(a)}(z)\, f^{(b)}(w)&\sim  e^{-\pi i s_{ab}}\, \varphi^{a\Leftarrow b}(\Delta)^{-1}\,  f^{(b)}(w)\, f^{(a)}(z)   \;,\\
\left[e^{(a)}(z),f^{(b)}(w) \right\} &\sim -  \delta_{a,b} \frac{\psi^{(a)}(z)-\psi^{(b)}(w)}{z-w}  \;,
\end{aligned}
\end{tcolorbox}
\noindent where throughout the paper, ``$\simeq$" means equality up to $z^n w^{m\geq 0}$ terms,  ``$\sim$" means equality up to $z^{n\geq 0} w^{m}$ and $z^{n} w^{m\geq 0}$ terms, and we define 
\begin{equation}
\Delta\equiv z-w \;.
\end{equation}
The super-commutator $[e^{(a)}(z),f^{(b)}(w)\}$ is defined as
\begin{equation}
[e^{(a)}(z),f^{(b)}(w)\}\equiv e^{(a)}(z)\,f^{(b)}(w) - e^{\pi i s_{ab}} f^{(b)}(w)\,e^{(a)}(z)\,.
\end{equation}
The exponent $s_{ab}$ in the $e$-$e$, $f$-$f$, and $e$-$f$ relations captures the (mutual) statistics of the $e^{(a)}$ and $f^{(b)}$ generators.
In particular, $s_{aa}$ captures the self-statistics of the $e^{(a)}$ and $f^{(a)}$ generators.\footnote{
In \cite{Li:2020rij}, $s_{aa}$ is denoted as $|a|$.}
When $s_{aa}=0$ (resp.\  $s_{aa}=1$), $e^{(a)}$ and $f^{(a)}$ are bosonic (resp.\ fermionic).
The Cartan generators $\psi^{(a)}$ are always bosonic.

In order for the $e$-$e$ and $f$-$f$ equations to hold for arbitrary pairs of $a$ and $b$, we require the reciprocity condition\footnote{
This is a generalization of the choices made in \cite{Li:2020rij}, \cite{Galakhov:2021xum}, and \cite{Galakhov:2021vbo}, in order to incorporate more general (mutual) statistics among generators.
}
\begin{equation}\label{eq:Reciprocity}
e^{\pi i (s_{ab}+s_{ba})} \varphi^{a\Leftarrow b}(u) \varphi^{b\Leftarrow a}(-u)=1 \,,
\end{equation}
which gives 
\begin{equation}\label{eq:stab}
N_{ab}\equiv \frac{1}{2}(	(s_{ab}+s_{ba})+(t_{ab}+t_{ba})+(|a\rightarrow b|+|b\rightarrow a|) 
)\in \mathbb{Z}\,.
\end{equation}
Later on, when defining the representation, we will also need
\begin{equation}\label{eq:Naa}
N_{aa}=1 \quad \textrm{mod}\quad 2\,,   
%\,,\qquad  \textrm{and} \qquad 
\end{equation}
and 
\begin{equation}\label{eq:sab}
s_{ab}=s_{ba} \quad \textrm{mod}\quad 2\,.
\end{equation}
Once $s_{ab}$ is given, we can choose $t_{ab}$ in the definition of the bonding factor \eqref{eq:BondingFactorDef} to satisfy these constraints.
A simple choice is
\begin{equation}
\begin{aligned}
&s_{aa}=|a\rightarrow a|+1
\,,\quad s_{ab}=s_{aa}s_{bb}\,,\qquad t_{ab}=|a\rightarrow b| (|a\rightarrow b|-|b\rightarrow a|)\,.
\end{aligned}
\end{equation}
With this choice, $e^{(a)}$ and $f^{(a)}$ are bosonic (resp.\ fermionic) when the number of self-loops at $a$, namely $|a\rightarrow a|$, is odd (resp.\ even); the super-commutator $[e^{(a)}(z),f^{(b)}(w)\}$ denotes an anti-commutator $\{ e^{(a)}(z),f^{(b)}(w)\}$ when both $a$ and $b$ are fermionic, and a commutator $[e^{(a)}(z),f^{(b)}(w)]$ otherwise.
In this paper, we do not restrict the choice of $s_{ab}$ and $t_{ab}$, as long as they satisfy the constraints \eqref{eq:stab}, \eqref{eq:Naa}, and \eqref{eq:sab}.
\bigskip

The field relations in \eqref{eq:QuadraticFields} capture the quadratic relations of the quiver Yangian in a concise manner. 
But sometimes, one needs the relations in terms of the modes $(e^{(a)}_j,\psi^{(a)}_j,f^{(a)}_j)$, for example when comparing with other known algebras or defining the co-products. 
To derive the relations in terms of modes, one can simply substitute the mode expansion \eqref{eq:ModeExpansions} into the field relations \eqref{eq:QuadraticFields}, expand the bonding factors $\varphi^{a\Leftarrow b}(z-w)$, and finally extract the terms $\sim$ $z^{-n-1}w^{-m-1}$ \cite{Li:2020rij}: 
\begin{tcolorbox}[ams align]
\label{eq:QuadraticModes}
\begin{aligned}
\left[\psi^{(a)}_n \, , \, \psi^{(b)}_m\right]&=0 \;,\\
\sum^{|b\rightarrow a|}_{k=0}(-1)^{|b\rightarrow a|-k} \, \sigma^{b\rightarrow a}_{|b\rightarrow a|-k}\,  [\psi^{(a)}_n\, e^{(b)}_m]_k &=\sum^{|a\rightarrow b|}_{k=0} \sigma^{a\rightarrow b}_{|a\rightarrow b|-k}\, [ e^{(b)}_m\, \psi^{(a)}_n]^k  \;,\\
 \sum^{|b\rightarrow a|}_{k=0}(-1)^{|b\rightarrow a|-k} \, \sigma^{b\rightarrow a}_{|b\rightarrow a|-k}\,  [e^{(a)}_n\, e^{(b)}_m]_k &=e^{\pi i s_{ab}}\sum^{|a\rightarrow b|}_{k=0} \sigma^{a\rightarrow b}_{|a\rightarrow b|-k}\, [ e^{(b)}_m\, e^{(a)}_n]^k  \;,\\
 \sum^{|a\rightarrow b|}_{k=0} \sigma^{a\rightarrow b}_{|a\rightarrow b|-k}\, [\psi^{(a)}_n\, f^{(b)}_m]_k &= \sum^{|b\rightarrow a|}_{k=0}(-1)^{|b\rightarrow a|-k} \, \sigma^{b\rightarrow a}_{|b\rightarrow a|-k}\,  [ f^{(b)}_m\, \psi^{(a)}_n]^k  \;,\\
\sum^{|a\rightarrow b|}_{k=0} \sigma^{a\rightarrow b}_{|a\rightarrow b|-k}\, [f^{(a)}_n\, f^{(b)}_m]_k &=e^{-\pi i s_{ab}} \sum^{|b\rightarrow a|}_{k=0}(-1)^{|b\rightarrow a|-k} \, \sigma^{b\rightarrow a}_{|b\rightarrow a|-k}\,  [ f^{(b)}_m\, f^{(a)}_n]^k  ,\\
\left[e^{(a)}_n\, , \, f^{(b)}_m \right\}&=\delta^{a,b}\,\psi^{(a)}_{n+m} \;,
\end{aligned}
\end{tcolorbox}
\noindent where for the $\psi^{(a)}_n$ modes, $n\in \mathbb{Z}$, and for the $e^{(a)}_n$ and $f^{(a)}_n$ modes, $n\in\mathbb{Z}_{\geq 0}$.
Here $\sigma^{a\rightarrow b}_k$ denotes the $k^{\textrm{th}}$ elementary symmetric sum of the set $\{h_I\}$ with $I\in \{a\rightarrow b\}$,
and  we have defined the shorthand notations
\begin{equation}\label{eq-ABn}
\begin{aligned}
\left[A_n\, B_m\right]_k& \equiv \sum^{k}_{j=0} (-1)^j\,\tbinom{k}{j} \,A_{n+k-j}\, B_{m+j}\quad \textrm{and} \quad 
%\qquad \\
\left[B_m\,A_n\right]^k \equiv \sum^{k}_{j=0} (-1)^j\,\tbinom{k}{j} \,B_{m+j}\, A_{n+k-j}\,.
\end{aligned}
\end{equation}

The quadratic relations, \eqref{eq:QuadraticFields} in terms of fields or \eqref{eq:QuadraticModes} in terms of modes, were already defined in \cite{Li:2020rij} for quivers with superpotentials that come from toric CY$_3$'s.
In this paper, we will show that for an arbitrary quiver, this set of relations still defines a quiver Yangian, with a natural action on a set of representations to be defined from the quiver.
The only difference is that for general quivers, the representations will not be described by 3D crystals, but will instead have the structure of a poset.

In contrast to the quadratic relations, the higher order relations are not universal and need to be determined in a case-by-case or at least class-by-class manner. 
We will explain the method in Sec.\ \ref{ssec:HigherOrderRelations} after showing how to construct representations of the quiver Yangians.

\subsubsection{BPS quivers}
The definition of quiver Yangian in Sec.\ \ref{ssec:QYalgebra} applies to general quivers. 
It is particularly interesting when the quiver are the BPS quivers, namely the quivers (with potential) that captures the BPS spectrum of quantum field theory or string theory. 
The expectation is that in such a setup, the quiver Yangian captures the (spherical part of the) BPS algebra.
Interesting examples include:
\begin{itemize}
\item 2-acyclic quivers, namely the quivers with no cycle of length-$1$ or $2$.
They appear as BPS quivers of 4D $\mathcal{N}=2$ theories from wrapping M5-branes on Riemann surfaces \cite{Alim:2011ae,Alim:2011kw} and can be obtained directed from the ideal triangulations of the Riemann surfaces \cite{fomin2007clusteralgebrastriangulatedsurfaces,LabardiniFragoso2008,labardinifragoso2009quiverspotentialsassociatedtriangulated,Cecotti:2011rv}.
The simplest subclass are $m$-Kronecker quivers, see Sec.\ \ref{ssec:Kronecker}, e.g.\ the $m=1$ case corresponds to the $A_2$ quiver (the BPS quiver of the simplest Argyres-Douglas theory) and the $m=2$ case corresponds to the BPS quiver of the pure $SU(2)$ theory.\footnote{For further study on BPS algebras based on this class of quivers see \cite{Gaiotto:2024fso}.}
\item Toric CY$_3$ quivers (with potentials). These are the BPS quivers of the 4D $\mathcal{N}=2$ theory from type IIA string on toric Calabi-Yau threefolds, where the $\frac{1}{2}$-BPS sector are from the 1D6-D2-D0 bound states wrapping holomorphic cycles of the toric CY$_3$ \cite{Li:2020rij,Galakhov:2020vyb}. 
They have the property that the information of the quiver and potential can be uniquely repackaged into a 2D periodic quiver whose faces correspond to terms in the potential; such a 2D periodic quiver can then be uplifted to the 3D colored crystal that gives rise to the melting crystals that furnish the representations of the corresponding quiver Yangian.
\item Triple quivers, namely the quivers with potentials that are obtained by ``tripling" (finite undirected) graphs, where one first ``doubles" the graph by turning each link into a pair of arrows in opposite directions, then adds a self-loop to each vertex, and finally define appropriate potential. An important subclass is McKay quiver of ADE type, see Sec.\ \ref{ssec:ADE}. 
\item Symmetric quivers without potential. They appear in the knot-quiver correspondence \cite{Ekholm:2018eee,Ekholm:2019lmb}.
\end{itemize}
We will discuss the first three classes of examples in Sec.\ \ref{sec:BPSInvariants} and the last one in Sec.\ \ref{sec:KnotQuiver}.

\section{Representations for arbitrary quiver}
\label{sec:Representation}

The quiver Yangians for the toric CY$_3$ quiver were bootstrapped from their action on BPS crystals, namely, by demanding that the sets of BPS crystals furnish representations of the quiver Yangian. 
However, for arbitrary quivers, the  representations are not yet known.

In this section, we will construct representations of the quiver Yangians for arbitrary quivers.
This generalizes the similar story for the quivers with potentials from toric CY$_3$'s \cite{Szendroi,Mozgovoy:2008fd} to \textit{arbitrary} quivers with potentials.
However, unlike the case of toric CY$_3$'s, the states in these representations will not have the structure of 3D crystals.
Instead, the representation will be described by the poset of finite co-dimensional ideals of the Jacobian algebra of $(Q,W)$.

\subsection{General properties of highest weight representations of quiver Yangians}

One quiver Yangian $\textsf{Y}(Q,W)$ (or its trigonometric, elliptic, and generalized cohomological versions) can have multiple representations, and different representations are related to different ``framings", labeled by $\sharp$, of the same quiver with potential $(Q,W)$, which will be defined below.
The resulting ``framed quiver" is denoted as $({}^{\sharp}Q, {}^{\sharp}W)$, and its corresponding representation as ${}^{\sharp}\mathcal{P}_{(Q,W)}$. 

We will focus on highest weight representations ${}^{\sharp}\mathcal{P}_{(Q,W)}$ of $\textsf{Y}(Q,W)$, which are specified by their ground states $|{}^{\sharp}\Pi_0\rangle$, satisfying
\begin{equation}\label{eq:GroundStateDef}
f^{(a)}(z)\, |{}^{\sharp}\Pi_0\rangle =0 \qquad \textrm{and} \qquad \psi^{(a)}(z)\, |{}^{\sharp}\Pi_0\rangle= {}^{\sharp}\psi_{0}^{(a)}(z)\, |^{\sharp}\Pi_0\rangle
\end{equation}
for all $a\in Q_0$, and ${}^{\sharp}\psi_{0}^{(a)}(z)$ is a rational function that only depends on the framing $\sharp$, see below.
A level-$d$ state $|^{\sharp}\Pi_{d}\rangle$ of a representation ${}^{\sharp}\mathcal{P}_{(Q,W)}$ can be obtained by applying the raising operators $e^{(a)}(z)$ $d$ times to the ground state $|{}^{\sharp}\Pi_0\rangle$:
\begin{equation}
|^{\sharp}\Pi_{d}\rangle \sim e^{(a_1)}(z_1)\,  e^{(a_2)}(z_2) \cdots  e^{(a_{d})}(z_{d})  |^{\sharp}\Pi_0\rangle\,.
\end{equation}
The goal of this section is to first describe the states $ |^{\sharp}\Pi_0\rangle$ in terms of the framed quiver $({}^{\sharp}Q, {}^{\sharp}W)$, and then to determine the action of $(\psi^{(a)}(z), e^{(a)}(z), f^{(a)}(z))$ on any $|^{\sharp}\Pi_{d}\rangle$ so that ${}^{\sharp}\mathcal{P}_{(Q,W)}=\{|^{\sharp}\Pi_{d}\rangle\}$ indeed furnishes a representation of the quiver Yangian $\textsf{Y}(Q,W)$ defined in \eqref{eq:QuadraticFields}. 

\subsubsection{Review: Jacobian algebra of quiver with potential % and its ideals
}
\label{sssec:JacobianAlgebra}

To describe the states directly in terms of the quiver data, we will need the language of the path algebra and Jacobian algebra of a quiver (with potential), which we briefly review here. 
For more details see e.g.\ \cite{Ginzburg:2006fu, bocklandt2008graded, Szendroi,Mozgovoy:2008fd} or the textbook \cite{Kirillov}.

The \textit{path algebra} $\mathbb{C}Q$ of a quiver $Q$ is an algebra generated (over $\mathbb{C}$) by arrows in the quiver diagram, with multiplication defined by concatenating paths. 
Namely, a path $\mathfrak{p}$ in $\mathbb{C}Q$ is a sequence of $n$ arrows $I_i$:
\begin{equation}
 I_{1}\cdot I_2\cdot \ldots \cdot  I_n
\end{equation} satisfying $t(I_i)=s(I_{i+1})$. 
We define the source and the target of the path $\mathfrak{p}$ to be 
\begin{equation}
s(\mathfrak{p})=s(I_1) \qquad \textrm{and} \qquad  t(\mathfrak{p})=t(I_n)\,,
\end{equation}
and use the notation $\mathfrak{p}^{a\rightarrow b}$ to denote a path that starts at vertex $a$ and ends at vertex $b$.
The product $\mathfrak{p}^{a\rightarrow b}\cdot \mathfrak{q}^{c \rightarrow d}$ of two paths $\mathfrak{p}^{a\rightarrow b}$ and $\mathfrak{q}^{c \rightarrow d}$ is nonzero only if $b=c$.
In addition, sometimes we label the path only by its ending vertex:
\begin{equation}\label{eq:pathanotation}
\mathfrak{p}^{(b)} \equiv \mathfrak{p}^{a\rightarrow b} \,.
\end{equation}
Finally, in $\mathbb{C}Q$ there are also trivial paths (or zero-length paths) $I^{(a)}$ for each vertex $a\in Q_0$.
They are idempotent:
\begin{equation}\label{eq:Idempotent}
  (I^{(a)})^n=I^{(a)} \qquad \textrm{for } n\in \mathbb{N}\,,
\end{equation}
and thus can be used as projection operators. 

\medskip

For a quiver $Q$ with potential $W$, we need to quotient out all the relations $\{\partial_I W=0\}$ given by the potential, and consider the factor algebra, called \textit{Jacobian algebra}, $J_{(Q,W)}$:
\begin{equation}\label{eq:JacobianAlgebraDef}
   J_{(Q,W)}\equiv \mathbb{C}Q/\partial_I W\,.
\end{equation}
Namely, each relation $\partial_I W=0$ gives rise to an equivalence relation between different paths. For example, if the arrow $I_1$ appears in $W$ as
\begin{equation}
    W\ni I_1\cdot I_2 \cdot I_3 -I_1\cdot I_4\cdot I_5 \cdot I_6\,,
\end{equation}
then the relation $\partial_I W=0$ defines the path equivalence
\begin{equation}
 I_2 \cdot I_3 \simeq I_4\cdot I_5 \cdot I_6   \,.
\end{equation}
The Jacobian algebra $J_{(Q,W)}$ consists of all the paths $\in \mathbb{C}Q$ up to path equivalence defined by $W$.
$J_{(Q,W)}$ is also called the quiver-potential algebra; and when 
$(Q,W)$ is related to a Calabi-Yau threefold, $J_{(Q,W)}$ is also called Calabi-Yau-$3$ algebra \cite{Ginzburg:2006fu}.\footnote{
This  is not to be confused with the quiver Yangians, toroidal, and elliptic algebras based on the quiver with potential that we study in this paper.
}
Finally, using a trivial path $I^{(a)}$ as a projector, we can project to a subalgebra \cite{Szendroi,Mozgovoy:2008fd}:
\begin{equation}\label{eq:JaDef1}
 J^{(a)}_{(Q,W)} \equiv  I^{(a)}  \cdot J_{(Q,W)}\,,
\end{equation}
which consists of all the paths $\mathfrak{p}^{a\rightarrow *}\in J_{(Q,W)}$ that start from the vertex $a$ and are defined modulo the relations $\partial W=0$.\footnote{Note that this construction was used in 
\cite{Szendroi,Mozgovoy:2008fd} for the toric CY$_3$ cases; however, it is also applicable to general quivers.}

\subsubsection{Framing of quiver and Jacobian algebra for framed quiver}
\label{sssec:FramedQWandReps}

Different representations of the same quiver algebra are specified by different ``framings" of the same $(Q,W)$, defined as follows.

First, we extend the quiver $Q$ by adding a ``framing vertex", denoted by ``$\infty$", and a number of additional arrows $\tilde{I}$ between the framing vertex $\infty$ and some existing vertices $\mathfrak{a}_{i}, \mathfrak{b}_{j} \in Q_0$, called ``framed vertices".\footnote{
In this paper, we use the mathfrak font to label framed vertices.
}
The extended quiver so constructed is called the ``framed quiver" and denoted as ${}^{\sharp}Q$:
\begin{equation}\label{eq:FramedQuiver}
{}^{\sharp}Q=(^{\sharp}Q_0,{}^{\sharp}{Q}_1,s,t): \qquad \begin{cases}{}^{\sharp}Q_0=Q_0\cup \{\infty \}  \\
{}^{\sharp}Q_1= Q_1\cup \{\infty \rightarrow \mathfrak{a}_{i}\} \cup \{\mathfrak{b}_{j}\rightarrow \infty\} \,.
\end{cases}
\end{equation}

The simplest type of  framings, known as the \textit{canonical framings} and denoted as $({}^{(\mathfrak{a})}Q,{}^{(\mathfrak{a})}W)$, have only one arrow from the framing vertex $\infty$ to one framed vertex $\mathfrak{a}$ and no arrow back to $\infty$, with the potential remaining unchanged:\footnote{
For IIA string on CY$_3$, the canonical framing corresponds to having one D6-brane wrapping the entire CY$_3$ (introduced to compactify the moduli space) and with the stability condition at the NCDT chamber of the moduli space, see Sec.\ \ref{sec:BPSInvariants}.} 
\begin{equation}\label{eq:CanonicalQW}
{}^{(\mathfrak{a})}Q =\infty \longrightarrow Q\,, \quad \textrm{i.e.} \quad  
\begin{cases}
{}^{(\mathfrak{a})}Q_0=Q_0\cup \{\infty \}  \\
{}^{(\mathfrak{a})}Q_1= Q_1\cup \{\tilde{I}^{\infty \rightarrow \mathfrak{a}}\} 
\end{cases} 
\qquad \textrm{and} \quad 
{}^{(\mathfrak{a})}W=W    \,.
\end{equation}
In this paper, effectively this new arrow plays the same role as the zero-length path defined in \eqref{eq:Idempotent}, so for our purpose we will often rewrite $\tilde{I}^{\infty \rightarrow \mathfrak{a}}$ as:
\begin{equation}\label{eq:Iinftyto0}
\tilde{I}^{\infty \rightarrow \mathfrak{a}} = I^{(\mathfrak{a})} \,.
\end{equation}
Throughout the paper, we will use $\tilde{I}^{\infty \rightarrow \mathfrak{a}}$ and $I^{(\mathfrak{a})}$ interchangeably. 

Borrowing intuition from the $\mathcal{W}$ algebras,  we will call the representations that correspond to the canonical framings, denoted as ${}^{(\mathfrak{a})}\mathcal{P}_{(Q,W)}$ for $\mathfrak{a}\in Q_0$, the \textit{vacuum-like representations} of $\textsf{Y}(Q,W)$, since the growth of their characters are among the slowest.
A priori, there are $|Q_0|$ different vacuum-like representations for the same algebra $\textsf{Y}(Q,W)$; however, the number of inequivalent vacuum-like representations is less than $|Q_0|$ when there are non-trivial automorphisms of the quiver.\footnote{
Which one of these vacuum-like representation deserves to be called \textit{the vacuum representation} depends on the quiver in question and even on the physical context.
}
Later in the paper, we sometimes fix one framed vertex and label it by $\mathfrak{a}=0$, and call the corresponding representation ${}^{(0)}\mathcal{P}_{(Q,W)}$ the  \textit{vacuum representation}.

\medskip

For other, more complicated framings, the representation ${}^{\sharp}\mathcal{P}_{(Q,W)}$ can be constructed by superimposing ``positive" and ``negative" ${}^{(\mathfrak{a})}\mathcal{P}_{(Q,W)}$ representations (with possibly different $\mathfrak{a}$'s), see Sec.\ \ref{ssec:NonVacuumReps}.
As a result, the new arrows (to and from the framing vertex $\infty$) are related to the existing arrows in ${}^{(0)}Q_1$; more precisely, they are related to paths (up to equivalence) $\in J^{(0)}_{(Q,W)}$ via:
\begin{equation}\label{eq:tildeI}
\tilde{I}^{\infty \rightarrow \mathfrak{a}_i} = I^{(0)}\cdot \mathfrak{p}^{0\rightarrow  \mathfrak{a}_i} 
\qquad \textrm{and}  \qquad    
(I^{(0)}\cdot \mathfrak{p}^{0\rightarrow \mathfrak{b}_j}) \cdot \tilde{I}^{ \mathfrak{b}_i\rightarrow \infty} =1
\end{equation}
where we have chosen a particular $\mathfrak{a}$ such that ${}^{\sharp}\mathcal{P}_{(Q,W)} \subset {}^{(\mathfrak{a})}\mathcal{P}_{(Q,W)}$ and then set $\mathfrak{a}=0$, for more details see \eqref{eq:Subspace}.
For later convenience, we define 
\begin{equation}\label{eq:hatI}
\hat{I}^{\infty \rightarrow \mathfrak{b}_i}    =I^{(0)}\cdot \mathfrak{p}^{0\rightarrow \mathfrak{b}_j}\,.
\end{equation}
Note that $\hat{I}^{\infty \rightarrow \mathfrak{b}_i} $ is not a framing arrow, but forms a closed loop with the framing arrow $\tilde{I}^{ \mathfrak{b}_i\rightarrow \infty}$.
Recall that each arrow $I\in Q_1$ is assigned a weight $h_I$, subject to the constraints \eqref{eq:LoopConstraints} coming from the potential $W$ \eqref{eq:Potential}.
The weight of a new arrow $\tilde{I}$ to or from the framing vertex $\infty$ will also acquire a weight $h_{\tilde{I}}$, given by 
\begin{equation}\label{eq:weightnewarrows}
h_{\tilde{I}^{\infty \rightarrow \mathfrak{a}_i}} =\sum_{I\in \mathfrak{p}^{0\rightarrow \mathfrak{a}_i}}    h_I
\qquad  \textrm{and} \qquad 
h_{\tilde{I}^{ \mathfrak{b}_i\rightarrow \infty}} 
=- \sum_{I\in \mathfrak{p}^{0\rightarrow \mathfrak{b}_i}} h_I   \,,
\end{equation}
where the sum is over all the arrows $I$ in the path and we have used $h(I^{(0)})=0$.
Finally, to enforce the relation between the existing weights $\{h_{I}\}$ and the new weights $\{h_{\tilde{I}}\}$ \eqref{eq:weightnewarrows}, the potential also needs to be supplemented by additional terms involving the new arrows:
\begin{equation}
{}^{\sharp}W=W+\sum\textrm{monomials involving both $\{I\}$ and $\{\tilde{I}\}$}\,.
\end{equation}
As we will see in Sec.\ \ref{ssec:NonVacuumReps}, the choice of framing $\sharp$ \eqref{eq:FramedQuiver} on a quiver with potential $(Q,W)$, together with an appropriate new arrow assignment \eqref{eq:tildeI}, defines a representation ${}^{\sharp}\mathcal{P}_{(Q,W)}$ of the quiver algebra $\textrm{A}(Q,W)$.

\bigskip 

The Jacobian algebra $J_{(Q,W)}$ in \eqref{eq:JacobianAlgebraDef} was defined for the unframed $(Q,W)$.
In this paper, we define a version for the framed quiver:
\begin{equation}\label{eq:JsharpDef}
J^{\sharp}_{(Q,W)}
=\left( \bigcup_{ \{\mathfrak{a}_i \} } \tilde{I}^{\infty \rightarrow \mathfrak{a}_i}\cdot J_{(Q,W)} \right) 
\boldsymbol{\big{\backslash}} 
\left(\bigcup_{\{\mathfrak{b}_j\}} \hat{I}^{\infty\rightarrow \mathfrak{b}_j }\cdot J_{(Q,W)}\right)\,,
\end{equation}
where $\tilde{I}^{\infty \rightarrow \mathfrak{a}_i}$ and $\hat{I}^{ \infty\rightarrow\mathfrak{b}_j}$ are defined in  \eqref{eq:tildeI} and \eqref{eq:hatI}, respectively.
In particular, for the simplest canonical framings, it reduces to the definition \eqref{eq:JaDef1}, which already appeared in \cite{Szendroi,Mozgovoy:2008fd}:
\begin{equation}\label{eq:JaDef2}
J^{(a)}_{(Q,W)}
=\tilde{I}^{\infty\rightarrow a}\cdot J_{(Q,W)} =I^{(a)}\cdot J_{(Q,W)}\,,
\end{equation}
which is simply a projection of $J_{(Q,W)}$ using the idempotent element $I^{(a)}$.

\subsubsection{General properties of highest weight representations}
\label{sssec:GeneralPropertyRep}

We consider the highest weight representation ${}^{\sharp}\mathcal{P}_{(Q,W)}$ associated to the framed quiver $({}^{\sharp}Q,{}^{\sharp}W)$.
First of all, for general quivers, the constraint $\partial W=0$ can involve linear constraints among more than two terms.
In this paper, we will focus on those framing $\sharp$ such that in the resulting representation, for each state of the representation, the constraints $\partial_{I} W=0$ (if $W$ is present) always eliminate all but one path with the same weight.\footnote{For example, the vacuum representation of the quiver Yangian for the affine $D_4$ (see Sec.\ \ref{ssec:ADE}) will not be included, see Appx.\ \ref{appssec:onframing}.}  
First, note that this constraint on framings does not concern the case without potential, for which one can define non-commutative addition/subtraction to resolve higher-order poles, see Appx. \ref{appssec:resolving}. 
Second, this condition is imposed within each state of the representations: different states (at different levels) can have different paths with the same weight. 
Lastly, by ``eliminating all but one" we mean that there is only one linearly independent path after the constraints $\partial_{I} W=0$ are imposed.

Such a highest weight representation has the following properties.
\begin{enumerate}
\item Each element ${}^{\sharp}\Pi_{d}$ in ${}^{\sharp}\mathcal{P}$ consists of a set of $d$ paths, up to path equivalence defined by $W$, namely:\footnote{
The left superscript ${\sharp}$ of $^{\sharp}\Pi_{d}$ is to emphasize that it is a state in the representation ${}^{\sharp}\mathcal{P}_{(Q,W)}$. 
We sometimes omit it when there is no risk of confusion. 
}
\begin{equation}\label{eq:PiDef1}
{}^{\sharp}\Pi_{d}=\{ \mathfrak{p}^{(a_1)}_1, \mathfrak{p}^{(a_2)}_2, \dots, \mathfrak{p}^{(a_{d})}_{d}\}
\,,\quad \textrm{with}\quad 
\mathfrak{p}^{(a_i)}_i  \in J^{\sharp}_{(Q,W)} \,, 
\end{equation}
where the label $a_i$ denotes the end vertex of the path $\mathfrak{p}_i$, as defined in  \eqref{eq:pathanotation}.\footnote{
Note that ${}^{\sharp}\Pi_{d}$ is a set, namely, the order of the paths in it does not matter.
}
Then analogous to the case of toric CY$_3$'s,\footnote{The toric Calabi-Yau-three quiver is special in that each constraint $\partial_I W=0$ itself always involve precisely two terms, therefore one does not need to restrict the framings in order to use the algorithm below.} we can view each path $\mathfrak{p}^{(a)}$ (up to path equivalence) in ${}^{\sharp}\Pi_{d}$ as an ``atom" of color $a$ (denoted as $\sqbox{$a$}$):
\begin{equation}\label{eq:AtomDef}
\textrm{atom }\sqbox{$a$} \equiv \mathfrak{p}^{(a)} \textrm{ up to path equivalence } \in J^{\sharp}_{(Q,W)} \,,
\end{equation}
and ${}^{\sharp}\Pi_{d}$ as a ``compound" with $d$ atoms -- i.e.\ not necessarily a crystal.
Hence, we sometimes also write 
\begin{equation}\label{eq:PiDef2}
{}^{\sharp}\Pi_{d}=\{ \sqboxs{$a_1$}_1, \sqboxs{$a_2$}_2, \dots, \sqboxs{$a_{d}$}_{d}\}\,.
\end{equation}
For later convenience, we define the subset ${}^{\sharp}\Pi^{{a}} \subset {}^{\sharp}\Pi$, consisting of only atoms of color $a$:
\begin{equation}\label{eq:piaDef}
{}^{\sharp}\Pi^{(a)}=\{ \mathfrak{p}^{(a)}_1, \mathfrak{p}^{(a)}_2, \dots\} = \{ \sqbox{$a$}_1, \sqbox{$a$}_2, \dots, \} \subset {}^{\sharp}\Pi\,.
\end{equation}

The weight of a path or atom is defined as:
\begin{equation}\label{eq:weightPathAtom}
h(\mathfrak{p}^{(a)})=h(\sqbox{$a$})=\sum_{J\in \mathfrak{p}^{(a)}}    h_J\,,
\end{equation}
where the sum is over all arrows $J$ in the path.
Although it is usually possible to deduce the color of the atom once its weight is given, we sometimes explicitly attach the color to the weight in notations like $\sum_{J\in \mathfrak{p}^{(a)}}    (h_J)^{(a)}$.
Finally, for later convenience, we define the set of weights:
\begin{equation}\label{eq:WofPi}
\mathcal{W}({}^{\sharp}\Pi_{d})
=\{ h(\sqboxs{$a_1$}_1)^{(a_1)}, h(\sqboxs{$a_2$}_2)^{(a_2)}, \dots, h( \sqboxs{$a_{d}$}_{d})^{(a_2)}\}\,,
\end{equation}
where the superscripts, although redundant, make some later formulae more transparent.

For the case of toric CY$_3$'s, ${}^{\sharp}\Pi_{d}$ has the structure of a 3D crystal.
However, this structure is lost for general quivers; instead, ${}^{\sharp}\Pi_{d}$ is uniquely specified once we specify all the paths in it, see \eqref{eq:PiDef1}.
Finally, we will often call $d$, the number of atoms in $\Pi_{d}$, the ``level", since it also reflects the level at which
we reach $\Pi_{d}$ in the iterative definition of the representation. 

\item The ground state (or level-0 state) of the representation ${}^{\sharp}\Pi_{0}$ is the empty set:
\begin{equation}\label{eq:PiGroundState}
{}^{\sharp}\Pi_{0}=\{\varnothing\} \,.
\end{equation}

\item The highest weight representation ${}^{\sharp}\mathcal{P}_{(Q,W)}$ has the structure of a poset, i.e.\ a set whose elements have partial orders.
Namely, for certain pairs of elements $(\Pi,\Pi')$, one can define an order $\Pi < \Pi' $; and the term ``partial" means that not all pairs are comparable: there can exist pairs $(\Pi,\Sigma)$ with neither $\Pi < \Sigma$ nor $\Sigma < \Pi$.
The (partial) order for two states in the representation ${}^{\sharp}\mathcal{P}_{(Q,W)}$ is defined as follows:
\begin{equation}
{}^{\sharp}\Pi_{d} \subset {}^{\sharp}\Pi'_{d'} 
\qquad \Longrightarrow \qquad  
{}^{\sharp}\Pi_{d} < {}^{\sharp}\Pi'_{d'} \,,
\end{equation}
which implies $d <d'$.
\end{enumerate}

\subsection{Construction of representations from framed quivers}
\label{ssec:PosetFromQW}

Now we explain how to construct the poset representation ${}^{\sharp}\mathcal{P}_{(Q,W)}$ from a framed quiver and potential $({}^{\sharp}Q,{}^{\sharp}W)$, namely, how to determine the elements ${}^{\sharp}\Pi$ within ${}^{\sharp}\mathcal{P}_{(Q,W)}$.
(We remind that we focus on those framing $\sharp$ such that in the resulting representation, for each state of the representation, the constraints $\partial_{I} W=0$ (if $W$ is present) always eliminate all but one path with the same weight.)
This can be done recursively from level-0.
Before we go into details, let us first outline the general strategy.

\subsubsection{Ansatz for action}

We will use the quiver Yangians of Sec.\ \ref{sec:QuiverYangians} to construct these representations.
However, they will also serve as representations of the toroidal and elliptic quiver algebras (and those for the generalized cohomologies) of Appx.\ \ref{appsec:TrigElliptic}.
A quiver Yangian $\textsf{Y}(Q,W)$ has three types of generators: raising, Cartan, and lowering operators, denoted by $(e^{(a)}(z),\psi^{(a)}(z),f^{(a)}(z))$, %denoting (raising, Cartan, lowering) operators, 
see  \eqref{eq:epsif} and \eqref{eq:ModeExpansions}.
We will start with the ground state $|^{\sharp}\Pi_{0}\rangle$ defined by \eqref{eq:GroundStateDef}, and apply the raising operator $e^{(a)}(z)$ to it repeatedly to generate higher level states $|^{\sharp}\Pi_{d}\rangle$.
For this, we first need to determine the action of $(e^{(a)}(z),\psi^{(a)}(z),f^{(a)}(z))$ on any state $|\Pi\rangle$.\footnote{
When the formula doesn't depend on the framing $\sharp$ or level $d$ in an explicit manner, we often omit this information when there is no risk of confusion.
}
We choose the following ansatz:
\begin{equation}\label{eq:psiefAnsatz}
    \begin{aligned}
\psi^{(a)}(z)|\Pi\rangle&= \Psi_{\Pi}^{(a)}(z)|\Pi\rangle \;,\\
e^{(a)}(z)|\Pi\rangle &=\sum_{\sqbox{$a$} \,\in \,\textrm{Add}(\Pi)} 
 \frac{\llbracket \Pi\rightarrow \Pi+\sqbox{$a$}\rrbracket}{z-h(\sqbox{$a$})}|\Pi+\sqbox{$a$}\rangle \;,\\
f^{(a)}(z)|\Pi\rangle &=\sum_{\sqbox{$a$}\, \in\, \textrm{Rem}(\Pi)}
\frac{\llbracket \Pi\rightarrow \Pi-\sqbox{$a$}\rrbracket}{z-h(\sqbox{$a$})}|\Pi-\sqbox{$a$}\rangle \;.
\end{aligned}
\end{equation}

Formally, this ansatz looks exactly the same as the one for the toric CY$_3$'s in \cite{Li:2020rij}. 
The crucial difference for a general quiver is in the definition of Add$(\Pi)$ (resp.\ Rem$(\Pi)$) in the action of $e^{(a)}(u)$ (resp.\ $f^{(a)}(u)$).
For toric CY$_3$'s, each $|\Pi\rangle$ has the structure of a certain 3D crystal, which can be uplifted from the periodic quiver $\tilde{Q}$ defined in \eqref{eq:periodicQDef}. 
The action of $e^{(a)}(u)$ (resp.\ $f^{(a)}(u)$) then adds (resp.\ removes) an atom such that the resulting state $|\Pi+\sqbox{$a$}\rangle$ (resp.\ $|\Pi-\sqbox{$a$}\rangle$) is again an ``allowed" 3D crystal state. 
For a general quiver, first of all, one cannot define a corresponding periodic quiver.
Instead, one needs to first construct all the states $|\Pi_{d}\rangle$ from $(^{\sharp}Q,^{\sharp}W)$ iteratively.

\medskip

Let us now explain the ansatz \eqref{eq:psiefAnsatz}.
First of all, every element $\Pi$ of the poset ${}^{\sharp}\mathcal{P}_{(Q,W)}$ is an eigenstate of all the Cartan generators:
\begin{equation}\label{eq:CartanAction}
\begin{aligned}
\psi^{(a)}(z)|\Pi\rangle&= \Psi_{\Pi}^{(a)}(z)|\Pi\rangle \;,
\end{aligned}
\end{equation}
with the eigenvalue function taking the form 
\begin{equation}\label{eq:ChargeFunctionDef}
\Psi^{(a)}_{\Pi}(z) ={}^{\sharp}\psi^{(a)}_{0}(z) \prod_{b\in Q_0} \prod_{\sqbox{$b$} \in \Pi } \varphi^{a\Leftarrow b}(z-h(\sqbox{$b$}))   \,,
\end{equation}
which will often be called the ``charge function".
There are two types of contributions to  \eqref{eq:ChargeFunctionDef}:
\begin{enumerate}
\item The first factor ${}^{\sharp}\psi^{(a)}_{0}(z)$ is the $\psi^{(a)}(z)$ eigenvalue of the ground state $|{}^{\sharp}\Pi_0\rangle$ (see \eqref{eq:GroundStateDef}) and captures the contribution from the ground state $|{}^{\sharp}\Pi_0\rangle$.
\item  The factor $\varphi^{a\Leftarrow b}(z-h(\sqbox{$b$}))$ gives the contribution from each atom $\sqbox{$b$}\in \Pi$, where  $\varphi^{a\Leftarrow b}$ is the bonding factor defined in \eqref{eq:BondingFactorDef} from the quiver $Q$, and $h(\sqbox{$b$})$ is the weight of the atom $\sqbox{$b$}$ defined in \eqref{eq:weightPathAtom}.
Note that all the atoms of all the colors contribute to the charge function of color $a$ --- not just those with color $a$ --- hence the total number of $\varphi$ factors in \eqref{eq:ChargeFunctionDef} is $|\Pi|$. 
\end{enumerate}

\medskip

The $\psi^{(a)}(z)$ eigenvalue of the state $|\Pi\rangle$  completely fixes the action of the raising and lowering operators, $e^{(a)}(z)$ and $f^{(a)}(z)$, on it. 
\begin{enumerate}
\item First of all, the poles of $ \Psi_{\Pi}^{(a)}(z)$ determine the set Add$(\Pi)$ (resp.\ Rem$(\Pi)$) for the action of $e^{(a)}(z)$ (resp.\ $f^{(a)}(z)$) as follows.
For each $ \Psi^{(a)}_{\Pi}(z)$, with $a=1,2,\dots, |Q_0|$, we take all its poles, assign them with color $a$, and define
\begin{equation}
\mathcal{S}^{(a)}(\Pi) \equiv \{p^{(a)}_1,\dots p^{(a)}_{m_{a}}\}\,,
\end{equation}
where $m_a$ is the total number of poles with color $a$ for the element $\Pi$. 
Then we take the union of $\mathcal{S}^{(a)}(\Pi)$ for all colors $a\in Q_0$:
\begin{equation}\label{eq:SofPiDef}
\mathcal{S}(\Pi) \equiv \bigcup_{a\in Q_0} \mathcal{S}^{(a)}(\Pi)\,.
\end{equation}
This set of poles $\mathcal{S}(\Pi)$ can be divided into two sets:
\begin{itemize}
\item The poles that coincide with the weights of the atoms inside $\Pi$:
\begin{equation}\label{eq:RofPiDef}
\mathcal{R}(\Pi) \equiv \mathcal{\W}(\Pi) \cap \mathcal{S}(\Pi)\,,
\end{equation}
where $\mathcal{W}(\Pi)$ was defined in \eqref{eq:WofPi}.
They will correspond to the weights of the atoms that can be removed from the state $|\Pi\rangle$ by the lowering operators $f^{(a)}(z)$.
\item The remaining ones:
\begin{equation}\label{eq:AofPiDef}
\mathcal{A}(\Pi) \equiv  \mathcal{S}(\Pi)\setminus  \mathcal{R}(\Pi)\,,
\end{equation}
will correspond to the weights of the atoms that can be added to the state $|\Pi\rangle$ by the raising operators $e^{(a)}(z)$.
\end{itemize}
\item The set of atoms that can be removed from $\Pi$ are those that are already inside $\Pi$ and whose weights coincide with the elements in $\mathcal{R}(\Pi)$: 
\begin{equation}\label{eq:RemPi}
\textrm{Rem}(\Pi)=\bigcup_{a\in Q_0}\left\{\sqbox{$a$}\in \Pi \, \middle|\, h(\sqbox{$a$})\in \mathcal{R}(\Pi)\right\}      \,.
\end{equation}
Given an initial state $|\Pi\rangle \in \mathcal{P}$ at level $d$, the lowering operator $f^{(a)}(z)$ takes it to a linear superposition of states at level $d-1$, 
each obtained by removing an atom $\sqbox{$a$}$ from the initial state $\Pi$, denoted by
\begin{equation}
|\Pi-\sqbox{$a$} \rangle\equiv |\Pi \backslash \{\sqbox{$a$}\} \rangle\,.
\end{equation}
\item Similarly, the set of atoms that can be added to $\Pi$ are those that can be reached from at least one existing atom in $\Pi$ via a single arrow $\in Q_1$, and whose weight coincide with an element in $\mathcal{A}(\Pi)$:
\begin{equation}\label{eq:AddPi}
\textrm{Add}(\Pi)=\bigcup_{b\in Q_0} \bigcup_{a\in Q_0}\left\{\sqbox{$a$}=\sqbox{$b$}\cdot I^{b\rightarrow a} \, \middle| \,\sqbox{$b$}\in \Pi \textrm{ and } h(\sqbox{$a$})\in \mathcal{A}(\Pi) \right\}   \,.
\end{equation}
Given an initial state
$|\Pi\rangle\in \mathcal{P}$ at level $d$, the raising operator $e^{(a)}(z)$ takes it to a linear superposition of states at level $d+1$, each obtained by adding an atom $\sqbox{$a$}$ to the initial state $\Pi$, denoted by
\begin{equation}
|\Pi+\sqbox{$a$} \rangle\equiv |\Pi \cup \{\sqbox{$a$}\} \rangle\,.
\end{equation}
\item In addition, the eigenvalue $ \Psi_{\Pi}^{(a)}(z)$ determines the coefficients of the $e^{(a)}(z)$ and $f^{(a)}(z)$ actions in \eqref{eq:psiefAnsatz}:
\begin{equation}\label{eq:EFCoefficientsSim}
\llbracket \Pi\rightarrow \Pi\pm \sqbox{$a$}\rrbracket \sim \left(\textrm{Res}_{u=h(\sqbox{$a$}) }\Psi^{(a)}_{\Pi}(u)\right)^{\frac{1}{2}}\,,
\end{equation}
where the residue ensures that the action is null if the weight of the atom $\sqbox{$a$}$ is not among the poles of $\Psi_{\Pi}(z)$. 
We will fix the precise expression later in Sec.\ \ref{ssec:Bootstrap}, since it is not important for the construction of the representation ${}^{\sharp}\mathcal{P}_{(Q,W)}$.     
\end{enumerate}

\subsubsection{Recursive construction of representations from framed quivers}
\label{sssec:RepresentationConstruction}

Now we can apply the action of the raising operators $e^{(a)}(z)$ on the ground state $|{}^{\sharp}\Pi_{0}\rangle$ recursively to obtain all the states $|{}^{\sharp}\Pi_{d}\rangle$ in the representation ${}^{\sharp}\mathcal{P}_{(Q,W)}$.

\medskip
\noindent \textbf{Ground state (level 0).}
The ground state of the representation ${}^{\sharp}\mathcal{P}_{(Q,W)}$ corresponds to the first element of the poset ${}^{\sharp}\mathcal{P}_{(Q,W)}$, which is the empty set $^{\sharp}\Pi_0=\{\varnothing\}$, see definition  \eqref{eq:PiGroundState}. 
The ground state is labeled as level-0.

Although $^{\sharp}\Pi_0$ is the empty set, i.e.\ with no atom, it specifies which representation of a given quiver Yangian $\textsf{Y}(Q,W)$ we are considering, via its eigenvalue of $\psi^{(a)}(z)$, see definition \eqref{eq:GroundStateDef}. 
Its eigenvalue function ${}^{\sharp}\psi_{0}^{(a)}(z)$ is determined by the framing $\sharp$ of the quiver as follows. 
Each arrow $\tilde{J}$ from the framing vertex $\infty$ to a framed vertex $\mathfrak{a}$ contributes a factor $(z-h(\tilde{J}^{\infty\rightarrow\mathfrak{a}}))$ to the denominator of ${}^{\sharp}\psi_{0}^{(a)}(z)$, whereas each arrow $\tilde{I}$ from a framed vertex $\mathfrak{a}$ back to the framing node $\infty$ contributes a factor $(z+h(\tilde{I}^{\mathfrak{a}\rightarrow\infty}))$ to the numerator of ${}^{\sharp}\psi_{0}^{(a)}(z)$:
\begin{equation}\label{eq:varphi0Def}
{}^{\sharp}\psi_{0}^{(a)}(z) 
=\frac{\prod_{\tilde{I}
\in \{\mathfrak{a} \rightarrow \infty\}} (z+h(\tilde{I}))}{\prod_{\tilde{J}
\in\{\infty \rightarrow \mathfrak{a}\}} (z-h(\tilde{J}))}
=\frac{\prod^{\mathfrak{s}^{(a)}_-}_{\beta=1} (z-\mathtt{z}^{(a)}_{\beta})}{\prod^{\mathfrak{s}^{(a)}_{+}}_{\alpha=1} (z-\mathtt{p}^{(a)}_{\alpha})}\,,
\end{equation}
where in the last step we change notation to highlight the fact that the weights of the arrows to and from the framing vertex $\infty$ give rise to the zeros and poles of the ground state charge function ${}^{\sharp}\psi_{0}^{(a)}(z)$, respectively:
\begin{equation}
\{\mathtt{p}^{(a)}_{\alpha}\}\equiv \bigcup_{\tilde{J} 
\in \{\infty \rightarrow \mathfrak{a}\} } \{ h(\tilde{J}) \} \qquad \textrm{and} \qquad \{\mathtt{z}^{(a)}_{\beta}\}\equiv\bigcup_{\tilde{I} 
\in \{\mathfrak{a} \rightarrow \infty\} }
\{- h(\tilde{I} )\}\,.
\end{equation}
Furthermore,  the number of arrows to and from $\infty$ are labeled by $\mathfrak{s}^{(a)}_{-}\equiv |\mathfrak{a}\rightarrow \infty|$ and $\mathfrak{s}^{(a)}_{+}\equiv |\infty\rightarrow \mathfrak{a}|$, respectively.
Finally, the choice of the framing is not completely arbitrary. 
We will explain the constraints on the choice later in Sec.\ \ref{ssec:NonVacuumReps}. 

\medskip

\noindent\textbf{Level $d$ to level $d+1$.}
The procedure of going from level $d$ to  level $d+1$ is as follows:
\begin{enumerate}
\item  Assume that we know the set of states $\{{}^{\sharp}\Pi_{d}\}$ at level $d$. 
Each state ${}^{\sharp}\Pi_{d}$ contains a set of paths $\in J^{\sharp}_{(Q,W)} $ as in \eqref{eq:PiDef1}, or equivalently a set of atoms as in \eqref{eq:PiDef2}.
We first compute its weight $\mathcal{W}({}^{\sharp}\Pi_{d})$ defined in \eqref{eq:WofPi}.
Then we compute its charge function $ \Psi^{(a)}_{{}^{\sharp}\Pi_{d}}(z)$ for all $a\in Q_0$ using the definition \eqref{eq:ChargeFunctionDef}. There are in total $|Q_0|$ such functions $ \Psi^{(a)}_{{}^{\sharp}\Pi_{d}}(z)$ for each state ${}^{\sharp}\Pi_{d}$.

\item For each ${}^{\sharp}\Pi_{d}$, use $\Psi^{(a)}_{{}^{\sharp}\Pi_{d}}(z)$ to determine its $\mathcal{S}({}^{\sharp}\Pi_{d})$, $\mathcal{R}({}^{\sharp}\Pi_{d})$, and $\mathcal{A}({}^{\sharp}\Pi_{d})$, defined in \eqref{eq:SofPiDef}, \eqref{eq:RofPiDef}, and \eqref{eq:AofPiDef}, respectively.
Then for each ${}^{\sharp}\Pi_{d}$, use its $\mathcal{A}({}^{\sharp}\Pi_d)$ and the information of the quiver $(Q,W)$ to determine the set of atoms Add$({}^{\sharp}\Pi_d)$ that can be added, defined in \eqref{eq:AddPi}. 

\item Within the set $\textrm{Add}({}^{\sharp}\Pi_{d})$, each atom $\sqbox{$a$}_{*}$ for some $a\in Q_0$  then gives rise to a new state at level $d+1$, defined by adding this atom $\sqbox{$a$}_{*}$ to the set ${}^{\sharp}\Pi_{d}$:
\begin{equation}
\begin{aligned}
{}^{\sharp}\Pi_{d}&=\{ \sqboxs{$a_1$}_1, \sqboxs{$a_2$}_2, \dots, \sqboxs{$a_{d}$}_{d}\} 
\quad\longrightarrow \quad  
{}^{\sharp}\Pi_{d+1}=\{ \sqboxs{$a_1$}_1, \sqboxs{$a_2$}_2, \dots, \sqboxs{$a_{d}$}_{d}, \sqboxs{$a$}_* \}  \,.
\end{aligned}
\end{equation}
Each element ${}^{\sharp}\Pi_{d}$ gives rise to  $|\textrm{Add}({}^{\sharp}\Pi_{d})|$  new elements ${}^{\sharp}\Pi_{d+1}$ at level $d+1$ in the poset ${}^{\sharp}\mathcal{P}$.
\item Finally, since different elements at level $d$, ${}^{\sharp}\Pi_{d}$ and ${}^{\sharp}\Pi'_{d}$, might give rise to the same element at level $d+1$, we take the union of all such ${}^{\sharp}\Pi_{d+1}$ \textit{without multiplicity} to be the set of elements at level $d+1$. 
For this step, we need to check the equivalence of two atoms by examining whether their corresponding paths are path equivalent.
\end{enumerate}
With the information on the ground state and the procedure of determining the set of states at level $d+1$ once the set of states at level $d$ is given, we can determine all the elements $\Pi$ inside the poset $\mathcal{P}$ recursively.
The partition function of the poset $\mathcal{P}$ is then given by
\begin{equation}\label{eq:Zpartitionfunction}
\mathcal{Z}_{\mathcal{P}}( \boldsymbol{x} ) = \sum_{\Pi \in \mathcal{P}} \, \boldsymbol{x}^{|\Pi|}\,,
\end{equation}
where $\boldsymbol{x}$ and $\boldsymbol{x}^{|\Pi|}$ are shorthand notations defined as
\begin{equation}
\boldsymbol{x} \equiv ( x_0,x_1,\dots x_{|Q_0|-1}) \qquad \textrm{and} \qquad  \boldsymbol{x}^{|\Pi|}\equiv \prod_{a\in Q_0} (x_a)^{|\Pi^{(a)}|}\,,
\end{equation}
where $\Pi^{(a)}$ is the subset of all atoms of color $a$ within $\Pi$, defined in \eqref{eq:piaDef}, and $|\Pi^{(a)}|$ is its size.
The vector $(|\Pi^{(0)}|,|\Pi^{(1)}|,\dots, |\Pi^{(|Q|_0-1)}|)$ will be mapped to the \textit{dimension vector} $\boldsymbol{d}$ of the quiver, see \eqref{eq:dvDef}. 

\medskip 

Finally, we have a few detailed examples illustrating this poset-generating procedure later in the paper, e.g., for the toric CY$_3$ quivers in Appx.\ \ref{appssec:toricCY3ex}, the affine Yangian of A-D-E type in Appx.\ \ref{appssec:PosetAn},  Sec.\ \ref{sssec:VacRepDn}, and Appx.\  \ref{appssec:PosetE678}, respectively, the $m$-loop quiver in Sec.\ \ref{sssec:mloopVacuumRep}, and finally the quiver for the trefoil knot (via the knot-quiver correspondence) in Appx.\ \ref{appsec:TFquiver}.
The interested readers may consult those first before continuing.

\subsubsection{Example: from level 0 to level 1}

Now let us apply this four-step procedure to obtain the set of states at level $1$ from the ground state.
This step is universal for all framed quivers $({}^{\sharp}Q,{}^{\sharp}W)$, namely, it doesn't depend on the details of  $({}^{\sharp}Q,{}^{\sharp}W)$.
\begin{enumerate}
\item First of all, since ${}^{\sharp}\Pi_0=\{\varnothing \}$, we have 
\begin{equation}
\mathcal{W}({}^{\sharp}\Pi_0)=\{\varnothing \}\,.
\end{equation}
By definition \eqref{eq:ChargeFunctionDef}, the charge function of the ground state is just
\begin{equation}
\Psi^{(a)}(z)
={}^{\sharp}\psi_{0}^{(a)}(z)   \,,
\end{equation}
given in \eqref{eq:varphi0Def}.
\item By definition \eqref{eq:SofPiDef}, the set of poles for the state ${}^{\sharp}\Pi_0$ is then 
\begin{equation}
\mathcal{S}({}^{\sharp}\Pi_0) =\bigcup_{a\in Q_0} \{\mathtt{p}^{(a)}_{\alpha} \}\,.
\end{equation}
By definition \eqref{eq:RofPiDef}, the set of removing poles is empty:
\begin{equation}\label{eq:RemVacuum}
\mathcal{R}({}^{\sharp}\Pi_0)
=\{\varnothing\}    \,.
\end{equation}
By definition \eqref{eq:AofPiDef}, the set of adding poles is then 
\begin{equation}\label{eq:AddVacuumPole}
\mathcal{A}({}^{\sharp}\Pi_0)
=\mathcal{S}({}^{\sharp}\Pi_0) 
=\bigcup_{a\in Q_0} \{p^{(a)}_{\alpha} \}   \,,
\end{equation}
and the set of atoms that can be added is
\begin{equation}\label{eq:AddVacuumAtom}
\textrm{Add}({}^{\sharp}\Pi_0)
=\bigcup_{\{\infty \rightarrow \mathfrak{a}\}} \{\tilde{J}^{\infty\rightarrow \mathfrak{a}} \} 
=\bigcup_{\{\infty \rightarrow \mathfrak{a}\}}\{I^{(0)}\cdot \mathfrak{p}^{0\rightarrow  \mathfrak{a}}\}
\end{equation}
by definition \eqref{eq:AddPi}, where the last step will be explained later in Sec.\  \ref{ssec:NonVacuumReps} and Appx.\ \ref{appssec:Method1}, in particular see \eqref{eq:starterDef}, \eqref{eq:pauser+Def}, and  \eqref{eq:Jinftytoa}.

\item Each path $\tilde{J}^{\infty\rightarrow \mathfrak{a}}$ in $\textrm{Add}({}^{\sharp}\Pi_0)$ then gives rise to a new state at level $1$:
\begin{equation}\label{eq:StatesLevel1}
{}^{\sharp}\Pi_{1}
=\{I^{(0)}\cdot \mathfrak{p}^{0\rightarrow  \mathfrak{a}}\}\,.
\end{equation}
There are in total $\sum_{a\in Q_0} \mathfrak{s}^{(a)}_{+}$ of them.

\item Finally, since all these states are from the same initial state ${}^{\sharp}\Pi_0$, they are all distinct. 
Therefore, there are in total $\sum_{a\in Q_0} \mathfrak{s}^{(a)}_{+}$ distinct states at level $1$.

\end{enumerate}

\subsubsection{Example: from level 1 to level 2 for the vacuum-like representations}

Starting from level 1, the detailed procedure will depend on the quiver in question. 
We will demonstrate the procedure above when we study concrete examples.
For now, let us just briefly explain the step from level 1 to level 2 for the vacuum-like representations, namely those that correspond to the canonically framed quivers.

\medskip

A vacuum-like representation corresponds to a canonically framed quiver $({}^{(\mathfrak{a})}Q,\\{}^{(\mathfrak{a})}W)$, where there is only one arrow from the framing vertex $\infty$ to the framed vertex $\mathfrak{a}$, see \eqref{eq:CanonicalQW}. 
Therefore there is only one state
\begin{equation}\label{eq:Pi1Vacuum}
{}^{(\mathfrak{a})}\Pi_{1}=\{\tilde{I}^{\infty \rightarrow \mathfrak{a}}\}=\{I^{(\mathfrak{a})}\}
\end{equation} at level 1.
Let us now apply the four-step procedure to obtain the states at level 2.
\begin{enumerate}
\item First of all, from \eqref{eq:Pi1Vacuum}, we have 
\begin{equation}
\mathcal{W}({}^{(\mathfrak{a})}\Pi_{1})
=\{(0)^{(\mathfrak{a})}\}    \,,
\end{equation}
where we have set $h(\tilde{I}^{\infty \rightarrow \mathfrak{a}})=0$ and attached the color for clarity.
By definition \eqref{eq:ChargeFunctionDef}, the charge functions of ${}^{(\mathfrak{a})}\Pi_{1}$ are 
\begin{equation}
\Psi^{(b)}(z)
=\left(\frac{1}{z}\right)^{\delta_{b,\mathfrak{a}}} \cdot \varphi^{b\Leftarrow \mathfrak{a}}(u) 
\qquad \textrm{with} \quad
\varphi^{b\Leftarrow \mathfrak{a}}(u)        
=\frac{\prod_{I\in \{b \rightarrow \mathfrak{a}\}}(z+h_I)}{\prod_{J\in\{\mathfrak{a}\rightarrow b\} }(z-h_J)}   \,,
\end{equation}
for all $b\in Q_0$.
\item By definition \eqref{eq:SofPiDef}, the set of poles for the state $\Pi_1$ is then \begin{equation}
\mathcal{S}({}^{(\mathfrak{a})}\Pi_1) 
=\{(0)^{(\mathfrak{a})}\}
\cup \bigcup_{b\in Q_0}
\bigcup_{J\in \{\mathfrak{a}\rightarrow b\}}\{(h_J)^{(b)}\}\,.
\end{equation}
By definition \eqref{eq:RofPiDef}, the set of removing poles is:
\begin{equation}\label{eq:RemLevel1}
\mathcal{R}({}^{(\mathfrak{a})}\Pi_1)
=\{(0)^{(\mathfrak{a})}\}\,,
\end{equation}
which corresponds to the single atom of the state ${}^{(\mathfrak{a})}\Pi_1$.
By definition \eqref{eq:AofPiDef}, the set of adding poles is then 
\begin{equation}\label{eq:AddLevel1Pole}
\mathcal{A}({}^{(\mathfrak{a})}\Pi_1)
=\bigcup_{b\in Q_0} \bigcup_{J\in \{\mathfrak{a}\rightarrow b\}} \{ (h_J)^{(b)}\}\,,
\end{equation}
with $b=t(J)$, and the set of atoms that can be added is
\begin{equation}\label{eq:AddLevel1Atom}
\textrm{Add}({}^{(\mathfrak{a})}\Pi_1)
=\bigcup_{b\in Q_0} \bigcup_{J\in \{\mathfrak{a} \rightarrow b\}}  \{I^{(\mathfrak{a})}\cdot J^{\mathfrak{a}\rightarrow b}\}    \,,
\end{equation}
by definition \eqref{eq:AddPi}.

\item Each path $I^{(\mathfrak{a})}\cdot J^{\mathfrak{a}\rightarrow b}$ in $\textrm{Add}({}^{(\mathfrak{a})}\Pi_1)$ then gives rise to a new state at level $2$:
\begin{equation}\label{eq:StatesLevel2}
{}^{(\mathfrak{a})}\Pi_{2,J}=\{I^{(\mathfrak{a})},\, I^{(\mathfrak{a})}\cdot J^{\mathfrak{a}\rightarrow b}\}\,.
\end{equation}
There are in total $\sum_{b\in Q_0}|\mathfrak{a}\rightarrow b|$ of them.

\item Finally, since all these states are from the same initial state ${}^{(\mathfrak{a})}\Pi_1$, they are all distinct, therefore, there are in total $\sum_{b\in Q_0}|\mathfrak{a}\rightarrow b|$ distinct states at level $2$ in the vacuum-like representation ${}^{(\mathfrak{a})}\mathcal{P}_{(Q,W)}$.
\end{enumerate}

\subsection{Posets as quiver Yangian representations}
\label{ssec:Bootstrap}

In this subsection, we show that the poset ${}^{\sharp}\mathcal{P}_{(Q,W)}$ constructed in Sec.\ \ref{ssec:PosetFromQW} is indeed a representation of the quiver Yangian $\textsf{Y}(Q,W)$, defined in Sec.\ \ref{ssec:QYalgebra}.

Recall that the procedure of generating the poset representation from a given quiver with potential $(Q,W)$ was inspired by the ansatz \eqref{eq:psiefAnsatz}.
In the poset-generating procedure, we really only need three pieces of information from the ansatz \eqref{eq:psiefAnsatz}: 
\begin{enumerate}
\item Each element $\Pi$ in the poset ${}^{\sharp}\mathcal{P}_{(Q,W)}$ is an eigenfunction of $\psi^{(a)}(z)$, for all $a\in Q_0$. 
The eigenvalue $\Psi^{(a)}_{\Pi}(z)$ is a rational function, given by \eqref{eq:ChargeFunctionDef}.
\item For a given element $\Pi$, applying the lowering operators $f^{(a)}(z)$ on it gives a state that is a linear combination of states with one of the removable atoms (defined in \eqref{eq:RemPi}) removed.
\item For a given element $\Pi$, applying the raising operators $e^{(a)}(z)$ on it gives a state that is a linear combination of states with one of the addable atoms (defined in \eqref{eq:AddPi}) added.
\end{enumerate}
Note that the details of the coefficients of the $e^{(a)}(z)$ and $f^{(a)}(z)$ actions were not crucial for the poset-generating procedure of Sec.\ \ref{sssec:RepresentationConstruction}, as long as they are non-zero for the relevant final states that $e^{(a)}(z)$ and $f^{(a)}(z)$ are supposed to create.

\bigskip

In this subsection, we derive the precise form of $\Psi^{(a)}_{\Pi}(z)$ and 
$\llbracket \Pi\rightarrow \Pi\pm \sqbox{$a$}\rrbracket$, by demanding that the action \eqref{eq:psiefAnsatz} reproduces the quadratic algebraic relation \eqref{eq:QuadraticFields}.
Note that although in constructing the poset, we have assumed the form of $\Psi^{(a)}_{\Pi}(z)$ as in \eqref{eq:ChargeFunctionDef}, we can actually determine it from scratch.
Moreover, we will give a systematic derivation of the signs in the coefficients $\llbracket \Pi\rightarrow \Pi\pm \sqbox{$a$}\rrbracket$.\footnote{
The sign choice here differs from \cite{Li:2020rij,Galakhov:2020vyb,Galakhov:2022uyu} in various ways, see below.
}

Let us now check the algebraic relations in \eqref{eq:QuadraticFields} by applying them on an arbitrary initial state $\Pi$ and using the ansatz \eqref{eq:psiefAnsatz}.

\medskip

\noindent \textbf{$\boldsymbol{\psi}$-$\boldsymbol{\psi}$ relations.} In the action \eqref{eq:psiefAnsatz}, since any state $\Pi$ is an eigenstate of $\psi^{(a)}(z)$ for all $a\in Q_0$, we have
\begin{equation}
\psi^{(a)}(z)\, \psi^{(b)}(w)\, |\Pi\rangle =  \psi^{(b)}(w)\, \psi^{(a)}(z)\, |\Pi\rangle   \,.
\end{equation}
Since this equation is true for any $\Pi$, we have the first equation of \eqref{eq:QuadraticFields}.

\medskip

\noindent \textbf{$\bm{\psi}$-$\bm{e}$ and $\bm{\psi}$-$\bm{f}$ relations.} To check the $\psi$-$e$ equation in \eqref{eq:QuadraticFields}, apply both sides of the equation on any initial state $\Pi$ to obtain
\begin{equation}\label{eq:psie2sides}
\begin{aligned}
& \sum_{\sqbox{$b$} \,\in \,\textrm{Add}(\Pi)} 
\frac{\llbracket \Pi\rightarrow \Pi+ \sqbox{$b$} \rrbracket}{w- h(\sqbox{$b$}) }\, \Psi^{(a)}_{\Pi+\sqbox{$b$}  }(z)\, |\Pi+\sqbox{$b$} \rangle    \\
\simeq \varphi^{a\Leftarrow b}(z-w)\,&\sum_{\sqbox{$b$} \,\in \,\textrm{Add}(\Pi)} 
\frac{\llbracket \Pi\rightarrow \Pi+ \sqbox{$b$}\rrbracket}{w-h(\sqbox{$b$})}\, \Psi^{(a)}_{\Pi}(z)\, |\Pi+\sqbox{$b$}\rangle  \,.
\end{aligned}
\end{equation}
Recall that ``$\simeq$" means equality up to regular terms in $w$, and because of the factor  $\frac{1}{w-h(\sqbox{$b$})}$ on both sides of the equation, the equation holds as long as  
\begin{equation}\label{eq:Psiforpsi-e}
\Psi^{(a)}_{\Pi+\sqbox{$b$}}(z) =  \varphi^{a\Leftarrow b} (z-h(\sqbox{$b$}))\, \Psi^{(a)}_{\Pi}(z) \,.
\end{equation}
Similarly, from the $\psi$-$f$ equation, we derive the constraint
\begin{equation}\label{eq:Psiforpsi-f}
\Psi^{(a)}_{\Pi-\sqbox{$b$}}(z) =  \varphi^{a\Leftarrow b} (z-h(\sqbox{$b$}))^{-1}\, \Psi^{(a)}_{\Pi}(z) \,.
\end{equation}
The solution of $\Psi^{(a)}_{\Pi}(u)$ to these two equations is
\begin{equation}\label{eq:ChargeFunction1}
\Psi^{(a)}_{\Pi}(z) %={}^{\sharp}\psi^{(a)}_{0}(z) 
\sim \prod_{b\in Q_0} \prod_{ \sqbox{$b$} \in \Pi } \varphi^{a\Leftarrow b}(z-h(\sqbox{$b$}))\,,
\end{equation}
which agrees with the definition of $\Psi^{(a)}_{\Pi}(z)$ in \eqref{eq:ChargeFunctionDef}; the prefactor ${}^{\sharp}\psi^{(a)}_{0}(z)$ in \eqref{eq:ChargeFunctionDef} is a function that cannot be fixed from the constraints \eqref{eq:Psiforpsi-e} and \eqref{eq:Psiforpsi-f} and it captures the contribution from the ground state of the representation, see \eqref{eq:varphi0Def}.

\medskip

\noindent \textbf{$\bm{e}$-$\bm{e}$, $\bm{f}$-$\bm{f}$, $\bm{e}$-$\bm{f}$   relations.} First, applying both sides of the $e$-$e$ relation in \eqref{eq:QuadraticFields} on an arbitrary initial state $\Pi$, and using the $e$ action from the ansatz \eqref{eq:psiefAnsatz}, we have
\begin{equation}\label{eq:ee2sides}
\begin{aligned}
 & \qquad \sum_{\sqbox{$b$} \,\in \,\textrm{Add}(\Pi)} \sum_{\sqbox{$a$} \,\in \,\textrm{Add}(\Pi+\sqbox{$b$})} 
 \frac{\llbracket \Pi\rightarrow \Pi+ \sqbox{$b$} \rightarrow \Pi+ \sqbox{$b$} +\sqbox{$a$}\rrbracket}{(z-h(\sqbox{$a$}))(w-h(\sqbox{$b$}))}\, |\Pi+\sqbox{$b$}+\sqbox{$a$}\rangle    \\
 &\sim e^{\pi i s_{ab}} \varphi^{a\Leftarrow b}(z-w) \times\\ &\qquad \sum_{\sqbox{$a$} \,\in \,\textrm{Add}(\Pi)} \sum_{\sqbox{$b$} \,\in \,\textrm{Add}(\Pi+\sqbox{$a$})} 
 \frac{\llbracket \Pi\rightarrow \Pi+ \sqbox{$a$} \rightarrow \Pi+ \sqbox{$a$} +\sqbox{$b$}\rrbracket}{(z-h(\sqbox{$a$}))(w-h(\sqbox{$b$}))}\, |\Pi+\sqbox{$a$}+\sqbox{$b$}\rangle   \,,
\end{aligned}
\end{equation}
where we have defined the shorthand notation
\begin{equation}
\llbracket \Pi_1 \rightarrow \Pi_2 \rightarrow \Pi_3 \rrbracket \,  \equiv \llbracket \Pi_1 \rightarrow \Pi_2 \rrbracket \,  \llbracket \Pi_2 \rightarrow \Pi_3 \rrbracket \,.
\end{equation}
Since ``$\sim$" means that the equation holds only up to regular terms in $z^{n\geq 0}w^{m \geq 0}$, we reach the constraint\footnote{
For the details of this derivation, see \cite[Sec.\ 6.5.4]{Li:2020rij}.
}
\begin{equation}\label{eq:constraintfromee}
\frac{\llbracket \Pi\rightarrow \Pi+ \sqbox{$b$} \rightarrow \Pi+ \sqbox{$b$} +\sqbox{$a$}\rrbracket}{\llbracket \Pi\rightarrow \Pi+ \sqbox{$a$} \rightarrow \Pi+ \sqbox{$a$} +\sqbox{$b$}\rrbracket} =e^{\pi i s_{ab}} \varphi^{a\Leftarrow b}(h(\sqbox{$a$})-h(\sqbox{$b$})) \,.
\end{equation}
Similarly, from the $f$-$f$ equation, we have
\begin{equation}\label{eq:constraintfromff}
\frac{\llbracket \Pi\rightarrow \Pi- \sqbox{$b$} \rightarrow \Pi- \sqbox{$b$} -\sqbox{$a$}\rrbracket}{\llbracket \Pi\rightarrow \Pi- \sqbox{$a$} \rightarrow \Pi- \sqbox{$a$} -\sqbox{$b$}\rrbracket} =e^{-\pi i s_{ab}} \varphi^{a\Leftarrow b}(h(\sqbox{$a$})-h(\sqbox{$b$}))^{-1}\,.
\end{equation}
Finally, from the $e$-$f$ relation, we have\footnote{
For the details of this derivation, see \cite[Sec.\ 6.5.3]{Li:2020rij}.
}
\begin{equation}\label{eq:constraintfromef}
\begin{aligned}
\llbracket \Pi\rightarrow \Pi- \sqbox{$b$} \rightarrow \Pi- \sqbox{$b$} +\sqbox{$a$}\rrbracket -e^{\pi i s_{ab}} \llbracket \Pi\rightarrow \Pi+\sqbox{$a$} \rightarrow \Pi+ \sqbox{$a$} -\sqbox{$b$}\rrbracket &\\
= \delta_{a,b} \delta_{h(\sqbox{$a$}),h(\sqbox{$b$})} \textrm{Res}_{u=h(\sqbox{$a$})}\Psi^{(a)}_{\Pi}(u) &\,,
\end{aligned}
\end{equation}
which gives two constraints, depending on whether the atoms created by $e$ is the same as the one removed by $f$. 
If they are different, we have
\begin{equation}\label{eq:constraintfromef1}
\frac{\llbracket \Pi\rightarrow \Pi- \sqbox{$b$} \rightarrow \Pi- \sqbox{$b$} +\sqbox{$a$}\rrbracket}{\llbracket \Pi\rightarrow \Pi+ \sqbox{$a$} \rightarrow \Pi +\sqbox{$a$} -\sqbox{$b$}\rrbracket} =e^{\pi i s_{ab}} \,,
\end{equation}
and otherwise
\begin{equation}\label{eq:constraintfromef2}
\begin{aligned}
\llbracket \Pi\rightarrow \Pi- \sqbox{$a$} \rightarrow \Pi \rrbracket &= \textrm{Res}_{u=h(\sqbox{$a$})}\Psi^{(a)}_{\Pi}(u)\\
\llbracket \Pi\rightarrow \Pi+ \sqbox{$a$} \rightarrow \Pi\rrbracket&=-e^{-\pi i s_{ab}} \textrm{Res}_{u=h(\sqbox{$a$})}\Psi^{(a)}_{\Pi}(u)
\end{aligned}
\end{equation}
Note that in eq.\ \eqref{eq:constraintfromef} and in the subsequent equations involving ``Res", we have assumed that $u=h(\sqbox{$a$})$ is a simple pole of $\Psi^{(a)}_{\Pi}(u)$.
As we will show later, when the weights $h_I$ are generic, each atom $\sqbox{$a$}$ corresponds to a simple pole in the charge function $\Psi^{(a)}_{\Pi}$, responsible for both its addition to and removal from $\Pi$.
However, there can be degenerate situations where the poles for two or more atoms have the same value, resulting in ``accidental" double or higher-order poles in the charge function.
When this happens, we need to resolve the higher-order pole first before evaluating the residue. 
(If the n$^{\textrm{th}}$-order pole happens to be pure, the condition \eqref{eq:constraintfromef} is simply modified into 
\begin{equation}\label{eq:constraintfromefhigher}
\begin{aligned}
\llbracket \Pi\rightarrow \Pi- \sqbox{$b$} \rightarrow \Pi- \sqbox{$b$} +\sqbox{$a$}\rrbracket -e^{\pi i s_{ab}} \llbracket \Pi\rightarrow \Pi+\sqbox{$a$} \rightarrow \Pi+ \sqbox{$a$} -\sqbox{$b$}\rrbracket &\\
= \delta_{a,b} \delta_{h(\sqbox{$a$}),h(\sqbox{$b$})} (\textrm{lim}_{u\rightarrow h(\sqbox{$a$}) }(u- h(\sqbox{$a$}))^n\Psi^{(a)}_{\Pi}(u) )&\,,
\end{aligned}
\end{equation}
and the actions of $e$ and $f$ can be easily modified to preserve the algebraic relations, and no resolution of the pole is needed.)
We will explain these points in more detail in Appendix \ref{appsec:poleresolution}.

Now we can solve for the coefficients $\llbracket \Pi\rightarrow \Pi\pm  \sqbox{$a$} \rrbracket$ from these constraints.
First of all, the consistency between the two equations in \eqref{eq:constraintfromef2} imposes \eqref{eq:Naa}.
Then the coefficients $\llbracket \Pi\rightarrow \Pi\pm  \sqbox{$a$} \rrbracket$ can be solved by 
\begin{equation}\label{eq:EFsolutions}
\llbracket \Pi\rightarrow \Pi\pm \sqbox{$a$}\rrbracket = e^{\frac{\pi i }{4} (t_{aa}+|a\rightarrow a|)} 
\epsilon(\sqbox{$a$},\Pi)\left( \textrm{Res}_{u=h(\sqbox{$a$})}\Psi^{(a)}_{\Pi}(u) \right)^{\frac{1}{2}}\,,
\end{equation}
where $ \epsilon(\sqbox{$a$},\Pi)$ is a phase factor defined by
\begin{equation}\label{eq:epsilonDef}
\epsilon(\sqbox{$a$},\Pi)= \prod_{\sqbox{$b$}\in \Pi}v(\sqbox{$a$},\sqbox{$b$})	
\end{equation}
with
\begin{equation}\label{eq:vDef}
v(\sqbox{$a$},\sqbox{$b$})=
\begin{cases}
\begin{aligned}
1\qquad \qquad \qquad \qquad \qquad \qquad 
\qquad \qquad 
 &\qquad \sqbox{$a$}=\sqbox{$b$}\\	
\textrm{exp}(i\pi (N_{ab}+\frac{s_{ab}-s_{ba}}{2}) \frac{1+\textrm{sgn}(\sqbox{$a$},\sqbox{$b$})}{2}) &\qquad \sqbox{$a$}\neq \sqbox{$b$}
\end{aligned}
\end{cases}
\end{equation}
where $N_{ab}$ is defined in \eqref{eq:stab} and $\textrm{sgn}(\sqbox{$a$},\sqbox{$b$})$ is an arbitrary sign function that orders all the atoms in the representations, and we also require \eqref{eq:sab}.
The prescription for choosing the branches for the square root is given as follows. 
First, in view of the recursive relations \eqref{eq:Psiforpsi-e}, the signs of $ \Psi^{(a)}_{\Pi}(z)^{\frac{1}{2}}$ can be recursively determined such that
\begin{equation}\label{eq:sqrt1}
\Psi^{(a)}_{\Pi+\sqbox{$b$}}(z) ^{\frac{1}{2}}=  \varphi^{a\Leftarrow b} (z-h(\sqbox{$b$}))^{
\frac{1}{2}} \,  \Psi^{(a)}_{\Pi}(z)^{\frac{1}{2}}\,.
\end{equation}
Secondly, since $\Psi^{(a)}_{\Pi}(u)$ is a rational function of $u$ with only simple poles (after resolution of accidental higher-order poles), we have
\begin{equation}
\textrm{Res}_{u=h(\sqbox{$a$})}\Psi^{(a)}_{\Pi+\sqbox{$b$}}(u) =
\varphi^{a\Leftarrow b}(h(\sqbox{$a$})-h(\sqbox{$b$})) \cdot
\textrm{Res}_{u=h(\sqbox{$a$})}\Psi^{(a)}_{\Pi}(u)  
\end{equation}
when $h(\sqbox{$a$})$ is a pole of $\Psi^{(a)}_{\Pi}(u)$.
We will choose the branch of the square roots such that 
\begin{equation}\label{eq:Reshalf}
\left(\textrm{Res}_{u=h(\sqbox{$a$})}\Psi^{(a)}_{\Pi+\sqbox{$b$}}(u) \right)^{\frac{1}{2}}=
\varphi^{a\Leftarrow b}(h(\sqbox{$a$})-h(\sqbox{$b$}))^{\frac{1}{2}} \cdot
\left(\textrm{Res}_{u=h(\sqbox{$a$})}\Psi^{(a)}_{\Pi}(u)\right)^{\frac{1}{2}}  
\end{equation}
Lastly, the branches of  $\varphi^{a\Leftarrow b} (u)^{\frac{1}{2}}$ are chosen such that 
\begin{equation}\label{eq:varphihalf1}
\varphi^{a\Leftarrow b} (u)^{\frac{1}{2}}
=e^{\frac{i\pi}{2}(t_{ab}+|a\rightarrow b|)}\left(\frac{\prod_{I\in \{a\rightarrow b\}} (-u-h_I)}{\prod_{J\in \{b\rightarrow a\}} (u-h_J)}\right)^{\frac{1}{2}}\,,
\end{equation}
which gives
\begin{equation}\label{eq:varphi0half}
\varphi^{a\Leftarrow a} (0)^{\frac{1}{2}}
=e^{\frac{i\pi}{2}(t_{aa}+|a\rightarrow a|)}\,.
\end{equation}
This method of fixing the signs of the square root allows for general statistical factors $s_{ab}$ and $t_{ab}$ (subject to the constraints \eqref{eq:stab}, \eqref{eq:Naa}, and \eqref{eq:sab}).

With this prescription for the signs, one can check that the solution \eqref{eq:EFsolutions} indeed satisfies \eqref{eq:constraintfromee}, \eqref{eq:constraintfromff}, and \eqref{eq:constraintfromef}.
For \eqref{eq:constraintfromee}, \eqref{eq:constraintfromff}, and the $\sqbox{$a$}\neq \sqbox{$b$}$ case of \eqref{eq:constraintfromef} to hold, we have used \eqref{eq:Reshalf} and \eqref{eq:varphihalf1}.
For the remaining case, namely when the $e^{(a)}$ and $f^{(a)}$ actions correspond to the same pole, we have used \eqref{eq:Reshalf}, \eqref{eq:varphi0half}, and the constraints \eqref{eq:Naa} and \eqref{eq:sab}.

To conclude this subsection, we summarize the action of the quiver Yangian on any state
\begin{tcolorbox}[ams equation]
\label{eq:ActiononRep}
\begin{aligned}
\vspace{-2in}
\psi^{(a)}(z)|\Pi\rangle&= \Psi_{\Pi}^{(a)}(z)|\Pi\rangle \;,\\
e^{(a)}(z)|\Pi\rangle &=\sum_{\sqbox{$a$} \,\in \,\textrm{Add}(\Pi)} 
 \frac{e^{\frac{\pi i }{4} (t_{aa}+|a\rightarrow a|)}\epsilon(\sqbox{$a$},\Pi)\left( \textrm{Res}_{u=h(\sqbox{$a$})}\Psi^{(a)}_{\Pi}(u) \right)^{\frac{1}{2}}}{z-h(\sqbox{$a$})}\,|\Pi+\sqbox{$a$}\rangle \;,\\
f^{(a)}(z)|\Pi\rangle &=\sum_{\sqbox{$a$}\, \in\, \textrm{Rem}(\Pi)}
 \frac{e^{\frac{\pi i }{4} (t_{aa}+|a\rightarrow a|)}\epsilon(\sqbox{$a$},\Pi)\left( \textrm{Res}_{u=h(\sqbox{$a$})}\Psi^{(a)}_{\Pi}(u) \right)^{\frac{1}{2}}}{z-h(\sqbox{$a$})}\,|\Pi-\sqbox{$a$}\rangle \;,
\end{aligned}
\end{tcolorbox} 
\noindent with $\Psi^{(a)}_{\Pi}(u)$ defined in \eqref{eq:ChargeFunctionDef} and $\epsilon$ given in \eqref{eq:epsilonDef} and \eqref{eq:vDef}.

\subsection{Non-vacuum representations and shifted quiver Yangians}
\label{ssec:NonVacuumReps}

The previous three subsections apply to all representations of quiver Yangians. 
For a given $(Q,W)$, a representation ${}^{\sharp}\mathcal{P}_{(Q,W)}$ of $\textsf{Y}(Q,W)$ is specified by its ground state charge function ${}^{\sharp}\psi_{0}^{(a)}(z)$ in \eqref{eq:varphi0Def},
whose information can also be encoded in the framing $\sharp$ of $(Q,W)$ as explained in Sec.\ \ref{sssec:FramedQWandReps}. 
Once the ${}^{\sharp}\psi_{0}^{(a)}(z)$ is given, one can apply the procedure of Sec.\ \ref{sssec:RepresentationConstruction} to generate all the states $\Pi_{d}$ in the representation.

We have seen that there are vacuum-like representations, which are special in that they have the simplest ground state charge functions
\begin{equation}\label{eq:vacuumGS1}
{}^{\sharp}\psi^{(a)}_{\textrm{vac}}(z) = \left(\frac{1}{z}\right)^{\delta_{a,\mathfrak{b}}} \,,
\end{equation}
where $\mathfrak{b}$ is the framed vertex in the canonical framing \eqref{eq:CanonicalQW}.
In this subsection, we will explain how to determine ${}^{\sharp}\psi_{0}^{(a)}(z)$ for non-vacuum representations, namely, how to fix their corresponding framings.

\subsubsection{Constructions of non-vacuum representations from vacuum-like representation}

First of all, the choice of the ground state charge function ${}^{\sharp}\psi_{0}^{(a)}(z)$ is not arbitrary: if there is no relation between the poles and zeros in ${}^{\sharp}\psi_{0}^{(a)}(z)$, the resulting representation would be isomorphic to a trivial tensor product of $\sum_{\mathfrak{b}\in Q_0}|\infty \rightarrow \mathfrak{b}|$ vacuum-like representations, since there would be no interaction among these poles and zeros.
To construct non-vacuum representations that are not of this type, we need to constrain the choice of the rational function ${}^{\sharp}\psi_{0}^{(a)}(z)$.

We generalize the idea of \cite{Galakhov:2021xum} from toric CY$_3$'s to arbitrary quivers.\footnote{
For the case of toric CY$_3$'s, to construct a generic representation for a given quiver Yangian, the starting point is the shape of the crystals ${}^{\sharp}\mathcal{C}$ \cite{Galakhov:2021xum}.
For a generic quiver that doesn't come from a toric CY$_3$, the representations do not have a geometric realization in terms of 3D crystals. 
However, the logic of obtaining the generic non-vacuum representations either as a subspace of the vacuum-like representation or by superimposing positive and negative vacuum-like representations still holds.
} 
There are two ways to view a non-vacuum representation ${}^{\sharp}\mathcal{P}_{(Q,W)}$ in terms of the vacuum-like representations ${}^{(\mathfrak{a})}\mathcal{P}_{(Q,W)}$, corresponding to two equivalent methods of determining its  ground state charge function ${}^{\sharp}\psi^{(a)}(u)$.\footnote{
Although  these two methods  are equivalent, which one is easier to use depends on the representation, see later.
}
Consider both ${}^{(\mathfrak{a})}\mathcal{P}$ and ${}^{\sharp}\mathcal{P}$ as some ``compound" of atoms with fixed internal structure given by the quiver $(Q,W)$.
\begin{enumerate}
\item The first way to view ${}^{\sharp}\mathcal{P}$ is as a superposition of ``positive" and ``negative"  ${}^{(\mathfrak{a})}\mathcal{P}$'s, with possibly different $\mathfrak{a}$.
A (positive) ${}^{(\mathfrak{a})}\mathcal{P}$ corresponds to the a vacuum-like representation, with the $^{\sharp}\psi_0$ taking the form \eqref{eq:vacuumGS1}, namely, it starts with a single atom of color $\mathfrak{a}$. 
A negative ${}^{(\mathfrak{a})}\mathcal{P}$ has exactly the same shape as the positive one, but whose atoms ``cancel" those in the positive one. 
Placing multiple positive and negative ${}^{(\mathfrak{a})}\mathcal{P}$, with possibly different $\mathfrak{a}\in Q_0$, at different locations (which are uniquely specified by the locations of their leading atoms), can then create  ${}^{\sharp}\mathcal{P}$.
\item The second way to view  ${}^{\sharp}\mathcal{P}$ is as a ``sub-compound" of ${}^{(\mathfrak{a})}\mathcal{P}$ for some fixed $\mathfrak{a}$.
Namely, they share the same internal structure, but the boundary of ${}^{\sharp}\mathcal{P}$ lies inside ${}^{(\mathfrak{a})}\mathcal{P}$. 
(For this method to work, $\mathfrak{a}$ needs to be chosen such that the ${}^{\sharp}\mathcal{P}$ to be constructed can indeed be embedded in ${}^{(\mathfrak{a})}\mathcal{P}$.\footnote{
In this computation, once this $\mathfrak{a}$ is chosen, we usually relabel it as $\mathfrak{a}=0$.
})
The ground state of ${}^{\sharp}\mathcal{P}$ then corresponds to some excited state of ${}^{(\mathfrak{a})}\mathcal{P}$.
\end{enumerate}
In both views, the important point is that all the non-vacuum representations ${}^{\sharp}\mathcal{P}$ share the same internal ``chemical" structure as the vacuum-like representations ${}^{(\mathfrak{a})}\mathcal{P}$, determined by $(Q,W)$, but differ in their outside shapes, namely their boundaries.

\medskip

Given a quiver, assume that we have already constructed the vacuum-like representations ${}^{(\mathfrak{a})}\mathcal{P}$ of the quiver Yangian, using the procedure of Sec.\ \ref{sssec:RepresentationConstruction}.
We can then construct an arbitrary non-vacuum representation ${}^{\sharp}\mathcal{P}$ of this quiver Yangian (that satisfies the property that for each state of the representation, the constraints $\partial_{I} W=0$ (if $W$ is present) always eliminate all but one path with the same weight), imitating the procedure of \cite{Galakhov:2021xum}. 
There are two alternative constructions, corresponding to the two views above, for their details see Appx.\ \ref{appssec:Method1} and Appx.\ \ref{appssec:Method2}, respectively, and for a discussion on the relative advantages of the two methods in different situations see Appx.\ \ref{appssec:Comparison2methods}.
The end result is the ground state charge function (defined in \eqref{eq:varphi0Def}) of the non-vacuum representation ${}^{\sharp}\mathcal{P}$, given in \eqref{eq:varphi0M1} for Method 1 and in \eqref{eq:varphi0M22} for Method 2.
The representation given by \eqref{eq:varphi0Def} is a representation of the shifted quiver Yangian with shift (see definition \eqref{eq:ModeExpansions})
\begin{equation}\label{eq:shift}
\mys^{(a)}\equiv\mys^{(a)}_{+}-\mys^{(a)}_{-}\,.
\end{equation}

\subsubsection{Relation to framed quiver}

There is a one-to-one correspondence between a representation given by \eqref{eq:varphi0Def} and a framed quiver.
This is easier to explain in the language of Method 1 in Appx. \ref{appssec:Method1}: each starter or pauser+ of color $a$ corresponds to an arrow from the framing vertex $\infty$ to the framed vertex $\mathfrak{a}$:\footnote{
For the definitions of the starters, the pauser$-$, the pauser$+$, and the stoppers, see \eqref{eq:starterDef}, \eqref{eq:pauser-Def}, \eqref{eq:pauser+Def}, and \eqref{eq:stopperDef}, respectively.
}
\begin{equation}\label{eq:Jinftytoa}
\{\tilde{J}^{\infty\rightarrow \mathfrak{a}}\}=\{\sqbox{$\mathfrak{a}$}^{\textrm{starter}}\} \cup \{\sqbox{$\mathfrak{a}$}^{\textrm{pauser+}}\}    
\end{equation}
with weight 
\begin{equation}
h_{\tilde{J}^{\infty\rightarrow \mathfrak{a}}}=h(\sqbox{$\mathfrak{a}$}^{\textrm{starter/pauser+}} )\,,
\end{equation}
and each stopper or pauser- of color $a$ corresponds to an arrow from the framed vertex $\mathfrak{a}$ back to the framing vertex $\infty$:
\begin{equation}
\{\tilde{I}^{\mathfrak{a}\rightarrow \infty}\}=\{\sqbox{$\mathfrak{a}$}^{\textrm{stopper}}\} \cup \{\sqbox{$\mathfrak{a}$}^{\textrm{pauser-}}\}    
\end{equation}
with weight 
\begin{equation}
h_{\tilde{I}^{\mathfrak{a}\rightarrow \infty}}=-h(\sqbox{$\mathfrak{a}$}^{\textrm{stopper/pauser-}} )\,.
\end{equation}
One also needs to supplement the potential with additional terms to account for the relations among these poles and zeros and the weights of the arrows in the unframed quiver.
In terms of the definition \eqref{eq:BondingFactorDef}, the ground state $ {}^{\sharp}\psi^{(a)}_0(u)$ can also be written as 
\begin{equation}\label{eq:psi0Framing}
   {}^{\sharp}\psi^{(a)}_0(u)=\varphi^{\mathfrak{a}\Leftarrow \infty }(u) \,,
\end{equation}
where we assume $s_{\infty a}=s_{a\infty}=0$.

\subsubsection{Non-vacuum representations as subspaces of vacuum-like representations}
\label{sssec:Subspace}

In both methods, we have implicitly used the fact that as a vector space, a non-vacuum representation ${}^{\sharp}\mathcal{P}_{(Q,W)}$ is a subspace of the vacuum-like representation ${}^{(\mathfrak{a})}\mathcal{P}_{(Q,W)}$ for some fixed $\mathfrak{a}$, which we relabel as $\mathfrak{a}=0$ and accordingly call ${}^{(0)}\mathcal{P}_{(Q,W)}$ the vacuum representation within this discussion. 
Namely, we have 
\begin{equation}\label{eq:Subspace}
{}^{\sharp}\mathcal{P}_{(Q,W)}
\subset {}^{(0)}\mathcal{P}_{(Q,W)}
\simeq J^{(0)}_{(Q,W)}\,.
\end{equation}
This is particularly evident from Method 2 of  constructing ${}^{\sharp}\mathcal{P}$.
Starting from the vacuum representation ${}^{(0)}\mathcal{P}$, Method 2 implements two operations to extract this subspace.
The first one is the ``truncation from above" at step 1, done by placing ``stoppers" at various positions to stop the representation from growing once we reach there.  
The second one is the ``truncation from below" at step 3, done by simply converting an excited state of ${}^{(0)}\mathcal{P}$ into the ground state of ${}^{\sharp}\mathcal{P}$. 

The observation \eqref{eq:Subspace} will be important in showing that a level-$d$ state ${}^{\sharp}\Pi_{d}$ in the non-vacuum representation ${}^{\sharp}\mathcal{P}$ also maps one-to-one to a codimension-$d$ ideal of $J^{(0)}_{(Q,W)}$, see Sec.\ \ref{sssec:IdealNonVacuum}.
In particular, a level-$d$ state ${}^{\sharp}\Pi_{d}$ in ${}^{\sharp}\mathcal{P}$ maps to a level-$(d+d_{\sharp})$ state ${}^{(0)}\Pi_{d+d_{\sharp}}$ in ${}^{(0)}\mathcal{P}$, where $d_{\sharp}$ denotes the number of atoms in ${}^{(0)}\Pi_{\sharp}$, the excited state in ${}^{(0)}\mathcal{P}$ that is used to define the representation ${}^{\sharp}\mathcal{P}$ in Method 2, see \eqref{eq:PsiPi1} and \eqref{eq:Step3Method2} and the discussion about them. 
(Note that it is possible for $d_{\sharp}$ to be infinite.)

Recall from Method 2 that there are two ways to view the ground state of ${}^{\sharp}\mathcal{P}$. 
As the ground state of ${}^{\sharp}\mathcal{P}$, it is the empty set; but as an excited state of the vacuum representation ${}^{(0)}\mathcal{P}$, it contains $d_{\sharp}$ atoms:
\begin{equation}
\begin{cases}
\begin{aligned}
{}^{\sharp}\Pi_{\textrm{g.s.}}
&= \{\varnothing\} \qquad \qquad \qquad \qquad \quad \, \textrm{viewed as ground state of } {}^{\sharp}\mathcal{P}\\
{}^{(0)}\Pi_{\sharp}
&=\{I^{(0)}, I^{(0)}\cdot I^{0\rightarrow a}, \dots\} \qquad  \textrm{viewed as excited state of } {}^{(0)}\mathcal{P}\,.
\end{aligned}
\end{cases}
\end{equation}
Now consider a level $d$ state of ${}^{\sharp}\mathcal{P}$, obtained as
\begin{equation}
|{}^{\sharp}\Pi_{d}\rangle\sim e^{(a_1)}(z_1)\dots e^{(a_{d})}(z_{d})|{}^{\sharp}\Pi_{\textrm{g.s.}}\rangle 
\quad \in {}^{\sharp}\mathcal{P} \,,
\end{equation}
which has $d$ atoms.
Taking its union with the set ${}^{(0)}\Pi_{\sharp}$, we obtain a state ${}^{(0)}\Pi_{d+d_{\sharp}}$ in the (possibly truncated) vacuum representation ${}^{(0)}\mathcal{P}$: 
\begin{equation}\label{eq:2ChargeFunctions}
{}^{(0)}\Pi_{d+d_{\sharp}} 
\equiv  {}^{\sharp}\Pi_{d} 
\cup 
{}^{(0)}\Pi_{\sharp}  
\quad \in 
{}^{(0)}\mathcal{P} \,.
\end{equation}
The two states ${}^{\sharp}\Pi_{d}$ and ${}^{(0)}\Pi_{d+d_{\sharp}}$ share the same charge function, up to a factor coming from removing poles from ${}^{(0)}\Pi_{\sharp}$
\begin{equation}
\begin{aligned}\label{eq:2ChargeFunctions1}
\Psi^{(a)}_{{}^{\sharp}\Pi_{d}}(z)
&={}^{\sharp}\psi_{0}^{(a)}(z)
\cdot \prod_{b\in Q_0} \prod_{\sqbox{$b$}\in {}^{\sharp}\Pi_{d} } \varphi^{a\Leftarrow b}(z-h(\sqbox{$b$})) \\   
&=\psi^{(a)}_0(z)
\cdot \prod_{b\in Q_0} \prod_{\sqbox{$b$} \in {}^{(0)}\Pi_{d+d_{\sharp}} }   \varphi^{a\Leftarrow b}(z-h(\sqbox{$b$}))
\cdot  \prod_{\sqbox{$a$}\in \textrm{Rem}({}^{(0)}\Pi_{\sharp})}(z-h(\sqbox{$a$})) \\
&=\Psi^{(a)}_{{}^{(0)}\Pi_{d+d_{\sharp}}}(z) \cdot  \prod_{\sqbox{$a$}\in \textrm{Rem}({}^{(0)}\Pi_{\sharp})}(z-h(\sqbox{$a$})) 
\end{aligned}
\end{equation}
due to  \eqref{eq:varphi0M22}.
This fact will be crucial in Sec.\ \ref{sssec:IdealNonVacuum}.

\subsection{States in the representation and finite codimensional ideals of \texorpdfstring{$J{(Q,W)}$}{J(Q,W)}}
\label{ssec:PathAlgebralIdeals}

In this subsection, we show that in the representation constructed via the iterative procedure of Sec.\ \ref{sssec:RepresentationConstruction}, each level-$d$ state ${}^{\sharp}\Pi_{d}$ in the representation ${}^{\sharp}\mathcal{P}$ maps one-to-one to a codimension-$d$ ideal of $J^{\sharp}_{(Q,W)}$, defined in \eqref{eq:JsharpDef}.
It is easiest to first show this for the vacuum-like representations and then for all the non-vacuum ones, since the latter can be obtained as superpositions of the former.

\subsubsection{Vacuum-like representations}

We will first prove that for a vacuum-like representation ${}^{(\mathfrak{a})}\mathcal{P}_{(Q,W)}$, each level-$d$ state ${}^{(\mathfrak{a})}\Pi_{d}$ maps one-to-one to a codimension-$d$ ideal of the projected Jacobian algebra $J^{(\mathfrak{a})}_{(Q,W)}$, defined in \eqref{eq:JaDef2}.
(In fact, for a vacuum-like representation, we have a stronger statement that  each level-$d$ state ${}^{(\mathfrak{a})}\Pi_{d}$ maps one-to-one to a codimension-$d$ ideal of the full Jacobian algebra $J_{(Q,W)}$, defined in \eqref{eq:JacobianAlgebraDef}.\footnote{
The stronger claim implies the weaker one because ${}^{(\mathfrak{a})}\mathcal{P}_{(Q,W)} \simeq J^{(a)}_{(Q,W)}\subset J_{(Q,W)}$.
})
The proof takes two steps:
\begin{enumerate}
\item We first show that the prescription, in particular eq.\ \eqref{eq:AddPi}, of the representation-generating procedure ensures an Adding Rule:
\begin{equation}\label{eq:AddingRule}
\begin{aligned}
\textrm{Adding Rule: } \quad &\textrm{In order to add a new atom $\sqbox{$a$}$ to a state $\Pi$, all its}\\
&\textrm{precursor atoms $\sqbox{$b$}$ must be already present in $\Pi$.}
\end{aligned}
\end{equation}
where the  definition of the ``precursor" will be given momentarily.
We emphasize that this Adding Rule is automatically enforced by \eqref{eq:AddPi}, since otherwise the corresponding adding pole would not be present in $\Psi^{(a)}_{\Pi}(u)$.
In proving this, we will also see that the adding pole for each atom is always a simple pole:
\begin{equation}\label{eq:SimplePole}
\begin{aligned}
\textrm{Rule of simple pole: } \quad &\textrm{Each addable atom $\sqbox{$a$}$ to the state $\Pi$ }\\ 
&\textrm{corresponds to a \textit{simple} pole in $\Psi^{(a)}_{\Pi}(u)$} \,. 
\end{aligned}
\end{equation}
We emphasize that this is not to say that the charge function $\Psi^{(a)}_{\Pi}(u)$ has only simple poles, but merely that each atom gives rise to a first-order pole; and an $n^{\textrm{th}}$-order pole in  $\Psi^{(a)}_{\Pi}(u)$ corresponds to $n$ different atoms with the same weight.\footnote{
	For such an example, see the $m$-loop quiver in Sec.\ \ref{ssec:mloop}.
} 
Similar to the Adding Rule, this rule is also automatically enforced by \eqref{eq:AddPi}.

\item We then show that the Adding Rule implies that each state generated this way corresponds to a finite codimensional ideal of the Jacobian algebra $J_{(Q,W)}$.
\end{enumerate}

To prove the Adding Rule \eqref{eq:AddingRule}, let us first explain the definition of the precursor. 
An atom $\sqbox{$a$}$ corresponds to a path $\in J^{(0)}_{(Q,W)}$, namely up to the path equivalence imposed by the potential $W$.
Now, if we do not impose the path equivalence, then these paths are different elements in the path algebra $\mathbb{C}Q$.
Suppose there are $N\geq 1$ paths $\in \mathbb{C}Q$ that correspond to the same atom $\sqbox{$a$}$, with  $N-1$ path equivalence conditions among them.
Among these N paths, there can be $n\leq N$ distinct atoms $\sqboxs{$b_i$}$ right before $\sqbox{$a$}$:\footnote{We emphasize that these $n$ distinct atoms need not have distinct colors, namely, the color $b_i$ might coincide.}
\begin{equation}\label{eq:npaths}
\mathfrak{p}^{0\rightarrow b_i}_{i,j} \cdot I^{b_i\rightarrow a}    \qquad i=1,\dots,n\,, \quad 
j=1,\dots, N_i
\end{equation} 
where we have explicitly written out the last arrow of each such path, and $N_i$ is the number of paths that lead to $\sqboxs{$b_i$}$, with $\sum^n_{i=1} N_i=N$.
The $n$ atoms that sit right before $\sqbox{$a$}$ on these $n$ paths are:
\begin{equation}\label{eq:precursors}
\sqboxs{$b_i$}=\mathfrak{p}^{0\rightarrow b_i}_{i,j} \qquad i=1,\dots,n\,, \quad \forall j=1,\dots, N_i\,,
\end{equation}
and they are defined as the \textit{precursors} of the atom $\sqbox{$a$}$ in the representation ${}^{\sharp}\mathcal{P}_{(Q,W)}\simeq J^{\sharp}_{(Q,W)}$.
They satisfy
\begin{equation}\label{eq:btoa}
\sqboxs{$b_i$}\cdot I^{b_i\rightarrow a} =\sqbox{$a$} \qquad \Longrightarrow \qquad h(\sqboxs{$b_i$})+h(I^{b_i\rightarrow a}) =h(\sqbox{$a$})\,,    \qquad i=1,\dots,n\,.
\end{equation}

\medskip

Next, we explain how the path equivalence is implemented. 
Let us first specialize to a vacuum-like representation ${}^{(\mathfrak{a})}\mathcal{P}_{(Q,W)} \simeq J^{(a)}_{(Q,W)}$.
Let us relabel $\mathfrak{a}=0$ for this discussion and later for the proof of the Adding Rule for vacuum-like representations. 

First of all, it is the path equivalence among the $n$ paths leading from the $n$ distinct precursors $\sqbox{$b$}$ to the atom $\sqbox{$a$}$ in question, rather than any path equivalence among the paths that lead to any of the precursors $\sqbox{$b_i$}$, that plays a role in the argument. 
Therefore, we should focus on the number of the precursors $n$, rather than the total number $N$ of different paths that lead to the atom $\sqbox{$a$}$.

When $n=1$, there is only one precursor $\sqbox{$b$}$:
%the atom $\sqbox{$a$}$ corresponds to a unique path $\in J^{(0)}_{(Q,W)}$, namely, :
\begin{equation}\label{eq:pathequivalenceFign1}
\begin{array}{c}
\begin{tikzpicture}[scale=0.9]
%\node%[state]  
%[regular polygon, regular polygon sides=4, draw=blue!50, very thick, fill=blue!10] (a1) at (0,0)  {$1$};
\node[vertex,minimum size=0.5mm,font=\footnotesize] (zero) at (0, 1.5)  {$0$};
%\node[vertex,minimum size=0.5mm,font=\footnotesize] (ci) at (0,0)  {$c_{i}$};
\node[vertex,minimum size=0.5mm,font=\footnotesize] (b) at (0,0)  {$b$};
%\node[vertex,minimum size=0.5mm,font=\footnotesize] (bj) at (2,-1)  {$b_{i+1}$};
\node[vertex,minimum size=0.5mm,font=\footnotesize] (a) at (0,-1)  {$a$};
%\draw[fill=white] (a1) (0,0) circle (0.5) ;
%\node (f) at (-2,0) {$\infty$};
%\draw[thick, ->] (f) -- (a1);
%\path[thick, -{Latex[length=2mm, width=1.5mm]},every loop/.append style=-{Latex[length=2mm, width=1.5mm]}] 
\draw[->,very thick] (zero) --  (b);
\draw[->] (b) --  (a) 
;
\end{tikzpicture}
\end{array} 
\end{equation}
with a single arrow from $\sqbox{$b$}$ to  $\sqbox{$a$}$, and there is no relevant path equivalence involved.
(Here and in the following, the thin arrows denote single arrows $I\in Q_1$ whereas the thick ones denote either a path that might consist of more than one arrow or multiple paths --- each might contain more than one arrow --- that are all path equivalent.)
%The atom $\sqbox{$b$}$ that is immediately before $\sqbox{$a$}$ is its unique precursor.

Now, suppose there are $n\geq 2$ precursors $\sqboxs{$b_i$}$ leading to the atom $\sqbox{$a$}$. 
For each $\sqboxs{$b_i$}$, there could be multiple paths going from the origin to it, labeled by $j=1,\dots, N_i$ in \eqref{eq:npaths} and \eqref{eq:precursors}.
First of all, since the focus is on atom $\sqboxs{$a$}$ now, we do not need to consider the path equivalence among these $N_i$ paths, and can instead view them as one path. 
Namely, effectively we are considering $n$ (instead of $N$) paths  $\sqboxs{$0$}\mathbf{\rightarrow} \sqboxs{$b_i$} \rightarrow\sqboxs{$a$}$, with $n-1$ path equivalence conditions among them.
Then, we can always label the $\sqboxs{$b$}$'s and $\sqboxs{$c$}$'s such that any two neighboring paths, with $\sqboxs{$b_i$}$ and $\sqboxss{$b_{i+1}$}$ as precursors of $\sqbox{$a$}$, share a common atom $\sqboxs{$c_{i}$}$ --- we take the one that is closest to $\sqbox{$a$}$:
\begin{equation}\label{eq:2ctobtoa}
I^{(0)}\cdots \mathfrak{p}^{c_{i}\rightarrow b_i} \cdot I^{b_i\rightarrow a}  \qquad \textrm{and} \qquad I^{(0)}\cdots \mathfrak{p}^{c_{i}\rightarrow b_{i+1}} \cdot I^{b_{i+1}\rightarrow a} \qquad i=1,\dots,n-1\,.
\end{equation}
We emphasize that these atoms $\sqboxs{$c_i$}$ need not be dinstinct (let alone their colors $c_i\in Q_0$) --- it is possible for them to be all identical.
Then, since the $i^{\textrm{th}}$ and $(i+1)^{\textrm{th}}$ paths are equivalent, there must exist an arrow $I^{a\rightarrow c_i}\in Q_1$ in order to implement this path equivalence:
\begin{equation}\label{eq:pathequivalenceFig}
\begin{array}{c}
\begin{tikzpicture}[scale=0.8]
%\node%[state]  
%[regular polygon, regular polygon sides=4, draw=blue!50, very thick, fill=blue!10] (a1) at (0,0)  {$1$};
\node[vertex,minimum size=0.5mm,font=\footnotesize] (zero) at (0,1.5)  {$0$};
\node[vertex,minimum size=0.5mm,font=\footnotesize] (ci) at (0,0)  {$c_{i}$};
\node[vertex,minimum size=0.5mm,font=\footnotesize] (bi) at (-2,-1)  {$b_i$};
\node[vertex,minimum size=0.1mm,font=\tiny] (bj) at (2,-1)  {$b_{i+1}$};
\node[vertex,minimum size=0.5mm,font=\footnotesize] (a) at (0,-2)  {$a$};
%\draw[fill=white] (a1) (0,0) circle (0.5) ;
%\node (f) at (-2,0) {$\infty$};
%\draw[thick, ->] (f) -- (a1);
%\path[thick, -{Latex[length=2mm, width=1.5mm]},every loop/.append style=-{Latex[length=2mm, width=1.5mm]}] 
\draw[->,very thick] (zero) --  (ci);
\draw[->,very thick] (ci) --  (bi); %{$I$} 
\draw[->,very thick] (ci) --  (bj) ;
\draw[->] (bi) --  (a) ;
\draw[->] (bj) --  (a) ;
\draw[->] (a) --  (ci) 
;
\end{tikzpicture}
\end{array} 
\end{equation}
Namely, the potential $W$ needs to contain
\begin{equation}
W\ni \mathfrak{p}^{c_{i}\rightarrow b_i}\cdot I^{b_i\rightarrow a}\cdot I^{a\rightarrow c_{i}}- \mathfrak{p}^{c_{i}\rightarrow b_{i+1}}\cdot I^{b_{i+1}\rightarrow a}\cdot I^{a\rightarrow c_{i}}
\end{equation}
in order for the relation 
\begin{equation}
0=\partial_{I^{a\rightarrow c_{i}}}W=  \mathfrak{p}^{c_{i}\rightarrow b_i}\cdot I^{b_i\rightarrow a}- \mathfrak{p}^{c_{i}\rightarrow b_{i+1}}\cdot I^{b_{i+1}\rightarrow a}
\end{equation}
to exist and enforce the path equivalence between the two paths in \eqref{eq:2ctobtoa}.
The relation also implies
\begin{equation}\label{eq:looprelationabc}
h(\sqboxs{$c_i$})-h(I^{a\rightarrow c_i})
=h(\sqboxs{$b_i$})+h(I^{b_{i}\rightarrow a})
=h(\sqboxss{$b_{i+1}$})+h(I^{b_{i+1}\rightarrow a}) \,.
\end{equation}
Note that although the $n-1$ colors $c_i$ need not be distinct, however, these $n-1$ arrows $I^{a\rightarrow c_i}$ are distinct.
In summary, the existence of $n$ equivalent paths, which can be considered as describing one atom $\sqbox{$a$}$, means that there must exist $n-1$ arrows from $a$ to $c_i\in Q_0$, with $i=1,\dots,n-1$.

\medskip

Now we are ready to prove the Adding Rule \eqref{eq:AddingRule} and the fact that each atom $\sqbox{$a$}$ that is addable to a state $\Pi$ must correspond to a \textit{simple} pole in $\Psi^{(a)}_{\Pi}(u)$.
Recall that the case of $n=1$ is different from the case of $n\geq 2$ in that the former does not require path equivalence. 
We will first explain the $n=1$ case, which is the easiest, and then prove the general $n\geq 2$ case.
We will also illustrate the cases with $n=2,3$ in Appx.\ \ref{appsec:AddingRulen=23}.

\medskip

\noindent $\boldsymbol{n=1}$. 
When there is only one path in $\mathbb{C}Q$ that corresponds to $\sqbox{$a$}$, i.e., \eqref{eq:npaths} with $n=1$ as shown in \eqref{eq:pathequivalenceFig}, obviously one first has to add the unique precursor $\sqbox{$b$}$ before one can add $\sqbox{$a$}$. 
Hence, the Adding Rule \eqref{eq:AddingRule} is trivially true. 
It is also easy to see why the adding pole is simple in this case.
In the representation-generation procedure, the adding pole that one uses to add $\sqbox{$a$}$ is from 
\begin{equation}
\Psi^{(a)}_{\Pi}(u) \ni \varphi^{a\Leftarrow b} (u-h(\sqbox{$b$}))\ni \frac{1}{u-h(I^{b\rightarrow a})-h(\sqbox{$b$})}= \frac{1}{u-h(\sqbox{$a$})}\,,    
\end{equation}
where $\Pi$ is any state that contains the path \eqref{eq:npaths} (with $n=1$) leading to the atom $\sqbox{$a$}$, but not the atom $\sqbox{$a$}$ itself; and in the last step we have used \eqref{eq:btoa}.\footnote{
Note that this pole $h(\sqbox{$a$})$ could not have come from other previous atoms in the path, otherwise one would have already added $\sqbox{$a$}$ in an earlier step in the iterative procedure. 
}

\medskip

\noindent \textbf{General $\bm{n}$}. 
The diagram that shows $n$ paths $\in \mathbb{C}Q$ leading to the atom $\sqbox{$a$}$ consists of multiple subgraphs \eqref{eq:pathequivalenceFig} as building blocks. 
In assembling these building blocks, the crucial information is whether some of the atoms $\sqboxs{$c_i$}$'s are identical. 
Assuming that among these $n-1$ $\sqboxs{$c_i$}$, there are $m\leq n-1$ different ones, then the final graph will only show these $m$ $\sqboxs{$c_i$}$'s.
Let us label the number of paths from $\sqboxs{$c_i$}$ to the different
$\sqboxs{$b_j$}$'s by $t_i$;  the set of $t_i$, with $i=1,\dots,m$, satisfy
\begin{equation}
t_i \geq 2 \qquad \textrm{and} \qquad \left(\sum^{m}_{i=1} t_i \right) -(m-1)=n   \,.
\end{equation} 
When all $t_i=2$, then $m=n-1$ and all the $\sqboxs{$c_i$}$ are distinct.
Finally, since the $t_i$ paths from $\sqboxs{$c_i$}$ to some $\sqboxs{$b_j$}$ and then to $\sqbox{$a$}$ are all equivalent, there must be $t_i-1$ arrows from $a\rightarrow c_i\in Q_i$, $I^{a\rightarrow c_i}_{k}$, with $k=1,\dots,t_1-1$, that account for these equivalences --- note that there could be more arrows from $a$ to $c_i$ in the quiver diagram, but they are not involved in these path equivalences. 

After this preparation, let us now show that if all the atoms $\sqboxs{$b_j$}$, with $j=1,\dots, n$, and all the $\sqboxs{$c_i$}$, with $i=1,\dots,m$, are already inside $\Pi$, then there is precisely a simple pole that can serve as the adding pole of $\sqbox{$a$}$.
(Note that all the $\sqboxs{$c_i$}$ are necessarily in $\Pi$ if all the $\sqboxs{$b_j$}$ are in.)
Each $\sqboxs{$b_j$}$, with $j=1,\dots, n$, if in $\Pi$, would contribute a factor
\begin{equation}\label{eq:polefrombi}
\Psi^{(a)}_{\Pi}(u) \ni \varphi^{a\Leftarrow b_j}(u- h(\sqboxs{$b_j$}) )     \ni \frac{1}{u -h(I^{b_j\rightarrow a})-h(\sqboxs{$b_j$}) } =\frac{1}{u-h(\sqbox{$a$})  }\,, \quad j=1,\dots, n\,.
\end{equation} 
On the other hand, each  $\sqboxs{$c_i$}$, with $i=1,\dots, m$, if in $\Pi$, would contribute a factor to $\Psi^{(a)}_{\Pi}(u)$
\begin{equation}\label{eq:zerofromc}
\Psi^{(a)}_{\Pi}(u) \ni \varphi^{a\Leftarrow c_i}(u- h(\sqboxs{$c_i$}))     \ni \prod^{t_i-1}_{k=1} (u +h(I^{a\rightarrow c_i}_k) -h(\sqboxs{$c_i$}))=(u-h(\sqbox{$a$}))^{t_i-1}\,,\quad i=1,\dots, m\,.
\end{equation}
The net number of poles at $h(\sqbox{$a$})$ is then
\begin{equation}
n- \sum^m_{i=1}(t_i-1)=    1\,.
\end{equation}
Namely, with all the $n$ $\sqbox{$b$}$'s and $m$ $\sqbox{$c$}$'s present, we have precisely a simple pole to add $\sqbox{$a$}$.
Now we just need to show that by removing any of the $\sqboxs{$b_i$}$, the degree of the pole at $h(\sqbox{$a$})$ can only decrease.

First, suppose we try to remove $\sqboxs{$b_j$}$ without doing anything else, this would change the degree by $-1$ --- the pole at $h(\sqbox{$a$})$ would disappear.
Then, one might hope to also remove some of the $\sqbox{$c$}$'s to remove their associated zeros.
For example, once we remove $\sqboxs{$b_j$}$, we are allowed to remove a $\sqboxs{$c_i$}$ that is a precursor of  $\sqboxs{$b_j$}$, which would change the degree by $+(t_i+1)$.
However, once we remove $\sqboxs{$c_i$}$, it would also remove all the other $\sqbox{$b$}$'s that are connected to $\sqboxs{$c_i$}$ down the path, and there are $t_i-1$ such $\sqbox{$b$}$'s, each of which changes the degree of the pole by $-1$.
In total, the change of the degree of the pole is
\begin{equation}
(-1)+(t_i-1)+ (t_i-1)\times (-1)=-1   \,. 
\end{equation}

Therefore, if we remove some $\sqboxs{$b_j$}$ from the $\Pi$ configuration that has all the $n$ $\sqbox{$b$}$'s and $m$ $\sqbox{$c$}$'s, the pole at $h(\sqbox{$a$})$ would necessarily disappear. 
To summarize, in order to add $\sqbox{$a$}$ to $\Pi$, all its precursors have to be present, and the adding pole is a simple\footnote{We remind that it can happen that a different atom gives rise to different simple pole but with the same value, in which case the pole need to be resolved, see Appendix \ref{appsec:poleresolution}.} pole in $\Psi^{(a)}_{\Pi}(u)$.
This completes our proof for the Adding Rule \eqref{eq:AddingRule} and the Rule of Simple Pole \eqref{eq:SimplePole} for the vacuum-like representations. 
To illustrate the proof, we also provide examples with $n=2,3$ in Appx.\ \ref{appsec:AddingRulen=23}

\bigskip

The Adding Rule \eqref{eq:AddingRule} implies that each level-$d$ state ${}^{(\mathfrak{a})}\Pi_{d}$ in a  vacuum-like representation ${}^{(\mathfrak{a})}\mathcal{P}_{(Q,W)}\simeq J^{(a)}_{(Q,W)}$ maps one-to-one to a codimension-$d$ ideal of the Jacobian algebra $J^{(a)}_{(Q,W)}$.
Here an ideal $\mathcal{I}\subset J^{(a)}_{(Q,W)}$ is defined as a subset of $J^{(a)}_{(Q,W)}$ that satisfies
\begin{equation}
(I^{(a)}\cdot\mathfrak{p}^{a\rightarrow b} )\in \mathcal{I} 
\quad \textrm{and} \quad 
\forall \,\, \mathfrak{p}^{b\rightarrow c} \in J_{(Q,W)}
\qquad \Longrightarrow \quad (I^{(a)}\cdot\mathfrak{p}^{a\rightarrow b})\cdot \mathfrak{p}^{b\rightarrow c} \in \mathcal{I} \,.
\end{equation}
Consider the complement of ${}^{(\mathfrak{a})}\Pi_d$ in  $J^{(a)}_{(Q,W)}$:
\begin{equation}\label{eq:PiComplementVac}
\overline{{}^{(\mathfrak{a})}\Pi_{d}}\equiv J^{(a)}_{(Q,W)}\backslash {}^{(\mathfrak{a})}\Pi_d \,.
\end{equation}
One can see that $\overline{{}^{(\mathfrak{a})}\Pi_{d}}\subset J^{(a)}_{(Q,W)}$ is an ideal of $J^{(a)}_{(Q,W)}$, using the Adding Rule \eqref{eq:AddingRule}. 
For any $(I^{(a)}\cdot\mathfrak{p}^{a\rightarrow b})\in \overline{{}^{(a)}\Pi_{d}}$ and any $\mathfrak{q}^{b\rightarrow c}\in J_{(Q,W)}$, we have
\begin{align}
(I^{(a)}\cdot \mathfrak{p}^{a\rightarrow b})
\in \overline{{}^{(a)}\Pi_{d}}
&\,\, \Rightarrow \,\,
(I^{(a)}\cdot\mathfrak{p}^{a\rightarrow b}) \notin {}^{(a)}\Pi_{d}\\
&\,\, \Rightarrow \,\,
(I^{(a)}\cdot\mathfrak{p}^{a\rightarrow b})\cdot \mathfrak{p}^{b\rightarrow c} \notin {}^{(a)}\Pi_{d}
\,\, \Rightarrow \,\, 
(I^{(a)}\cdot\mathfrak{p}^{a\rightarrow b})\cdot \mathfrak{p}^{b\rightarrow c} \in \overline{{}^{(a)}\Pi_{d}}\,,\nonumber
\end{align}
where in the second arrow we have used the Adding Rule.
Namely, we can view $(I^{(a)}\cdot\mathfrak{p}^{a\rightarrow b})\cdot \mathfrak{p}^{b\rightarrow c}$ as an atom to be added and $(I^{(a)}\cdot\mathfrak{p}^{a\rightarrow b})$ as one of its precursors (or a precursor of a precursor, or even further upstream), then if the latter is not present in ${}^{(a)}\Pi_{d}$, the former cannot be either, due to the Adding Rule. 
In other words, if a path $I^{(a)}\cdot\mathfrak{p}^{a\rightarrow b}$ is not inside ${}^{(a)}\Pi_{d}$, then extending it by multiplying it by an arbitrary path $\mathfrak{q}^{b\rightarrow c}\in J_{(Q,W)}$ still will not make it belong to ${}^{(a)}\Pi_{d}$.\footnote{
This is a generalization of the argument for the toric CY$_3$ case considered in \cite{Szendroi,Mozgovoy:2008fd}.
}
Therefore, each level-$d$ state ${}^{(\mathfrak{a})}\Pi_{d}$ in a vacuum-like representation ${}^{(\mathfrak{a})}\mathcal{P}_{(Q,W)}\simeq J^{(a)}_{(Q,W)}$ maps one-to-one to a codimension-$d$ ideal $\overline{{}^{(\mathfrak{a})}\Pi_{d}}\equiv J^{(a)}_{(Q,W)}\backslash {}^{(\mathfrak{a})}\Pi_d$ of the projected Jacobian algebra $J^{(a)}_{(Q,W)}$.

Finally, we mention that for the vacuum-like representations, a stronger argument holds:
each level-$d$ state ${}^{(\mathfrak{a})}\Pi_{d}$ in a  vacuum-like representation ${}^{(\mathfrak{a})}\mathcal{P}_{(Q,W)}\simeq J^{(a)}_{(Q,W)}$ maps one-to-one to a codimension-$d$ ideal $\overline{{}^{(\mathfrak{a})}\Pi_{d}}\equiv J_{(Q,W)}\backslash {}^{(\mathfrak{a})}\Pi_d$ of the full Jacobian algebra $J_{(Q,W)}$.
To see this, one can just rerun the argument above almost verbatim, only replacing $J^{(\mathfrak{a})}_{(Q,W)}$ by $J_{(Q,W)}$ in \eqref{eq:PiComplementVac}.

\subsubsection{Non-vacuum representation}
\label{sssec:IdealNonVacuum}
We have just shown that for a vacuum-like  representation ${}^{(\mathfrak{a})}\mathcal{P}_{(Q,W)}\simeq J^{(a)}_{(Q,W)}$, there is an Adding Rule 
\eqref{eq:AddingRule}, which  implies that for each of its level-$d$ state ${}^{(\mathfrak{a})}\Pi_{d}$, its complement $\overline{{}^{(\mathfrak{a})}\Pi_{d}}\equiv J^{(a)}_{(Q,W)}\backslash {}^{(\mathfrak{a})}\Pi_d$ is a codimension-$d$ ideal  of the projected Jacobian algebra $J^{(a)}_{(Q,W)}$. 
A corollary of the Adding Rule is the Rule of Simple Pole \eqref{eq:SimplePole}.
It is easy to generalize these statements to the non-vacuum representations, by combining the argument for the vacuum-like representations and the fact that a non-vacuum representation ${}^{\sharp}\Pi_{d}$ is a subspace of some vacuum-like representation, labeled by ${}^{(0)}\mathcal{P}$, see \eqref{eq:Subspace}.

\medskip

Let us first show that the Adding Rule \eqref{eq:AddingRule} still applies to a non-vacuum representation ${}^{\sharp}\mathcal{P}$. 
Suppose we want to add an atom $\sqbox{$a$}$ to a level-$d$ state ${}^{\sharp}\Pi_{d}$ in ${}^{\sharp}\mathcal{P}$. 
Recall from Sec.\ \ref{sssec:Subspace} that ${}^{\sharp}\Pi_{d}$ corresponds to the state ${}^{(0)}\Pi_{d+d_{\sharp}}$ in  ${}^{(0)}\mathcal{P}$:
\begin{equation}
{}^{\sharp}\Pi_{d} \in {}^{\sharp}\mathcal{P}   
\quad \longrightarrow \quad 
{}^{(0)}\Pi_{d+d_{\sharp}}={}^{\sharp}\Pi_{d}\cup {}^{(0)}\Pi_{\sharp}\in {}^{(0)}\mathcal{P} 
\end{equation}
and the two states share the same charge function up to a factor coming from removing poles from ${}^{(0)}\Pi_{\sharp}$, see \eqref{eq:2ChargeFunctions1}, which means that we can check the charge function of ${}^{(0)}\Pi_{d+d_{\sharp}}$ to see if it contains the adding pole for $\sqbox{$a$}$ in order to add $\sqbox{$a$}$ to ${}^{\sharp}\Pi_{d}$, namely:
\begin{equation}\label{eq:2adding}
\sqbox{$a$}\in \textrm{Add}({}^{(0)}\Pi_{d+d_{\sharp}})   \quad \Longleftrightarrow \quad  \sqbox{$a$}\in \textrm{Add}({}^{\sharp}\Pi_{d})\,.
\end{equation}
The condition for $\sqbox{$a$}\in \textrm{Add}({}^{(0)}\Pi_{d+d_{\sharp}})$ is that all the precursors of $\sqbox{$a$}$ in ${}^{(0)}\mathcal{P}$:  %$\{\sqbox{$b$_i}\}$, 
\begin{equation}\label{eq:PreaDef}
{}^{(0)}\textrm{Pre}(\sqbox{$a$})   \equiv \left\{\sqbox{$b$}\,\big|\,\sqbox{$b$} \textrm{ is a precursor of }\sqbox{$a$} \textrm{ in } {}^{(0)}\mathcal{P}  \right\}
\end{equation}
have to be already in ${}^{(0)}\Pi_{d+d_{\sharp}}$, namely
\begin{equation}\label{eq:Adding0}
{}^{(0)}\textrm{Pre}(\sqbox{$a$})\subset {}^{(0)}\Pi_{d+d_{\sharp}} \quad \Longleftrightarrow \quad \sqbox{$a$}\in \textrm{Add}({}^{(0)}\Pi_{d+d_{\sharp}})  \,.
\end{equation}
These precursors can be divided into two classes, those in ${}^{(0)}\Pi_{\sharp}$ and those that are not, and only the second class are those precursors of $\sqbox{$a$}$ that are also in the representation ${}^{\sharp}\mathcal{P}$:
\begin{equation}\label{eq:sharpPreaDef}
\begin{aligned}
{}^{\sharp}\textrm{Pre}(\sqbox{$a$})   
&\equiv \left\{\textrm{precursors of }\sqbox{$a$} \textrm{ in } {}^{\sharp}\mathcal{P} \right\}\\
&\equiv \left\{\sqbox{$b$}\,\big|\,\sqbox{$b$} \textrm{ is a precursor of }\sqbox{$a$} \textrm{ in } {}^{(0)}\mathcal{P} \,\, \& \,\, \sqbox{$b$}  \notin {}^{(0)}\Pi_{\sharp} \right\}  \,.
\end{aligned}\end{equation}
The following two conditions are identical:
\begin{equation}\label{eq:2Pre}
\begin{aligned}
& {}^{(0)}\textrm{Pre}(\sqbox{$a$})\subset {}^{(0)}\Pi_{d+d_{\sharp}} \quad \Longleftrightarrow \quad {}^{\sharp}\textrm{Pre}(\sqbox{$a$})\subset {}^{\sharp}\Pi_{d}    \,.
\end{aligned}
\end{equation}
Combining \eqref{eq:2adding},  \eqref{eq:Adding0}, and \eqref{eq:2Pre}, we see that the Adding Rule for the vacuum-like representations implies the Adding Rule for the representation ${}^{\sharp}\mathcal{P}$ immediately:
\begin{flalign}\label{eq:sharpAddingRule}
& {}^{\sharp}\textrm{Pre}(\sqbox{$a$})\subset {}^{\sharp}\Pi_{d} 
\,\,\, \Leftrightarrow \,\,\, 
{}^{(0)}\textrm{Pre}(\sqbox{$a$})\subset {}^{(0)}\Pi_{d+d_{\sharp}} 
\,\,\, \Leftrightarrow\,\,\, \sqbox{$a$}\in \textrm{Add}({}^{(0)}\Pi_{d+d_{\sharp}})  
\,\,\, \Leftrightarrow \,\,\,  \sqbox{$a$}\in \textrm{Add}({}^{\sharp}\Pi_{d}) \,.
\end{flalign}

Similar to the case for the vacuum-like representations, the Adding Rule \eqref{eq:sharpAddingRule} implies that each level-$d$ state ${}^{\sharp}\Pi_{d}$ in the non-vacuum representation ${}^{\sharp}\mathcal{P}_{(Q,W)}$ maps one-to-one to a codimension-$d$ ideal $J^{\sharp}_{(Q,W)}\backslash {}^{\sharp}\Pi_{d}$ of  $J^{\sharp}_{(Q,W)}$, defined in \eqref{eq:JsharpDef}.
Here an ideal $\mathcal{I}\subset J^{\sharp}_{(Q,W)}$ is defined as a subset of $J^{\sharp}_{(Q,W)}$ that satisfies
\begin{equation}
(I^{(0)}\cdot\mathfrak{p}^{a\rightarrow b} )\in \mathcal{I} 
\quad \textrm{and} \quad 
\forall \,\, \mathfrak{p}^{b\rightarrow c} \in J_{(Q,W)}
\qquad \Longrightarrow \quad (I^{(0)}\cdot\mathfrak{p}^{a\rightarrow b})\cdot \mathfrak{p}^{b\rightarrow c} \in \mathcal{I} \,.
\end{equation}
Consider the complement of ${}^{\sharp}\Pi_d$ in  $J^{\sharp}_{(Q,W)}$:
\begin{equation}\label{eq:PiComplementNonVac}
\overline{{}^{\sharp}\Pi_{d}}\equiv J^{\sharp}_{(Q,W)}\backslash {}^{\sharp}\Pi_d \,.
\end{equation}
One can see that $\overline{{}^{\sharp}\Pi_{d}}\subset J^{\sharp}_{(Q,W)}$ is an ideal of $J^{\sharp}_{(Q,W)}$, using the Adding Rule for the representation ${}^{\sharp}\mathcal{P}$ \eqref{eq:sharpAddingRule}. 
For any $(I^{(0)}\cdot \mathfrak{p}^{a\rightarrow b})\in \overline{{}^{\sharp}\Pi_{d}}$ and any $\mathfrak{q}^{c\rightarrow d}\in J_{(Q,W)}$, we have
\begin{align}
(I^{(0)}\cdot \mathfrak{p}^{a\rightarrow b})
\in \overline{{}^{\sharp}\Pi_{d}}
&\,\, \Rightarrow \,\,
(I^{(0)}\cdot\mathfrak{p}^{a\rightarrow b}) \notin {}^{\sharp}\Pi_{d}\\
&\,\, \Rightarrow \,\,
(I^{(0)}\cdot\mathfrak{p}^{a\rightarrow b})\cdot \mathfrak{p}^{b\rightarrow c} \notin {}^{\sharp}\Pi_{d}
\,\, \Rightarrow \,\, 
(I^{(0)}\cdot\mathfrak{p}^{a\rightarrow b})\cdot \mathfrak{p}^{b\rightarrow c} \in \overline{{}^{\sharp}\Pi_{d}}\,,\nonumber
\end{align}
where in the second arrow we have used the Adding Rule  \eqref{eq:sharpAddingRule}. 

Finally, note that unlike the vacuum-like representations,  the analogous stronger statement doesn't hold for the non-vacuum representation.
Namely, for a level-$d$ state ${}^{\sharp}\Pi_{d}$ in a  non-vacuum representation ${}^{\sharp}\mathcal{P}_{(Q,W)}\simeq J^{\sharp}_{(Q,W)}$, its complement in $J_{(Q,W)}$, 
$J_{(Q,W)}\backslash {}^{\sharp}\Pi_d$, is not an ideal of the  Jacobian algebra $J_{(Q,W)}$.

\subsection{Representations in terms of modes}
\label{ssec:RepMode}

So far we have described the quiver Yangian representation as spanned by finite codimensional ideals of the Jacobian algebra.
Now we show how to construct quiver Yangian representations by applying repeatedly the raising operator on the ground state $|\varnothing\rangle_{\sharp}$.
The procedure is as follows:
\begin{enumerate}
	\item Given a framing $\sharp$, with $N^{(a)}=|\infty\rightarrow a|\geq 0$ arrows from the framing node $\infty$ to vertex $a$, the ground state $|\varnothing\rangle_{\sharp}$ satisfies\footnote{If there are arrows from vertex $a$ back to $\infty$, they would also impose constraints on the ground state $|\varnothing\rangle_{\sharp}$.  }
	\begin{equation}\label{eq:GSdef}
		e^{(a)}_{n\geq N^{(a) }}|\varnothing\rangle_{\sharp}=0 \,.
	\end{equation} 
	For the NCDT chamber, which is the focus of the current paper, the ground state $|\varnothing\rangle_{\sharp}$ is defined by
\begin{equation}\label{eq:GSdefNCDT}
e^{(\mathfrak{a})}_{n\geq 1}|\varnothing\rangle_{\sharp}=0
\,, \quad
\qquad e^{(b)}_{n\geq 0}|\varnothing\rangle_{\sharp}=0
	\end{equation} 
	where $\mathfrak{a}$ is the framed vertex and $b$ are the remaining ones. 
	When there are also arrows from $a$ back to the framing node $\infty$, there are loops involving $\infty$. 
	A loop of length $\ell$ involving  $\infty$ corresponds to a condition on $|\varnothing\rangle_{\sharp}$ with order-$(\ell-1)$ relation applying on $|\varnothing\rangle_{\sharp}$.
	Together with the condition \eqref{eq:GSdef} imposed by the arrows from $\infty$ to $a$, they give the definition of the ground state $|\varnothing\rangle_{\sharp}$.

	\item Now one can construct the representation by applying the modes $e^{(a)}_n$'s on $|\varnothing\rangle_{\sharp}$ iteratively. 
	This is similar but different to the construction of the representation in terms of posets of ideals by applying $e^{(a)}(z)$ iteratively (see Sec.\ \ref{ssec:PosetFromQW}).
	\begin{enumerate}
		\item The level-$0$ state is the  ground state  $|\varnothing\rangle_{\sharp}$.
		\item At each level, first apply $e^{(a)}_n$ with all $a\in Q_0$ and all $n\geq 0$ on all the states of the current level, then remove the redundant states by using the quadratic and higher order relations to shuffle $e^{(a)}_n$'s around and imposing the conditions on $|\varnothing\rangle_{\sharp}$ (both simple ones like \eqref{eq:GSdef} and higher order ones if present).
		The quardratic relations are given in \eq{eq:QuadraticModes}, a loop of length $\ell$ corresponds to order-$\ell$ operator relations among relevant $e^{(a)}_n$'s, and finally a loop of length $\ell$ involving $\infty$ corresponds to order-$(\ell-1)$ operator relations among relevant $e^{(a)}_n$'s acting on the ground state.
		
	\end{enumerate}
	
	\item At each level, one can compare the result with the counting obtained by the procedure of Sec.\ \ref{ssec:PosetFromQW} to ensure that one has indeed obtained the correct number of states.
	
\end{enumerate} 
This procedure is to be contrasted with the one given in Sec.\ \ref{ssec:PosetFromQW}, where the representation is described in terms of ideal poset.
In that procedure, each state is given by a set of paths that start with $p^{\infty \rightarrow \mathfrak{a}\rightarrow \dots}$, and at each level, one determines all the states of the next level by looking at the pole structure of the charge functions of the states of the current level.
In addition, while the procedure in Sec.\ \ref{ssec:PosetFromQW} doesn't require knowing the algebraic relations of the quiver Yangian, the current method requires the knowledge of all relations, quadratic and higher order ones.\footnote{Sometimes there is no higher order relation, e.g.\ when there is no loop.}
See Sec.\ \ref{ssec:HigherOrderRelations} on how to determine higher order relations.

\subsection{Higher order relations of quiver Yangians}
\label{ssec:HigherOrderRelations}

Recall that the quadratic relations of the quiver Yangian have a universal form for all quivers; however, the higher order relations need to be determined in a case-by-case or at least class-by-class manner. 
There are two methods to determine the higher order relations. 

One is the method based on the representations, using the fact that there are two ways to compute the characters of a representation of the quiver Yangian \cite{Li:2020rij}.
\begin{enumerate}
\item Directly counting the finite co-dimensional ideals of the Jacobian algebra $J^{\sharp}_{(Q,W)}$. 
\item Counting the independent states created by applying the modes of the raising operators $e^{(a)}_n$ on the ground state of the representation, as explained in Sec.\ \ref{ssec:RepMode}. 
There are three effects that reduces the number of independent states (at each level). 
\begin{enumerate}
\item Constraint from the framing, i.e.\ the condition \eqref{eq:GSdefNCDT}.
\item Quandratic relations.
\item Higher order relations.
\end{enumerate}
\end{enumerate}
Now, at each level $\ell$, compare the number $n_1$ of states computed via method-1 with the number $n_2$ computed via method-2 after using only (a), (b), and (c) with all the higher order relations determined from previous levels.
If there are $n_1 > n_2$ redundant states, it means that there are $n_1-n_2$ new higher order relations among the $e^{(a)}_m$ modes that we can determine at this level.
Doing this level by level can then fix the higher order relations (in terms of modes) order by order.\footnote{ 
Very often, after a few steps one can guess the higher order relations in terms of fields and then check whether they provide all the higher order relations (in terms of modes).}
(After obtaining these relations among the $e^{(a)}_m$ modes, we impose parallel higher order relations among $f^{(a)}_m$ modes.)
See \cite{Li:2020rij} for the detailed procedure and for examples of determining Serre relations in affine Yangian of $\mathfrak{gl}_n$.
See also \cite[Sec.\ 2.1.1]{Gaberdiel:2018nbs} for another method of obtaining the conjectured form of higher order relations.

The second method of determining the higher order relations is to use the (conjectured)\footnote{This was shown in \cite{Rapcak:2018nsl,Rapcak:2020ueh} for affine Yangian of $\mathfrak{gl}_1$ and $\mathfrak{gl}_{1|1}$, and conjectured to apply for all toric Calabi-Yau-three quivers there and in \cite{Galakhov:2021vbo}.
For more general quivers, comparing the higher order relations obtained this way with those using the mode computation, one can test this conjecture for general quivers.} isomorphism between the subalgebra generated by $e^{(a)}$'s and the spherical part of equivariant CoHA for the same quiver. 
It would be interesting to compare the results from these two methods.
We leave a systematic study of this for future work.

\section{Counting refined BPS invariants as character of quiver Yangians}
\label{sec:BPSInvariants}

The quiver Yangian can be defined for an arbitrary quiver and potential, and it is fairly easy to compute its characters (at least as a power series using Mathematica).
On the other hand, when the quiver and potential satisfy some conditions, one can define the Donaldson-Thomas (DT) invariants of the quiver; and the generating function of the framed DT invariants count the BPS states of the corresponding supersymmetric quiver quantum mechanics, which are effective theories describing D-brane bound-states in the corresponding geometry.

The characters of the quiver Yangians representations are known to reproduce the classical DT invariants of the quiver, which correspond to the unrefined BPS invariants, for quivers from toric CY$_3$'s \cite{Li:2020rij, Mozgovoy:2020has}. 
In this section, we show that this generalizes to arbitrary quivers.
More importantly,  we will show how to refine the definition of the characters of the quiver Yangians representations to taken into account the refinement in the framed BPS invariants, which are related to the motivic DT invariants, for more general quivers. 
Since it is easy to compute the refined characters of the quiver Yangians representations for general quivers, the quiver Yangians provide an alternative method of computing the motivic DT invariants and of counting BPS states for these systems.

In this paper, we focus on two main classes of quivers:
\begin{itemize}
\item BPS quivers in 4D $\mathcal{N}=2$ theories, which can have non-trivial superpotentials. 
\item Quivers in the knot-quiver correspondence, which are symmetric quivers without potentials.
\end{itemize}

In this section, we focus on the first class, and study the second class in Sec.\ \ref{sec:KnotQuiver}.
The first class includes the 4D $\mathcal{N}=2$  theories \cite{Cecotti:2011rv,Alim:2011ae,Alim:2011kw} for which the BPS quivers are two-acyclic and 4D $\mathcal{N}=2$ theories from compactifying type IIA string on Calabi-Yau threefold $X$.
For the two-acyclic cases, we can refine the characters using the physical spin-grading of the $e^{(a)}_n$ generators; we have also checked that for simple examples, these results can be reproduced by computing in terms of ideal poset, using rather simple prescriptions. 
For the Calabi-Yau quivers, the definition of refined BPS invariants is often not unique and depends, e.g.\ on the choice of a  certain subtorus of the CY$_3$. 
We have adopted a simple and universal choice for all quivers in this class; and we have also discussed other choices and compared with results in the literature.

\subsection{Review: BPS spectrum in 4D \texorpdfstring{$\mathcal{N}=2$}{N=2} gauge theory and Donaldson-Thomas invariants of quiver}

We first give a lightening review of Donaldson-Thomas invariants of quivers and their relation to the BPS spectra of 4D $\mathcal{N}=2$ theories. 

\subsubsection{BPS quiver}

For a general 4D $\mathcal{N}=2$ (i.e.\ with $8$ supercharges) gauge theory, the spectrum of BPS particles can be captured by the so-called BPS quiver.\footnote{
Note that there exists 4D $\mathcal{N}=2$ theories that do not have any  quiver description for their BPS spectrum (at any point of the moduli space), e.g.\ the theory obtained from wrapping M5-branes on an unpunctured higher genus ($g\geq 3$) Riemann surface \cite{Gaiotto:2009hg}.
It was conjectured in \cite{Cecotti:2011rv} that a 4D $\mathcal{N}=2$ theory whose BPS phases form a dense subspace of $S^1$ cannot have a BPS quiver description. 
We will not consider these types of %4D $\mathcal{N}=2$ 
theories in this paper.
}
Consider the worldvolume theory of the $\frac{1}{2}$-BPS particle (i.e.\ with $4$ supercharges): it is given by an $\mathcal{N}=4$ quiver quantum mechanics \cite{Denef:2002ru}. 
The quiver and potential reviewed in Sec.\ \ref{ssec:QuiverPotential} correspond to the following entities in the quiver quantum mechanics:
\begin{itemize}
\item Each vertex $a\in Q_0$ corresponds to a $U(N_a)$ gauge factor.
\item Each arrow $I$ corresponds to a chiral field that transforms as the bi-fundamental with respect to the two gauge factors at the source $a=s(I)$ and the target $b=t(I)$
of the arrow $\overline{U(N_a)}\times U(N_b)$.
\item The potential in Sec.\ \ref{ssec:QuiverPotential} is just  the superpotential of the $\mathcal{N}=4$ quiver quantum mechanics, and thus has to satisfy certain properties. 
\item The constraints \eqref{eq:LoopConstraints} come from the global symmetries of the quiver quantum mechanics.\footnote{
There might also exist gauge symmetries, which can impose further constraints on the charges $\{h_I\}$. 
Since this has to be analyzed in a case-by-case manner, in this paper we will only impose the constraints \eqref{eq:LoopConstraints} from the potential \eqref{eq:Potential}.
}
\end{itemize}
 
We summarize a few important classes of 4D $\mathcal{N}=2$ theories whose BPS spectra can be described by such a quiver quantum mechanics.
\begin{itemize}

\item For the orbifold type 4D $\mathcal{N}=2$ field theories, the quiver and superpotential can be obtained using fractional branes \cite{Douglas:1996sw,Diaconescu:1997br,Diaconescu:1999dt,Douglas:2000ah,Douglas:2000qw}.

\item  Among the 4D $\mathcal{N}=2$ field theories from type IIA string theory on non-compact CY$_3$'s, the quiver and superpotential can be obtained via the brane tiling technique when the CY$_3$ is toric \cite{Hanany:2005ve, Franco:2005rj, Franco:2005sm}; on the other hand, for those non-toric Calabi-Yau singularities that can be partially resolved by blowing up a complex surface, one can also obtain the quiver and superpotential using the method of exceptional collections \cite{Herzog:2003zc, Herzog:2004qw, Aspinwall:2004bs,Hanany:2006nm}.

\item For the 4D $\mathcal{N}=2$ field theories from the IR limit of wrapping $N$ M5-branes on punctured Riemann surfaces (the so-called Class $\mathcal{S}$ theories \cite{Gaiotto:2009we}), the quiver and potential can be obtained by studying the ideal triangulation of the Riemann surface \cite{Cecotti:2012gh,Alim:2011ae,Alim:2011kw}.

\item For the 4D $\mathcal{N}=2$ field theories from geometric engineering (i.e.\ type II string theory on local singularities of CY$_3$'s) \cite{Katz:1996fh,Katz:1996th}, the quiver and potential can be obtained by considering the vacua and structures of kinks of their dual 2D $\mathcal{N}=2$ theories in  the 4D/2D correspondence \cite{Cecotti:2010fi}.

\end{itemize}
Obviously one theory can fall into multiple categories in the list above, and therefore can have more than one method to have its quiver and superpotential determined.  

\subsubsection{Donaldson-Thomas invariants and wall-crossing formula}

The $\frac{1}{2}$-BPS spectrum of a 4D $\mathcal{N}=2$ theory can be captured by the refined BPS index\footnote{
Note that we have factored out the contribution from the universal half-hypermultiplet, which transforms as $2(0,0)\oplus (\frac{1}{2},0)$ and tracks the center-of-mass DOF in $\mathbb{R}^3$ --- otherwise the definition would be $\overline{\underline{\Omega}}_{\gamma}(y, u)\equiv -\frac{1}{2}\textrm{Tr}_{\mathcal{H}'_{BPS, \gamma}}(-y)^{2J'_3}(2J'_3)^2$, where both $\mathcal{H}'_{BPS, \gamma}$ and $J'_3$ are for the theory before the universal half-hypermultiplet is factored out,  see \cite{Denef:2007vg}.
} 
\begin{equation}\label{eq:RefinedIndexDef}
\overline{\underline{\Omega}}^{\textrm{ref}}_{\gamma}(y, u)\equiv -\frac{1}{2}\textrm{Tr}_{\mathcal{H}_{BPS, \gamma}}(-y)^{2J_3}\,,
\end{equation}
where $\mathcal{H}_{BPS, \gamma}$ is the Hilbert space of BPS particles with electromagnetic charge $\gamma\in \Gamma$, with $\Gamma\sim \mathbb{Z}^{2r}$ being the charge lattice of the theory and equipped with the symplectic form $\langle \cdot,\cdot\rangle$, and $J_3$ is the generator of the rotation group $Spin(3)$. 
The variable $u$ labels the point in the Coulomb branch moduli space $\mathcal{B}$ (of complex dimension $r$), and it enters the r.h.s.\ of \eqref{eq:RefinedIndexDef} through the fact that the BPS particles contributing to \eqref{eq:RefinedIndexDef} satisfy $M=|Z_{\gamma}(u)|$, where $Z_{\gamma}(u)\in \mathbb{C}$ is the central charge.
Finally, in the unrefined limit $y\rightarrow 1$, the index \eqref{eq:RefinedIndexDef} reduces to the unrefined BPS index (also called $2^{\textrm{nd}}$ helicity supertrace):
\begin{equation}\label{eq:2ndHelicitySupertrace}
\overline{\underline{\Omega}}^{\textrm{unref}}_{\gamma}(u)\equiv \lim_{y\rightarrow 1}\overline{\underline{\Omega}}^{\textrm{ref}
}_{\gamma}(y, u)=-\frac{1}{2}\textrm{Tr}_{\mathcal{H}_{BPS, \gamma}}(-1)^{2J_3}\,,
\end{equation}
which loses the information on the flavor charges.

The index $\overline{\underline{\Omega}}^{\textrm{ref}}_{\gamma}(y, u)$ (hence also $\overline{\underline{\Omega}}^{\textrm{unref}}_{\gamma}(u)$) only depends on $u$ in a discrete manner. 
Given $\gamma$, there exist  ``walls of marginal stability" (of real co-dimension $1$)
in the Coulomb branch moduli space $\mathcal{B}$, satisfying $\gamma=\gamma_1+\gamma_2$ (for linearly independent $\gamma_{1,2}$) and $\textrm{arg}Z_{\gamma_1}(u)=\textrm{arg}Z_{\gamma_2}(u)$.
These walls of marginal stability divide $\mathcal{B}$ into multiple chambers.
The index $\overline{\underline{\Omega}}^{\textrm{ref}}_{\gamma}(y, u)$ is constant within each chamber but jumps when $u$ crosses a wall of marginal stability, because that is when a particle of charge $\gamma$ can decay into two particles with charges $\gamma_{1}$ and $\gamma_2$ or vice versa.
Hence, we will also denote the index $\overline{\underline{\Omega}}^{\textrm{ref}}_{\gamma}(y,u)$ by $\overline{\underline{\Omega}}^{\textrm{ref}}_{\gamma, \theta}(y)$, where $\theta$ labels different chambers in the K\"ahler moduli space.\footnote{
Here $\theta$ corresponds to the stability parameter in the mathematics literature.
}

\medskip

The $\overline{\underline{\Omega}}^{\textrm{ref}}_{\gamma, \theta}(y)$ in \eqref{eq:RefinedIndexDef} is the ``framed" BPS index, to be distinguished from the unframed version $\Omega^{\textrm{ref}}_{\gamma,\theta}(y)$.\footnote{
For example, for the type IIA string theory on CY$_3$, the unframed index counts the BPS D2-D0 bound states whereas the framed one counts the BPS bound states of D2-D0 with 1 D6-brane wrapping the entire CY$_3$.
}
The latter appears directly in the wall-crossing formula of Kontsevich and Soibelman \cite{Kontsevich:2008fj} that captures the intricate wall-crossing structure of the 4D $\mathcal{N}=2$ theories.
Let us briefly summarize it.
In the unrefined case, consider a coordinate system $X_{\gamma}$ on the complex torus $T_u\equiv \Gamma^{*}\otimes_{\mathbb{Z}}\mathbb{C}^{\times}$ that obeys $X_{\gamma}X_{\gamma'}=X_{\gamma+\gamma'}$.
Moreover, define a function $\sigma(\gamma)\equiv (-1)^{\langle \gamma^{e}, \gamma^{m}\rangle}$ where $\gamma^{e}$ and $\gamma^{m}$ are the electric and magnetic part of the charge $\gamma$ ($=\gamma^{e}+\gamma^{m}$).\footnote{
The factor $\sigma(\gamma)$ is introduced to take into account the fact that the fermion number of the bound state of two BPS particles with charges $\gamma$ and $\gamma'$ is shifted by $\langle \gamma, \gamma'\rangle$.
}
Define the operators $e_{\gamma}\equiv \sigma(\gamma)\, X_{\gamma}$, which generate an associative algebra with multiplication $e_{\gamma}\,  e_{\gamma'}  =(-1)^{ \langle \gamma , \gamma' \rangle } \, e_{\gamma+\gamma'}$.
For the refined case, uplifting the definition above to  the quantum torus $T_u$ and the corresponding operator $\hat{e}_{\gamma}$, which is related to $e_{\gamma}$ via  $\lim_{y\rightarrow 1} (y^2-1)\hat{e}_{\gamma}=e_{\gamma}$ (for a detailed definition see \cite{Kontsevich:2008fj}), the associative algebra now becomes $\hat{e}_{\gamma}\,  \hat{e}_{\gamma'}  =(-y)^{ \langle \gamma , \gamma' \rangle } \, \hat{e}_{\gamma+\gamma'}$.
Now define the operator
\begin{equation}
\mathcal{K}_{\gamma}\equiv \textrm{exp}\left(\sum^{\infty}_{n=1} \frac{1}{n^2}\, \hat{e}_{n\, \gamma}
\right)
\end{equation}
and the product 
\begin{equation}\label{eq:StackyA}
\mathcal{A}\equiv \prod^{}_{\gamma} (\mathcal{K}_{\gamma})^{\Omega_{\gamma,\theta}(y)}\,,
\end{equation}
where $\Omega_{\gamma,\theta}(y)$ are the unframed DT invariants,\footnote{
By comparing the wall-crossing behaviors of these BPS indices with those of the Donaldson-Thomas (DT) invariants of the BPS quiver $(Q,W)$, it was shown by \cite{Gaiotto:2010okc} that the unrefined BPS index $\Omega^{\textrm{unref}}_{\gamma,\theta}$ %in \eqref{eq:2ndHelicitySupertrace} 
corresponds to the unrefined DT invariants (also called numerical or classical DT invariants) in  \cite{Kontsevich:2008fj}, and then by \cite{Dimofte:2009bv} that  the refined BPS index  $\Omega_{\gamma,\theta}(y)$ corresponds to the motivic DT invariants (also called quantum DT invariants) in \cite{Kontsevich:2008fj} (with identification $y=\mathbb{L}^{\frac{1}{2}}$).
(Note that in the cases where the refinement choice is not unique, one should distinguish between refined BPS indices and motivic DT invariants, for recent discussions see e.g.\ \cite{Descombes:2021snc,Cirafici:2021yda}. However we will not systematically study the refinement choice in this paper.)
} 
and  the product is taken in the order  (from right to left) in which the phase of the central charge $Z$ increases.
The ``wall-crossing theorem" states that $\mathcal{A}(\boldsymbol{x})$, defined in \eqref{eq:StackyA}, is independent of the stability parameter $\theta$ \cite{Kontsevich:2008fj}.\footnote{
Physically, this corresponds to the fact that the hyperk\"ahler metric of the BPS moduli space of the 4D theory on $\mathbb{R}^3\times S^1$ is continuous \cite{Gaiotto:2010okc}.
}
The KS wall-crossing formula \eqref{eq:StackyA} then offers a way to compute the unframed DT invariants $\Omega_{\gamma,\theta}(y)$ once we know their values in one chamber of the moduli space.
It was conjectured that the integrability of $\Omega_{\gamma, \theta}(y)$ in one chamber together with the  wall-crossing formula \eqref{eq:StackyA} 
guarantee its integrality in all other chambers \cite{Kontsevich:2008fj}.

\medskip

In this section, we consider the situation where the $\frac{1}{2}$-BPS spectrum of the 4D $\mathcal{N}=2$ theory is given by a quiver (called BPS quiver).
Each vertex $a\in Q_0$ corresponds to an elementary hypermultiplet, with charge $\gamma_a$; any BPS particle (with Im$Z_{\gamma}>0$)\footnote{Those with Im$Z_{\gamma}<0$ are anti-particles.} can be decomposed in terms of these elementary ones: $\gamma=\sum_{a\in Q_0}d_a \gamma_a$ with $d_a\geq 0$.
The Dirac pairings of basic charges are related to the arrows of the quiver by $\langle \gamma_a, \gamma_b\rangle=|a\rightarrow b|-|b\rightarrow a|$.

Now consider the moduli space of the (semi)stable representation of the quiver $Q$, with dimension vector $\{d_a\}$. (A quiver representation with dimension vector $\{d_a\}$ consists of the collection of vector space $V_a$ of dimension $d_a$ for each vertex $a\in Q_0$, together with linear maps $\phi_{I^{a\rightarrow b}}\in \textrm{Hom}(V_a,V_b)$ for each arrow $I^{a\rightarrow b}\in Q_1$. A quiver representation $R$ is stable (resp.\ semistable) if any sub-representation $S\subset R$  satisfies $\textrm{arg}Z(S)< \textrm{arg}Z(R)$ (resp.\  $\textrm{arg}Z(S)\leq \textrm{arg}Z(R)$).)
The DT-invariant $\Omega_{\gamma,\theta}(y)$ counts the (compactly-supported) de Rham cohomolgy of quiver representations with dimension vector $\{d_a\}$, where $\gamma=\sum_a d_a \gamma_a$, and stability parameter $\theta$, which is related to the central charge by $\theta\sim -\textrm{Re}Z$ and corresponds to the FI parameter.
See e.g.\ \cite{Kirillov} for more on quiver representation and \cite{Kontsevich:2008fj,Kontsevich:2010px} for DT invariants for quivers.

\medskip

In this paper, we would like to use the character of the quiver Yangian to compute the generating function  of the refined BPS indices $\overline{\underline{\Omega}}_{\gamma,\theta}(y)$ in the so-called non-commutative DT (NCDT) chamber: 
\begin{equation}\label{eq:GFrefinedBPS}
\mathcal{Z}_{\textrm{NC}}^{\textrm{ref}}(y,\boldsymbol{x})= \sum_{\boldsymbol{d}\in \mathbb{N}^{|Q_0|}\backslash \{0\}} \overline{\underline{\Omega}}^{\textrm{NC}}_{\,\boldsymbol{d}}(y)\, \boldsymbol{x}^{\boldsymbol{d}}\,,
\end{equation}
where we have already translated into the BPS quiver description.
Here
\begin{equation}\label{eq:dvDef}
\boldsymbol{d}\equiv  (d_0, d_1, \dots d_{|Q|_0-1})  
%\qquad \textrm{and} \qquad 
%\boldsymbol{x}\equiv (x_0, x_1, \dots x_{|Q|_0-1}) 
\end{equation}
is the dimension vector of the NC-(semi)stable representation of the quiver and 
\begin{equation}\label{eq:xDef}
\boldsymbol{x}\equiv (x_0, x_1, \dots x_{|Q|_0-1}) 
\end{equation}
are the fugacities of the $|Q_0|$ vertices.
The unrefined version of \eqref{eq:GFrefinedBPS} is
\begin{equation}\label{eq:GFBPS}
\mathcal{Z}^{\textrm{unref}}_{\textrm{NC}}(\boldsymbol{x})
\equiv \lim_{y\rightarrow 1}\,  \mathcal{Z}_{\textrm{NC}}^{\textrm{ref}}(y,\boldsymbol{x})
=\sum_{\boldsymbol{d}\in \mathbb{N}^{|Q_0|}\backslash \{0\}} \overline{\underline{\Omega}}^{\textrm{NC}}_{\,\boldsymbol{d}}(1)\, \boldsymbol{x}^{\boldsymbol{d}}\,.
\end{equation}

\subsubsection{Simplification for symmetric quivers}

When the quiver is symmetric, the DT invariant is independent of the stability parameter.
Thus one can consider the motivic generating series  \cite{Kontsevich:2008fj}:
\begin{equation}\label{eq:GDef}
\boldsymbol{G}(y,\boldsymbol{x})=\textrm{PE}\left[-\frac{\boldsymbol{\Omega}(y,\boldsymbol{x})}{y-y^{-1}}\right]\,,
\end{equation}
where P.E.\ stands for the plethystic exponential:
\begin{equation}\label{eq:PEDef}
\textrm{PE}[f(x)]\equiv \textrm{exp}\left(\sum^{\infty}_{k=1}\frac{f(x^k)}{k}\right)\,,
\end{equation} 
and $\boldsymbol{\Omega}(y,\boldsymbol{x})$ is the generating function of the  unframed motivic DT invariants $\Omega_{\boldsymbol{d}}(y)$:
\begin{equation}\label{eq:Omega}
\boldsymbol{\Omega}(y,\boldsymbol{x})=\sum_{\boldsymbol{d}\in \mathbb{N}^{|Q_0|}_0\backslash \{0\}} \Omega_{\boldsymbol{d}}(y) \, \boldsymbol{x}^{\boldsymbol{d}}
=\sum_{\boldsymbol{d}\in \mathbb{N}^{|Q_0|}_0\backslash \{0\}} \sum_{n=0}(-1)^{|\boldsymbol{d}|+n+1}\Omega_{\boldsymbol{d},n} \, \boldsymbol{x}^{\boldsymbol{d}}\, y^n\,,
\end{equation}
where $|\boldsymbol{d}|\equiv \sum_{a}d_a$ and $\Omega_{\boldsymbol{d},n}\in \mathbb{Z}$.
When the quiver is symmetric, the NCDT partition function is related to  the unframed DT invariants $\Omega_{\boldsymbol{d}}(y)$ via \cite{Morrison:2011rz}
\begin{equation}\label{eq:ZNCsymmetric}
\mathcal{Z}^{\textrm{ref}}_{\textrm{NC}}(y,\boldsymbol{x})=\hat{\texttt{S}}_{-\boldsymbol{f}}\cdot \textrm{PE}\left[-\frac{\sum_{\boldsymbol{d}} (y^{2\boldsymbol{f}\cdot \boldsymbol{d}}-1)\, \Omega_{\boldsymbol{d}}(y)\, \boldsymbol{x}^{\boldsymbol{d}}}{y-y^{-1}}\right]
\end{equation}
with the operator
\begin{equation}
\hat{\texttt{S}}_{-\boldsymbol{f}}: \qquad\boldsymbol{x}^{\boldsymbol{d}}\rightarrow (-y)^{-\boldsymbol{f}\cdot \boldsymbol{d}}\, \boldsymbol{x}^{\boldsymbol{d}}\,.
\end{equation}
One can rewrite the expression \eqref{eq:ZNCsymmetric} as 
\begin{equation}
\mathcal{Z}^{\textrm{ref}}_{\textrm{NC}}(y,\boldsymbol{x})=\widehat{\textrm{Sgn}}_0\cdot \textrm{PE}\left[-\frac{\sum_{\boldsymbol{d}} (y^{\boldsymbol{f}\cdot \boldsymbol{d}}-y^{-\boldsymbol{f}\cdot \boldsymbol{d}})\, \Omega_{\boldsymbol{d}}(y)\, \boldsymbol{x}^{\boldsymbol{d}}}{y-y^{-1}}\right]
\end{equation}
with 
\begin{equation}
\widehat{\textrm{Sgn}}_0: \qquad x_{\mathfrak{a}}\rightarrow -x_{\mathfrak{a}}\,,
\end{equation}
when the framing vector $\boldsymbol{f}=e_{\mathfrak{a}}$ with $\mathfrak{a}\in Q_0$ being the framed vertex.
(In this section, most of the time $\mathfrak{a}=0$.)
Conversely, if we can compute $\mathcal{Z}_{\textrm{NC}}(y,\boldsymbol{x})$, we can easily obtain
\begin{equation}\label{eq:tildeOmega}
\widetilde{\boldsymbol{\Omega}}_{\textrm{NC}}(y,\boldsymbol{x}) \equiv   \sum_{\boldsymbol{d}\in \mathbb{N}^{|Q_0|}_0\backslash \{0\}}(y^{\boldsymbol{f}\cdot \boldsymbol{d}}-y^{-\boldsymbol{f}\cdot \boldsymbol{d}})\, 
\Omega_{\boldsymbol{d}}(y)\, \boldsymbol{x}^{\boldsymbol{d}}
=-(y-y^{-1})\cdot \,\textrm{PLog}\left[\widehat{\textrm{Sgn}}_0\cdot\boldsymbol{\mathcal{Z}}^{\textrm{ref}}_{\textrm{NC}}(y,\boldsymbol{x}) \right]\,,
\end{equation}
where \textrm{PLog} (the plethystic log)  is the inverse function of P.E.\ defined in \eqref{eq:PEDef}:
\begin{equation}
\textrm{PLog}[g(x)]\equiv \sum^{\infty}_{n=1}\frac{\mu(n)}{n} \log{g(x^n)}\,,
\end{equation}
where $\mu(n)$ is the M\"obius function.
From \eqref{eq:tildeOmega}, one can then obtain the unframed motivic DT invariant for each dimension vector $\boldsymbol{d}$.

\subsection{Refining the characters of the quiver Yangians}
\label{ssec:Refining}

For the 4D $\mathcal{N}=2$ theory from type IIA string theory on a toric CY$_3$, the unrefined characters of the quiver Yangian reproduce the generating function \eqref{eq:GFBPS} of the classical DT invariants. 
In particular, the vacuum character corresponds to the generating series of  DT invariants in the non-commutative DT (NCDT) chamber \cite{Li:2020rij}, and non-vacuum characters correspond to the other chambers (separated by walls of the second kind \cite{Kontsevich:2008fj}) and open DT invariants \cite{Galakhov:2021xum}.
For a general quiver, we conjecture that the computation of the framed numerical DT invariants is still given by counting finite co-dimensional ideals of the Jacobian algebra $J_{(Q,W)}$, which 
as explained in Sec.\ \ref{ssec:PathAlgebralIdeals} is captured by the unrefined vacuum character of the quiver Yangian. 
Now we would like to uplift to the refined case, namely we want to compute the refine characters of the vacuum representations and relate them to the motivic DT invariants.

\medskip

Recall that there are two ways to describe the quiver Yangian representations.
The first one is in terms of finite codimensional ideals of the Jacobian algebra, and can be constructed using the procedure in Sec.\ \ref{ssec:PosetFromQW}.
The second one is by applying repeatedly the raising operator on the ground state $|\varnothing\rangle_{\sharp}$, as explained in Sec.\ \ref{ssec:RepMode}. 

\medskip

Let's first consider the second description since the refinement in this case is more straightforward.
In this description of the quiver Yangian representations, the refinement is given by assigning the (spin) grading to the raising operators $e^{(a)}_n$'s. 
Namely, once we obtain all the states, in the form of $\prod e^{(a)}_n |\varnothing\rangle_{\sharp}$, one can associate a grading to each such a state once we define the grading of each raising operator $e^{(a)}_n$ and the grading of the product.
For example, for the 2-acyclic BPS quivers of the 4D $\mathcal{N}=2$ theories, it is natural to assign
\begin{equation}
e^{(a)}_n:\qquad y^{2n+1}
\end{equation} 
for all $a\in Q_0$ and define the grading of the product $e^{(a)}_ne^{(b)}_m$ by
\begin{equation}
e^{(a)}_ne^{(b)}_m:\qquad y^{2n+2m+2+n(a,b)}
\end{equation} 
where $n(a,b)$ is determined by the Dirac pairing of the charges corresponding to the vertices $a$ and $b$.
A priori, different classes of quivers could require different definition of the grading, and need to be  fixed case by case.

\medskip

Below, we will use this method to refine the quiver Yangian characters for a few examples. 
However, since there are a few advantages of describing the quiver Yangian representations in terms of ideals of Jacobian algebra, e.g.\ the construction is universal for all quivers, and it is very easy to incorporate the loops (and hence the superpotentials) and arbitrary framings, we would like to explore an alternative way to define the refinement of the characters of the quiver Yangian representation that is more adapted to this description.

For the purpose of computing refined DT invariants, it is enough to choose a particular $\mathfrak{a}$ such that the corresponding vacuum-like representation is the largest among all vacuum-like representations for the same quiver.
We then relabel it as $\mathfrak{a}=0$, and call the corresponding ${}^{(0)}\mathcal{P}_{(Q,W)}$ the vacuum representation, as in Sec.\ \ref{sec:Representation}.
Indeed, in this section, we have already implicitly made this choice, i.e.\ in this section we will only consider the vacuum representation ${}^{(0)}\mathcal{P}_{(Q,W)}$.

\medskip

The unrefined character of the representation $\mathcal{P}$ of a quiver Yangian $\mathrm{Y}(Q,W)$ considered in \cite{Li:2020rij,Galakhov:2021xum} is just the partition function \eqref{eq:Zpartitionfunction}, which we reproduce here: 
\begin{equation}\label{eq:ZunrefinedP}
\boldsymbol{\mathcal{Z}}^{\textrm{unref}}_{\mathcal{P}}(\boldsymbol{x})\equiv
\sum_{\Pi\in \mathcal{P}} \boldsymbol{x}^{|\Pi|} \qquad \textrm{with} \quad
\boldsymbol{x}^{|\Pi|}\equiv \prod_{a\in Q_0}\, (x_a)^{|\Pi|_a} \,,
\end{equation}
where $|\Pi|_a$ denotes the number of atoms of color $a$ in the state $\Pi$.
The refined version is 
\begin{equation}\label{eq:ZrefinedP}
\boldsymbol{\mathcal{Z}}^{\textrm{ref}}_{\mathcal{P}}(y,\boldsymbol{x})\equiv
\sum_{\Pi\in \mathcal{P}} y^{\upsilon(\Pi)}\, \boldsymbol{x}^{|\Pi|}\,,
\end{equation}
where $\upsilon(\Pi)$ measures the total $2J_3$ value of the state $\Pi$; from now on we will also sometimes call this the $y$-charge. 

\medskip

We would like to determine, for a given quiver with potential $(Q,W)$, 
a prescription of $v(\Pi)$ such that the refined character \eqref{eq:ZrefinedP} for the vacuum representation ${}^{(0)}\mathcal{P}$ can reproduce the generating function of the refined BPS index \eqref{eq:GFrefinedBPS} up to signs:
\begin{equation}\label{eq:ZcrystaltoZNC}
	\boldsymbol{\mathcal{Z}}^{\textrm{ref}}_{{}^{(0)}\mathcal{P}}(y,\boldsymbol{x})\equiv
	\sum_{\Pi\in {}^{(0)}\mathcal{P}} y^{\upsilon(\Pi)}\, \boldsymbol{x}^{|\Pi|}
	\quad \longrightarrow \quad   \mathcal{Z}^{\textrm{ref}}_{\textrm{NC}}(y,\boldsymbol{x})=\sum_{\Pi\in {}^{(0)}\mathcal{P}} \textrm{sign}(\Pi)y^{\upsilon(\Pi)}\, \boldsymbol{x}^{|\Pi|} \,,
\end{equation}
with 
\begin{equation}\label{eq:signPi}
	\textrm{sign}(\Pi)\equiv (-1)^{|\Pi|_0+\langle |\Pi| , |\Pi| \rangle} \,,
\end{equation}
where $|\Pi|$ is the vector $(|\Pi|_0, \dots, |\Pi|_{|Q_0|-1})$ with $|\Pi|_{a}$ being the number of atoms of color $a$ in the state $\Pi$ and $|\Pi|$ is mapped to the dimension vector $\boldsymbol{d}$ \eqref{eq:dvDef}; $\langle\cdot ,\cdot \rangle$ is the Ringel form of the quiver $Q$ \cite{Mozgovoy:2008fd}:
\begin{equation}
	\langle U, V\rangle = \sum_{a\in Q_0} U_a V_a-\sum_{I^{a\rightarrow b}\in Q_1} U_a V_b\,.
\end{equation}
One can rewrite the relation \eqref{eq:ZcrystaltoZNC} as
\begin{equation}
	\mathcal{Z}^{\textrm{ref}}_{\textrm{NC}}(y,\boldsymbol{x})=\hat{\textbf{S}}\cdot \boldsymbol{\mathcal{Z}}^{\textrm{ref}}_{{}^{(0)}\mathcal{P}}(y,\boldsymbol{x})\,,
\end{equation}
where the operation $\hat{\textbf{S}}$ inserts a $\textrm{sign}(\Pi)$ \eqref{eq:signPi} for each term $\Pi$ in the expansion of the vacuum character $\boldsymbol{\mathcal{Z}}^{\textrm{ref}}_{{}^{(0)}\mathcal{P}}(y,\boldsymbol{x})$.

For the symmetric quiver, the sign change in \eqref{eq:ZcrystaltoZNC} can be absorbed into a change of variables \cite{Mozgovoy:2008fd}:
\begin{equation}\label{eq:SgnSym}
	\textrm{symmetric quiver}: \qquad
	\mathcal{Z}^{\textrm{ref}}_{\textrm{NC}}(y,\boldsymbol{x}) 
	=\hat{\textbf{S}}\cdot \boldsymbol{\mathcal{Z}}^{\textrm{ref}}_{{}^{(0)}\mathcal{P}}(y,\boldsymbol{x})
	= \boldsymbol{\mathcal{Z}}^{\textrm{ref}}_{{}^{(0)}\mathcal{P}}(y, \widehat{\textrm{Sgn}'} \,\boldsymbol{x})\,,
\end{equation}
where the operator $\widehat{\textrm{Sgn}'}$  acts on the set of fugacities
$\boldsymbol{x}$ by
\begin{equation}\label{eq:Soperation}
	%\begin{aligned}
	\widehat{\textrm{Sgn}'} =  \hat{S}_0\cdot  \hat{S}_1: \qquad 
	\begin{cases}
		\begin{aligned}
			&\hat{S}_0 : \qquad x_a\rightarrow -x_a \quad \textrm{for } a\in Q_0\,, a\neq 0 \\
			&\hat{S}_1 : \qquad x_a\rightarrow -x_a \quad \textrm{for } a\in Q_0 \,\textrm{ s.t. } I^{a\rightarrow a}\in Q_1 \,.    
		\end{aligned}    
	\end{cases}
	%\end{aligned}
\end{equation} 
Therefore, for symmetric quivers, the $\widetilde{\boldsymbol{\Omega}}^{\textrm{NC}}(y,\boldsymbol{x}) $ defined in \eqref{eq:tildeOmega} can be computed via
\begin{equation}\label{eq:tildeOmega1}
	\begin{aligned}
		\widetilde{\boldsymbol{\Omega}}^{\textrm{NC}}(y,\boldsymbol{x}) 
		&=-(y-y^{-1})\cdot \textrm{PLog}\left[ \boldsymbol{\mathcal{Z}}^{\textrm{ref}}_{{}^{(0)}\mathcal{P}}(y,\widehat{\textrm{S}}_{\textrm{total}} \cdot\boldsymbol{x})\right]
	\end{aligned}
\end{equation}
with 
\begin{equation}\label{eq:SgnTotal}
	\widehat{\textrm{S}}_{\textrm{total}}= \widehat{\textrm{Sgn}_0}\cdot \widehat{\textrm{Sgn}'}\,.
\end{equation}
From this result one can then read off the unframed refined DT invariants $\Omega_{\boldsymbol{d}}(y)$ from the definition of $\widetilde{\boldsymbol{\Omega}}^{\textrm{NC}}(y,\boldsymbol{x}) 
$ in \eqref{eq:tildeOmega}.

To study the refined version \eqref{eq:ZcrystaltoZNC}, the main task is to define the function $\upsilon(\Pi)$ for a given quiver. 
One natural prescription is that the $y$-charge of a state $\Pi$ is given by the sum of the $y$-charges of all the atoms in $\Pi$: 
\begin{equation}\label{eq:upsilonAssumption}
\upsilon(\Pi)\equiv \sum_{a\in Q_0}\sum_{\sqbox{$a$}\in \Pi} \upsilon(\sqbox{$a$}) \,.
\end{equation}
We will assume \eqref{eq:upsilonAssumption} for examples in this paper. 
If we adopt this assumption, the only remaining task is to determine the function $\upsilon(\sqbox{$a$})$ for a given quiver. 

\smallskip

Recall that an atom $\sqbox{$a$}$ is equivalent to a path $\mathfrak{p}^{\infty\rightarrow a}$, hence the refinement prescription consists of the following three steps:
\begin{enumerate}
\item Define a function $\Upsilon$ on  each arrow $I\in Q_1$ to keep track of its $y$-charge.
The detailed assignment $\Upsilon(I)$ will depend on the individual quiver $(Q,W)$.
\item Since an atom $\sqbox{$a$}$ is equivalent to a path $\mathfrak{p}^{\infty\rightarrow a}$, it is natural to define 
\begin{equation}\label{eq:UpsilonAtom}
\Upsilon(\sqbox{$a$})\equiv \sum_{I\in \mathfrak{p}^{\infty\rightarrow a}} \Upsilon(I)\,.
\end{equation}
\item Determine $\upsilon(\sqbox{$a$})$ as an (integer) function of  $\Upsilon(\sqbox{$a$})$.  
\end{enumerate}
In this procedure, there are two places, i.e.\ step 1 and step 3, where a choice needs to be made, resulting in different refinements. 
The choice of the function in step 3 is the important one and depends on the types of quivers, whereas the choice in step 1 is more technical than structural.
%It turns out that the choice in step 3 is different for the BPS quivers of this section and the knot quivers of Sec.\ \ref{sec:KnotQuiver}. 
%Let us focus on the former case for now. 

%Finally, in this paper, we have studied refinements of the characters of quiver Yangians, for various examples of the BPS quivers of 4D $\mathcal{N}=2$ theories and knot quivers, which match known results computed by other methods. 
%A more systematic study will be presented in future work. 

\medskip

The refinement for the 4D $\mathcal{N}=2$ theories comes from the massive little group $\mathfrak{su}(2)$, namely, turning on $y$ means turning on the graviphoton background in 4D.  
As we will see, for a pair of vertices $a,b\in Q_0$, the set of arrows from $a$ to $b$, $\{a \rightarrow b\}$, decomposes into a few spin-$J$ representations.
Note that this decomposition is not unique, and different decompositions correspond to different ways to refine the character.
Although the choice is to be made in a case-by-case manner, in general we will choose a  decomposition that respects some symmetries of the quiver and its vacuum representation, to be explained later. 
Once we fix such a choice, it then determines the function $\Upsilon(I)$  to be :
\begin{equation}\label{eq:UpsilonJ3}
\Upsilon(I) = 2 \cdot J_3 \textrm{ value of }  I\,,    
\end{equation}
up to immaterial permutations of arrows within the same representation and possibly overall shifts.

\medskip

The are different possible prescriptions in step 3, depend on type of quivers.
\begin{itemize}
\item For simple examples of two-acyclic BPS quivers and knot-quivers that we studied in this paper, we find
\begin{equation}\label{eq:upsilonKQsec4}
\begin{aligned}
\upsilon(\sqbox{$a$})=\delta_{a\in Q_0'}\cdot\Upsilon(\sqbox{$a$})    \qquad  \textrm{with} \quad \Upsilon(\sqbox{$a$})= \sum_{I\in \mathfrak{p}^{\infty\rightarrow a}} \Upsilon(I) \,,
\end{aligned}
\end{equation}
where $\delta_a=1$ when $a$ is in the subset $Q_0'\subset Q_0$ to be determined, otherwise 0,
together with \eqref{eq:UpsilonAtom} and \eqref{eq:UpsilonJ3}, possibly with overall shifts.
For more complicated quivers, there could be systematic shifts that depend on the level in  \eqref{eq:UpsilonAtom} and \eqref{eq:UpsilonJ3}, and even in \eqref{eq:upsilonAssumption}.

\item For Calabi-Yau quivers, recall that for this class, there exists a family of refinement, so the prescription is not unique. 
One simple, symmetric (under $y\rightarrow y^{-1}$), and universal choice for step 3 is: 
\begin{equation}\label{eq:upsilonAtomType1}
\upsilon(\sqbox{$0$})=\textrm{sign}(\Upsilon(\sqbox{$0$})) \qquad \textrm{and} \qquad
\upsilon(\sqbox{$a$})=0 \quad  \textrm{for} \,\,\,  a\in Q_0\,, a\neq 0\,,
\end{equation}
where $0\in Q_0$ is the framed vertex. 
However, since the symmetry $y\rightarrow y^{-1}$ is a priori not respected in the computation of motivic DT invariants, below we will also discuss other refinement choices to match some existing results in the literature. 
\end{itemize}

\subsection{Example: quiver with one vertex and no loop}
\label{ssec:m0loop}

Let us give further evidence for our assertions by considering some examples.
The simplest quiver is the quiver with one vertex and no arrow.
By definition \eqref{eq:BondingFactorDef}, the bonding factor is simply
\begin{equation}\label{eq:varphim0}
\begin{aligned}
&\varphi^{0\Leftarrow 0}(u)=1\,,
\end{aligned}
\end{equation}
where we have chosen $t_{00}=1$; and we will also choose $s_{00}=1$.
Plugging these into the definition \eqref{eq:QuadraticFields} then gives the quiver Yangian for this quiver:
\begin{equation}\label{eq:m0Field}
\begin{aligned}
\psi(z)\, \psi(w)&=   \psi(w)\, \psi(z)  \;,\\
\,\psi(z) \, e(w)&\simeq\, e(w)\,\psi(z) \,, \\
\,e(z)\, e(w)&\sim -\, e(w)\, e(z) \,,\\
\,\psi(z)\, f(w)&\simeq \, f(w)\, \psi(z)\,, \\
\,f(z)\, f(w)&\sim -\, f(w)\, f(z) \,,\\
\{e(z)\, , \, f(w)\}&\sim -\frac{\psi(z)-\psi(w)}{z-w} \,.
\end{aligned}
\end{equation}
The relations in terms of the modes are
\begin{equation}\label{eq:m0Mode}
	\begin{aligned}
		&[\psi_{j}\,,\, \psi_{k}]  = [\psi_{j}\,,e_{k}]=[\psi_{j}\,,\, f_{k}]= 0 \,,\\
		&\{e_{j}\,,\, e_{k}\} = \{f_{j}\,,\, f_{k}\} =0\,, \qquad \{e_{j}\,,\, f_{k}\}=\psi_{j+k}\,.
	\end{aligned}    
\end{equation}

We can now consider the representation that corresponds to the NCDT chamber, with one arrow from $\infty$ to the single vertex.
In fact, we can also easily consider the more general case with $m\geq 0$ arrows from $\infty$ to the vertex, and as we will see its structure is very similar to the $m$-Kronecker quiver. 
From \eqref{eq:varphi0Def}, its charge function is
\begin{equation}
{}^{\sharp}\psi_{0}(z)=\prod^{m}_{i=1}\frac{1}{z-h_i}   \,,
\end{equation}
with $h_{i}$ the weight of the $i^{\textrm{th}}$ arrow.
Applying the state generating algorithm, we obtain
\begin{align}\label{eq:mframingVacuumRep}
\textrm{level-}0:&\qquad \{\varnothing\} \nonumber\\
\textrm{level-}1:&\qquad 
\{ \mathfrak{p}^{}_1\} \,,\quad  
\{\mathfrak{p}^{}_2\} \,,\quad  \dots \quad 
\{ \mathfrak{p}^{}_m\}\nonumber \\
\textrm{level-}2:&\qquad 
\{ \mathfrak{p}^{}_1,\, \mathfrak{p}^{}_2\} \,,
\quad  
\{ \mathfrak{p}^{}_1,\, \mathfrak{p}^{}_3\} \,,
\,\,  \dots \,\,  
\{ \mathfrak{p}^{}_{m-1},\, \mathfrak{p}^{}_m\} \nonumber\\
	\dots \dots &\nonumber\\
	\textrm{level-}m:&\qquad 
\{ \mathfrak{p}^{}_1,\,\mathfrak{p}^{}_2,\,\dots \mathfrak{p}^{}_m\} \,,
\end{align}
where $\mathfrak{p}^{}_i$ denotes the $i^\textrm{th}$ arrow from $\infty$ to the vertex. 

Alternatively in terms of modes, the representation can also be described by all possible products
$e_{n_i}e_{n_{i-1}}\cdots e_{0}$ acting on the ground state $|\varnothing\rangle_m$, with $m-1\geq n_{j}>n_{j-1}\geq 0$, where we have used the condition \eqref{eq:GSdef} on $|\varnothing\rangle_m$ and the $e-e$ relation \eqref{eq:m0Mode}.

Both these descriptions give the unrefined character
\begin{equation}\label{eq:ZA1funref}
	\boldsymbol{\mathcal{Z}}(x)=(1+x)^m\,.
\end{equation}
To refine the character, in terms of modes, if we use $e_n\sim y^{2n+1}$, we have
\begin{equation}\label{eq:ZA1fref}
	\boldsymbol{\mathcal{Z}}(y,x)=\prod^{m-1}_{k=0}(1+y^{2k+1}x)=(-y x;y^2)_{m}\,,
\end{equation}
which can be reproduced using the prescription \eqref{eq:upsilonKQsec4} with 
\begin{equation}
	\Upsilon(I_i)=2i-1\,, \qquad i=1,2,\dots, m
\end{equation}
which is related to \eqref{eq:UpsilonJ3} by an overall shift of $m$.

\subsection{Example: Kronecker quivers}
\label{ssec:Kronecker}

Let us now consider the $m$-Kronecker quiver, which consists of two vertices, $0$ and $1$, and $m$ arrows from $0$ to $1$: 
\begin{equation}
Q=\begin{array}{c}
\begin{tikzpicture}[scale=1]
\node[vertex,minimum size=0.5mm,font=\footnotesize] %[regular polygon, regular polygon sides=4, draw=blue!50, very thick, fill=blue!10] 
(a0) at (-2,0)  {$0$};
\node[vertex,minimum size=0.5mm,font=\footnotesize] (a1) at (2,0)  {$1$};
\draw[->>>]  node[below] {$I_{i=1,2,\dots,m}$} (a0) -- (a1) ;
;
\end{tikzpicture}
\end{array} 
\end{equation}
Since there is no closed loop in $Q$, $W=0$, hence there is no relation between the weights of the arrows $h_i$, $i=1,2,\dots, m$.
By definition \eqref{eq:BondingFactorDef}, the bonding factors are
\begin{equation}\label{eq:varphiKronecker}
\begin{aligned}
&\varphi^{0\Leftarrow 0}(u)=e^{i\pi t_{00}}
\,, \qquad
\varphi^{1\Leftarrow 1}(u)=e^{i\pi t_{11}}\,, \,\, \\
&
\varphi^{0\Leftarrow 1}(u)=e^{i\pi t_{01}}\prod^{m}_{i=1}(u+h_i)\,, \qquad \varphi^{1\Leftarrow 0}(u)=e^{i\pi t_{10}}\frac{1}{\prod^{m}_{i=1}(u-h_i)}\,.
\end{aligned}
\end{equation}
Plugging \eqref{eq:varphiKronecker} into the definition \eqref{eq:QuadraticFields} then gives the quiver Yangian for the $m$-Kronecker quiver.
One natural choice of the statistic factors is
\begin{equation}
\begin{aligned}
&t_{00}=t_{11}=t_{01}=0\,, \quad t_{10}=1\,, \qquad s_{00}=s_{11}=s_{10}=s_{01}=1\,.
\end{aligned}
\end{equation} 

\subsubsection{Vacuum representations}

The vacuum representation for the $m$-Kronecker quiver can then be constructed iteratively using the procedure of Sec.\ \ref{sssec:RepresentationConstruction}.
\begin{align}\label{eq:KroneckerVacuumRep}
\textrm{level-}0:&\qquad \{\varnothing\}\\
\textrm{level-}1:&\qquad \{I^{(0)}\}\nonumber\\
\textrm{level-}2:&\qquad 
\{I^{(0)},\, \mathfrak{p}^{0\rightarrow 1}_1\} \,,\quad  
\{I^{(0)},\, \mathfrak{p}^{0\rightarrow 1}_2\} \,,\quad  \dots \quad 
 \{I^{(0)},\, \mathfrak{p}^{0\rightarrow 1}_m\}\nonumber \\
\textrm{level-}3:&\qquad 
\{I^{(0)},\, \mathfrak{p}^{0\rightarrow 1}_1,\, \mathfrak{p}^{0\rightarrow 1}_2\} \,,
\quad  
\{I^{(0)},\, \mathfrak{p}^{0\rightarrow 1}_1,\, \mathfrak{p}^{0\rightarrow 1}_3\} \,,
\,\,  \dots \,\,  
\{I^{(0)},\, \mathfrak{p}^{0\rightarrow 1}_{m-1},\, \mathfrak{p}^{0\rightarrow 1}_m\} \nonumber\\
\dots \dots &\nonumber\\
\textrm{level-}(m+1):&\qquad 
\{I^{(0)},\, \mathfrak{p}^{0\rightarrow 1}_1,\,\mathfrak{p}^{0\rightarrow 1}_2,\,\dots \mathfrak{p}^{0\rightarrow 1}_m\} \,,\nonumber
\end{align}
where we have defined the shorthand notation 
\begin{equation}
\mathfrak{p}^{0\rightarrow 1}_{i}\equiv I^{(0)}\cdot I^{0\rightarrow 1}_i\,.
\end{equation}
We see that the representation shares a similar structures as the representation \eqref{eq:mframingVacuumRep} of the single vertex quiver with $m$ framing arrows.
Counting these states then gives the unrefined vacuum character
\begin{equation}
\boldsymbol{\mathcal{Z}}^{(m)}_{\textrm{Kronecker}}(\boldsymbol{x})=1+x_0 \, \sum^{m}_{n=0} \binom{m}{n} x_1^n\,.
\end{equation}

\medskip

Since this is a quiver without any loop, therefore no superpotential and hence no higher order relations, we can also easily describe the representation in terms of $e^{(a)}_n$ acting on the ground state $|\varnothing\rangle$.
One can simply consider all states $\prod e^{(a)}_n |\varnothing\rangle$ and then remove dependent ones by using the mode version of the quadratic relations.
We list the $1+2^m$ states in the vacuum representation for the first few $m$'s.
\begin{enumerate}
\item $m=1$
\begin{equation}\label{eq:KroStates1}
|\varnothing\rangle 
\,, \qquad e^{(0)}_0|\varnothing\rangle
\,, \qquad e^{(1)}_0 e^{(0)}_0|\varnothing\rangle\,.
\end{equation}
\item $m=2$
\begin{equation}\label{eq:KroStates2}
|\varnothing\rangle 
\,, \qquad e^{(0)}_0|\varnothing\rangle
\,, \qquad e^{(1)}_0 e^{(0)}_0|\varnothing\rangle
\,, \qquad e^{(1)}_1 e^{(0)}_0|\varnothing\rangle
\,, \qquad e^{(1)}_1e^{(1)}_0e^{(0)}_0|\varnothing\rangle\,.
\end{equation}
\item $m=3$
\begin{equation}\label{eq:KroStates3}
\begin{aligned}
&|\varnothing\rangle 
\,, \quad e^{(0)}_0|\varnothing\rangle
\,, \quad e^{(1)}_0 e^{(0)}_0|\varnothing\rangle
\,, \quad e^{(1)}_1 e^{(0)}_0|\varnothing\rangle
\,, \quad e^{(1)}_2 e^{(0)}_0|\varnothing\rangle
\,, \quad \\ &e^{(1)}_1e^{(1)}_0e^{(0)}_0|\varnothing\rangle\,
\,, \quad e^{(1)}_2 e^{(1)}_0e^{(0)}_0|\varnothing\rangle
\,, \quad e^{(1)}_2 e^{(1)}_1e^{(0)}_0|\varnothing\rangle
\,, \quad e^{(1)}_2 e^{(1)}_1e^{(1)}_0e^{(0)}_0|\varnothing\rangle\,.
\end{aligned}
\end{equation}
\end{enumerate}

\subsubsection{Refinement}

Let us first compute the refined characters by considering the representation in terms of $e^{(a)}_n$ acting on the vacuum. 
With $e^{(0)}_n\sim e^{(1)}_n\sim y^{2n+1}$ and
\begin{equation} 
e^{(0)}_n e^{(0)}_{n'} \sim e^{(1)}_n e^{(1)}_{n'}\sim  y^{2(n+n'+1)}
\,,\qquad 
e^{(0)}_n e^{(1)}_{n'} \sim y^{2m} e^{(1)}_n e^{(0)}_{n'}\sim  y^{2(n+n'+1)+m}\,.
\end{equation}
we obtain the refined character
\begin{equation}\label{eq:ZrefKroNew}
\boldsymbol{\mathcal{Z}}^{(m)}_{\textrm{Kronecker}}(y,\boldsymbol{x})=1+yx_0\, \sum^{m}_{n=0} x_1^n \left(\sum^m_{i_1=1} y^{2 i_1-m-1} \sum^{m}_{i_2=i_1+1} y^{2 i_2-m-1} \dots \sum^{m}_{i_n=i_{n-1}+1}  y^{2 i_n-m-1} \right)\,.
\end{equation}
For example, the states in \eqref{eq:KroStates1}, \eqref{eq:KroStates2}, \eqref{eq:KroStates3} reproduce  precisely
\begin{align}\label{eq:ZmKro}
	\boldsymbol{\mathcal{Z}}^{(1)}_{\textrm{Kronecker}}(y, \boldsymbol{x})&=1+yx_0+yx_0\, x_1\,;\\
	\boldsymbol{\mathcal{Z}}^{(2)}_{\textrm{Kronecker}}(y, \boldsymbol{x})&=1+yx_0+(y^2+1)\, x_0\, x_1+yx_0\, x_1^2\,;\nonumber\\
	\boldsymbol{\mathcal{Z}}^{(3)}_{\textrm{Kronecker}}(y, \boldsymbol{x})&=1+yx_0+(y^3+y+\frac{1}{y})\, x_0\, x_1 +(y^3+y+\frac{1}{y})\, x_0\, x_1^2+yx_0\, x_1^3\,.\nonumber
\end{align}

Let us now compute the refine character using the representation in terms of ideal posets.  
To define the refinement, we choose the prescription \eqref{eq:upsilonKQsec4} with $Q'_0=\{0,1\}$ and with the  $m$ arrows transforming as a spin-$(\frac{m-1}{2})$ of $\mathfrak{su}(2)$, therefore we can assign the $\Upsilon$ charges to these arrows as 
\begin{equation}
\Upsilon(I_i)=2 i-m-1\qquad i=1,2,\dots, m
\end{equation}
by formula \eqref{eq:UpsilonJ3}. 

Note that we are still free to assign the $\Upsilon$ charges to the framing arrow $I^{\infty \rightarrow 0}$.
If we choose $\Upsilon(I^{(0)})=0$, then the refined character of the vacuum representation \eqref{eq:KroneckerVacuumRep} is 
\begin{equation}\label{eq:ZrefKro}
\boldsymbol{\mathcal{Z}}^{(m)}_{\textrm{Kronecker}}(y,\boldsymbol{x})=1+x_0\, \sum^{m}_{n=0} x_1^n \left(\sum^m_{i_1=1} y^{2 i_1-m-1} \sum^{m}_{i_2=i_1+1} y^{2 i_2-m-1} \dots \sum^{m}_{i_n=i_{n-1}+1}  y^{2 i_n-m-1} \right)\,.
\end{equation}
which is related to \eqref{eq:ZrefKro} by simply rescaling $x_0\rightarrow y x_0$.
%We list the result for the first few $m$'s:
%\begin{align}
%\boldsymbol{\mathcal{Z}}^{(1)}_{\textrm{Kronecker}}(y, \boldsymbol{x})&=1+x_0+x_0\, x_1\,;\\
%\boldsymbol{\mathcal{Z}}^{(2)}_{\textrm{Kronecker}}(y, \boldsymbol{x})&=1+x_0+(y+\frac{1}{y})\, x_0\, x_1+x_0\, x_1^2\,;\nonumber\\
%\boldsymbol{\mathcal{Z}}^{(3)}_{\textrm{Kronecker}}(y, \boldsymbol{x})&=1+x_0+(y^2+1+\frac{1}{y^2})\, x_0\, x_1 +(y^2+1+\frac{1}{y^2})\, x_0\, x_1^2+x_0\, x_1^3\,;\nonumber\\
%\boldsymbol{\mathcal{Z}}^{(4)}_{\textrm{Kronecker}}(y, \boldsymbol{x})&=1+x_0
%+(y^3+y+\frac{1}{y}+\frac{1}{y^3})\, x_0\, x_1  
%+(y^4+y^2+2+\frac{1}{y^2}+\frac{1}{y^4})\, x_0\, x_1^2\nonumber\\
%&\qquad\qquad
%+(y^3+y+\frac{1}{y}+\frac{1}{y^3})\, x_0\, x_1^3
%+x_0\, x_1^4\,.\nonumber
%\end{align}
The refined character $\boldsymbol{\mathcal{Z}}^{(m)}_{\textrm{Kronecker}}(y, \boldsymbol{x})$ is related to the partition function in the NC chamber by
\begin{equation}
\boldsymbol{\mathcal{Z}}^{(m)}_{\textrm{Kronecker}}(y, \boldsymbol{x})  =\widehat{\textrm{Sgn}}\cdot   \mathcal{Z}^{(m)}_{\textrm{NC}}(y, \boldsymbol{x})  
\end{equation}
with 
\begin{equation}
\widehat{\textrm{Sgn}}: \qquad x_0\rightarrow x_0,\, \qquad x_1\rightarrow (-1)^{m-1} x_1    \,,
\end{equation}
reproducing the motivic DT invariants for Kronecker quivers in the NC chamber, see e.g.\ \cite[Sec.\ 3]{Mozgovoy:2020has}.

\medskip

Note that it is also easy to generalize to more complicated framings. 
For example, consider the line defect that corresponds to the representation $[N_0, N_1]=[1,-1]$, where $N_{a}=|\infty \rightarrow a|$, for the $1$-Kronecker quiver. 
Note that there is now a loop $\infty\rightarrow 0\rightarrow 1\rightarrow \infty$.
Its ground state charge functions are
\begin{equation}\label{eq:psi01m1}
\psi^{(0)}_0(z)=\frac{1}{z}
\qquad \textrm{and} \qquad 
\psi^{(1)}_0(z)=z-h \,,    
\end{equation}
giving the refined character
\begin{equation}\label{eq:Z1m1}
\boldsymbol{\mathcal{Z}}(y,x)= 1+y  x_0 \,,
\end{equation}
where the second term corresponds to $e^{(1)}_0|\varnothing\rangle$.
Note that unlike the first line in  \eqref{eq:ZmKro}, which corresponds to states in \eqref{eq:KroStates1}, the character \eqref{eq:Z1m1} terminates earlier due to the zero in $\psi^{(1)}_0(z)$ in \eqref{eq:psi01m1}.
In terms of the mode relations, the loop $\infty\rightarrow 0\rightarrow 1\rightarrow \infty$ effectively imposes $e^{(1)}_0e^{(0)}_0|\varnothing\rangle=0$.
We can uplift $x_{0,1}$ to variables in the quantum torus:
\begin{equation}
\boldsymbol{\mathcal{Z}}(y,X)=1+y  X_{\gamma_0}  \,,
\end{equation}
with $X_{\gamma_0}X_{\gamma_1}=yX_{\gamma_0+\gamma_1}$, then after multiplied by $X_{-\gamma_0-\gamma_1}$ from right, it reproduces the correct framed degeneracy $X_{-\gamma_0-\gamma_1}+X_{-\gamma_1}$ \cite{Cordova:2013cea}.
It is straightforward to generalize this computation to more complicated framings and quivers, for more examples see \cite{Gaiotto:2024fso}.

\subsection{Example: quivers from toric \texorpdfstring{CY$_3$'s}{CY3s} without compact 4-cycles}

For toric CY$_3$'s, since the quiver Yangians and their vacuum representations were already defined (see \cite{Li:2020rij}), we will only focus on refining their vacuum characters and computing the refined DT invariants from them.\footnote{
In \cite{Li:2020rij}, the vacuum-like representations were all loosely called the vacuum representations.
} 
In Appx.\ \ref{appssec:toricCY3ex}, we explain how to construct these crystal representations using the procedure of Sec.\ \ref{sssec:RepresentationConstruction}.
In this section, we study the case when the  toric CY$_3$'s have no compact $4$-cycles; the case with compact $4$-cycles will be considered in the next subsection.

\subsubsection{Refining the vacuum characters of affine Yangians of \texorpdfstring{$\mathfrak{g}$}{g}}
\label{sssec:Refineghat}

There are three classes of toric CY$_3$'s without compact $4$-cycles: $\mathbb{C}^2/\mathbb{Z}_{n+1}\times \mathbb{C}$, $xy=z^m w^n$, and $\mathbb{C}^2/(\mathbb{Z}_2\times \mathbb{Z}_2)\times \mathbb{C}$; for their quivers, see  \cite[Sec.\ 8.2]{Li:2020rij},  \cite[Sec.\ 8.3]{Li:2020rij}, and \cite[Sec.\ 6]{Noshita:2021ldl}, respectively.
The corresponding quiver Yangians are the affine Yangians of $\mathfrak{g}=\mathfrak{h}\oplus \mathfrak{u}(1)$, with the Lie (super)algebra $\mathfrak{g}$ being $\mathfrak{gl}_{n+1}$, $\mathfrak{gl}_{m|n}$, and $D(2,1|\alpha)$, respectively.
These quivers are all symmetric, with the number of vertices given by
\begin{equation}
|Q_0|=\textrm{rank}(\mathfrak{g})=\textrm{rank}(\mathfrak{h})+1\,,
\end{equation}
and after imposing both loop constraints and vertex constraints, there are two parameters $\texttt{h}_{1,2}$, corresponding to the two equivariant parameters of toric CY$_3$'s.

\medskip

As we will show, the vacuum character of the affine Yangian of $\mathfrak{g}=\mathfrak{h}\oplus \mathfrak{u}(1)$ is\footnote{
The subscript ${}^{(0)}\mathcal{P}$ is to emphasize that the result is computed directly by counting configurations in the representation ${}^{(0)}\mathcal{P}$, whereas the subscript $\mathfrak{g}$ is to emphasize the dependence on the algebra $\mathfrak{g}$.
}
\begin{align}\label{eq:VacuumCharacterAffineADEProduct}
&\boldsymbol{\mathcal{Z}}^{\textrm{unref}}_{{}^{(0)}\mathcal{P}}(\boldsymbol{x}) =  
\boldsymbol{\mathcal{Z}}^{\textrm{unref}}_{\mathfrak{g}}(\boldsymbol{Q},\texttt{x}) \\
&=M(\texttt{x})^{r(\mathfrak{g})} \, \prod_{\boldsymbol{\beta} \in \Delta_{+}}\prod^{\infty}_{i=1} \prod^{\infty}_{j=1}\frac{1}{(1-\textrm{sgn}(\boldsymbol{\beta})\boldsymbol{Q}^{\boldsymbol{\beta}}\texttt{x}^{i+j-1})^{\textrm{sgn}(\boldsymbol{\beta})}\, (1-\textrm{sgn}(\boldsymbol{\beta})\boldsymbol{Q}^{-\boldsymbol{\beta}}\texttt{x}^{i+j-1})^{\textrm{sgn}(\boldsymbol{\beta})}}  \,, \nonumber
\end{align}
where 
\begin{equation}
\boldsymbol{x}\equiv(x_0,x_1,\dots,x_{r(\mathfrak{h})})
\end{equation}
are the fugacities for the $\textrm{rank}(\mathfrak{g})=\textrm{rank}(\mathfrak{h})+1$ colors, $r(\mathfrak{h})\equiv \textrm{rank}(\mathfrak{h})$, and $r(\mathfrak{g})\equiv \textrm{rank}(\mathfrak{g})$; $\boldsymbol{Q}$ and $\boldsymbol{Q}^{\boldsymbol{\beta}}$ are defined as
\begin{equation}\label{eq:QQbeta}
 \boldsymbol{Q}\equiv (x_1,x_2\dots,x_{r(\mathfrak{h})})
 \qquad \textrm{and} \qquad  \boldsymbol{Q}^{\boldsymbol{\beta}}\equiv \prod^{r(\mathfrak{h})}_{i=1} x^{\beta_i}_{i}\,,
\end{equation}
where $\boldsymbol{\beta}\in\Delta_{+}$ are the positive roots (in the $\alpha$-basis) of $\mathfrak{h}$, with $\textrm{sgn}(\boldsymbol{\beta})=1$ and $-1$ for bosonic and fermionic roots, respectively. 
The scalar $\texttt{x}$ is defined as
\begin{equation}\label{eq:qQdef}
\texttt{x}\equiv x_0 \,  \boldsymbol{Q}^{\mathbf{a}} =x_0\prod^{r(\mathfrak{h})}_{i=1} x^{a_i}_{i} \,,
 \end{equation}
with $\mathbf{a}$ being the Kac label of $\mathfrak{h}$.
Finally, $M(\texttt{x})$ is the MacMahon function \begin{equation}\label{eq:MacM}
\begin{aligned}
M(\texttt{x})&\equiv \prod^{\infty}_{k=1}\frac{1}{(1-\texttt{x}^k)^k}
%=\prod^{\infty}_{i=1}\prod^{\infty}_{j=1}\frac{1}{1-\texttt{x}^{i+j-1}}\,. 
=1+x+3\,x^2+6\, x^3+13\, x^4 +24\, x^5+\mathcal{O}(x^6)\,.
\end{aligned}
\end{equation}
We have checked the vacuum character formula \eqref{eq:VacuumCharacterAffineADEProduct} when $\mathfrak{g}$ belongs to  the three cases above, and also when $\mathfrak{g}=\mathfrak{h}\oplus \mathfrak{u}(1)$ with $\mathfrak{h}$ being a finite ADE Lie algebra (see Sec.\ \ref{ssec:ADE}).
Later in Sec.\ \ref{ssec:ADE}, we will also provide a derivation \eqref{eq:VacuumCharacterAffineADEProduct} based on the (conjectured) relation between the affine Yangian of $\mathfrak{g}$ and $\mathfrak{g}$ matrix extended $\mathcal{W}_{1+\infty}$ algebras.

\medskip

Before we refine the character \eqref{eq:VacuumCharacterAffineADEProduct}, first note that the infinite product form \eqref{eq:VacuumCharacterAffineADEProduct} can be rewritten as a (generalized) plethystic exponential\footnote{
Note that the infinite product form or the plethystic exponential form do not exist for CY$_3$ quivers with compact $4$-cycles.
}
\begin{equation}\label{eq:ZPEz}
\boldsymbol{\mathcal{Z}}^{\textrm{unref}}_{\mathfrak{g}}(\boldsymbol{Q},\texttt{x})=\textrm{PE}^{\textrm{gen}}\left[\mathfrak{z}^{\textrm{unref}}_{\mathfrak{g}}(\boldsymbol{Q},\texttt{x})\right] \,,
\end{equation}
where $\textrm{PE}^{\textrm{gen}}$ generalizes \eqref{eq:PEDef} and handles both ``bosonic" and ``fermionic" variables:
\begin{equation}\label{eq:PEgen}
\textrm{PE}^{\textrm{gen}}\left[f(x_{\textrm{B}},x_{\textrm{F}})\right]\equiv \textrm{exp}\left(\sum^{\infty}_{k=1}\frac{f( (x_{\textrm{B}})^k, (-x_{\textrm{F}})^k)}{k}\right)\,.
\end{equation} 
Here $x_{\textrm{B}}$ and $x_{\textrm{F}}$ stand for the collection of bosonic and fermionic variables, respectively, and they are the fugacities of the bosonic and fermionic vertices in $Q$, which have odd and even number of self-loops, respectively.\footnote{
Note that here we have specialized to the choice $t_{aa}=0$ in \eqref{eq:BondingFactorDef}, to match with the convention in   \cite{Li:2020rij}.
} 
(For the designation of bosonic and fermionic vertices for Lie superalgebras, see e.g.\  \cite[Sec.\ 8.3]{Li:2020rij}.)
Finally, the function $\mathfrak{z}_{\mathfrak{g}}(\boldsymbol{Q},\texttt{x})$ counts the ``single-particle" spectrum and takes the factorized form
\begin{equation}\label{eq:PLogZADE}
\begin{aligned}
\mathfrak{z}^{\textrm{unref}}_{\mathfrak{g}}(\boldsymbol{Q},\texttt{x})
&=m(\texttt{x})\, \chi_{\mathfrak{g}}(\boldsymbol{Q})\,,
\end{aligned}
\end{equation}
where $m(\texttt{x})$ is the plethystic log of the MacMahon function $M(\texttt{x})$ in \eqref{eq:MacM}:
\begin{equation}\label{eq:mdef}
m(\texttt{x})=\textrm{PLog}\left[M(\texttt{x})\right]=\frac{\texttt{x}}{(1-\texttt{x})^2} \,,
\end{equation}
and $\chi_{\mathfrak{g}}(\boldsymbol{Q})$ is  the character of the adjoint representation of $\mathfrak{g}$: 
\begin{equation}\label{eq:chi}
\chi_{\mathfrak{g}}(\boldsymbol{Q})=\textrm{rank}(\mathfrak{g})+\sum_{\boldsymbol{\beta} \in \Delta}\textrm{sgn}(\boldsymbol{\beta})\boldsymbol{Q}^{\boldsymbol{\beta}}\,,
\end{equation}
where $\boldsymbol{\beta}\in\Delta$ runs over the set of all (non-zero) roots (in the $\alpha$-basis) of the finite ADE algebra $\mathfrak{h}$.
Finally, since the sign operator $\widehat{\textrm{S}}_{\textrm{total}}$  defined in \eqref{eq:SgnTotal} is just\footnote{
Here we have used the fact that $\widehat{\textrm{S}}_{\textrm{total}}$  defined in \eqref{eq:SgnTotal} is $x_a\rightarrow -x_a$ only when the vertex $a$ has no self-loop in the quiver $Q$, whose statistic factor $S_{aa}$ is then $|a\rightarrow a|+1=1$, hence $a$ is fermionic.
Since for the toric CY$_3$ quivers, a vertex $a$ can only have zero, one, or three self-loops, the other scenario $x_a\rightarrow x_a$ happens when the vertex $a$ has one or three self-loops, with makes its statistic factor $S_{aa}=|a\rightarrow a|+1=0$ mod $2$, hence $a$ is bosonic.
} 
\begin{equation}
\widehat{\textrm{S}}_{\textrm{total}}: \qquad x_{\textrm{B}} \rightarrow x_{\textrm{B}}   
\quad \textrm{and} \quad 
x_{\textrm{F}} \rightarrow -x_{\textrm{F}}  
\end{equation}
for the affine Yangian of $\mathfrak{g}$, we can rewrite the relation  \eqref{eq:ZPEz} as:
\begin{equation}\label{eq:ZPEz1}
\boldsymbol{\mathcal{Z}}^{\textrm{unref}}_{\mathfrak{g}}(\boldsymbol{Q},\texttt{x})=\textrm{PE}^{\textrm{gen}}\left[\mathfrak{z}^{\textrm{unref}}_{\mathfrak{g}}(\boldsymbol{Q},\texttt{x})\right] 
=\widehat{\textrm{S}}_{\textrm{total}} \cdot \textrm{PE}\left[\mathfrak{z}^{\textrm{unref}}_{\mathfrak{g}}(\boldsymbol{Q},\texttt{x})\right]\,.
\end{equation}
This will be useful later for the connection to the unframed DT invariants.
Note that the sign operation $\widehat{\textrm{S}}_{\textrm{total}}$
ensures that all the coefficients in $\boldsymbol{\mathcal{Z}}^{\textrm{unref}}_{{}^{(0)}\mathcal{P}}(\boldsymbol{x})=\boldsymbol{\mathcal{Z}}^{\textrm{unref}}_{\mathfrak{g}}(\boldsymbol{Q},\texttt{x})$ are positive.

\medskip

Now we define the refinement of \eqref{eq:ZPEz}.
It is easier to first refine the single-particle partition function $\mathfrak{z}^{\textrm{unref}}_{\mathfrak{g}}(\boldsymbol{Q},\texttt{x})\rightarrow \mathfrak{z}^{\textrm{ref}}_{\mathfrak{g}}(y, \boldsymbol{Q},\texttt{x})$ directly and then compute
\begin{equation}\label{eq:ZPEzref}
\boldsymbol{\mathcal{Z}}^{\textrm{ref}}_{\mathfrak{g}}(y, \boldsymbol{Q},\texttt{x})=\textrm{PE}^{\textrm{gen}}\left[\mathfrak{z}^{\textrm{ref}}_{\mathfrak{g}}(y, \boldsymbol{Q},\texttt{x})\right]=\widehat{\textrm{S}}_{\textrm{total}} \cdot \textrm{PE}\left[\mathfrak{z}^{\textrm{ref}}_{\mathfrak{g}}(y, \boldsymbol{Q},\texttt{x})\right]\,, 
\end{equation}
where the parameter $y$ is bosonic.
Using \eqref{eq:ZPEzref}, we see that  the $\tilde{\Omega}(y,\boldsymbol{x})$ defined in \eqref{eq:tildeOmega} is 
\begin{equation}\label{eq:tildeOmegaSym1}
\widetilde{\boldsymbol{\Omega}}(y,\boldsymbol{x})=-(y-y^{-1})\textrm{PLog}\left[\widehat{\textrm{S}}_{\textrm{total}} \cdot\boldsymbol{\mathcal{Z}}^{\textrm{ref}}_{\mathfrak{g}}(y,\boldsymbol{x})\right]=-(y-y^{-1})\, \mathfrak{z}^{\textrm{ref}}_{\mathfrak{g}}(y, \boldsymbol{Q},\texttt{x})
\end{equation}
from which one obtains the final generating function of the refined DT invariants in these cases.\footnote{
Note that this fixes $\boldsymbol{\Omega}(y,\boldsymbol{x})$ only up to terms that are independent of $x_0$.
}

We will focus on the refinement that preserves the factorized form, i.e.\
\begin{equation}\label{eq:PLogZrefADE}
\begin{aligned}
\mathfrak{z}^{\textrm{unref}}_{\mathfrak{g}}(\boldsymbol{Q},\texttt{x})\quad\longrightarrow\quad \mathfrak{z}^{\textrm{ref}}_{\mathfrak{g}}(y, \boldsymbol{Q},\texttt{x})
&=m^{\textrm{ref}}(y,\texttt{x})\, \chi^{\textrm{ref}}_{\mathfrak{g}}(y, \boldsymbol{Q})\,.
\end{aligned}
\end{equation}
Namely, we will not touch the adjoint character $\chi_{\mathfrak{g}}(\boldsymbol{Q})$.
The refinement of the MacMahon function was already studied in \cite{Iqbal:2007ii, Dimofte:2009bv}.
There is in fact a one-parameter family of refinements \cite{Dimofte:2009bv}:\footnote{
This formula is related to the one in \cite[below eq.\ (2.19)]{Dimofte:2009bv} by $x=q_1q_2$ and $y=\frac{q_1}{q_2}$.
}
\begin{equation}\label{eq:MacMref}
M^{\textrm{ref}}_{\delta}(y, \texttt{x})=\prod^{\infty}_{i=1}\prod^{\infty}_{j=1}\frac{1}{1-y^{i-j+\delta}\,\texttt{x}^{i+j-1}} 
\end{equation}
which correspond to different conventions for the refined DT invariant, see later.
The refinement in \cite{Iqbal:2007ii} corresponds to $\delta=1$ whereas the one in  \cite{Dimofte:2009bv} corresponds to  $\delta=0$.
The plethystic log of the refined MacMahon function $M^{\textrm{ref}}_{\delta}(y, \texttt{x})$ \eqref{eq:MacMref} is
\begin{equation}\label{eq:mrefDef}
m^{\textrm{ref}}_{\delta}(y, \texttt{x})
=\textrm{PLog}\left[M^{\textrm{ref}}_{\delta}(y, \texttt{x})\right]
=y^{\delta}\frac{1}{y-y^{-1}}\left(\frac{\texttt{x}\, y}{1- \texttt{x}\, y}-\frac{\texttt{x}/y}{1-\texttt{x}/y}\right) \,.
\end{equation}
One can check that in the unrefined limit $y\rightarrow 1$ together with $\delta\rightarrow 0$, it reduces to $m(\texttt{x})$ \eqref{eq:mdef}.
The $\tilde{\Omega}(y,\boldsymbol{x})$ \eqref{eq:tildeOmegaSym1} is then 
\begin{equation}\label{eq:tildeOmegaSym2}
\widetilde{\boldsymbol{\Omega}}(y,\boldsymbol{x})
=-y^{\delta}\left(\sum^{\infty}_{n=1} \, (y^{n}-y^{-n})\,\texttt{x}^n \right) \chi^{\textrm{ref}}_{\mathfrak{g}}(y, \boldsymbol{Q})\,,
\end{equation}
from which one obtains the final generating series of the unframed refined DT invariants in these cases:
\begin{equation}\label{eq:OmegaSym1}
\boldsymbol{\Omega}(y,\boldsymbol{x})=-y^{\delta}\frac{\texttt{x}}{1-\texttt{x}}\, \chi^{\textrm{ref}}_{\mathfrak{g}}(y, \boldsymbol{Q})\,.
\end{equation}

\medskip

As we will see, one refinement that allows for a particularly simple description at the level of the representation (for all $\mathfrak{g}$) is given by
\begin{equation}\label{eq:SimpleRef}
m^{\textrm{ref}}(y, \texttt{x})=    m^{\textrm{ref}}_{\delta=0}(y, \texttt{x})\qquad \textrm{and}\qquad \chi^{\textrm{ref}}_{\mathfrak{g}}(y, \boldsymbol{Q}) =\chi_{\mathfrak{g}}(\boldsymbol{Q})\,,
\end{equation}
which leads to

\vspace{-1em}
{\footnotesize
\begin{align}\label{eq:VacuumCharacterAffineADEProductRef}
&\boldsymbol{\mathcal{Z}}^{\textrm{ref}}_{{}^{(0)}\mathcal{P}}(\boldsymbol{x}) =  
\boldsymbol{\mathcal{Z}}^{\textrm{ref}}_{\mathfrak{g}}(\boldsymbol{Q},\texttt{x}) \\
&=M(y, \texttt{x})^{r(\mathfrak{g})} \, \prod_{\boldsymbol{\beta} \in \Delta_{+}}\prod^{\infty}_{i=1} \prod^{\infty}_{j=1}\frac{1}{(1-\textrm{sgn}(\boldsymbol{\beta})y^{i-j}\boldsymbol{Q}^{\boldsymbol{\beta}}\texttt{x}^{i+j-1})^{\textrm{sgn}(\boldsymbol{\beta})}\, (1-\textrm{sgn}(\boldsymbol{\beta})y^{i-j}\boldsymbol{Q}^{-\boldsymbol{\beta}}\texttt{x}^{i+j-1})^{\textrm{sgn}(\boldsymbol{\beta})}}  
  \,.  \nonumber
\end{align}
}\normalsize
\noindent
To compare with other conventions in the literature, we will also discuss more complicated refinements for individual cases.
With the choice \eqref{eq:SimpleRef}, the $\boldsymbol{\Omega}(y,\boldsymbol{x})$  \eqref{eq:OmegaSym1} is further simplified into\footnote{
Note that this fixes $\boldsymbol{\Omega}(y,\boldsymbol{x})$ only up to terms that are independent of $x_0$.
}
\begin{equation}\label{eq:OmegaSym2}
\boldsymbol{\Omega}(y,\boldsymbol{x})=-\frac{\texttt{x}}{1-\texttt{x}}\, \chi_{\mathfrak{g}}(\boldsymbol{Q})\,.
\end{equation}

\subsubsection{\texorpdfstring{Affine Yangian of  $\mathfrak{gl}_{n+1}$, refined colored plane partition, and motivic DT invariants of $\mathbb{C}^2/\mathbb{Z}_{n+1}\times \mathbb{C}$}{Affine Yangian of gln+1, refined colored plane partition, and motivic DT invariants of C2/Zn+1 C}}
\label{sssec:gln}

For the toric CY$_3$'s $\mathbb{C}^2/\mathbb{Z}_{n+1}\times \mathbb{C}$, the quiver Yangians are the affine Yangians of $\mathfrak{gl}_{n+1}$, and the vacuum representation is given by the colored plane partitions. 
Let us first review the $\mathbb{C}^3$ case and then study the general $n$ case. 

\subsubsubsection{\texorpdfstring{$n=0$}{n=0} and refined plane partitions}
\label{sssec:gl1}

For the simplest toric CY$_3$, which is $\mathbb{C}^3$, the quiver  is
\begin{equation}\label{fig:C3quiver}
Q=\begin{array}{c}\begin{tikzpicture}[scale=0.8]
\node[vertex,minimum size=0.5mm,font=\footnotesize] %[regular polygon, regular polygon sides=4, draw=blue!50, very thick, fill=blue!10] 
(a1) at (0,0)  {$0$};
\path[->] 
(a1) edge [in=90, out=150, loop, thin, above left,font=\footnotesize] node {$I_3$} ()
(a1) edge [in=210, out=270, loop, thin, below left,font=\footnotesize] node {$I_1$} ()
(a1) edge [in=330, out=30, loop, thin, right,font=\footnotesize] node {$I_2$} ()
;
\end{tikzpicture}
\end{array} \quad \textrm{and}\quad W=\textrm{Tr}[I_1 I_2I_3-I_1I_3 I_2]
\end{equation}
with constraint $h_1+h_2+h_3=0$.
The corresponding quiver Yangian is the affine Yangian of $\mathfrak{gl}_1$ (see \cite[Sec.\ 5]{Li:2020rij}), which was first defined by \cite{Tsymbaliuk:2014fvq}.
The vacuum representation corresponds to the canonically framed quiver defined in \eqref{eq:CanonicalQW} and is spanned by all the plane partitions (with trivial asymptotics) \cite{Prochazka:2015deb,Datta:2016cmw,Gaberdiel:2017dbk}.
The character of the affine Yangian of $\mathfrak{gl}_1$ is given by the MacMahon function \eqref{eq:MacM}:
\begin{equation}\label{eq:MacMexp}
\begin{aligned}
M(x)&%\equiv \prod^{\infty}_{k=1}\frac{1}{(1-x^k)^k}
=\prod^{\infty}_{i=1}\prod^{\infty}_{j=1}\frac{1}{1-x^{i+j-1}}
=1+x+3\,x^2+6\, x^3+13\, x^4 +\mathcal{O}(x^5)\,.
\end{aligned}
\end{equation}

\medskip

As we mentioned before, the refinement of the MacMahon function was already studied in \cite{Iqbal:2007ii, Dimofte:2009bv}, and there is in fact a one-parameter family of refinements \cite{Dimofte:2009bv} given by \eqref{eq:MacMref}.
We will choose the refinement with $\delta=0$ (as in \cite{Dimofte:2009bv}) since it gives a character that is symmetric under $y\rightarrow \frac{1}{y}$.
To realize this, we set the three arrows $I_{1,2,3} \in Q_1$ to transform as the adjoint representation ($j=1$) under the $\mathfrak{su}(2)$. 
Therefore, by formula \eqref{eq:UpsilonJ3}, we have 
\begin{equation}\label{eq:Upsilongln}
\Upsilon(I_1)=2\,,\qquad \Upsilon(I_2)=-2\,,\qquad  \Upsilon(I_3)=0\,.
\end{equation}
Next, the prescription \eqref{eq:UpsilonAtom} and \eqref{eq:upsilonAtomType1} leads to the $y$-charge of an atom:
\begin{equation}\label{eq:upsilonAtomgl1F1}
\upsilon(\sqbox{$0$})=\textrm{sgn}\left(\textrm{number of }I_1\in \sqbox{$0$} - \textrm{number of }I_2\in \sqbox{$0$} \right)    \,.
\end{equation}
Since in this case, the representation has the geometric meaning of a plane partition, we can also label each atom by its 3D coordinate $(x_1,x_2,x_3)$, with the origin atom at $(0,0,0)$.
Therefore the formula \eqref{eq:upsilonAtomgl1F1} can be also be written as
\begin{equation}\label{eq:upsilonAtomgl1F2}
\upsilon(\sqbox{$0$})=\textrm{sgn}\left(x_1-x_2\right)   \,. 
\end{equation}
Namely, we divide the plane partitions  by half using the $x_1=x_2$ line, and the atoms with $x_1>x_2$ (resp.\ $x_1<x_2$) have charge $+1$ (resp.\ $-1$), and those that sit on the line have charge $0$.
Namely, an atom has $2 J_3$ value $(1,0,-1)$, depending on which side of the line it is on.

With this prescription, one can then directly compute the refined vacuum character by definition \eqref{eq:ZrefinedP}, and this leads to
\begin{align}\label{eq:ZMacMrefdelta0}
\displaystyle
&\boldsymbol{\mathcal{Z}}^{\textrm{ref}}_{{}^{(0)}\mathcal{P}}(y,x)=M^{\textrm{ref}}_{\delta=0}(y, \texttt{x})=\prod^{\infty}_{i=1}\prod^{\infty}_{j=1}\frac{1}{1-y^{i-j}x^{i+j-1}}  \\
&=\scalebox{0.85}{$1+x+(y+1+\tfrac{1}{y})\, x^2+(y^2+y+2+\tfrac{1}{y}+\tfrac{1}{y^2})\, x^3+(y^3+2y^2+2y+3+\tfrac{2}{y}+\tfrac{2}{y^2}+\tfrac{1}{y^3})\, x^4+\mathcal{O}(x^5)$} \,. \nonumber
\end{align}
When $y\rightarrow 1$, it reduces to the MacMahon function \eqref{eq:MacM}.
One can check that it agrees with the general formula \eqref{eq:VacuumCharacterAffineADEProduct} with $\mathfrak{h}=A_0$.

Applying the change of signs \eqref{eq:SgnSym}, we obtain the generating function of the framed refined DT invariants in the NCDT chamber:
\begin{equation}
\mathcal{Z}^{\textrm{ref}}_{\textrm{NC}}(y,x)=\boldsymbol{\mathcal{Z}}_{{}^{(0)}\mathcal{P}}^{\textrm{ref}}(y,-x)
= \sum_{d=0}\overline{\underline{\Omega}}^{\textrm{NC}}_{\,d}(y)\, x^{d}\,.
\end{equation}
Using the fact that the operation defined in \eqref{eq:SgnTotal} is 
$\widehat{\textrm{S}}_{\textrm{total}}=1$ in this case, and taking the plethystic log of $\boldsymbol{\mathcal{Z}}^{\textrm{ref}}_{{}^{(0)}\mathcal{P}}(y,x)$, we then have
\begin{equation}
\tilde{\Omega}(y,x)=-(y-y^{-1})\,\textrm{PLog}\left[\boldsymbol{\mathcal{Z}}^{\textrm{ref}}_{{}^{(0)}\mathcal{P}}(y,x)\right]=-\sum^{\infty}_{n=1} \, (y^{n}-y^{-n})\,x^n \,,
\end{equation}
from which one obtains the generating series of the unframed motivic DT invariants:
\begin{equation}
\boldsymbol{\Omega}(y,x)=- \frac{x}{1-x}\,.
\end{equation}
Note that for general values of $\delta$, the refinement \eqref{eq:MacMref} corresponds to\footnote{
This is consistent with the result with the choice $\delta=1$ in \cite{Iqbal:2007ii} and the choice $\delta=3$ in \cite{Mozgovoy:2020has}.
} 
\begin{equation}
\boldsymbol{\Omega}_{\delta}(y,x)=-y^{\delta}  \frac{x}{1-x} \,. 
\end{equation}

\subsubsubsection{General \texorpdfstring{$n$}{n} and refined colored plane partitions}

For the toric CY$_3$'s $\mathbb{C}^2/\mathbb{Z}_{n+1}\times \mathbb{C}$, the quiver is given by the tripling of the Dynkin diagram of the affine $A_n$ algebra:
\begin{equation}\label{eq:AffineAnQuiver}
\begin{array}{c}
 \begin{tikzpicture}[
block/.style={
circle, draw, minimum size={width("$\tiny{n\!\!-\!\!1}$")+0pt},
font=\small},scale=0.8]
\node[vertex,minimum size=0.5mm,font=\scriptsize] (a0) at (0,2) {0};
\node[vertex,minimum size=0.5mm,font=\scriptsize] (a1) at (-4,0) {1};
\node[vertex,minimum size=0.5mm,font=\scriptsize] (a2) at (-2,0) {2};
%\vertex (a0) at (0,2) {0};
%\vertex (a1) at (-4,0) {1};
%\vertex (a2) at (-2,0) {2};
\node (adots) at (0,0) {$\dots$};
\node[vertex,minimum size=0.1mm,font=\tiny] (an-1) at (2,0) {$n\!\!-\!\!1$};
\node[vertex,minimum size=0.5mm,font=\scriptsize] (an) at (4,0) {$n$};
%\vertex (an-1) at (2,0) {$n-1$};
%\vertex (an) at (4,0) {$n$};

\path[-{Latex[length=2mm, width=1.5mm]},every loop/.append style=-{Latex[length=2mm, width=1.5mm]}] 

%%%%%%%%%%%%%%%%%%%%%
(a0) edge [in=60, out=120, loop, thin, above,font=\scriptsize] node {$C_0$} ()
(a1) edge [in=150, out=210, loop, thin, left,font=\scriptsize] node {$C_1$} ()
(an) edge [in=330, out=30, loop, thin, right,font=\scriptsize] node {$C_n$} ()
(a2) edge [in=300, out=240, loop, thin, below,font=\scriptsize] node {$C_2$} ()
 (an-1) edge [in=300, out=240, loop, thin, below,font=\scriptsize] node {$C_{n-1}$} ()

%%%%%%%%%%%%%%%%%%%%%%%%%%%%
(a1) edge   [thin, bend left,font=\scriptsize]  node [above] {$B_0$} (a0) 
(a0) edge   [thin,font=\scriptsize]  node [right]{$A_0$} (a1)
%%%%%%%%%%%%%%%%%%%%%%%%%%%%
(a1) edge   [thin,font=\scriptsize]  node [above] {$A_1$} (a2) 
(a2) edge   [thin, bend left,font=\scriptsize]  node [below]{$B_1$} (a1)
%%%%%%%%%%%%%%%%%%%%%%%%%%%%
(a2) edge   [thin,font=\scriptsize]  node [above] {$A_2$} (adots) 
(adots) edge   [thin, bend left,font=\scriptsize]  node [below]{$B_2$} (a2)
%%%%%%%%%%%%%%%%%%%%%%%%%%%%
(adots) edge   [thin,font=\scriptsize]  node [above] {$A_{n-2}$} (an-1) 
(an-1) edge   [thin, bend left,font=\scriptsize]  node [below]{$B_{n-2}$} (adots)
%%%%%%%%%%%%%%%%%%%%%%%%%%%%
(an-1) edge   [thin,font=\scriptsize]  node [above] {$A_{n-1}$} (an) 
(an) edge   [thin, bend left,font=\scriptsize]  node [below]{$B_{n-1}$} (an-1)
%%%%%%%%%%%%%%%%%%%%%%%%%%%%
(an) edge   [thin,font=\scriptsize]  node [left] {$A_{n}$} (a0) 
(a0) edge   [thin,bend left,font=\scriptsize]  node [above]{$B_n$} (an);
\end{tikzpicture}
\end{array}
\end{equation}
The weight assignment is
\begin{equation}
h(A_i)=h_1\,, \qquad h(B_i)=h_2\, \qquad h(C_i)=h_3 \qquad i=0,1,\dots, n\,,
\end{equation}
with the constraint $h_1+h_2+h_3=0$.
The superpotential is 
\begin{equation}
W=\sum^{n}_{i=0}\textrm{Tr}[A_i B_i C_i-B_{i-1}A_{i-1}C_i] \,.
\end{equation}
For the corresponding quiver Yangians and their crystal representations see \cite[Sec.\ 8.2]{Li:2020rij}; we have also shown in Appx.\ \ref{appssec:PosetAn} how to reproduce the crystal representation using the procedure of Sec.\ \ref{sssec:RepresentationConstruction}.

The states in the vacuum representation can be described by plane partitions colored by the color function $c\in \mathbb{Z}/(n+1)\mathbb{Z}$ 
\begin{equation}
    c(x_1,x_2,x_3)=x_1-x_2 \quad \textrm{mod }n+1\,.
\end{equation}
Namely,  the box at the origin (with coordinate $(0,0,0)$) has color $0$; and if a box at the coordinate $(x_1,x_2,x_3)$ has color $a$, then the colors of the three boxes that are its nearest neighbors along the positive directions are :
\begin{equation}\label{eq:PPColorScheme}
(x_1+1,x_2,x_3): \,\, a+1\,, \quad  (x_1,x_2+1,x_3): \,\, a-1\,, \quad   (x_1,x_2,x_3+1): \,\, a\,, 
\end{equation}
with the convention that $0-1=n$.
For example, for $A_3$, there are four vertices $a=0,1,2,3$, corresponding to the four colors of the boxes in the plane partition:
\begin{equation}
A_3: \qquad 
    0:\quad  
\begin{tikzpicture}[scale=0.5]
\pgfmathsetmacro{\cubex}{1}
\pgfmathsetmacro{\cubey}{1}
\pgfmathsetmacro{\cubez}{1}
\draw[black,fill=blue!50] (0,0,0) -- ++(-\cubex,0,0) -- ++(0,-\cubey,0) -- ++(\cubex,0,0) -- cycle;
\draw[black,fill=blue!50] (0,0,0) -- ++(0,0,-\cubez) -- ++(0,-\cubey,0) -- ++(0,0,\cubez) -- cycle;
\draw[black,fill=blue!50] (0,0,0) -- ++(-\cubex,0,0) -- ++(0,0,-\cubez) -- ++(\cubex,0,0) -- cycle;
\end{tikzpicture} \qquad 
1:\quad 
\begin{tikzpicture}[scale=0.5]
\pgfmathsetmacro{\cubex}{1}
\pgfmathsetmacro{\cubey}{1}
\pgfmathsetmacro{\cubez}{1}
\draw[black,fill=red!50] (0,0,0) -- ++(-\cubex,0,0) -- ++(0,-\cubey,0) -- ++(\cubex,0,0) -- cycle;
\draw[black,fill=red!50] (0,0,0) -- ++(0,0,-\cubez) -- ++(0,-\cubey,0) -- ++(0,0,\cubez) -- cycle;
\draw[black,fill=red!50] (0,0,0) -- ++(-\cubex,0,0) -- ++(0,0,-\cubez) -- ++(\cubex,0,0) -- cycle;
\end{tikzpicture} \qquad 
2:\quad 
\begin{tikzpicture}[scale=0.5]
\pgfmathsetmacro{\cubex}{1}
\pgfmathsetmacro{\cubey}{1}
\pgfmathsetmacro{\cubez}{1}
\draw[black,fill=green!50] (0,0,0) -- ++(-\cubex,0,0) -- ++(0,-\cubey,0) -- ++(\cubex,0,0) -- cycle;
\draw[black,fill=green!50] (0,0,0) -- ++(0,0,-\cubez) -- ++(0,-\cubey,0) -- ++(0,0,\cubez) -- cycle;
\draw[black,fill=green!50] (0,0,0) -- ++(-\cubex,0,0) -- ++(0,0,-\cubez) -- ++(\cubex,0,0) -- cycle;
\end{tikzpicture} \qquad 
3:\quad 
\begin{tikzpicture}[scale=0.5]
\pgfmathsetmacro{\cubex}{1}
\pgfmathsetmacro{\cubey}{1}
\pgfmathsetmacro{\cubez}{1}
\draw[black,fill=yellow!50] (0,0,0) -- ++(-\cubex,0,0) -- ++(0,-\cubey,0) -- ++(\cubex,0,0) -- cycle;
\draw[black,fill=yellow!50] (0,0,0) -- ++(0,0,-\cubez) -- ++(0,-\cubey,0) -- ++(0,0,\cubez) -- cycle;
\draw[black,fill=yellow!50] (0,0,0) -- ++(-\cubex,0,0) -- ++(0,0,-\cubez) -- ++(\cubex,0,0) -- cycle;
\end{tikzpicture} 
\end{equation}
An example of a plane partition configuration colored with the scheme \eqref{eq:PPColorScheme} is as follows
\begin{equation}
\begin{tikzpicture}[scale=0.5]
\pgfmathsetmacro{\cubex}{1}
\pgfmathsetmacro{\cubey}{1}
\pgfmathsetmacro{\cubez}{1}
\draw[->] (-1,-1,-1) -- ++(4+\cubex,0,0); 
\draw[->] (-1,-1,-1) -- ++(0,4+\cubey,0); 
\draw[->] (-1,-1,-1) -- ++(0,0,4+\cubez);
\draw[black,fill=blue!50] (0,0,0) -- ++(-\cubex,0,0) -- ++(0,-\cubey,0) -- ++(\cubex,0,0) -- cycle;
\draw[black,fill=blue!50] (0,0,0) -- ++(0,0,-\cubez) -- ++(0,-\cubey,0) -- ++(0,0,\cubez) -- cycle;
\draw[black,fill=blue!50] (0,0,0) -- ++(-\cubex,0,0) -- ++(0,0,-\cubez) -- ++(\cubex,0,0) -- cycle;
\draw[black,fill=red!50] (0,0,1) -- ++(-\cubex,0,0) -- ++(0,-\cubey,0) -- ++(\cubex,0,0) -- cycle;
\draw[black,fill=red!50] (0,0,1) -- ++(0,0,-\cubez) -- ++(0,-\cubey,0) -- ++(0,0,\cubez) -- cycle;
\draw[black,fill=red!50] (0,0,1) -- ++(-\cubex,0,0) -- ++(0,0,-\cubez) -- ++(\cubex,0,0) -- cycle;
\draw[black,fill=green!50] (0,0,2) -- ++(-\cubex,0,0) -- ++(0,-\cubey,0) -- ++(\cubex,0,0) -- cycle;
\draw[black,fill=green!50] (0,0,2) -- ++(0,0,-\cubez) -- ++(0,-\cubey,0) -- ++(0,0,\cubez) -- cycle;
\draw[black,fill=green!50] (0,0,2) -- ++(-\cubex,0,0) -- ++(0,0,-\cubez) -- ++(\cubex,0,0) -- cycle;
\draw[black,fill=yellow!50] (0,0,3) -- ++(-\cubex,0,0) -- ++(0,-\cubey,0) -- ++(\cubex,0,0) -- cycle;
\draw[black,fill=yellow!50] (0,0,3) -- ++(0,0,-\cubez) -- ++(0,-\cubey,0) -- ++(0,0,\cubez) -- cycle;
\draw[black,fill=yellow!50] (0,0,3) -- ++(-\cubex,0,0) -- ++(0,0,-\cubez) -- ++(\cubex,0,0) -- cycle;
\draw[black,fill=blue!50] (0,0,4) -- ++(-\cubex,0,0) -- ++(0,-\cubey,0) -- ++(\cubex,0,0) -- cycle;
\draw[black,fill=blue!50] (0,0,4) -- ++(0,0,-\cubez) -- ++(0,-\cubey,0) -- ++(0,0,\cubez) -- cycle;
\draw[black,fill=blue!50] (0,0,4) -- ++(-\cubex,0,0) -- ++(0,0,-\cubez) -- ++(\cubex,0,0) -- cycle;

\draw[black,fill=yellow!50] (1,0,0) -- ++(-\cubex,0,0) -- ++(0,-\cubey,0) -- ++(\cubex,0,0) -- cycle;
\draw[black,fill=yellow!50] (1,0,0) -- ++(0,0,-\cubez) -- ++(0,-\cubey,0) -- ++(0,0,\cubez) -- cycle;
\draw[black,fill=yellow!50] (1,0,0) -- ++(-\cubex,0,0) -- ++(0,0,-\cubez) -- ++(\cubex,0,0) -- cycle;
\draw[black,fill=green!50] (2,0,0) -- ++(-\cubex,0,0) -- ++(0,-\cubey,0) -- ++(\cubex,0,0) -- cycle;
\draw[black,fill=green!50] (2,0,0) -- ++(0,0,-\cubez) -- ++(0,-\cubey,0) -- ++(0,0,\cubez) -- cycle;
\draw[black,fill=green!50] (2,0,0) -- ++(-\cubex,0,0) -- ++(0,0,-\cubez) -- ++(\cubex,0,0) -- cycle;
\draw[black,fill=red!50] (3,0,0) -- ++(-\cubex,0,0) -- ++(0,-\cubey,0) -- ++(\cubex,0,0) -- cycle;
\draw[black,fill=red!50] (3,0,0) -- ++(0,0,-\cubez) -- ++(0,-\cubey,0) -- ++(0,0,\cubez) -- cycle;
\draw[black,fill=red!50] (3,0,0) -- ++(-\cubex,0,0) -- ++(0,0,-\cubez) -- ++(\cubex,0,0) -- cycle;
\draw[black,fill=blue!50] (4,0,0) -- ++(-\cubex,0,0) -- ++(0,-\cubey,0) -- ++(\cubex,0,0) -- cycle;
\draw[black,fill=blue!50] (4,0,0) -- ++(0,0,-\cubez) -- ++(0,-\cubey,0) -- ++(0,0,\cubez) -- cycle;
\draw[black,fill=blue!50] (4,0,0) -- ++(-\cubex,0,0) -- ++(0,0,-\cubez) -- ++(\cubex,0,0) -- cycle;

\draw[black,fill=yellow!50] (1,1,0) -- ++(-\cubex,0,0) -- ++(0,-\cubey,0) -- ++(\cubex,0,0) -- cycle;
\draw[black,fill=yellow!50] (1,1,0) -- ++(0,0,-\cubez) -- ++(0,-\cubey,0) -- ++(0,0,\cubez) -- cycle;
\draw[black,fill=yellow!50] (1,1,0) -- ++(-\cubex,0,0) -- ++(0,0,-\cubez) -- ++(\cubex,0,0) -- cycle;
\draw[black,fill=yellow!50] (1,2,0) -- ++(-\cubex,0,0) -- ++(0,-\cubey,0) -- ++(\cubex,0,0) -- cycle;
\draw[black,fill=yellow!50] (1,2,0) -- ++(0,0,-\cubez) -- ++(0,-\cubey,0) -- ++(0,0,\cubez) -- cycle;
\draw[black,fill=yellow!50] (1,2,0) -- ++(-\cubex,0,0) -- ++(0,0,-\cubez) -- ++(\cubex,0,0) -- cycle;
\draw[black,fill=green!50] (2,1,0) -- ++(-\cubex,0,0) -- ++(0,-\cubey,0) -- ++(\cubex,0,0) -- cycle;
\draw[black,fill=green!50] (2,1,0) -- ++(0,0,-\cubez) -- ++(0,-\cubey,0) -- ++(0,0,\cubez) -- cycle;
\draw[black,fill=green!50] (2,1,0) -- ++(-\cubex,0,0) -- ++(0,0,-\cubez) -- ++(\cubex,0,0) -- cycle;
\draw[black,fill=red!50] (3,0,0) -- ++(-\cubex,0,0) -- ++(0,-\cubey,0) -- ++(\cubex,0,0) -- cycle;
\draw[black,fill=red!50] (3,0,0) -- ++(0,0,-\cubez) -- ++(0,-\cubey,0) -- ++(0,0,\cubez) -- cycle;
\draw[black,fill=red!50] (3,0,0) -- ++(-\cubex,0,0) -- ++(0,0,-\cubez) -- ++(\cubex,0,0) -- cycle;
\draw[black,fill=blue!50] (4,0,0) -- ++(-\cubex,0,0) -- ++(0,-\cubey,0) -- ++(\cubex,0,0) -- cycle;
\draw[black,fill=blue!50] (4,0,0) -- ++(0,0,-\cubez) -- ++(0,-\cubey,0) -- ++(0,0,\cubez) -- cycle;
\draw[black,fill=blue!50] (4,0,0) -- ++(-\cubex,0,0) -- ++(0,0,-\cubez) -- ++(\cubex,0,0) -- cycle;

\draw[black,fill=blue!50] (0,1,0) -- ++(-\cubex,0,0) -- ++(0,-\cubey,0) -- ++(\cubex,0,0) -- cycle;
%\draw[black,fill=blue!50] (0,1,0) -- ++(0,0,-\cubez) -- ++(0,-\cubey,0) -- ++(0,0,\cubez) -- cycle;
%\draw[black,fill=blue!50] (0,1,0) -- ++(-\cubex,0,0) -- ++(0,0,-\cubez) -- ++(\cubex,0,0) -- cycle;
\draw[black,fill=blue!50] (0,2,0) -- ++(-\cubex,0,0) -- ++(0,-\cubey,0) -- ++(\cubex,0,0) -- cycle;
%\draw[black,fill=blue!50] (0,2,0) -- ++(0,0,-\cubez) -- ++(0,-\cubey,0) -- ++(0,0,\cubez) -- cycle;
%\draw[black,fill=blue!50] (0,2,0) -- ++(-\cubex,0,0) -- ++(0,0,-\cubez) -- ++(\cubex,0,0) -- cycle;
\draw[black,fill=blue!50] (0,3,0) -- ++(-\cubex,0,0) -- ++(0,-\cubey,0) -- ++(\cubex,0,0) -- cycle;
\draw[black,fill=blue!50] (0,3,0) -- ++(0,0,-\cubez) -- ++(0,-\cubey,0) -- ++(0,0,\cubez) -- cycle;
\draw[black,fill=blue!50] (0,3,0) -- ++(-\cubex,0,0) -- ++(0,0,-\cubez) -- ++(\cubex,0,0) -- cycle;
\end{tikzpicture}
\end{equation}
The set of plane partitions colored in this manner is precisely the colored crystals for the toric CY$_3$ $(\mathbb{C}^2/\mathbb{Z}_{n+1})\times \mathbb{C}$.

\bigskip

As in all cases, the state-generating procedure can be easily implemented in Mathematica.
Using it one can compute the vacuum character to arbitrary high order; we give the first four levels:\footnote{
This explicit expansion is for generic $n$; for smaller $n$, one just replaces the index $i$ in $x_i$ by $i$ mod $n+1$.
}
\begin{align}\label{eq:ZAnexpansion}
&\boldsymbol{Z}^{\textrm{unref}}_{{}^{(0)}\mathcal{P}}(\boldsymbol{x})=1+x_0+(x_0^2+x_0x_1+x_0x_n) \\
&\qquad\qquad+(x_0^3+x_0^2 x_1+x^2_0 x_n+x_0 x_1x_2+x_0 x_1x_n +x_0 x_{n-1}x_n)\nonumber\\
&\qquad\qquad+(x_0^4+x_0^3 x_1+x_0^3 x_n+ x_0^2x_1^2+x_0^2x_n^2+x_0^2x_1 x_2+2x_0^2x_1 x_n+x_0^2x_{n-1}x_n \nonumber\\
&\qquad\qquad +x_0x_1x_2x_3+x_0x_1x_2x_n+x_0x_1x_{n-1}x_n+x_0x_{n-2}x_{n-1}x_n)+\mathcal{O}(\boldsymbol{x}^5) \,. \nonumber
%   +x_0^5+x_6 x_0^4+x_6^2
%   x_0^3+x_3 x_6 x_0^3+x_3
%   x_6^2 x_0^2+x_2 x_3 x_6
%   x_0^2+x_3 x_4 x_6 x_0^2+x_1
%   x_2 x_3 x_6 x_0+x_2 x_3 x_4
%   x_6 x_0+x_3 x_4 x_5 x_6
%   x_0,x_0^6+x_6 x_0^5+x_6^2
%   x_0^4+x_3 x_6 x_0^4+x_6^3
%   x_0^3+x_3 x_6^2 x_0^3+x_2
%   x_3 x_6 x_0^3+x_3 x_4 x_6
%   x_0^3+x_3^2 x_6^2 x_0^2+x_2
%   x_3 x_6^2 x_0^2+x_3 x_4
%   x_6^2 x_0^2+x_1 x_2 x_3 x_6
%   x_0^2+x_2 x_3 x_4 x_6
%   x_0^2+x_3 x_4 x_5 x_6
%   x_0^2+x_2 x_3^2 x_4 x_6
%   x_0+x_1 x_2 x_3 x_4 x_6
%   x_0+x_2 x_3 x_4 x_5 x_6
%   x_0,x_0\right\}
\end{align}
With Mathematica we have checked that the expansion obeys the closed-form formula  \eqref{eq:VacuumCharacterAffineADEProduct} with $\mathfrak{h}=A_n$, which has 
\begin{equation}\label{eq:QxAn}
\boldsymbol{Q}\equiv (x_1,x_2\dots,x_{n})\,, \qquad
\boldsymbol{Q}^{\boldsymbol{\beta}}\equiv \prod^{n}_{i=1} x^{\beta_i}_{i}\,, \qquad 
\texttt{x}\equiv \prod^{n}_{i=0} x_{i} \,,
\end{equation}
where $\boldsymbol{\beta}\in\Delta_{+}$ are the positive roots (in the $\alpha$-basis) of $A_n$, and $\textrm{sgn}(\boldsymbol{\beta})=1$ for all $\boldsymbol{\beta}$. 
One can check that if we set all $x_a=x$, for $a=0,1,\dots,n$, the character \eqref{eq:VacuumCharacterAffineADEProduct} for $\mathfrak{h}=A_n$ reduces to the MacMahon function:
\begin{equation}
\boldsymbol{\mathcal{Z}}^{\textrm{unref}}_{\mathfrak{gl}_{n+1}}(\boldsymbol{x})|_{x_a=x} =M(x) \,,
\end{equation}
which is actually a highly non-trivial formula.

\medskip

We choose a similar refinement as for the BPS quiver for the $\mathbb{C}^3$ case in \eqref{eq:Upsilongln}:
\begin{equation}
\Upsilon(I^{a\rightarrow a+1})=2\,, \qquad \Upsilon(I^{a\rightarrow a-1})=-2\,, \qquad \Upsilon(I^{a\rightarrow a})=0\,. \qquad 
\end{equation}
Using \eqref{eq:upsilonAtomType1} and the same argument around \eqref{eq:upsilonAtomgl1F1}, we have
\begin{equation}\label{eq:upsilonAtomAn}
\upsilon(\sqbox{$0$})=\textrm{sgn}\left(\textrm{x}_1-\textrm{x}_2\right)   \qquad \textrm{and} \qquad \upsilon(\sqbox{$a$})=0 \quad \textrm{for } a\neq 0\,,
\end{equation}
where $(\textrm{x}_1,\textrm{x}_2,\textrm{x}_3)$ is the coordinate of the atom of color $0$. 
The resulting refined vacuum character is \eqref{eq:VacuumCharacterAffineADEProductRef} with $\mathfrak{h}=A_n$, with variables given by \eqref{eq:QxAn}.
In the expansion of \eqref{eq:VacuumCharacterAffineADEProductRef}, the first level at which $y$ shows up is $n+2$.
Applying the change of signs \eqref{eq:SgnSym} on \eqref{eq:VacuumCharacterAffineADEProductRef} with $\mathfrak{h}=A_n$, we have the generating function of the framed refined DT invariants in the NCDT chamber:
\begin{equation}\label{eq:ZrefNCAn}
\mathcal{Z}^{\textrm{ref}}_{\textrm{NC}}(y,\boldsymbol{x})=\boldsymbol{\mathcal{Z}}_{{}^{(0)}\mathcal{P}}^{\textrm{ref}}(y,\boldsymbol{x})|_{x_0\rightarrow -x_0}
= \sum_{d=0}\overline{\underline{\Omega}}^{\textrm{NC}}_{\,d}(y)\, x^{d}\,.
\end{equation}
Using the fact that the operation defined in \eqref{eq:SgnTotal} is 
$\widehat{\textrm{S}}_{\textrm{total}}=1$ in this case, and taking the plethystic log of $\boldsymbol{\mathcal{Z}}^{\textrm{ref}}_{{}^{(0)}\mathcal{P}}(y,x)$, we find
\begin{equation}
\tilde{\Omega}(y,\boldsymbol{x})=-(y-y^{-1})\,\textrm{PLog}\left[\boldsymbol{\mathcal{Z}}^{\textrm{ref}}_{{}^{(0)}\mathcal{P}}(y,\boldsymbol{x})\right]=-\left(\sum^{\infty}_{n=1} \, (y^{n}-y^{-n})\,\texttt{x}^n  \right) \cdot \chi_{\mathfrak{gl}_{n+1}}(\boldsymbol{Q})\,,
\end{equation}
from which one obtains the generating series of the unframed motivic DT invariants:
\begin{equation}
\boldsymbol{\Omega}(y,x)= -\frac{\texttt{x}}{1-\texttt{x}}\cdot \chi_{\mathfrak{gl}_{n+1}}(\boldsymbol{Q})\,,
%\sum^{\infty}_{n=1} x^n
\end{equation}
with variables related to $\boldsymbol{x}$ by  \eqref{eq:QxAn}.
Note that unlike the framed refined DT invariants computed by the expansion of \eqref{eq:ZrefNCAn}, it is independent of $y$.
This is a feature of the particularly simple choice of the refinement. 

\subsubsection{Affine Yangian of \texorpdfstring{$\mathfrak{gl}_{n|m}$}{gln|m} and motive DT invariants of generalized conifolds}

For the toric CY$_3$'s given by the algebraic relations $xy=z^m w^n$ (the so-called generalized conifolds), the quiver Yangians are the affine Yangian of $\mathfrak{gl}_{n|m}$, see \cite[Sec.\ 8.3]{Li:2020rij}.

\begin{figure}
\begin{center}
\includegraphics[width=5 cm]{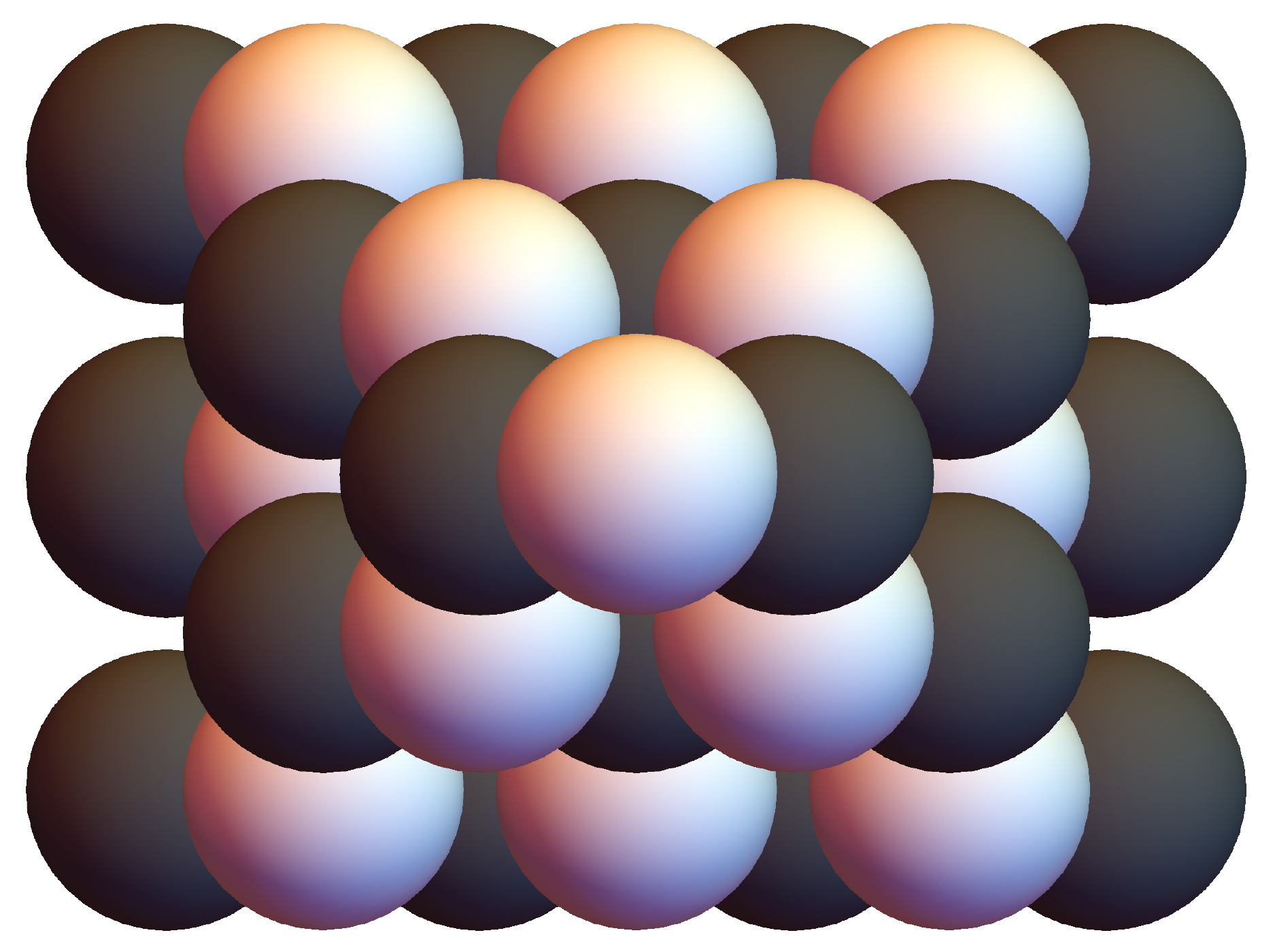}
\end{center}
\caption{The two-colored pyramid.}
\label{fig:Pyramid}
\end{figure}

Let us first consider the simplest case in this class, 
the resolved conifold, corresponding to $m=n=1$.
Its BPS quiver and superpotential are
\begin{equation}
Q=\begin{array}{c}
\begin{tikzpicture}[scale=1]
\node[vertex,minimum size=0.5mm,font=\footnotesize]  %[regular polygon, regular polygon sides=4, draw=blue!50, very thick, fill=blue!10] 
(a1) at (-2,0)  {$0$};
\node[vertex,minimum size=0.5mm,font=\footnotesize]  
%[regular polygon, regular polygon sides=4, draw=blue!50, very thick, fill=blue!10] 
(a2) at (2,0)  {$1$};
\path[->] 
(a1) edge[->>]  [thin, bend left,font=\footnotesize]  node [above] {$(A_1,h_1), \,\, (A_2,-h_1)$} (a2)
(a2) edge[->>]   [thin, bend left,font=\footnotesize]  node [below]{$(B_1,h_2), \,\, (B_2,-h_2)$} (a1);
\end{tikzpicture}
\end{array} \quad \textrm{and}\quad W=\textrm{Tr}[-A_1B_1A_2B_2+A_1B_2A_2B_1]\,.
\end{equation}
The corresponding quiver Yangian is the affine Yangian of $\mathfrak{gl}_{1|1}$ (see \cite[Sec.\ 8]{Li:2020rij}), whose vacuum representation is spanned by two-colored pyramids \cite{Szendroi}, see Fig.\ \ref{fig:Pyramid}, and its vacuum character is \cite{Young2007, bryan2010generating}\footnote{
Note that in our definition of characters directly from the crystal representation, each state counts as $1$, without signs, unlike in \cite{bryan2010generating}; our $(x_0,x_1)$ correspond to their $(q_0,-q_1)$.
}
\begin{align}\label{eq:Pyramid}
\boldsymbol{\mathcal{Z}}_{{}^{(0)}\mathcal{P}}^{\textrm{unref}}(\boldsymbol{x})&=M(x_0 x_1)^2\prod^{\infty}_{k=1}(1+x_1\, (x_0x_1)^{k})^k (1+\tfrac{1}{x_1}\, (x_0x_1)^k )^k\\
%&= M(x_0 x_1)^2\prod^{\infty}_{i=1}\prod^{\infty}_{j=1}(1+x_1\, (x_0x_1)^{i+j-1})(1+\tfrac{1}{x_1}\, (x_0x_1)^{i+j-1})\\
&=1+x_0+2\,x_0\, x_1+(4\, x_0^2 \, x_1+x_0\, x_1^2)+(2\, x_0^3 \, x_1+8\, x_0^2\, x_1^2)+\mathcal{O}(\boldsymbol{x}^5) \,,\nonumber
\end{align}
which agrees with the general formula \eqref{eq:VacuumCharacterAffineADEProduct} with specialization
\begin{equation}\label{eq:Qxgl11}
%\boldsymbol{Q}\equiv x_1\,, \qquad
\boldsymbol{Q}^{\boldsymbol{\beta}}= x_1\,, \qquad 
\textrm{sgn}(\boldsymbol{\beta})=-1\,, \qquad
\texttt{x}= x_0 x_1\,.
\end{equation}
With the substitution
\begin{equation}
x_0=-\frac{\texttt{x}}{Q} \qquad \textrm{and} \qquad x_1=-Q
\end{equation}
the vacuum character \eqref{eq:Pyramid} reproduces the unrefined BPS partition function 
\begin{equation}
\begin{aligned}
\mathcal{Z}^{\textrm{unref}}_{\textrm{NC}}(Q,\texttt{x})=\sum_{\beta, n\geq 0} \, \overline{\underline{\Omega}}_{\gamma_{\beta,n}}\,Q^{\beta}\,(-\texttt{x})^n 
=&M(\texttt{x})^2\prod^{\infty}_{k=1}(1-Q\,\texttt{x}^{k})^k (1-\tfrac{1}{Q}\,\texttt{x}^{k})^k \,, \\
%=&M(q)^2\prod^{\infty}_{i=1}\prod^{\infty}_{j=1}\frac{1}{(1-Qq^{i+j-1})(1-\frac{1}{Q}q^{i+j-1})}
\end{aligned}
\end{equation}
where $Q$ and $-\texttt{x}$ are the fugacities that count the D2 and D0 charges, respectively. 

\medskip

Again, there is a two-parameter family of refinements to  the pyramid counting, see \cite{Cirafici:2021yda}.
We choose the most symmetric one, which corresponds to the choice in which the two arrows from $0$ to $1$ both transform as spin-0 representations of $\mathfrak{su}(2)$ and thus acquire the $\Upsilon$-charge:
\begin{equation}\label{eq:conifoldA12}
\Upsilon(A_1)= 1 \qquad \textrm{and} \qquad \Upsilon(A_2)=-1\,,
\end{equation}
by \eqref{eq:UpsilonJ3};
and the two arrows from $1$ back to $0$  transform as a spin-$\frac{1}{2}$ representation of $\mathfrak{su}(2)$ and thus acquire the $\Upsilon$-charge:
\begin{equation}\label{eq:conifoldB12}
\Upsilon(B_1)= 1 \qquad \textrm{and} \qquad \Upsilon(B_2)=-1\,.
\end{equation}
Next, the prescription \eqref{eq:UpsilonAtom} and \eqref{eq:upsilonAtomType1} gives the $y$-charge of an atom as:
\begin{equation}\label{eq:upsilonAtomgl2F1}
\begin{aligned}
\upsilon(\sqbox{$0$})&=\textrm{sgn}\left(\textrm{\# of }A_1\in \sqbox{$0$} + \textrm{\# of }B_1\in \sqbox{$0$} -\textrm{\# of }A_2\in \sqbox{$0$}-\textrm{\# of }B_2\in \sqbox{$0$} \right)  \\
\upsilon(\sqbox{$1$})&=0
\end{aligned}
\end{equation}
Namely, we divide the pyramid in Fig.\ \ref{fig:Pyramid} along the diagonal line, and the atoms on the two different sides of the diagonal have $\pm 1$ charges, respectively, while those that sit on the diagonal have charge $0$.

With this prescription, one can directly compute the refine character and obtains

\vspace{-1em}
{\footnotesize
\begin{align}\label{eq:Pyramidref}
&\boldsymbol{\mathcal{Z}}_{{}^{(0)}\mathcal{P}}^{\textrm{ref}}(y,\boldsymbol{x})\equiv M(y, x_0 x_1)^2\prod^{\infty}_{i=1}\prod^{\infty}_{j=1}(1+y^{i-j}\,x_1\,(x_0x_1)^{i+j-1})(1+y^{i-j}\,x^{-1}_1\,(x_0x_1)^{i+j-1})\\
&=1+x_0+2\,x_0\, x_1+\left((y+2+\tfrac{1}{y})\, x_0^2 \, x_1+x_0\, x_1^2\right)+\left((y+\tfrac{1}{y})\, x_0^3 \, x_1+(2y+4+\tfrac{2}{y})\, x_0^2\, x_1^2\right)+\mathcal{O}(\boldsymbol{x}^5) 
\,, \nonumber
\end{align}
}\normalsize 
\noindent
which agrees with the general formula \eqref{eq:VacuumCharacterAffineADEProductRef} with specialization \eqref{eq:Qxgl11}.
(When $y\rightarrow 1$, \eqref{eq:Pyramidref} reduces to the unrefined pyramid counting function \eqref{eq:Pyramid}.)
In this case the $\widehat{\textrm{Sgn}'}$ defined in \eqref{eq:Soperation} is 
$(x_{0},x_{1})\rightarrow (x_{0},-x_{1})$, and therefore
\begin{equation}
\mathcal{Z}^{\textrm{ref}}_{\textrm{NC}}(y, \boldsymbol{x})
=\sum_{\boldsymbol{d}} \, \overline{\underline{\Omega}}^{\textrm{ref}}_{\,\boldsymbol{d}}(y)\,\boldsymbol{x}^{\boldsymbol{d}} 
=\boldsymbol{\mathcal{Z}}_{{}^{(0)}\mathcal{P}}^{\textrm{ref}}(y,x_0,-x_1) 
\end{equation}
gives the generating function of the framed refined DT invariants.
For the unframed ones, the general formula \eqref{eq:OmegaSym2} gives 
\begin{equation}\label{eq:OmegaConifold}
\begin{aligned}
\boldsymbol{\Omega}^{\textrm{ref}}(y,\boldsymbol{x})
=\sum_{\boldsymbol{d}} \Omega^{\textrm{ref}}_{\boldsymbol{d}}(y) \boldsymbol{x}^{\boldsymbol{d}}
&=-\frac{x_0x_1}{1-x_0\, x_1}\left(2- x_1-\frac{1}{x_1} \right) \\
&\simeq  \left( x_0+x_1-2x_0x_1 \right) \frac{1}{1-x_0\, x_1}\,,
\end{aligned}
\end{equation}
where we have added $-x_1$ in going to the second line to make the expression more symmetric; since $x_1$ is independent of $x_0$, this does not change $\tilde{\Omega}^{\textrm{ref}}(y,\boldsymbol{x})$ and hence not $\mathcal{Z}^{\textrm{ref}}_{\textrm{NC}}$. 
One can also directly take the plethystic log of $Z^{\textrm{ref}}(y,-\boldsymbol{x})$ to obtain $\widetilde{\boldsymbol{\Omega}}(y,\boldsymbol{x})$ defined in \eqref{eq:tildeOmega}, and then translate it into $\boldsymbol{\Omega}(y,\boldsymbol{x})$ in \eqref{eq:OmegaConifold}.

\medskip

Before we end the discussion of this example, we mention that there exist other deformations, which correspond to small numerical changes in  $\boldsymbol{\Omega}(y,\boldsymbol{x})$, but not to any structural changes in $\mathcal{Z}^{\textrm{ref}}_{\textrm{NC}}(y, \boldsymbol{x})$. 
For example, if we keep \eqref{eq:conifoldA12} and \eqref{eq:conifoldB12} but had chosen
\begin{equation}\label{eq:upsilonAtomgl2F1choice2}
\begin{aligned}
\upsilon(\sqbox{$0$})&=0\\
\upsilon(\sqbox{$1$})&=\textrm{sgn}\left(\textrm{\# of }A_1\in \sqbox{$1$} + \textrm{\# of }B_1\in \sqbox{$1$} -\textrm{\# of }A_2\in \sqbox{$1$}-\textrm{\# of }B_2\in \sqbox{$1$} \right)  \,,
\end{aligned}
\end{equation}
instead of \eqref{eq:upsilonAtomgl2F1}, then we would have gotten
\begin{equation}
\boldsymbol{\Omega}^{\textrm{ref}
}(y,\boldsymbol{x})=\left( (x_0+x_1)-(y+\tfrac{1}{y})\, x_0\, x_1 \right) \frac{1}{1-x_0\, x_1}\,,
\end{equation}
which differs from the convention in \cite{Mozgovoy:2020has} by replacing $y+\tfrac{1}{y}$ by $y+y^3$. 
This difference can be traced back to the ambiguity in defining the refined MacMahon function \eqref{eq:MacMref} \cite{Dimofte:2009bv} (see also \cite{Mozgovoy:2020has,Descombes:2021snc,Cirafici:2021yda} for a recent discussions on this).

\subsection{Example: quivers from toric \texorpdfstring{CY$_3$'s}{CY3s} with compact 4-cycles}

Let us consider a simple example of a toric CY$_3$ with a compact $4$-cycle: $\mathbb{C}^3/\mathbb{Z}_3$. 
Its quiver is
\begin{equation}\label{eq:C3Z3Quiver}
	\begin{array}{c}
		\begin{tikzpicture}[
			block/.style={
				circle, draw, minimum size={width("$\tiny{n-1}$")+0pt},
				font=\small}]
			\node[block,minimum size=0.5mm,font=\footnotesize] (a0) at (0,2) {0};
			\node[block,minimum size=0.5mm,font=\footnotesize] (a1) at (-2,0) {1};
			\node[block,minimum size=0.5mm,font=\footnotesize] (a2) at (2,0) {2};
			\path[-{Latex[length=2mm, width=1.5mm]},every loop/.append style=-{Latex[length=2mm, width=1.5mm]}] 
			%%%%%%%%%%%%%%%%%%%%%%%%%%%% 
			(a0) edge[->>>]   [thin,font=\footnotesize]  node [left]{$X^{(0)}_{1,2,3}$} (a1)
			%%%%%%%%%%%%%%%%%%%%%%%%%%%%
			(a1) edge [->>>]  [thin,font=\footnotesize]  node [below] {$X^{(1)}_{1,2,3}$} (a2) 
			(a2) edge[->>>]   [thin,font=\footnotesize]  node [right]{$X^{(2)}_{1,2,3}$} (a0);
		\end{tikzpicture}
	\end{array}
\end{equation}
with weight assignments
\begin{equation}
	h(X^{(a)}_i)=h_i\,, \qquad a=0,1,2\,, \quad i=1,2,3\,,
\end{equation}
subject to the constraint $h_1+h_2+h_3=0$.
The superpotential is 
\begin{equation}
	W=-\sum^3_{i,j,k=1}\epsilon^{ijk}\textrm{Tr}[X^{(0)}_i X^{(1)}_j X^{(2)}_k] \,,
\end{equation}
where $\epsilon^{ijk}$ is the rank-$3$ totally anti-symmetric tensor.
For the corresponding quiver Yangians and their crystal representations see \cite[Sec.\ 9.1]{Li:2020rij}.

The states in the vacuum representation can be described by plane partitions colored by a color function $c\in \mathbb{Z}/3\mathbb{Z}$ defined as
\begin{equation}\label{eq:ColorFunctionC3Z3}
	c(\textrm{x}_1,\textrm{x}_2 ,\textrm{x}_{3})=\textrm{x}_1+\textrm{x}_2 +\textrm{x}_3\quad \textrm{mod }3\,,
\end{equation}
where $(\textrm{x}_1,\textrm{x}_2 ,\textrm{x}_{3})$ is the coordinate of the box, with the box at the origin having coordinates $(0,0,0)$.
Namely, the three vertices $a=0,1,2$ correspond to three colors of the boxes in the plane partition:
\begin{equation}\label{eq:ColorFunctionC3Z3BRG}
	\mathbb{C}^3/\mathbb{Z}_3: \qquad 
	0:\quad  
	\begin{tikzpicture}[scale=0.5]
		\pgfmathsetmacro{\cubex}{1}
		\pgfmathsetmacro{\cubey}{1}
		\pgfmathsetmacro{\cubez}{1}
		\draw[black,fill=blue!50] (0,0,0) -- ++(-\cubex,0,0) -- ++(0,-\cubey,0) -- ++(\cubex,0,0) -- cycle;
		\draw[black,fill=blue!50] (0,0,0) -- ++(0,0,-\cubez) -- ++(0,-\cubey,0) -- ++(0,0,\cubez) -- cycle;
		\draw[black,fill=blue!50] (0,0,0) -- ++(-\cubex,0,0) -- ++(0,0,-\cubez) -- ++(\cubex,0,0) -- cycle;
	\end{tikzpicture} \qquad 
	1:\quad 
	\begin{tikzpicture}[scale=0.5]
		\pgfmathsetmacro{\cubex}{1}
		\pgfmathsetmacro{\cubey}{1}
		\pgfmathsetmacro{\cubez}{1}
		\draw[black,fill=red!50] (0,0,0) -- ++(-\cubex,0,0) -- ++(0,-\cubey,0) -- ++(\cubex,0,0) -- cycle;
		\draw[black,fill=red!50] (0,0,0) -- ++(0,0,-\cubez) -- ++(0,-\cubey,0) -- ++(0,0,\cubez) -- cycle;
		\draw[black,fill=red!50] (0,0,0) -- ++(-\cubex,0,0) -- ++(0,0,-\cubez) -- ++(\cubex,0,0) -- cycle;
	\end{tikzpicture} \qquad 
	2:\quad 
	\begin{tikzpicture}[scale=0.5]
		\pgfmathsetmacro{\cubex}{1}
		\pgfmathsetmacro{\cubey}{1}
		\pgfmathsetmacro{\cubez}{1}
		\draw[black,fill=green!50] (0,0,0) -- ++(-\cubex,0,0) -- ++(0,-\cubey,0) -- ++(\cubex,0,0) -- cycle;
		\draw[black,fill=green!50] (0,0,0) -- ++(0,0,-\cubez) -- ++(0,-\cubey,0) -- ++(0,0,\cubez) -- cycle;
		\draw[black,fill=green!50] (0,0,0) -- ++(-\cubex,0,0) -- ++(0,0,-\cubez) -- ++(\cubex,0,0) -- cycle;
	\end{tikzpicture} 
\end{equation}
An example of a plane partition configuration colored with the scheme \eqref{eq:ColorFunctionC3Z3BRG} is as follows
\begin{equation}
	\begin{tikzpicture}[scale=0.5]
		\pgfmathsetmacro{\cubex}{1}
		\pgfmathsetmacro{\cubey}{1}
		\pgfmathsetmacro{\cubez}{1}
		\draw[->] (-1,-1,-1) -- ++(4+\cubex,0,0); 
		\draw[->] (-1,-1,-1) -- ++(0,4+\cubey,0); 
		\draw[->] (-1,-1,-1) -- ++(0,0,4+\cubez);
		\draw[black,fill=blue!50] (0,0,0) -- ++(-\cubex,0,0) -- ++(0,-\cubey,0) -- ++(\cubex,0,0) -- cycle;
		\draw[black,fill=blue!50] (0,0,0) -- ++(0,0,-\cubez) -- ++(0,-\cubey,0) -- ++(0,0,\cubez) -- cycle;
		\draw[black,fill=blue!50] (0,0,0) -- ++(-\cubex,0,0) -- ++(0,0,-\cubez) -- ++(\cubex,0,0) -- cycle;
		\draw[black,fill=red!50] (0,0,1) -- ++(-\cubex,0,0) -- ++(0,-\cubey,0) -- ++(\cubex,0,0) -- cycle;
		\draw[black,fill=red!50] (0,0,1) -- ++(0,0,-\cubez) -- ++(0,-\cubey,0) -- ++(0,0,\cubez) -- cycle;
		\draw[black,fill=red!50] (0,0,1) -- ++(-\cubex,0,0) -- ++(0,0,-\cubez) -- ++(\cubex,0,0) -- cycle;
		\draw[black,fill=green!50] (0,0,2) -- ++(-\cubex,0,0) -- ++(0,-\cubey,0) -- ++(\cubex,0,0) -- cycle;
		\draw[black,fill=green!50] (0,0,2) -- ++(0,0,-\cubez) -- ++(0,-\cubey,0) -- ++(0,0,\cubez) -- cycle;
		\draw[black,fill=green!50] (0,0,2) -- ++(-\cubex,0,0) -- ++(0,0,-\cubez) -- ++(\cubex,0,0) -- cycle;
		\draw[black,fill=blue!50] (0,0,3) -- ++(-\cubex,0,0) -- ++(0,-\cubey,0) -- ++(\cubex,0,0) -- cycle;
		\draw[black,fill=blue!50] (0,0,3) -- ++(0,0,-\cubez) -- ++(0,-\cubey,0) -- ++(0,0,\cubez) -- cycle;
		\draw[black,fill=blue!50] (0,0,3) -- ++(-\cubex,0,0) -- ++(0,0,-\cubez) -- ++(\cubex,0,0) -- cycle;
		\draw[black,fill=red!50] (0,0,4) -- ++(-\cubex,0,0) -- ++(0,-\cubey,0) -- ++(\cubex,0,0) -- cycle;
		\draw[black,fill=red!50] (0,0,4) -- ++(0,0,-\cubez) -- ++(0,-\cubey,0) -- ++(0,0,\cubez) -- cycle;
		\draw[black,fill=red!50] (0,0,4) -- ++(-\cubex,0,0) -- ++(0,0,-\cubez) -- ++(\cubex,0,0) -- cycle;
		
		\draw[black,fill=red!50] (1,0,0) -- ++(-\cubex,0,0) -- ++(0,-\cubey,0) -- ++(\cubex,0,0) -- cycle;
		\draw[black,fill=red!50] (1,0,0) -- ++(0,0,-\cubez) -- ++(0,-\cubey,0) -- ++(0,0,\cubez) -- cycle;
		\draw[black,fill=red!50] (1,0,0) -- ++(-\cubex,0,0) -- ++(0,0,-\cubez) -- ++(\cubex,0,0) -- cycle;
		\draw[black,fill=green!50] (2,0,0) -- ++(-\cubex,0,0) -- ++(0,-\cubey,0) -- ++(\cubex,0,0) -- cycle;
		\draw[black,fill=green!50] (2,0,0) -- ++(0,0,-\cubez) -- ++(0,-\cubey,0) -- ++(0,0,\cubez) -- cycle;
		\draw[black,fill=green!50] (2,0,0) -- ++(-\cubex,0,0) -- ++(0,0,-\cubez) -- ++(\cubex,0,0) -- cycle;
		%\draw[black,fill=blue!50] (3,0,0) -- ++(-\cubex,0,0) -- ++(0,-\cubey,0) -- ++(\cubex,0,0) -- cycle;
		%\draw[black,fill=blue!50] (3,0,0) -- ++(0,0,-\cubez) -- ++(0,-\cubey,0) -- ++(0,0,\cubez) -- cycle;
		%\draw[black,fill=blue!50] (3,0,0) -- ++(-\cubex,0,0) -- ++(0,0,-\cubez) -- ++(\cubex,0,0) -- cycle;
		\draw[black,fill=red!50] (4,0,0) -- ++(-\cubex,0,0) -- ++(0,-\cubey,0) -- ++(\cubex,0,0) -- cycle;
		\draw[black,fill=red!50] (4,0,0) -- ++(0,0,-\cubez) -- ++(0,-\cubey,0) -- ++(0,0,\cubez) -- cycle;
		\draw[black,fill=red!50] (4,0,0) -- ++(-\cubex,0,0) -- ++(0,0,-\cubez) -- ++(\cubex,0,0) -- cycle;
		
		\draw[black,fill=green!50] (1,1,0) -- ++(-\cubex,0,0) -- ++(0,-\cubey,0) -- ++(\cubex,0,0) -- cycle;
		\draw[black,fill=green!50] (1,1,0) -- ++(0,0,-\cubez) -- ++(0,-\cubey,0) -- ++(0,0,\cubez) -- cycle;
		\draw[black,fill=green!50] (1,1,0) -- ++(-\cubex,0,0) -- ++(0,0,-\cubez) -- ++(\cubex,0,0) -- cycle;
		\draw[black,fill=blue!50] (1,2,0) -- ++(-\cubex,0,0) -- ++(0,-\cubey,0) -- ++(\cubex,0,0) -- cycle;
		\draw[black,fill=blue!50] (1,2,0) -- ++(0,0,-\cubez) -- ++(0,-\cubey,0) -- ++(0,0,\cubez) -- cycle;
		\draw[black,fill=blue!50] (1,2,0) -- ++(-\cubex,0,0) -- ++(0,0,-\cubez) -- ++(\cubex,0,0) -- cycle;
		\draw[black,fill=blue!50] (2,1,0) -- ++(-\cubex,0,0) -- ++(0,-\cubey,0) -- ++(\cubex,0,0) -- cycle;
		\draw[black,fill=blue!50] (2,1,0) -- ++(0,0,-\cubez) -- ++(0,-\cubey,0) -- ++(0,0,\cubez) -- cycle;
		\draw[black,fill=blue!50] (2,1,0) -- ++(-\cubex,0,0) -- ++(0,0,-\cubez) -- ++(\cubex,0,0) -- cycle;
		\draw[black,fill=blue!50] (3,0,0) -- ++(-\cubex,0,0) -- ++(0,-\cubey,0) -- ++(\cubex,0,0) -- cycle;
		\draw[black,fill=blue!50] (3,0,0) -- ++(0,0,-\cubez) -- ++(0,-\cubey,0) -- ++(0,0,\cubez) -- cycle;
		\draw[black,fill=blue!50] (3,0,0) -- ++(-\cubex,0,0) -- ++(0,0,-\cubez) -- ++(\cubex,0,0) -- cycle;
		\draw[black,fill=red!50] (4,0,0) -- ++(-\cubex,0,0) -- ++(0,-\cubey,0) -- ++(\cubex,0,0) -- cycle;
		\draw[black,fill=red!50] (4,0,0) -- ++(0,0,-\cubez) -- ++(0,-\cubey,0) -- ++(0,0,\cubez) -- cycle;
		\draw[black,fill=red!50] (4,0,0) -- ++(-\cubex,0,0) -- ++(0,0,-\cubez) -- ++(\cubex,0,0) -- cycle;
		
		\draw[black,fill=red!50] (0,1,0) -- ++(-\cubex,0,0) -- ++(0,-\cubey,0) -- ++(\cubex,0,0) -- cycle;
		%\draw[black,fill=blue!50] (0,1,0) -- ++(0,0,-\cubez) -- ++(0,-\cubey,0) -- ++(0,0,\cubez) -- cycle;
		%\draw[black,fill=blue!50] (0,1,0) -- ++(-\cubex,0,0) -- ++(0,0,-\cubez) -- ++(\cubex,0,0) -- cycle;
		\draw[black,fill=green!50] (0,2,0) -- ++(-\cubex,0,0) -- ++(0,-\cubey,0) -- ++(\cubex,0,0) -- cycle;
		%\draw[black,fill=blue!50] (0,2,0) -- ++(0,0,-\cubez) -- ++(0,-\cubey,0) -- ++(0,0,\cubez) -- cycle;
		%\draw[black,fill=blue!50] (0,2,0) -- ++(-\cubex,0,0) -- ++(0,0,-\cubez) -- ++(\cubex,0,0) -- cycle;
		\draw[black,fill=blue!50] (0,3,0) -- ++(-\cubex,0,0) -- ++(0,-\cubey,0) -- ++(\cubex,0,0) -- cycle;
		\draw[black,fill=blue!50] (0,3,0) -- ++(0,0,-\cubez) -- ++(0,-\cubey,0) -- ++(0,0,\cubez) -- cycle;
		\draw[black,fill=blue!50] (0,3,0) -- ++(-\cubex,0,0) -- ++(0,0,-\cubez) -- ++(\cubex,0,0) -- cycle;
		
	\end{tikzpicture}
\end{equation}
The set of plane partitions colored in this manner is precisely the colored crystals for the toric CY$_3$ $\mathbb{C}^3/\mathbb{Z}_3$.

\bigskip

As in all cases, the state-generating procedure can be easily implemented in Mathematica.
Using it one can compute the vacuum character to arbitrary high order; we give the first $5$ levels:
\begin{align}\label{eq:ZC3Z3expansion}
	\boldsymbol{\mathcal{Z}}^{\mathbb{C}^3/\mathbb{Z}_3,\textrm{unref}}_{{}^{(0)}\mathcal{P}}(\boldsymbol{x})=&1+x_0+3 x_0 x_1+(3 x_0 x_1^2+3x_0 x_1 x_2)+(3 x_0^2 x_1 x_2+ x_0x_1^3+9 x_0 x_1^2 x_2)\nonumber\\
	&+(9 x_0^2x_1^2x_2+6x_0x_1^3 x_2+9x_0x_1^2 x^2_2)+\mathcal{O}(\boldsymbol{x}^6)\,.
\end{align}
If we choose a similar refinement as for the BPS quiver for the $\mathbb{C}^3$ case \eqref{eq:Upsilongln}:
\begin{equation}
	\Upsilon(X^{(a)}_1)=2\,, \qquad \Upsilon(X^{(a)}_2)=-2\,, \qquad \Upsilon(X^{(a)}_3)=0\,,  \qquad a=0,1,2\,,
\end{equation}
then using \eqref{eq:upsilonAtomType1} and the same arguments as around \eqref{eq:upsilonAtomgl1F1}, we have
\begin{equation}\label{eq:upsilonAtomC3Z3}
	\upsilon(\sqbox{$0$})=\textrm{sgn}\left(\textrm{x}_1-\textrm{x}_2\right)   \qquad \textrm{and} \qquad \upsilon(\sqbox{$a$})=0 \quad \textrm{for } a=1,2\,,
\end{equation}
where $(\textrm{x}_1,\textrm{x}_2,\textrm{x}_3)$ is the coordinate of the atom of color $0$. 
The resulting refined vacuum character is 
\begin{equation}\label{eq:ZC3Z3expansionRef}
	\begin{aligned}
		\boldsymbol{\mathcal{Z}}^{\mathbb{C}^3/\mathbb{Z}_3,\textrm{ref}}_{{}^{(0)}\mathcal{P}}(y,
		\boldsymbol{x})=&1+x_0+3 x_0 x_1+(3 x_0 x_1^2+3x_0 x_1 x_2)\\
		&+\left((y+1+\tfrac{1}{y}) x_0^2 x_1 x_2+ x_0x_1^3+9 x_0 x_1^2 x_2\right)\\
		&+\left((3y+3+\tfrac{3}{y}) x_0^2x_1^2x_2+6x_0x_1^3 x_2+9x_0x_1^2 x^2_2\right)+\mathcal{O}(\boldsymbol{x}^6)\,.
	\end{aligned}
\end{equation}

\medskip

Now one might try to match with known result on motivic DT invariants, e.g.\ in \cite{Mozgovoy:2020has}.
However, note that we have adopted a refinement choice that respects the Poincar\'e duality symmetry $y\leftrightarrow \frac{1}{y}$.
On the other hand, the motivic DT invariants in e.g.\ \cite{Mozgovoy:2020has} do not have this symmetry since they capture cohomologies with compact support (on a non-compact algebraic variety).
One might first try to extract the part of the character that corresponds to $L^2$-cohomologies \cite{Lee:2016dbm,Duan:2020qjy,Mozgovoy:2020has} from their result, which does obey the Poincar\'e symmetry, and then compare to our result.
One can also explore other refinement choices that do not obey the Poincar\'e symmetry.
We leave these to future work.

\subsection{Examples: affine ADE quivers}
\label{ssec:ADE}

In this subsection, we consider the McKay quivers for the ADE groups, namely the triple quivers of the affine ADE Dynkin diagrams. 
The $A_n$ type corresponds to the toric CY$_3$'s $(\mathbb{C}^2/\mathbb{Z}_{n+1})\times \mathbb{C}$ (by the McKay correspondence), and hence has been considered in \cite{Li:2020rij}; we include type A here for comparison and also to treat the ADE class uniformly. 
The resulting quiver Yangians are the affine Yangians of $\mathfrak{g}$, where $\mathfrak{g}=\mathfrak{h}\oplus\mathfrak{u}(1)$ with $\mathfrak{h}$ the finite ADE algebras. 
We will construct their representations, and check that the characters of their vacuum representations reproduce the DT-invariants of the corresponding quiver.
Finally, we conjecture that the affine Yangian of $\mathfrak{g}$ is isomorphic to the universal enveloping algebra (UEA) of a $\mathfrak{g}$ matrix-extended $\mathcal{W}_{1+\infty}$ algebra, where $\mathfrak{g}=\mathfrak{h}\oplus\mathfrak{u}(1)$ with $\mathfrak{h}$ the ADE algebras.\footnote{ 
Note that in the ADE family, the $D_4$ case is special in that its Dynkin diagram has a 4-valent vertex (labeled by color $2$); as a result, in its vacuum representation there exist states where the constraint $\partial_{I^{2\rightarrow 2}} W=0$ is not strong enough to eliminate all but one path and the charge function has double poles; this requires us to only consider smaller representations (by adding additional framing arrows back to $\infty$) in which this problem doesn't occur, see more details in Appx.\ \ref{appssec:onframing}.} 

\subsubsection{Affine Yangian of ADE Lie algebras}

\subsubsubsection{McKay quivers from affine ADE Dynkin diagrams}

In general, the McKay correspondence refers to the study of smooth resolutions of orbifold singularities $\mathbb{C}^n/G$, for finite groups $G\subset SL(n,\mathbb{C})$, using the representation theory of $G$.\footnote{
One can even be more general and consider the cases where $\mathbb{C}^n$ is replaced by an $n$ complex-dimensional variety and $G$ by one of its finite automorphism groups \cite{Bridgeland:2001xf}.
}   
The original McKay correspondence states that the smooth resolutions of the simple surface singularities $\mathbb{C}^2/G$ for finite groups $G\subset SL(2,\mathbb{C})$ can be captured by the extended Dynkin diagrams of the simple Lie algebras $\mathfrak{h}$
\cite{McKay:1980}:
\begin{equation}
\begin{array}{c|ccccc}
 G\subset SL(2,C)  
% & \textrm{trivial}
 &\quad \mathbb{Z}_{n+1} \qquad
 & \qquad 2D_{2(n-2)} \qquad
 & \quad 2\mathcal{I}  \qquad
 & \qquad 2\mathcal{O} \qquad
& \qquad 2\mathcal{T}\qquad  \\
\hline
\mathfrak{h}      
& A_n 
& D_n
& \quad E_6
& \qquad E_7
& \qquad E_8
\end{array}
\end{equation}
where $2D_{2(n-2)}$, $2\mathcal{T}$, $2\mathcal{O}$ and $2\mathcal{I}$ are the binary dihedral, tetrahedral,  octahedral, and icosahedral groups (of order $2(n-2)$, 24, 48 and 120), respectively.

In general, a McKay quiver $Q(G,V)$, for $G$ a finite group and $V$ one of its finite linear representations,  is defined as follows. 
Each irreducible representation $\rho_i$ of $G$ corresponds to a vertex $a_i$. (We will always include the trivial representation $\rho_0$.)
The number of arrows $n_{ij}$ between $a_i$ and $a_j$ is determined by the decomposition
\begin{equation}
    V\otimes \rho_i = \oplus n_{ij} \, \rho_j\,.
\end{equation}
For the original McKay correspondence%of \cite{McKay:1980}
, $G\subset SU(2)$ and  $V\simeq\mathbb{C}^2$ is the defining representation, and the resulting McKay quivers for $\mathbb{C}^2/G$ are the ``doubling" of the corresponding extended Dynkin diagrams,
namely, for each connected pair $(a,b)$ of vertices in the extended Dynkin diagram, there is a pair of arrows $a\rightarrow b$ and $b\rightarrow a$ in the corresponding quiver \cite{McKay:1980}. 

\medskip

In this section, we consider the Calabi-Yau threefolds $(\mathbb{C}^2/G) \times \mathbb{C}$, with finite group $G\subset SL(2,\mathbb{C})\subset SL(3,\mathbb{C})$. 
The corresponding McKay quivers, namely $Q(G,V)$ with $G\subset SL(2,\mathbb{C})$ but $V\simeq \mathbb{C}^3$, are McKay quivers for $Q(G,\mathbb{C}^2)$ --- i.e.\ the doubling of the corresponding extended ADE Dynkin diagram --- with additional self-loops for each vertex, which are also called ``triple quivers" (or Ginzburg quivers) of the original extended ADE quiver.\footnote{Related to this case is when the quiver is the triple quiver of the (non-affine) ADE Dynkin diagram, then the resulting quiver Yangian is the yangian of $\mathfrak{g}$ in the Chevalley basis, with $\mathfrak{g}$ being the ADE Lie algebra.}

To summarize, the McKay quivers associated with ADE Lie algebras $\mathfrak{h}$ that we will consider in this section are defined as follows.
The set of vertices is
\begin{equation}
|Q_0|
=\textrm{rank}(\mathfrak{h})+1 
=\textrm{rank}(\mathfrak{g})\,,
\end{equation}
where $\mathfrak{h}$ is an ADE Lie algebra and $\mathfrak{g}=\mathfrak{h}\oplus\mathfrak{u}(1)$.
The set of arrows $Q_1$ is given by the doubling of the extended Dynkin diagram of $\mathfrak{h}$ and an additional self-loop for each vertex.
We give these quivers explicitly, together with the numbering of the vertices and the equivariant weight assignments of the arrows:
the affine A$_{n\geq 1}$ type triple quiver was given in \eqref{eq:AffineAnQuiver}, and the affine D$_{n\geq 5}$ and affine E$_{6,7,8}$ type triple quivers are:
\begin{equation}\label{eq:AffineDnQuiver}
D_{n\geq 5}:\qquad
\begin{array}{c}
\begin{tikzpicture}[
block/.style={
circle, draw, minimum size={width("$\tiny{n-2}$")+0pt},
font=\small}]
\node[vertex,minimum size=0.5mm,font=\scriptsize] (a0) at (-2,2) {0};
\node[vertex,minimum size=0.5mm,font=\scriptsize] (a1) at (-4,0) {1};
\node[vertex,minimum size=0.5mm,font=\scriptsize] (a2) at (-2,0) {2};
\node (adots) at (0,0) {$\dots$};
\node[vertex,minimum size=0.1mm,font=\tiny] (an-2) at (2,0) {$n\!\!-\!\!2$};
\node[vertex,minimum size=0.1mm,font=\tiny] (an-1) at (2,2) {$n\!\!-\!\!1$};
\node[vertex,minimum size=0.5mm,font=\scriptsize] (an) at (4,0) {$n$};

\path[-{Latex[length=2mm, width=1.0mm]},every loop/.append style=-{Latex[length=2mm, width=1.0mm]}] 
(a0) edge [in=60, out=120, loop, thin, above,font=\scriptsize] node {$h_3$} ()
(an-1) edge [in=60, out=120, loop, thin, above,font=\scriptsize] node {$h_3$} ()
(a1) edge [in=150, out=210, loop, thin, left,font=\scriptsize] node {$h_3$} ()
(an) edge [in=330, out=30, loop, thin, right,font=\scriptsize] node {$h_3$} ()
(a2) edge [in=300, out=240, loop, thin, below,font=\scriptsize] node {$h_3$} ()
 (an-2) edge [in=300, out=240, loop, thin, below,font=\scriptsize] node {$h_3$} ()

(a0) edge   [thin, bend left,font=\scriptsize]  node [right] {$h_1$} (a2) 
(a2) edge   [thin, bend left,font=\scriptsize]  node [left]{$h_2$} (a0)
(a2) edge   [thin,font=\scriptsize]  node [above] {$h_1$} (a1) 
(a1) edge   [thin, bend right,font=\scriptsize]  node [below]{$h_2$} (a2)
(a2) edge   [thin,font=\scriptsize]  node [above] {$h_1$} (adots) 
(adots) edge   [thin, bend left,font=\scriptsize]  node [below]{$h_2$} (a2)
(adots) edge   [thin,font=\scriptsize]  node [above] {$h_1$} (an-2) 
(an-2) edge   [thin, bend left,font=\scriptsize]  node [below]{$h_2$} (adots)
(an-2) edge   [thin,font=\scriptsize]  node [above] {$h_1$} (an) 
(an) edge   [thin, bend left,font=\scriptsize]  node [below]{$h_2$} (an-2)
(an-1) edge   [thin, bend left,font=\scriptsize]  node [right] {$h_1$} (an-2) 
(an-2) edge   [thin, bend left,font=\scriptsize]  node [left]{$h_2$} (an-1);
\end{tikzpicture}
\end{array}
\end{equation}
\begin{equation}\label{eq:AffineE6Quiver}
E_6:\qquad
\begin{array}{c}
\begin{tikzpicture}[scale=0.6]
%\node at (0,0) {$D_4$};
\node[vertex,minimum size=0.5mm,font=\scriptsize] (a0) at (0,4) {0};
\node[vertex,minimum size=0.5mm,font=\scriptsize] (a6) at (0,2) {6};
\node[vertex,minimum size=0.5mm,font=\scriptsize] (a3) at (0,0) {3};
\node[vertex,minimum size=0.5mm,font=\scriptsize] (a2) at (-2,0) {2};
\node[vertex,minimum size=0.5mm,font=\scriptsize] (a1) at (-4,0) {1};
\node[vertex,minimum size=0.5mm,font=\scriptsize] (a4) at (2,0) {4};
\node[vertex,minimum size=0.5mm,font=\scriptsize] (a5) at (4,0) {5};
%\draw[-] (a6) -- (a0);
%\draw[-] (a3) -- (a6);
%\draw[-] (a3) -- (a2);
%\draw[-] (a2) -- (a1);
%\draw[-] (a3) -- (a4);
%\draw[-] (a4) -- (a5);
\path[-{Latex[length=2mm, width=1.0mm]},every loop/.append style=-{Latex[length=2mm, width=1.0mm]}] 
(a1) edge [in=150, out=210, loop, thin, left,font=\scriptsize] node {$h_3$} ()

(a6) edge [in=150, out=210, loop, thin, left,font=\scriptsize] node {$h_3$} ()

(a5) edge [in=330, out=30, loop, thin, right,font=\scriptsize] node {$h_3$} ()

(a0) edge [in=60, out=120, loop, thin, above,font=\scriptsize] node {$h_3$} ()

(a2) edge [in=300, out=240, loop, thin, below,font=\scriptsize] node {$h_3$} ()
(a3) edge [in=300, out=240, loop, thin, below,font=\scriptsize] node {$h_3$} ()
(a4) edge [in=300, out=240, loop, thin, below,font=\scriptsize] node {$h_3$} ()

(a3) edge   [thin,font=\scriptsize]  node [above] {$h_1$} (a4) 
(a4) edge   [thin,font=\scriptsize]  node [above] {$h_1$} (a5) 
(a4) edge   [thin, bend left,font=\scriptsize]  node [below]{$h_2$} (a3)
(a5) edge   [thin, bend left,font=\scriptsize]  node [below]{$h_2$} (a4)
(a3) edge   [thin,font=\scriptsize]  node [above] {$h_1$} (a2) 
(a2) edge   [thin,font=\scriptsize]  node [above] {$h_1$} (a1) 
(a1) edge   [thin, bend right,font=\scriptsize]  node [below]{$h_2$} (a2)
(a2) edge   [thin, bend right,font=\scriptsize]  node [below]{$h_2$} (a3)

(a0) edge   [thin, bend left,font=\scriptsize]  node [right] {$h_1$} (a6) 
(a6) edge   [thin, bend left,font=\scriptsize]  node [right] {$h_1$} (a3) 

(a3) edge   [thin, bend left,font=\scriptsize]  node [left]{$h_2$} (a6)
(a6) edge   [thin, bend left,font=\scriptsize]  node [left]{$h_2$} (a0);
%\draw[<-{Latex[length=2mm, width=1.5mm]}] (c) edge[bend right=80] node[below] {$\sigma$} (d);
\end{tikzpicture}
\end{array}
\end{equation}
\begin{equation}\label{eq:AffineE7Quiver}
E_7:\qquad
\begin{array}{c}
\begin{tikzpicture}[scale=0.6]
%\node at (0,0) {$D_4$};
\node[vertex,minimum size=0.5mm,font=\scriptsize] (a0) at (-6,0) {0};
\node[vertex,minimum size=0.5mm,font=\scriptsize] (a7) at (0,2) {7};
\node[vertex,minimum size=0.5mm,font=\scriptsize] (a3) at (0,0) {3};
\node[vertex,minimum size=0.5mm,font=\scriptsize] (a2) at (-2,0) {2};
\node[vertex,minimum size=0.5mm,font=\scriptsize] (a1) at (-4,0) {1};
\node[vertex,minimum size=0.5mm,font=\scriptsize] (a4) at (2,0) {4};
\node[vertex,minimum size=0.5mm,font=\scriptsize] (a5) at (4,0) {5};
\node[vertex,minimum size=0.5mm,font=\scriptsize] (a6) at (6,0) {6};
%\draw[-] (a6) -- (a0);
%\draw[-] (a3) -- (a6);
%\draw[-] (a3) -- (a2);
%\draw[-] (a2) -- (a1);
%\draw[-] (a3) -- (a4);
%\draw[-] (a4) -- (a5);
\path[-{Latex[length=2mm, width=1.0mm]},every loop/.append style=-{Latex[length=2mm, width=1.0mm]}] 
(a0) edge [in=150, out=210, loop, thin, left,font=\scriptsize] node {$h_3$} ()

(a1) edge [in=300, out=240, loop, thin, below,font=\scriptsize] node {$h_3$} ()

(a7) edge [in=60, out=120, loop, thin, above,font=\scriptsize] node {$h_3$} ()

(a6) edge [in=330, out=30, loop, thin, right,font=\scriptsize] node {$h_3$} ()

(a2) edge [in=300, out=240, loop, thin, below,font=\scriptsize] node {$h_3$} ()
(a3) edge [in=300, out=240, loop, thin, below,font=\scriptsize] node {$h_3$} ()
(a4) edge [in=300, out=240, loop, thin, below,font=\scriptsize] node {$h_3$} ()
(a5) edge [in=300, out=240, loop, thin, below,font=\scriptsize] node {$h_3$} ()

(a0) edge   [thin, font=\scriptsize]  node [above] {$h_1$} (a1) 
(a1) edge   [thin, font=\scriptsize]  node [above] {$h_1$} (a2) 
(a2) edge   [thin, font=\scriptsize]  node [above] {$h_1$} (a3) 
(a3) edge   [thin,font=\scriptsize]  node [above] {$h_1$} (a4) 
(a4) edge   [thin,font=\scriptsize]  node [above] {$h_1$} (a5) 
(a5) edge   [thin,font=\scriptsize]  node [above] {$h_1$} (a6) 

(a3) edge   [thin,bend right,font=\scriptsize]  node [right] {$h_1$} (a7) 

(a6) edge   [thin, bend left,font=\scriptsize]  node [below]{$h_2$} (a5)
(a5) edge   [thin, bend left,font=\scriptsize]  node [below]{$h_2$} (a4)
(a4) edge   [thin, bend left,font=\scriptsize]  node [below]{$h_2$} (a3)
(a3) edge   [thin, bend left,font=\scriptsize]  node [below]{$h_2$} (a2)
(a2) edge   [thin, bend left,font=\scriptsize]  node [below]{$h_2$} (a1)
(a1) edge   [thin, bend left,font=\scriptsize]  node [below]{$h_2$} (a0)

(a7) edge   [thin, bend right,font=\scriptsize]  node [left]{$h_2$} (a3)
(a1) edge   [thin, bend left,font=\scriptsize]  node [below]{$h_2$} (a0);
%\draw[<-{Latex[length=2mm, width=1.5mm]}] (c) edge[bend right=80] node[below] {$\sigma$} (d);
\end{tikzpicture}
\end{array}
\end{equation}
\begin{equation}\label{eq:AffineE8Quiver}
E_8:\qquad
\begin{array}{c}
\begin{tikzpicture}[scale=0.6]
%\node at (0,0) {$D_4$};
\node[vertex,minimum size=0.5mm,font=\scriptsize](a0) at (10,0) {0};
\node[vertex,minimum size=0.5mm,font=\scriptsize] (a8) at (0,2) {8};
\node[vertex,minimum size=0.5mm,font=\scriptsize] (a3) at (0,0) {3};
\node[vertex,minimum size=0.5mm,font=\scriptsize] (a2) at (-2,0) {2};
\node[vertex,minimum size=0.5mm,font=\scriptsize] (a1) at (-4,0) {1};
\node[vertex,minimum size=0.5mm,font=\scriptsize] (a4) at (2,0) {4};
\node[vertex,minimum size=0.5mm,font=\scriptsize] (a5) at (4,0) {5};
\node[vertex,minimum size=0.5mm,font=\scriptsize] (a6) at (6,0) {6};
\node[vertex,minimum size=0.5mm,font=\scriptsize] (a7) at (8,0) {7};
%\draw[-] (a6) -- (a0);
%\draw[-] (a3) -- (a6);
%\draw[-] (a3) -- (a2);
%\draw[-] (a2) -- (a1);
%\draw[-] (a3) -- (a4);
%\draw[-] (a4) -- (a5);
\path[-{Latex[length=2mm, width=1.mm]},every loop/.append style=-{Latex[length=2mm, width=1.mm]}] 
(a1) edge [in=150, out=210, loop, thin, left, font=\scriptsize] node {$h_3$} ()

(a8) edge [in=60, out=120, loop, thin, above,font=\scriptsize] node {$h_3$} ()

(a7) edge [in=300, out=240, loop, thin, below, font=\scriptsize] node {$h_3$} ()

(a0) edge [in=330, out=30, loop, thin, right,font=\scriptsize] node {$h_3$} ()

(a2) edge [in=300, out=240, loop, thin, below, font=\scriptsize] node {$h_3$} ()
(a3) edge [in=300, out=240, loop, thin, below, font=\scriptsize] node {$h_3$} ()
(a4) edge [in=300, out=240, loop, thin, below, font=\scriptsize] node {$h_3$} ()
(a5) edge [in=300, out=240, loop, thin, below, font=\scriptsize] node {$h_3$} ()
(a6) edge [in=300, out=240, loop, thin, below, font=\scriptsize] node {$h_3$} ()

(a0) edge   [thin, font=\scriptsize]  node [above] {$h_1$} (a7) 
(a7) edge   [thin, font=\scriptsize]  node [above] {$h_1$} (a6) 
(a6) edge   [thin, font=\scriptsize]  node [above] {$h_1$} (a5) 
(a5) edge   [thin, font=\scriptsize]  node [above] {$h_1$} (a4) 
(a4) edge   [thin, font=\scriptsize]  node [above] {$h_1$} (a3)
(a3) edge   [thin, font=\scriptsize]  node [above] {$h_1$} (a2)
(a2) edge   [thin, font=\scriptsize]  node [above] {$h_1$} (a1)

(a1) edge   [thin, bend right, font=\scriptsize]  node [below]{$h_2$} (a2)
(a2) edge   [thin, bend right, font=\scriptsize]  node [below]{$h_2$} (a3)
(a3) edge   [thin, bend right, font=\scriptsize]  node [below]{$h_2$} (a4)
(a4) edge   [thin, bend right, font=\scriptsize]  node [below]{$h_2$} (a5)
(a5) edge   [thin, bend right, font=\scriptsize]  node [below]{$h_2$} (a6)
(a6) edge   [thin, bend right, font=\scriptsize]  node [below]{$h_2$} (a7)
(a7) edge   [thin, bend right, font=\scriptsize]  node [below]{$h_2$} (a0)
(a3) edge   [thin, bend right, font=\scriptsize]  node [right]{$h_1$} (a8)

(a8) edge   [thin, bend right, font=\scriptsize]  node [left] {$h_2$} (a3) ;
%\draw[<-{Latex[length=2mm, width=1.5mm]}] (c) edge[bend right=80] node[below] {$\sigma$} (d);
\end{tikzpicture}
\end{array}
\end{equation}
By construction, all these McKay quivers are non-chiral:
\begin{equation}
    \chi_{ab}\equiv |a\rightarrow b|-|b\rightarrow a|=0\,, \qquad \textrm{for any } a, b \in Q_0\,.
\end{equation}
The equivariant weight assignment of the arrows is determined as follows.
First,  in all the affine ADE quivers, there is no more than one arrow between any pair of vertices, i.e.\ $|a\rightarrow b|=0$ or $1$ for any $a,b\in Q_0$.
Hence  an arrow can be uniquely labeled by its starting point and end point as $I^{a\rightarrow b}$.
The potential $W$ of these McKay quivers contains terms with 
\begin{equation}\label{eq:WADEquivers}
W=\sum_{(a,b)} \textrm{Tr}[I^{a\rightarrow a} I^{a\rightarrow b}  I^{b\rightarrow a} - I^{b\rightarrow b} I^{b\rightarrow a}  I^{a\rightarrow b}]\,,
\end{equation}
where $(a,b)$ are two neighboring vertices in $Q$, namely they are connected by $I^{a\rightarrow b}$ and $I^{b\rightarrow a}$. The superpotential \eqref{eq:WADEquivers} then suggests the assignment of equivariant weights for these arrows with the constraint on $h_i$:
\begin{equation}
    h_1+h_2+h_3=0\,.
\end{equation}
Finally, we can use the shorthand notation to denote an atom (or a path):
\begin{equation}\label{eq:ADEshorthand}
\mathfrak{p}_{a}\equiv I^{(a)} \quad \textrm{and}\quad \mathfrak{p}_{a}^{a\rightarrow a_1\rightarrow\ldots \rightarrow a_n}\equiv I^{(a)}\cdot I^{a\rightarrow a_1}\cdot I^{a_1 \rightarrow a_2}\cdot \ldots \cdot I^{a_{n-1}\rightarrow a_n}   \,. 
\end{equation}

\subsubsubsection{Affine Yangian of ADE type}

Plugging in the McKay quivers \eqref{eq:AffineAnQuiver} -- \eqref{eq:AffineE8Quiver} into the general formula  \eqref{eq:QuadraticFields} of quiver Yangians then gives the affine Yangians of $\mathfrak{g}$, where $\mathfrak{g}=\mathfrak{h}\oplus\mathfrak{u}(1)$ with $\mathfrak{h}$ the ADE Lie algebras.
Although the Calabi-Yau threefold $(\mathbb{C}^2/G) \times \mathbb{C}$ is only toric when $G=\mathbb{Z}_{n+1}$, we can nevertheless construct the representations of these algebras using the procedure of Sec.\ \ref{sssec:RepresentationConstruction}.
The characters of the vacuum representations give the framed DT-invariants of the corresponding McKay quivers.\footnote{The $D_4$ case needs to be treated separately, see Appx.\ \ref{appsssec:PosetD4}.
}
Moreover, the structure of the characters of the vacuum representations suggest an isomorphism between the affine Yangian of $\mathfrak{g}$ and (the UEA of) the $\mathfrak{g}$-extended $\mathcal{W}_{1+\infty}$ algebras. 

\subsubsection{Vacuum representations for affine ADE quiver}
\label{sssec:VacRepDn}

In this subsection, we apply the procedure of Sec.\ \ref{sssec:RepresentationConstruction} to obtain the vacuum representations of the quiver Yangians for the affine D$_{n\geq 5}$-type triple quivers \eqref{eq:AffineDnQuiver}; we explain the computations for the affine D$_{4}$ case in Appx.\ \ref{appsssec:PosetD4}, the affine A-type case \eqref{eq:AffineAnQuiver} in Appx.\ \ref{appssec:PosetAn}, and the affine E-type cases \eqref{eq:AffineE6Quiver},  \eqref{eq:AffineE7Quiver}, and \eqref{eq:AffineE8Quiver} in Appx.\ \ref{appssec:PosetE678}.

\medskip

We begin by constructing the vacuum representation for the quiver Yangian from an affine $D_{n\geq 5}$-type triple quiver \eqref{eq:AffineDnQuiver}.
Since the corresponding CY$_3$ $(\mathbb{C}^2)/\Gamma\times \mathbb{C}$ with $\Gamma=2D_{2(n-2)}$ being the binary dihedral group of order $2(n-1)$ is not toric, this set of states does not have a crystal description.
The weight assignments of the arrows are
\begin{equation}
\begin{aligned}
h_1:& \qquad I^{0\rightarrow 2}\,, I^{2\rightarrow 1}\,, I^{n-1\rightarrow n-2}\,, I^{n-2\rightarrow n}\,, \quad  \textrm{and} \quad I^{i\rightarrow i+1}\,, \quad i=2,\dots, n-3\\
h_2:& \qquad I^{2\rightarrow 0}\,, I^{1\rightarrow 2}\,, I^{n-2\rightarrow n-1}\,, I^{n\rightarrow n-2}\,, \quad  \textrm{and} \quad I^{i\rightarrow i-1}\,, \quad i=3,\dots, n-2\\
h_3:& \qquad I^{a\rightarrow a}\,, \quad a=0,\dots, n \,.
\end{aligned}    
\end{equation}

\noindent \textbf{Level 0.} There is only one state at level 0: the ground state $\Pi_{0}=\{\varnothing\}$, which is the empty set. 
By definition \eqref{eq:ChargeFunctionDef}, the charge function of $\Pi_{0}$ is just the vacuum charge function, which can be read off from the framed quiver  by definition \eqref{eq:varphi0Def}:
\begin{equation}
\Psi^{(0)}_{\Pi_0}(z)=   \psi_{0}(z)=\frac{1}{z}\,, \qquad
\Psi^{(a)}_{\Pi_0}(z)=   1 \,\,\,  \textrm{with} \,\, a=1,2,\dots,n\,,\\
\end{equation}
which has only one pole at $z=0$ for color $0$.
Together with the fact that the state $\Pi_{0}$ is the empty set, by definition \eqref{eq:RemPi} and \eqref{eq:AddPi} (or equivalently from \eqref{eq:RemVacuum} and \eqref{eq:AddVacuumAtom}), we have
\begin{equation}
\textrm{Rem}(\Pi_{0})=\{ \varnothing \} 
\qquad \textrm{and} \qquad 
\textrm{Add}(\Pi_{0})=\{ \mathfrak{p}_0 \} \,.
\end{equation}

\noindent \textbf{Level 1.} From the single pole in $\textrm{Add}(\Pi_0)$, we obtain one state at level-1:
\begin{equation}
\Pi_1=\{
\mathfrak{p}_0
\}\,,
\end{equation}
by Step-4 of the procedure in Sec.\ \ref{sssec:RepresentationConstruction}, or directly from \eqref{eq:StatesLevel1}.
Its charge functions are
\begin{equation}
\begin{aligned}
&\Psi^{(0)}_{\Pi_1}(z)=\frac{1}{z}\cdot\frac{z+h_3}{z-h_3}\,, \qquad \Psi^{(2)}_{\Pi_1}(z)=\frac{z+h_2}{z-h_1}\,, \\
&\Psi^{(a)}_{\Pi_1}(z)=   1 \qquad \qquad \textrm{for} \quad a=1,3,\dots,n\,,
\end{aligned}
\end{equation}
by definition \eqref{eq:ChargeFunctionDef}.
Then by definition \eqref{eq:RemPi} and \eqref{eq:AddPi}, we have
\begin{equation}
\textrm{Rem}(\Pi_{1})=\{
\mathfrak{p}_0
\} \qquad \textrm{and} \qquad 
\textrm{Add}(\Pi_{1})=\{
\mathfrak{p}_0^{0\rightarrow 2} ,\, \mathfrak{p}_0^{0\rightarrow 0}
\} \,.
\end{equation}

\noindent \textbf{Level 2.} From the two atoms in $\textrm{Add}(\Pi_1)$, we obtain two states at level 2 
\begin{equation}
\Pi_{2,1}= \{
\mathfrak{p}_0 ,\, 
\mathfrak{p}_0^{0\rightarrow 2}
\} \qquad \textrm{and}\qquad  
\Pi_{2,2}=\{
\mathfrak{p}_0 ,\, 
\mathfrak{p}_0^{0\rightarrow 0}
\}\,.
\end{equation}

The charge functions of $\Pi_{2,1}$ are
\begin{equation}
\begin{aligned}
&\Psi^{(0)}_{\Pi_{2,1}}(z)=\frac{1}{z-h_3}\,,   \qquad \Psi^{(2)}_{\Pi_{2,1}}(z)=\frac{z+h_3-h_1}{z-h_1}\,,  \\
& \Psi^{(1)}_{\Pi_{2,1}}(z)=\Psi^{(3)}_{\Pi_{2,1}}(z)=\frac{z+h_2-h_1}{z-2h_1}   \,, \qquad
\Psi^{(a)}_{\Pi_{2,1}}(z)=   1\,, \,\,\, \textrm{for} \,\, a=4,\dots, n\,,
\end{aligned}
\end{equation}
which gives the set of removable and addable atoms as
\begin{equation}
\begin{aligned}
\textrm{Rem}(\Pi_{2,1})=\{
\mathfrak{p}_0^{0\rightarrow 2}
\} \quad \textrm{and} \quad
\textrm{Add}(\Pi_{2,1})= \{
\mathfrak{p}_0^{0\rightarrow 2\rightarrow 1} ,\, 
\mathfrak{p}_0^{0\rightarrow 2\rightarrow 3} ,\,
\mathfrak{p}_0^{0\rightarrow 0}  \}
\,,
\end{aligned}
\end{equation}
among which the addable atoms give rise to three states at level $3$:
\begin{equation}\label{eq:DnNewStates21}
\begin{aligned}
\{
\mathfrak{p}_0 ,\,
\mathfrak{p}_0^{0\rightarrow 2} ,\, \mathfrak{p}_0^{0\rightarrow 2 \rightarrow 1} 
\}\,,  \qquad 
\{
\mathfrak{p}_0 ,\,
\mathfrak{p}_0^{0\rightarrow 2},\, \mathfrak{p}_0^{0\rightarrow 2 \rightarrow 3}
\}\,,\qquad 
&\{
\mathfrak{p}_0 ,\,
\mathfrak{p}_0^{0\rightarrow 2} ,\, \mathfrak{p}_0^{0\rightarrow 0}
\}\,.
\end{aligned}
\end{equation}
Similarly, the charge functions of $\Pi_{2,2}$ are
\begin{equation} 
\begin{aligned}
&\Psi^{(0)}_{\Pi_{2,2}}(z)=\frac{z+h_3}{(z-h_3)(z-2h_3)}\,, \qquad  \Psi^{(2)}_{\Pi_{2,2}}(z)=\frac{z+h_2-h_3}{z-h_1}\,, \\
&\Psi^{(a)}_{\Pi_{2,2}}(z)=   1 \,\,\, \textrm{for} \,\, a=1,3,\dots,n\,,
\end{aligned}
\end{equation}
which gives the set of removable and addable atoms as
\begin{equation}
\begin{aligned}
\textrm{Rem}(\Pi_{2,2})=\{
\mathfrak{p}_0^{0\rightarrow 0}
\} \qquad \textrm{and} \qquad 
\textrm{Add}(\Pi_{2,2})= \{
\mathfrak{p}_0^{0\rightarrow 2} ,\, 
\mathfrak{p}_0^{0\rightarrow 0 \rightarrow 0}  \}\,,
\end{aligned}
\end{equation}
among which the addable atoms lead to two states at level $3$:
\begin{equation}\label{eq:DnNewStates22}
\begin{aligned}
&\{
\mathfrak{p}_0,\,
\mathfrak{p}_0^{0\rightarrow 0},\, 
\mathfrak{p}_0^{0\rightarrow 2}
\}  \qquad \textrm{and} \qquad
\{
\mathfrak{p}_0,\,
\mathfrak{p}_0^{0\rightarrow 0},\, 
\mathfrak{p}_0^{0\rightarrow 0 \rightarrow 0}
\}\,.
\end{aligned}
\end{equation}
Finally, among the 5 states in \eqref{eq:DnNewStates21} and \eqref{eq:DnNewStates22}, there are only $4$ distinct ones, namely, there are $4$ new states at level 3:
\begin{equation}
\begin{aligned}
&\Pi_{3,1}=\{
\mathfrak{p}_0 ,\,
\mathfrak{p}_0^{0\rightarrow 2} ,\, \mathfrak{p}_0^{0\rightarrow 2 \rightarrow 1} 
\}\,,\qquad \quad \,\,
\Pi_{3,2}=\{
\mathfrak{p}_0 ,\,
\mathfrak{p}_0^{0\rightarrow 2},\, \mathfrak{p}_0^{0\rightarrow 2 \rightarrow 3}
\}\,, \\
&\Pi_{3,3}=\{
\mathfrak{p}_0 ,\,
\mathfrak{p}_0^{0\rightarrow 2} ,\, \mathfrak{p}_0^{0\rightarrow 0}
\}\,,
\qquad \qquad \quad
\Pi_{3,4}=\{
\mathfrak{p}_0 ,\,
\mathfrak{p}_0^{0\rightarrow 0} ,\, 
\mathfrak{p}_0^{0\rightarrow 0 \rightarrow 0}
\}\,.\\
\end{aligned}
\end{equation}

One can repeat this procedure to construct the entire poset and count the vacuum characters to arbitrary high order.  
We show the Hasse diagram of the poset up to level 3:
\begin{equation}
\begin{array}{c}
\begin{tikzpicture}
\node (a0) at (0,0)  {$\{\varnothing\}$};
\node (a11) at (0,-1)  {$\{
\mathfrak{p}_0
\}$};
\node (a21) at (-3,-2)  {$\{
\mathfrak{p}_0 ,\, 
\mathfrak{p}_0^{0\rightarrow 2}
\}$};
\node (a22) at (3,-2)  {$\{
\mathfrak{p}_0 ,\, 
\mathfrak{p}_0^{0\rightarrow 0}
\}$};
\node (a31) at (-5,-3)  {$\{
\mathfrak{p}_0 ,\,
\mathfrak{p}_0^{0\rightarrow 2} ,\, \mathfrak{p}_0^{0\rightarrow 2 \rightarrow 1} 
\}$};
\node (a32) at (-1,-3)  {$\{
\mathfrak{p}_0 ,\,
\mathfrak{p}_0^{0\rightarrow 2},\, \mathfrak{p}_0^{0\rightarrow 2 \rightarrow 3}
\}$};
\node (a33) at (2.75,-3)  {$\{
\mathfrak{p}_0 ,\,
\mathfrak{p}_0^{0\rightarrow 2} ,\, \mathfrak{p}_0^{0\rightarrow 0}
\}$};
\node (a34) at (6.5,-3)  {$\{
\mathfrak{p}_0 ,\,
\mathfrak{p}_0^{0\rightarrow 0} ,\, 
\mathfrak{p}_0^{0\rightarrow 0 \rightarrow 0}
\}$};

%
%\node (dots1) at (-5,-4)  {$\vdots$};
%\node (dots1) at (-1,-4)  {$\vdots$};
%\node (dots1) at (3,-4)  {$\vdots$};
%\node (dots1) at (7,-4)  {$\vdots$};
%
\draw[->] (a0) -- (a11);
\draw[->] (a11) -- (a21);
\draw[->] (a11) -- (a22);
\draw[->] (a21) -- (a31);
\draw[->] (a21) -- (a32);
\draw[->] (a21) -- (a33);
\draw[->] (a22) -- (a33);
\draw[->] (a22) -- (a34);

\end{tikzpicture}
\end{array}
\end{equation}
and the vacuum character up to level 4:\footnote{This explicit expansion is for  $n\geq 6$; the expressions for $n=5$ is special.}
\begin{align}\label{eq:ZDnexpansion}
&\boldsymbol{\mathcal{Z}}^{D_n,\textrm{unref}}_{{}^{(0)}\mathcal{P}}(\boldsymbol{x})=1+x_0+(x_0^2+x_0x_2)+(x_0^3+x^2_0 x_2+x_0 x_1x_2+x_0 x_2x_3)\\
&\qquad \qquad
+(x_0^4+x_0^3 x_2+ x_0^2x_2^2+x_0^2x_1x_2+x_0^2x_2 x_3+x_0x_1 x_2x_3+x_0x_2x_3x_4)+\mathcal{O}(\boldsymbol{x}^5)\,. \nonumber
%   +x_0^5+x_6 x_0^4+x_6^2
%   x_0^3+x_3 x_6 x_0^3+x_3
%   x_6^2 x_0^2+x_2 x_3 x_6
%   x_0^2+x_3 x_4 x_6 x_0^2+x_1
%   x_2 x_3 x_6 x_0+x_2 x_3 x_4
%   x_6 x_0+x_3 x_4 x_5 x_6
%   x_0,x_0^6+x_6 x_0^5+x_6^2
%   x_0^4+x_3 x_6 x_0^4+x_6^3
%   x_0^3+x_3 x_6^2 x_0^3+x_2
%   x_3 x_6 x_0^3+x_3 x_4 x_6
%   x_0^3+x_3^2 x_6^2 x_0^2+x_2
%   x_3 x_6^2 x_0^2+x_3 x_4
%   x_6^2 x_0^2+x_1 x_2 x_3 x_6
%   x_0^2+x_2 x_3 x_4 x_6
%   x_0^2+x_3 x_4 x_5 x_6
%   x_0^2+x_2 x_3^2 x_4 x_6
%   x_0+x_1 x_2 x_3 x_4 x_6
%   x_0+x_2 x_3 x_4 x_5 x_6
%   x_0,x_0\right\}
\end{align}
We have checked that for $n\geq 5$, the vacuum character can be captured by the closed-form formula \eqref{eq:VacuumCharacterAffineADEProduct}, with specialization 
\begin{equation}\label{eq:QxDn}
\boldsymbol{Q}^{\boldsymbol{\beta}}\equiv \prod^{n}_{i=1} x^{\beta_i}_{i}\,, \qquad 
\textrm{sgn}(\boldsymbol{\beta})=1\,, \qquad
\texttt{x}\equiv x_0 x_1 x_{n-1}x_{n} \left( \prod^{n-2}_{i=2}x_i \right)^2\,,
\end{equation}
where 
$\boldsymbol{\beta}\in\Delta_{+}$ are the positive roots (in the $\alpha$-basis) of the $D_n$ algebra.
We will discuss the refinement of the vacuum character in Sec.\ \ref{ssec:RefinementADE}.

\medskip

Combining this result for the D$_{n \geq 5}$-type with those in Appx.\ \ref{appssec:PosetAn} for A-type and Appx.\ \ref{appssec:PosetE678} for E-type, we have checked that they all obey the closed-form formula \eqref{eq:VacuumCharacterAffineADEProduct} with all $\textrm{sgn}(\boldsymbol{\beta})=1$ and the Kac label $\mathbf{a}$ in the definition $\texttt{x}\equiv x_0 \,  \boldsymbol{Q}^{\mathbf{a}} $ being
\begin{align}\label{eq:KacLabelADE}
&%A_n: \quad
\mathbf{a}_{A_n}=(\underbrace{1,1,\dots,1}_\text{$n$})\,, \qquad
%D_n: \quad  
\mathbf{a}_{D_n}=(1,\underbrace{2,\dots,2}_\text{$n-3$},1,1)\,,\\
%E_6: \qquad  
&\mathbf{a}_{E_6}=(1,2,3,2,1,2)\,, \quad
%E_7: \qquad  &
\mathbf{a}_{E_7}=(2,3,4,3,2,1,2)\,, \quad
%E_8: \qquad  &
\mathbf{a}_{E_8}=(2,4,6,5,4,3,2,3)
\,.\nonumber
\end{align}

\subsubsection{Refined vacuum character and motivic DT invariants for affine ADE quivers}
\label{ssec:RefinementADE}

For the affine Yangian of $\mathfrak{g}$, where $\mathfrak{g}=\mathfrak{h}\oplus\mathfrak{u}(1)$ with $\mathfrak{h}$ a Lie algebra of ADE type, it is natural to generalize the refinement choice \eqref{eq:Upsilongln} used for the affine Yangian of $\mathfrak{gl}_{n+1}$.  
Namely, for the affine ADE family, we choose the refinement that the arrow with weight $h_{1,2,3}$ has the $\Upsilon$-charge $1$, $-1$, and $0$, respectively:
\begin{equation}
\begin{aligned}
&h(I)=h_1 \,\, \rightarrow \,\, \Upsilon(I^{a})=2\,, \quad     
h(I)=h_2 \,\, \rightarrow \,\, \Upsilon(I)=-2   \,, \quad
h(I)=h_3 \,\, \rightarrow \,\, \Upsilon(I)=0    \,.
\end{aligned}
\end{equation}
Then adopting the deformation \eqref{eq:UpsilonAtom} and \eqref{eq:upsilonAtomType1}, we obtain the refined vacuum character for the affine ADE quivers as the  general formula \eqref{eq:VacuumCharacterAffineADEProductRef} with $\textrm{sgn}(\beta)=1$.
In these cases, $\widehat{\textrm{S}}_{\textrm{total}}=1$, and therefore
\begin{equation}
\mathcal{Z}^{\textrm{ADE, ref}}_{\textrm{NC}}(y, \boldsymbol{x})
=\sum_{\boldsymbol{d}} \, \overline{\underline{\Omega}}^{\textrm{ADE, ref}}_{\,\boldsymbol{d}}(y)\,\boldsymbol{x}^{\boldsymbol{d}} 
=\boldsymbol{\mathcal{Z}}^{\textrm{ADE, ref}}_{{}^{(0)}\mathcal{P}}(y,\boldsymbol{x})
\end{equation}
gives the generating function of the framed refined DT invariants. 
Finally, the generating function of the unframed motivic DT invariants $\boldsymbol{\Omega}(y,\boldsymbol{x})$ is given by the general formula \eqref{eq:OmegaSym2} together with \eqref{eq:chi} with all $\textrm{sgn}(\boldsymbol{\beta})=1$.

\subsection{Isomorphism between affine Yangian of \texorpdfstring{$\mathfrak{g}$}{g} and UEA of \texorpdfstring{$\mathfrak{g}$-extended}{g-extended} \texorpdfstring{$\mathcal{W}_{1+\infty}$}{W1+infinity}   algebra}

We have seen that the refined vacuum character for the affine Yangians of $\mathfrak{g}$, where $\mathfrak{g}=\mathfrak{h}\oplus \mathfrak{u}(1)$ with $\mathfrak{h}$ being the ADE algebra or the super Lie algebra $\mathfrak{sl}_{n|m}$, is of an infinite product form \eqref{eq:VacuumCharacterAffineADEProductRef} and hence can be rewritten in terms of a generalized plethystic exponential \eqref{eq:ZPEzref}, which we reproduce here
\begin{equation}\label{eq:ZADEPE}
\boldsymbol{\mathcal{Z}}^{\textrm{ref}}_{\mathfrak{g}}(y, \boldsymbol{Q},\texttt{x})=\textrm{PE}^{\textrm{gen}}\left[\mathfrak{z}^{\textrm{ref}}_{\mathfrak{g}}(y, \boldsymbol{Q},\texttt{x})\right]\,,
\end{equation}
with $\mathfrak{z}^{\textrm{ref}}_{\mathfrak{g}}(y, \boldsymbol{Q},\texttt{x})$ given in \eqref{eq:PLogZrefADE}.
Physically, the plethystic formula \eqref{eq:ZADEPE} means that the vacuum character $\boldsymbol{\mathcal{Z}}^{\textrm{ref}}_{\mathfrak{g}}(y, \boldsymbol{Q},\texttt{x})$ can be viewed as the counting of the multi-particle spectrum of some system whose single-particle spectrum is counted by \linebreak
$\mathfrak{z}^{\textrm{ref}}_{\mathfrak{g}}(y, \boldsymbol{Q},\texttt{x})$ in  \eqref{eq:PLogZrefADE}.
In this subsection, we show that the formula \eqref{eq:ZADEPE} supports the conjecture that the affine Yangian of $\mathfrak{g}$ is isomorphic to the UEA of the $\mathfrak{g}$ matrix-extended  $\mathcal{W}_{1+\infty}$ algebra,\footnote{
For studies on the relation between affine Yangian of $\mathfrak{gl}_{m|n}$ and the $\mathfrak{gl}_{m|n}$-extended $\mathcal{W}$ algebra, see e.g.\ \cite{Ueda:2020yon,Bao:2022jhy}.} which in turn gives an $\mathcal{W}$ algebra interpretation of the plethystic formula  \eqref{eq:ZADEPE}.

\medskip

Let us first consider the case of the 
ordinary $\mathcal{W}_{1+\infty}$ algebra, whose UEA is already known to be isomorphic to the affine Yangian of $\mathfrak{gl}_1$ \cite{Prochazka:2015deb,Gaberdiel:2017dbk}.
The $\mathcal{W}_{1+\infty}$ algebra under consideration is the one with a spectrum containing one field $W^{(s)}(z)$ per spin $s=1,2,\dots, \infty$, where each field $W^{(s)}(z)$ is a scalar. 
The refined vacuum character of the $\mathcal{W}_{1+\infty}$ algebra should be the refined MacMahon function defined in \eqref{eq:MacMref}, which has the Plethystic exponential expression:
\begin{equation}\label{eq:ZmpW}
Z^{\mathcal{W}_{1+\infty}}_{\delta}(y, \texttt{x}) 
=M^{\textrm{ref}}_{\delta}(y, \texttt{x})
=\prod^{\infty}_{i=1}\prod^{\infty}_{j=1}\frac{1}{1-y^{i-j+\delta}\,\texttt{x}^{i+j-1}}
=\textrm{PE}[ m^{\textrm{ref}}_{\delta}(y, \texttt{x}) ]\,,
\end{equation}
with $m^{\textrm{ref}}_{\delta}(y, \texttt{x})$ given by \eqref{eq:mrefDef}, which should be interpreted as the single-particle character 
\begin{equation}\label{eq:zspW}
z^{\mathcal{W}_{1+\infty}}_{\textrm{s.p.}}(y,\texttt{x})  
= m^{\textrm{ref}}_{\delta}(y, \texttt{x})=\frac{y^{\delta}}{y-y^{-1}}\left(\frac{\texttt{x}\, y}{1- \texttt{x}\, y}-\frac{\texttt{x}/y}{1-\texttt{x}/y}\right) =\sum^{\infty}_{m=1}\, \texttt{x}^m\sum^{m}_{s=1}\,y^{2s-m-1+\delta}\,.
\end{equation}
Now let us give a $\mathcal{W}$ algebra derivation of the single-particle character   \eqref{eq:zspW}.

First of all, the fields of the $\mathcal{W}_{1+\infty}$ algebra have the mode expansions   
\begin{equation}
W^{(s)}(z)=\sum_{m\in Z} \frac{W^{(s)}_m}{z^{m+s}} \,.
\end{equation}
The states in the vacuum module are generated by repeatedly applying modes of the form $ W^{(s)}_{m\leq -s}$ on the vacuum $|0\rangle$.
For example, the states at the first few levels are:
\begin{equation}
\begin{aligned}
\textrm{level }0: \qquad &|0\rangle\,; \\
\textrm{level }1: \qquad &W^{(1)}_{-1}|0\rangle\,; \\
\textrm{level }2: \qquad &W^{(2)}_{-2}|0\rangle\,, \quad  W^{(1)}_{-2}|0\rangle\,, \quad (W^{(1)}_{-1})^2|0\rangle\,; \\
\textrm{level }3: \qquad &W^{(3)}_{-3}|0\rangle\,, \quad  W^{(2)}_{-3}|0\rangle\,, \quad W^{(1)}_{-3}|0\rangle\,, \quad
W^{(2)}_{-2} W^{(1)}_{-1}|0\rangle\,,\quad
W^{(1)}_{-2} W^{(1)}_{-1}|0\rangle\,, \quad 
(W^{(1)}_{-1})^3|0\rangle\,;\\
\vdots \qquad\quad &
\end{aligned}
\end{equation}
Each state $\prod_{i} W^{(s_i)}_{-m_i}|0\rangle$ contributes a term
\begin{equation}
\prod_{i} W^{(s_i)}_{-m_i}|0\rangle :\qquad y^{\sum_{i}(2s_i-m_i-1+\delta)}\texttt{x}^{\sum_i m_i}
\end{equation}
to the refined vacuum character of $\mathcal{W}_{1+\infty}$.
One can easily check that they indeed reproduce the refined character \eqref{eq:ZmpW}.

Among all these states, the single-particle spectrum consists of those states generated by only one $W^{(s)}_m$ mode, namely
\begin{equation}
\textrm{level }m: \qquad     W^{(s)}_{-m} |0\rangle \quad \textrm{for } s=1,2,\dots,m \,.
\end{equation}
Each state $W^{(s)}_{-m} |0\rangle$ contributes a term
\begin{equation}\label{eq:WsRef}
 W^{(s)}_{-m}|0\rangle :\qquad y^{2s-m-1+\delta}\texttt{x}^{ m}
\end{equation}
to the single-particle refined character, reproducing  \eqref{eq:zspW}.\footnote{
If we choose $y^{2s}\texttt{x}^{m}$ here, it would reproduce the convention in \cite{Iqbal:2007ii}.
}
The unrefined limit $y\rightarrow 1$ reproduces the wedge character of $\mathcal{W}_{1+\infty}$.
We comment that the $2 s$ in the exponent of $y$ in \eqref{eq:WsRef} is rather intriguing because normally the spin is not associated with any conserved charge in a $\mathcal{W}$ algebra.

\smallskip

Before we move on to the $\mathfrak{g}$-extended $\mathcal{W}_{1+\infty}$ algebra case, we caution that it does not seem easy to understand the Plethystic formula \eqref{eq:ZmpW} directly as a character of the affine Yangian of $\mathfrak{gl}_1$. 
If we regard the character \eqref{eq:ZmpW} as the generating function of the set of plane partitions, it is hard to see how to extract the ``single particle" piece from the plane partition counting. 
If we regard the character \eqref{eq:ZmpW} directly as the character of the affine Yangian of $\mathfrak{gl}_1$, namely, counted by the states generated by repeatedly applying the $e_{j\geq 0}$ generators on the vacuum $|0\rangle$, one has to use all the relations among the $e$'s, including the Serre relations, to reproduce the counting of the MacMahon function (see \cite[Sec.\ 5.3]{Li:2020rij}) --- then it is also hard to see how to extract the single-particle spectrum from this perspective.\footnote{
The difference is whether the particles are free or interacting: in terms of the $\mathcal{W}$ algebra, the modes are free, but in terms of the affine Yangian generators, they are interacting.
} 
Therefore, the map to the $\mathcal{W}_{1+\infty}$ algebra actually provides an easy and physical way to understand the plethystic formula \eqref{eq:ZmpW} with \eqref{eq:zspW}.
We will now generalize this to the affine Yangian of $\mathfrak{g}$, and try to understand the vacuum character of the affine Yangian of $\mathfrak{g}$ \eqref{eq:ZADEPE} from the $\mathcal{W}$ algebra perspective. 

The spectrum of the $\mathfrak{g}$-extended $\mathcal{W}_{1+\infty}$ algebra  contains one field $\mathbf{W}^{(s)}(z)$ per spin $s=1,2,\dots, \infty$, but now each field $\mathbf{W}^{(s)}(z)$ is a matrix that transforms in the adjoint representation of the Lie algebra $\mathfrak{g}$. 
For our current purpose, we consider the case when $\mathfrak{g}=\mathfrak{h}\oplus\mathfrak{u}(1)$ with $\mathfrak{h}$ being a ADE Lie algebra or the super Lie algebra $\mathfrak{sl}_{n|m}$. 

Similar to the original $\mathcal{W}_{1+\infty}$ algebra, the fields in the $\mathfrak{g}$-extended $\mathcal{W}_{1+\infty}$ algebra have the mode expansion   
\begin{equation}
\mathbf{W}^{(s)}(z)=\sum_{m\in Z} \frac{\mathbf{W}^{(s)}_m}{z^{m+s}}\,,
\end{equation}
where $\mathbf{W}^{(s)}_m$ are matrices transforming in the adjoint representation of $\mathfrak{g}$, and can be expanded as
\begin{equation}
\mathbf{W}^{(s)}_m =\sum_{\boldsymbol{\alpha}} W^{(s)}_{\boldsymbol{\alpha},m}  \mathcal{T}^{\boldsymbol{\alpha}}\,,
\end{equation}
where  $\mathcal{T}^{\boldsymbol{\alpha}}$ are the basis generators in the adjoint representation of $\mathfrak{g}$.
The full spectrum of the $\mathfrak{g}$-extended $\mathcal{W}_{1+\infty}$ algebra is then given by applying $W^{(s)}_{\boldsymbol{\alpha},m}$ on the vacuum repeatedly, with $\boldsymbol{\alpha}$ running through all the generators of $\mathfrak{g}$, $s=1,2,\dots, \infty$, and $m\leq -s$; whereas its single-particle spectrum is simply given by applying each such $W^{(s)}_{\boldsymbol{\alpha},m}$ only once on the vacuum:
\begin{equation}\label{eq:spSpectrumgW}
\textrm{level }m: \qquad     W^{(s)}_{\boldsymbol{\alpha}, -m} |0\rangle \quad \textrm{for } s=1,2,\dots,m \,.
\end{equation}
The refined character that counts these single-particle states in \eqref{eq:spSpectrumgW} is then 
\begin{equation}
z^{\mathfrak{g}-\mathcal{W}_{1+\infty}}_{\textrm{s.p.}}(y,\boldsymbol{Q},\texttt{x}) = z^{\mathcal{W}_{1+\infty}}_{\textrm{s.p.}}(y,\texttt{x}) \cdot \chi_{\mathfrak{g}}(\boldsymbol{Q})\,,
\end{equation}
where the first factor is the single-particle refined character of $\mathcal{W}_{1+\infty}$ \eqref{eq:zspW}, and $\chi_{\mathfrak{g}}(\boldsymbol{Q})$ is the character of the adjoint representation of $\mathfrak{g}$ \eqref{eq:chi}.
Therefore, \linebreak
$z^{\mathfrak{g}-\mathcal{W}_{1+\infty}}_{\textrm{s.p.}}(y,\boldsymbol{Q},\texttt{x}) $ precisely reproduces \eqref{eq:PLogZrefADE} with general $\delta$ and $\chi^{\textrm{ref}}_{\mathfrak{g}}(y, \boldsymbol{Q})=\chi^{\textrm{ref}}_{\mathfrak{g}}(\boldsymbol{Q})$.
Finally, translating the information on the even/odd roots to the boson/fermion properties of the vertices (for details see  \cite[Sec.\ 8.3]{Li:2020rij})), we have \eqref{eq:ZADEPE}.

\section{Quiver Yangians for symmetric quivers  and knot-quiver correspondence}
\label{sec:KnotQuiver}

In this section, we consider symmetric quivers without potential. 
One of the main motivations is that these are the quivers that appear in the knot-quiver correspondence \cite{Kucharski:2017poe,Kucharski:2017ogk}, which maps the generating  series of the colored HOMFLY-PT polynomials of \cite{Freyd1985,przytycki1988invariants} 
(and LMOV invariants of \cite{Ooguri:1999bv,Labastida:2000zp}) 
of a given knot $K$ to the motivic generating series of the DT invariants (and the motive DT invariants) of a corresponding quiver $Q_{K}$.
Using the quiver diagonalization of \cite{Jankowski:2022qdp}, these symmetric quivers can be decomposed into $m$-loop quivers. 
Focusing on these $m$-loop quivers and the trefoil knot quiver, we study their quiver Yangians and construct their vacuum representations; we then compute the refined vacuum characters and show that they reproduce the motivic generating series of the DT invariants of the corresponding quivers. 

\subsection{Review: knot-quiver correspondence}

\subsubsection{Motivic DT invariants for symmetric quivers without potential}
For a symmetric quiver, there is no wall-crossing phenomenon, therefore one can define the motivic generating series for the motivic DT invariants as:\footnote{
Note that in this section we adopt a convention that is more commonly used in the knot-quiver literature, which is slightly different from the one used in Sec.\ \ref{sec:BPSInvariants}, see \eqref{eq:GDef}.
}
\begin{equation}\label{eq:DDefKQ}
\boldsymbol{\mathcal{G}}(y,\boldsymbol{x})\equiv \textrm{PE}\left[
\frac{\boldsymbol{\Omega}(y,\boldsymbol{x})}{y^2-1 }\right]    
\end{equation}
with %$\boldsymbol{\Omega}(y,\boldsymbol{x})$ 
\begin{equation}\label{eq:Omega2}
\boldsymbol{\Omega}(y,\boldsymbol{x})=\sum_{\boldsymbol{d}\in \mathbb{N}^{|Q_0|}_0\backslash \{0\}} \Omega_{\boldsymbol{d}}(y) \, \boldsymbol{x}^{\boldsymbol{d}}
=\sum_{\boldsymbol{d}\in \mathbb{N}^{|Q_0|}_0\backslash \{0\}} \sum_{n=0}(-1)^{|\boldsymbol{d}|+n+1}\, \Omega_{\boldsymbol{d},n} \, \boldsymbol{x}^{\boldsymbol{d}}\, y^n\,, 
\end{equation}
with $\boldsymbol{x}=(x_0,\dots,x_{|Q|_0-1})$,  $d=(d_0,\dots, d_{|Q|_0-1})$, and $\boldsymbol{x}^{\boldsymbol{d}}\equiv\prod^{|Q|_0-1}_{a=0}x^{d_a}_a$.
When a symmetric quiver has no potential, its motivic generating series is known \cite{Kontsevich:2008fj,Kontsevich:2010px,Meinhardt2014,franzen2018semistable}:
\begin{equation}\label{eq:GQfromC}
\boldsymbol{\mathcal{G}}(y,\boldsymbol{x})=\sum_{\boldsymbol{d}} \frac{(-y)^{\boldsymbol{d}\cdot \boldsymbol{C}\cdot  \boldsymbol{d}}}{(y^2;y^2)_{\boldsymbol{d}}} \boldsymbol{x}^{\boldsymbol{d}}\,,
\end{equation}
where $\boldsymbol{C}$ is the adjacency matrix of the quiver (i.e.\ $\boldsymbol{C}_{ab}=|a\rightarrow b|$), and $(y^2;y^2)_{\boldsymbol{d}}\equiv \prod^{|Q|_0-1}_{a=0} (y^2;y^2)_{d_a}$ with the $q$-Pochhammer symbol $(z;q)_n\equiv \prod^{n-1}_{k=0}(1-z q^k)$. 
It was shown in \cite{efimov_2012} that the $\Omega_{\boldsymbol{d},n}$ (see definition in \eqref{eq:Omega2}) computed from \eqref{eq:GQfromC} are positive integers. 

\subsubsection{Knot-quiver correspondence}

The symmetric quivers without potential are interesting because they are the quivers that appear in the knot-quiver correspondence 
\cite{Kucharski:2017poe,Kucharski:2017ogk}, which we briefly review now.
For a knot $K\subset S^3$, consider the generating series
\begin{equation}\label{eq:PkGSDef}
\boldsymbol{P}^K(y,\alpha,x)\equiv \sum^{\infty}_{r=0} P^K_r(y,\alpha)x^r    \,,
\end{equation}
where $P^K_r(y,\alpha)$ is the HOMFLY-PT polynomial of the knot $K$ colored by the $r^{\textrm{th}}$ symmetric representation of $\mathfrak{u}(N)$, defined in \cite{Freyd1985, przytycki1988invariants} and generalizing both Jones polynomials and Alexander polynomials. 
$\boldsymbol{P}^K(y,\alpha,x)$ is related to the LMOV invariants $N^{K}_{i,j,r}$ (defined in \cite{Ooguri:1999bv,Labastida:2000zp}) by
\begin{equation}
 \boldsymbol{P}^K(y,\alpha,x) = \textrm{PE}\left[\frac{\boldsymbol{N}^K(y,\alpha,x)}{y^2-1}\right]  \,, 
\end{equation}
where $\boldsymbol{N}^K(y,\alpha,x)= \sum_{i,j,r}N^{K}_{i,j,r}y^i\, \alpha^j\, x^r$ is the generating series of the LMOV invariants.
The knot-quiver correspondence states that for a knot $K\subset S^3$, there exists a corresponding quiver $Q^K$ and a set of integers $\{A_a,Y_a\}$ for each $a\in Q_0$ such that 
\cite{Kucharski:2017poe,Kucharski:2017ogk}:\footnote{Note that there can be more than one quiver that corresponds to the same knot, see \cite{Kucharski:2017ogk,Ekholm:2018eee,Jankowski:2021flt}.}
\begin{equation}\label{eq:KQcorrespondence}
\boldsymbol{\mathcal{G}}(y, \boldsymbol{x})|_{x_a= y^{Y_a-\boldsymbol{C}_{aa}}\cdot\alpha^{A_a} \cdot x} = \boldsymbol{P}^{K}(y,\alpha,x) \,.
\end{equation}
Accordingly, the generating function of the motivic DT invariants of the quiver $Q$ (defined in \eqref{eq:Omega}) maps to the generating series of the LMOV invariants by:
\begin{equation}\label{eq:yOmegatoN}
\boldsymbol{\Omega}(y, \boldsymbol{x})|_{x_a= y^{Y_a-\boldsymbol{C}_{aa}}\cdot\alpha^{A_a} \cdot x} = \boldsymbol{N}^{K}(y,\alpha,x) \,.
\end{equation}
In the correspondence \eqref{eq:KQcorrespondence} (hence also \eqref{eq:yOmegatoN}), the substitution 
\begin{equation}\label{eq:KQvariables}
x_a \rightarrow y^{Y_a-\boldsymbol{C}_{aa}}\cdot\alpha^{A_a} \cdot x  
\end{equation}
collapses the vector $\{x_0,x_1,\dots, x_{|Q_0|-1}\}$ to a scalar; in this sense, $\boldsymbol{\mathcal{G}}(y, \boldsymbol{x})$ (resp.\ $ \boldsymbol{\Omega}(y, \boldsymbol{x})$) is a refined (by $\boldsymbol{d}$ grading) version of $\boldsymbol{P}^{K}(y,\alpha,x)$ (resp.\ $\boldsymbol{N}^{K}(y,\alpha,x)$).

The knot-quiver correspondence has been proven for rational knots \cite{Stosic:2017wno} and arborescent knots \cite{Stosic:2020xwn} and checked for numerous examples such as various torus knots and twisted knots, see e.g.\ \cite{Panfil:2018sis,Ekholm:2019lmb,Jankowski:2021flt}; it has already been generalized to toric CY$_3$'s (beyond the conifold) \cite{Panfil:2018faz,Kimura:2020qns} and to the knot complement \cite{Kucharski:2020rsp,Ekholm:2021irc}.\footnote{ 
However, note that the original knot-quiver correspondence \eqref{eq:KQcorrespondence}
needs to be slightly extended to deal with the situation when a disk (corresponding to a quiver vertex) wraps the $L_{K}$ more than once \cite{Ekholm:2018eee,Ekholm:2021gyu}, such as for the knots $9_{42}$ (see \cite[Sec.\ 6.1]{Ekholm:2018eee} and \cite[Sec.\ 6.2]{Ekholm:2021gyu}) and $10_{132}$ (see \cite[Sec.\ 6.3]{Ekholm:2021gyu}).
We will not discuss this complication in this paper.
}

\subsubsection{Physical realization of knot-quiver correspondence}

To understand the physical realization of the knot-quiver correspondence, let us first recall the relation between the knot invariants and the topological string.

The HOMFLY-PT polynomial of a knot $K$ can be evaluated in terms of the expectation value of a Wilson loop supported on $K\subset S^3$ in the $U(N)_k$ Chern-Simons theory in  $S^3$ \cite{Witten:1988hf}:
\begin{equation}
P^K_{\textrm{R}}(y,\alpha)=\langle W_{\textrm{R}}[K]\rangle_{S^3}\,,
\end{equation}
with parameters
\begin{equation}
y^2=e^{\frac{2\pi i}{N+k}} \qquad \textrm{and} \qquad \alpha=y^N \,,
\end{equation}
where $\textrm{R}$ labels the representation of $U(N)_k$.
The $U(N)_k$ Chern-Simons theory is in turn related to the open topological string with $e^{g_s}=y^2$ in  $T^{*}S^3$ with A-branes on the Lagrangian knot conormal $L_K$ \cite{Witten:1992fb}, and then via large-$N$ duality to the closed topological string on the resolved conifold $\textrm{X}$ with K\"ahler parameter $t=Ng_s$ and branes on $L_K$ \cite{Ooguri:1999bv}.

Embedding the topological string into M-theory on $\mathbb{R}^4\times \textrm{X}\times S^1$ with M2-branes wrapping holomorphic disks ending on M5-branes wrapping $L_{K}\times \mathbb{R}^2 \times S^1$, one obtains a 3D $\mathcal{N}=2$ Chern-Simons matter theory $T[L_{K}]$ on $\mathbb{R}^2\times S^1$,  whose BPS vortices are counted by the LMOV invariants \cite{Dimofte:2010tz,Terashima:2011qi,Dimofte:2011ju,Yagi:2013fda,Lee:2013ida,Cordova:2013cea}.
The generating series of the HOMFLY-PT polynomials of \eqref{eq:PkGSDef} can be reproduced by the BPS vortex partition function of $T[L_K]$, which can be obtained by taking a double scaling limit \cite{Fuji:2012nx,Ekholm:2018eee}: 
\begin{equation}\label{eq:DoubleScaling}
\begin{aligned}
\mathcal{Z}^{\textrm{open top.}}_{L_{K}}(y,\alpha,x)
=Z^{vortex}_{T[L_{K}]}(y,\alpha,x)
=&\boldsymbol{P}^K(y,\alpha,x) \\
\quad\xlongrightarrow[ y^{r}\rightarrow p]{g_s\rightarrow 0}\quad
&\int  \frac{dp}{p}\, e^{\frac{1}{g_s} (-W_{L_K}(\alpha,p)+\log x \log p+\dots) } \,,
\end{aligned}
\end{equation}
where $p$ can be viewed as the conjugate momentum of $x$, and $ W_{L_K}(\alpha,p)$ and $\log x \log p$ are the source term and the Chern-Simons terms of the 3D $\mathcal{N}=2$ theory $T[L_{K}]$ \cite{Fuji:2012nx,Ekholm:2018eee}, respectively.

As explained in \cite{Ekholm:2018eee}, since the knot-quiver correspondence involves a refinement of the Hilbert space by the $\boldsymbol{d}$ grading, see \eqref{eq:KQvariables}, one should apply a double scaling limit  directly to $\boldsymbol{\mathcal{G}}(y, \boldsymbol{x})$:
\begin{equation}\label{eq:DoubleScalingNew}
\boldsymbol{\mathcal{G}}(y, \boldsymbol{x})\quad \xlongrightarrow[ y^{2d_a}\rightarrow p_a]{g_s\rightarrow 0}\quad\int  \prod_a\frac{dp_a}{p_a}\, e^{\frac{1}{g_s} (-W_{Q_K}(\alpha,p)+\log x_a \cdot\log p_a+\dots) } \,,
\end{equation}
where $p_a$ can be viewed as the conjugate momentum of $x_a$ and $W_{Q_K}(\alpha,p)$ describes another 3D $\mathcal{N}=2$ Chern-Simons matter theory $T[Q_k]$ that is ``dual" to $T[L_K]$. 
The ingredients of $T[Q_K]$ are:
\begin{itemize}
\item Each vertex $a\in Q_0$ corresponds to a $U(1)$ gauge factor and a chiral field charged under the $U(1)$.
\item The number of arrows from $a$ to $b$, namely $\boldsymbol{}{C}_{ab}$, corresponds to the Chern-Simons coupling between the $U(1)$ at vertex $a$ and the $U(1)$ at vertex $b$.
\item The FI couplings are $x_a=x $, which are from the mixed Chern-Simons term between the $U(1)$ gauge factors and the dual topological symmetry.
\end{itemize}
The BPS vortices of  the 3D $\mathcal{N}=2$ theory $T[Q_K]$ can be described by a $\mathcal{N}=4$ quiver quantum mechanics 
\cite{Hwang:2017kmk}.
We conjecture that the quiver Yangian of $Q_K$ is precisely the BPS algebra of this theory. 

\subsection{Refining vacuum character of quiver Yangian}

It is beyond the scope of the current paper to derive the quiver Yangian of $Q_K$ directly as the BPS algebra describing the BPS vortices of  the 3D $\mathcal{N}=2$ theory $T[Q_K]$.
Instead, we will show, for some representative cases, that the refined characters of the vacuum-like representations of the quiver Yangian of $Q_K$ reproduce precisely the motivic DT invariants and hence the LMOV invariants of the knot $K$.
Similar to Sec.\ \ref{sec:BPSInvariants}, it is enough to consider a vacuum-like representation ${}^{(\mathfrak{a})}\mathcal{P}$ that is large enough to incorporate all the DT invariants; we can then label the vertices such that this $\mathfrak{a}=0$, and simply refer to ${}^{(0)}\mathcal{P}$ as the vacuum representation.
However, for the symmetric quivers in this section, we will consider all the vacuum-like representations and check that their refined characters all reproduce the correct motivic DT invariants.

For a given symmetric quiver without potential, it is straightforward to write down the quiver Yangian and construct its vacuum-like representation using the procedure of Sec.\ \ref{sssec:RepresentationConstruction}.
It is then easy to compute their unrefined character using \eqref{eq:ZunrefinedP}.

To refine the character as in  \eqref{eq:ZrefinedP}, for the quivers from the knot-quiver correspondence, we choose the prescription in step 3 of the refinement procedure (explained in Sec.\ \ref{ssec:Refining}) to be  
\begin{equation}\label{eq:upsilonKQ}
\begin{aligned}
\upsilon(\sqbox{$a$})=\delta_{a\in Q_0'}\cdot\Upsilon(\sqbox{$a$})    \qquad  \textrm{with} \quad \Upsilon(\sqbox{$a$})= \sum_{I\in \mathfrak{p}^{\infty\rightarrow a}} \Upsilon(I) \,,
\end{aligned}
\end{equation}
where $\delta_a=1$ when $a$ is in the subset $Q_0'\subset Q_0$ to be determined, otherwise 0.
This is an empirical prescription that works for the $m$-loop quivers (which are the building blocks of all symmetric quivers via quiver diagonalization) and the quiver from the trefoil knot.
We postpone the tasks of further checking and deriving \eqref{eq:upsilonKQ} to future work.

The next step is to check whether there exists a  $\Upsilon$-charge assignment for all arrows $I\in Q_1$ such that the refined character of a vacuum-like representation as defined in \eqref{eq:ZrefinedP} reproduces the motivic DT invariants computed via \eqref{eq:GQfromC}. 
In particular, this  refined character  should reproduce the generating function of the motivic DT invariants in the NC chamber up to signs:
\begin{equation}\label{eq:ZrefZNCKQ}
\boldsymbol{\mathcal{Z}}^{\textrm{ref}}_{{}^{(\mathfrak{a})}\mathcal{P}}(y,\boldsymbol{x})\equiv
\sum_{\Pi\in{}^{(\mathfrak{a})}\mathcal{P}} y^{\upsilon(\Pi)}\, \boldsymbol{x}^{|\Pi|} = \hat{\mathbf{S}}\cdot
\mathcal{Z}_{\textrm{NC},\mathfrak{a}}(y,\boldsymbol{x}) \,,
\end{equation}
where\footnote{
Note the slight difference from the convention \eqref{eq:tildeOmega} in Sec.\ \ref{sec:BPSInvariants}.
} 
\begin{equation}\label{eq:OmegaTildeKQ}
\mathcal{Z}_{\textrm{NC},\mathfrak{a}}(y,\boldsymbol{x})
=\textrm{PE}\left[\frac{\widehat{\boldsymbol{\Omega}}_{\mathfrak{a}}(y,\boldsymbol{x})}{y^2-1}\right]  \quad \textrm{with}\quad \widehat{\boldsymbol{\Omega}}_{\mathfrak{a}}(y,\boldsymbol{x})= \sum_{\boldsymbol{d}}  (y^{2\boldsymbol{f}_{\mathfrak{a}}\cdot \boldsymbol{d}}-1)\Omega_{\boldsymbol{d}}(y) \boldsymbol{x}^{\boldsymbol{d}}    \,.
\end{equation}
Here $\boldsymbol{f}_{\mathfrak{a}}$ is the framing vector $\textrm{e}_{\mathfrak{a}}$ and $\hat{\mathbf{S}}$ is an operator that changes the signs of $x_a$.
For the symmetric quivers considered in this section, we have checked that for all $\mathfrak{a}$, there exists a corresponding $\Upsilon$-charge assignment such that \eqref{eq:ZrefZNCKQ} holds.

\subsection{Quiver with one vertex and \texorpdfstring{$m$ loops}{m loops}}
\label{ssec:mloop}

In this subsection we study the simplest non-trivial example of a symmetric quiver without potential: a quiver with one vertex and $m$ loops.
As shown in \cite{Jankowski:2022qdp}, the $m$-loop quivers can be used as building blocks to construct arbitrary symmetric quivers without potential. 
We will write down their quiver Yangians, construct their vacuum representation, and determine a refinement prescription under which the refined vacuum characters indeed reproduces the motivic DT invariants computed from the motivic generating series \eqref{eq:GQfromC}. 

When $m=1$, the quiver is called Jordan quiver. 
It will be evident from the vacuum character that its quiver Yangian should be isomorphic to the universal enveloping algebra (UEA) of the Heisenberg algebra; and we will show this explicitly.

\subsubsection{\texorpdfstring{$m$-loop}{m-loop} quiver and its quiver Yangian}

Consider the quiver with one vertex (labeled $0$) and $m$ self-arrows:
\begin{equation}\label{eq:mloopQuiver}
Q=\begin{array}{c}
\begin{tikzpicture}[scale=1]
%\node%[state]  
%[regular polygon, regular polygon sides=4, draw=blue!50, very thick, fill=blue!10] (a1) at (0,0)  {$1$};
\vertex (a1) at (0,0)  {$0$};
%\draw[fill=white] (a1) (0,0) circle (0.5) ;
%\node (f) at (-2,0) {$\infty$};
%\draw[thick, ->] (f) -- (a1);
\path[thick, -{Latex[length=2mm, width=1.5mm]},every loop/.append style=-{Latex[length=2mm, width=1.5mm]}] 
%(a1) edge [in=90, out=150, loop, thin, above left] node {$h_3$} ()
%(a1) edge [in=210, out=270, loop, thin, below left] node {$h_1$} ()
(a1) edge [out=330, in=30, loop, thin, right] node {$I_1,I_2,\cdots, I_m$} ()
;
\end{tikzpicture}
\end{array} \qquad \textrm{and}\qquad W=0 \,.
\end{equation}
We label the $m$ self-arrows $I^{0\rightarrow0}$ by $I_{i}$ with $i=1,2,\dots, m$.
Let us assign to the arrow $I_i$ the weight $h_i$.
Since $W=0$, there is no relation between the $\{h_i\}$, so the quiver Yangian will have $m$ parameters.

Since there is only one vertex, there is only one bonding factor, namely the self-bonding factor $\varphi(z)=\varphi^{0\Leftarrow 0}(z)$ from the single vertex to itself, which can be read off from the quiver  \eqref{eq:mloopQuiver} by  definition
\eqref{eq:BondingFactorDef}:\footnote{
Since there is only one vertex, in this section, we will mostly drop the color label $0$, as well as arrow labels like $0\rightarrow 0$ and $0\Leftarrow 0$.
}
\begin{equation}\label{eq:varphimloopQuiver}
\varphi(u)\equiv \varphi^{0\Leftarrow 0}(z)=e^{i\pi t}\prod_{i=1}\frac{u+h_i}{u-h_i}    \,.
\end{equation}

The quiver Yangian defined by the $m$-loop quiver \eqref{eq:mloopQuiver} is then given by the general definition \eqref{eq:QuadraticFields} with $a=b=0$, and the bonding factor $\varphi(z)=\varphi^{0\Leftarrow 0}(z)$ \eqref{eq:varphimloopQuiver}:
\begin{equation}\label{eq:mloopQYQuadratic}
\begin{aligned}
\psi(z)\, \psi(w)&=   \psi(w)\, \psi(z)  \;,\\
\left(\prod^{m}_{i=1}(\Delta-h_i)\right)\,\psi(z) \, e(w)&\simeq e^{i\pi t} \left(\prod^{m}_{i=1}(\Delta+h_i)\right)\, e(w)\,\psi(z) \,, \\
\left(\prod^{m}_{i=1}(\Delta-h_i)\right)\,e(z)\, e(w)&\sim e^{i\pi (t+s)}\left(\prod^{m}_{i=1}(\Delta+h_i)\right)\, e(w)\, e(z) \,,\\
\left(\prod^{m}_{i=1}(\Delta+h_i)\right)\,\psi(z)\, f(w)&\simeq e^{-i\pi t} \left(\prod^{m}_{i=1}(\Delta-h_i)\right)\,f(w)\, \psi(z)\,, \\
\left(\prod^{m}_{i=1}(\Delta+h_i)\right)\,f(z)\, f(w)&\sim e^{-i\pi (t+s)} \left(\prod^{m}_{i=1}(\Delta-h_i)\right)\, f(w)\, f(z) \,,\\
e(z)f(w)-e^{i\pi s} f(w)e(z)&\sim -\frac{\psi(z)-\psi(w)}{z-w} \,,
\end{aligned}
\end{equation}
where we have used $\Delta=z-w$.
One convenient choice of statistic factors is
\begin{equation}\label{eq:stmloop}
t=0 \qquad \textrm{and} \qquad 
s=m+1\,.
\end{equation}

\subsubsection{Vacuum representation of \texorpdfstring{$m$-loop}{m-loop} quiver}
\label{sssec:mloopVacuumRep}

Let us consider the vacuum representation that is given by the canonical framing:
\begin{equation}\label{eq:FramedmloopQuiver}
^{(0)}Q=\begin{array}{c}
\begin{tikzpicture}[scale=1]
%\node%[state]  
%[regular polygon, regular polygon sides=4, draw=blue!50, very thick, fill=blue!10] 
\vertex (a1) at (0,0)  {$0$};
%\draw[fill=white] (a1) (0,0) circle (0.5) ;
\node (f) at (-2,0) {$\infty$};
\draw[thick, -{Latex[length=2mm, width=1.5mm]},every loop/.append style=-{Latex[length=2mm, width=1.5mm]}] (f) -- (a1); 
%\node[below] {$0$};
%\path[thick, ->] (f) -- (a1) [below] {$0$};
\path[thick, -{Latex[length=2mm, width=1.5mm]},every loop/.append style=-{Latex[length=2mm, width=1.5mm]}] 
%(a1) edge [in=90, out=150, loop, thin, above left] node {$h_3$} ()
%(a1) edge [in=210, out=270, loop, thin, below left] node {$h_1$} ()
(a1) edge [out=330, in=30, loop, thin, right] node {$I_1,I_2,\cdots, I_m$} ()
;
\end{tikzpicture}
\end{array}\qquad \textrm{and}\quad {}^{(0)} W=0 \,.
\end{equation}
We first construct the representation that corresponds to this framing using the procedure of Sec.\ \ref{sssec:RepresentationConstruction}.
Since there is only one vertex, namely one color, we can drop the color index in this subsection, and for simplicity we adopt the choice of statistic factors \eqref{eq:stmloop}.

\medskip

\noindent \textbf{Level 0.} There is only one state at level 0: the ground state $\Pi_{0}=\{\varnothing\}$, which is the empty set. 
By definition \eqref{eq:ChargeFunctionDef}, the charge function of $\Pi_{0}$, is just the vacuum charge function, which by definition \eqref{eq:varphi0Def} can be read off from the framed quiver \eqref{eq:FramedmloopQuiver}  to be
\begin{equation}
\Psi_{\Pi_0}(z)=   \psi_{0}(z)=\frac{1}{z} \,,
\end{equation}
which has only one pole at $z=0$.
Together with the fact that the state $\Pi_{0}$ is the empty set, we have
\begin{equation}
\textrm{Rem}(\Pi_{0})=\{\varnothing\} \qquad \textrm{and} \qquad \textrm{Add}(\Pi_{0})=\{I^{(0)} \}\,,
\end{equation}
by definition \eqref{eq:RemPi} and \eqref{eq:AddPi}.

\medskip

\noindent \textbf{Level 1.} From the single path (or atom) in $\textrm{Add}(\Pi_0)$, we obtain one state at level-1:
\begin{equation}
\Pi_1=\{I^{(0)}\}
\end{equation}
by Step-4 of the procedure in Sec.\ \ref{sssec:RepresentationConstruction}, or directly from \eqref{eq:StatesLevel1}.
Its charge function is
\begin{equation}
\Psi_{\Pi_1}(z)=\frac{1}{z}\cdot\prod^{m}_{i=1}\frac{z+h_i}{z-h_i}
\end{equation}
by definition \eqref{eq:ChargeFunctionDef}.
Then by definition \eqref{eq:RemPi} and \eqref{eq:AddPi}, we have
\begin{equation}
\textrm{Rem}(\Pi_{1})=\{I^{(0)}\} \qquad \textrm{and} \qquad \textrm{Add}(\Pi_{1})=\{I^{(0)}\cdot I_1,\dots,I^{(0)}\cdot I_m \} \,.
\end{equation}

\medskip

\noindent \textbf{Level-2.} From the $m$ atoms in $\textrm{Add}(\Pi_1)$, we obtain $m$ states at level-2:
\begin{equation}
\Pi_{2,i}=\{I^{(0)}, I^{(0)}\cdot I_{i}\} \qquad i=1,\dots,m\,,
\end{equation}
by Step-4 of the procedure in Sec.\ \ref{sssec:RepresentationConstruction}.
Their charge functions are
\begin{equation}
\Psi_{\Pi_{2,i}}(z)=\frac{1}{z}\cdot\prod^{m}_{j=1}\frac{z+h_j}{z-h_j} \cdot\prod^{m}_{k=1}\frac{z+h_k-h_i}{z-h_k-h_i}=\prod^{m}_{j=1}\frac{z+h_j}{z-h_j} \cdot\prod^{m}_{k=1, k\neq i}\frac{z+h_k-h_i}{z-h_k-h_i}\cdot\frac{1}{z-2 h_i}
\end{equation}
by definition \eqref{eq:ChargeFunctionDef}.
Then by definition \eqref{eq:RemPi} and \eqref{eq:AddPi}, we have
\begin{equation}
\begin{aligned}
\textrm{Rem}(\Pi_{2,i})&=\{I^{(0)}\cdot I_i\} \,, \\
\textrm{Add}(\Pi_{2,i})&=\{I^{(0)}\cdot I_{k\neq i}\, |\, k=1,\dots,m\}\cup  \{I^{(0)}\cdot I_{i}\cdot I_{k} \, |\, k=1,\dots,m\}\,,
\end{aligned}
\end{equation}
where the first factor in $\textrm{Add}(\Pi_{2,i})$ corresponds to  extending from the first atom (at level 1) by an arrow other than $I_i$, whereas the second factor corresponds to extending from the second atom (at level 2) by all possible arrows.

\smallskip

\noindent \textbf{Level $d$.} Repeating this argument recursively, we see that at level $d$, the set of states are given by all possible configurations with $d$ atoms sitting at the $d$ nodes of an $m$-ary tree, see also \cite{Reineke2011DegenerateCH}.
The number of such configurations is determined by a combinatorics reasoning. 
The unrefined character of the vacuum representation \eqref{eq:ZunrefinedP} in this case is
\begin{equation}\label{eq:ZmTree}
\boldsymbol{\mathcal{Z}}_{{}^{(0)}\mathcal{P}}^{(m)}(x)=\sum^{\infty}_{d=0} \mathbf{n}^{(m)}_{d} x^{d}     \,,
\end{equation}
where we have used $x$ to denote $x_0$ in this case, and $\mathbf{n}^{(m)}_{d}$ counts the number of states at level-$d$ (i.e.\ with $d$ atoms) in the vacuum representation of the $m$-loop quiver.
Since it has the interpretation that it counts the number of configurations with $d$ nodes in an $m$-ary tree, the character satisfies
\begin{equation}\label{eq:treeformula}
\boldsymbol{\mathcal{Z}}_{{}^{(0)}\mathcal{P}}^{(m)}(x)=1+x\, \left(\boldsymbol{\mathcal{Z}}_{{}^{(0)}\mathcal{P}}^{(m)}(x)\right)^m   \,. 
\end{equation}
Solving the constraint \eqref{eq:treeformula} with $\mathbf{n}^{(m)}_{d}$ as unknowns gives 
\begin{equation}
\mathbf{n}^{(m)}_{d}=\frac{1}{(m-1)d+1}\binom{m \,d}{d}   \,, 
\end{equation}
which agrees with \cite{Reineke2011DegenerateCH}.
For example, 
\begin{equation}\label{eq:Zm}
\begin{aligned}
\boldsymbol{\mathcal{Z}}_{{}^{(0)}\mathcal{P}}^{(0), \textrm{unref}}(x)&=1+x\,, \\
\boldsymbol{\mathcal{Z}}_{{}^{(0)}\mathcal{P}}^{(1), \textrm{unref}}(x)&=1+x+x^2+x^3+x^4+x^5+x^6+\mathcal{O}(x^7)=\frac{1}{1-x}  \,, \\
\boldsymbol{\mathcal{Z}}_{{}^{(0)}\mathcal{P}}^{(2),\textrm{unref}}(x)&=1+x+2\,x^2+ 5\,x^3+14\,x^4+42\,x^5+132\,x^6+\mathcal{O}(x^7)\,,\\
\boldsymbol{\mathcal{Z}}_{{}^{(0)}\mathcal{P}}^{(3),\textrm{unref}}(x)&=1+x+3\,x^2+ 12\,x^3+55\,x^4+273\,x^5 +1428\,x^6+\mathcal{O}(x^7)\,,\\
\end{aligned}
\end{equation}
as can be easily checked by hand.

\subsubsection{Refinement}

For the case of the $m$-loop quiver, the motivic generating series \eqref{eq:GDef} is simply 
\begin{equation}\label{eq:Gmloop}
\boldsymbol{G}^{(m)}(y,x)=\sum_{d} \frac{(-y)^{m\,d^2}}{(y^2;y^2)_{d}} x^{d}=\textrm{PE}\left[\frac{\boldsymbol{\Omega}^{(m)}(y,x)}{y^2-1}\right]\,.
\end{equation}
The generating function for the unframed motivic DT invariants $\boldsymbol{\Omega}^{(m)}(y,x)$ is computed by taking the plethystic log of \eqref{eq:Gmloop}; for example
\begin{equation}\label{eq:Omegam}
\begin{aligned}
\boldsymbol{\Omega}^{(0)}(y,x)&=- x\,;\\
\boldsymbol{\Omega}^{(1)}(y,x)&=y x\,;\\
\boldsymbol{\Omega}^{(2)}(y,x)&=-y^2 x + y^4 x^2 - y^8 x^3 + (y^{10} + y^{14}) x^4+\mathcal{O}(\boldsymbol{x}^5)\,;\\
\boldsymbol{\Omega}^{(3)}(y,x)&=y^3 x + y^8 x^2 + (y^{11} + y^{13} + y^{17}) x^3 \\
&\quad  + (y^{14} + y^{16} + 2 y^{18} + 
   y^{20} + 2 y^{22} + y^{24} + y^{26} + y^{30})x^4+\mathcal{O}(x^5)\,.\\
%\vdots\qquad&  
\end{aligned}
\end{equation}
The goal now is to find a refinement prescription on \eqref{eq:ZmTree} that reproduces the motivic DT invariants $\Omega^{(m)}_{d}$.
We have found that the prescription  
\begin{equation}\label{eq:upsilonmloop}
\upsilon(\sqbox{$0$})=  \Upsilon(\sqbox{$0$})= \sum_{I\in \mathfrak{p}^{0\rightarrow 0}}    \Upsilon(I)\,,
\end{equation}
which was summarized in  \eqref{eq:upsilonKQ} with $Q'_0=Q_0=\{0\}$, together with the $\Upsilon$-charge assignment on the arrows
\begin{equation}\label{eq:Upsilonmloop}
\Upsilon(I^{\infty\rightarrow 0})= m\qquad 
\textrm{and}\qquad \Upsilon(I_i)=2(i-1) \qquad i=1,2,\dots m    
\end{equation}
gives 
\begin{equation}
\boldsymbol{\mathcal{Z}}_{{}^{(0)}\mathcal{P}}^{(m),\textrm{ref}}(y,x)=\widehat{\textrm{Sgn}}_m\cdot \mathcal{Z}^{(m)}_{\textrm{NC}}(y,x) \,,
\end{equation}
where 
\begin{equation}\label{eq:OmegaTildeKQZm}
\mathcal{Z}^{(m)}_{\textrm{NC}}(y,x)=\textrm{PE}\left[\frac{\widehat{\boldsymbol{\Omega}}^{(m)}(y,x)}{y^2-1}\right]  \quad \textrm{with} \quad
\widehat{\boldsymbol{\Omega}}^{(m)}(y,\boldsymbol{x})= \sum_{d}  (y^{2 d}-1)\, \Omega^{(m)}_{d}(y) \, x^{d}   \,.
\end{equation}
Here the motivic DT invariants $\Omega^{(m)}_{d}(y)$ are the $y^d$ term in the $y$-expansion of  \eqref{eq:Omegam} and 
\begin{equation}
\widehat{\textrm{Sgn}}_m: \qquad x \rightarrow (-1)^{m-1}x\,.
\end{equation}
Note that the $\Upsilon$-charge on $I^{\infty\rightarrow 0}$ can also be absorbed
by a change of variable $x\rightarrow y^{-m}x$.
Namely, suppose we had adopted 
\begin{equation}
\Upsilon(I^{\infty\rightarrow 0})= 0\qquad 
\textrm{and}\qquad \Upsilon(I_i)=2(i-1) \qquad i=1,2,\dots m    
\end{equation}
instead of \eqref{eq:Upsilonmloop} and called the resulting refined character $\accentset{\circ}{\boldsymbol{\mathcal{Z}}}_{{}^{(0)}\mathcal{P}}^{(m),\textrm{ref}}
(y,x)$, then we have
\begin{equation}\label{eq:ZmloopRefDot}
 \accentset{\circ}{\boldsymbol{\mathcal{Z}}}_{{}^{(0)}\mathcal{P}}^{(m),\textrm{ref}}
(y,x)=\boldsymbol{\mathcal{Z}}_{{}^{(0)}\mathcal{P}}^{(m),\textrm{ref}}(y,x) |_{x\rightarrow y^{-m} x}\,,
\end{equation}
which is slightly easier to compute, and whose relation to $\mathcal{Z}^{(m)}_{\textrm{NC}}(y,x)$ is
\begin{equation}
\mathcal{Z}^{(m)}_{\textrm{NC}}(y,x)=\widehat{S}_m\cdot \accentset{\circ}{\boldsymbol{\mathcal{Z}}}_{{}^{(0)}\mathcal{P}}^{(m),\textrm{ref}}(y,x)\qquad \textrm{with}\quad \widehat{S}_m:\quad x\rightarrow (-y)^{m}(-x) \,.
\end{equation}
For example, 
\begin{align}\label{eq:Zrefm}
\accentset{\circ}{\boldsymbol{\mathcal{Z}}}_{{}^{(0)}\mathcal{P}}^{(0),\textrm{ref}}(y, x)=&1+x \,, \\
\accentset{\circ}{\boldsymbol{\mathcal{Z}}}_{{}^{(0)}\mathcal{P}}^{(1),\textrm{ref}}(y, x)=&1+x+x^2+x^3+x^4+\mathcal{O}(x^5) =\frac{1}{1-x}  \,, \\
\accentset{\circ}{\boldsymbol{\mathcal{Z}}}_{{}^{(0)}\mathcal{P}}^{(2),\textrm{ref}}(y, x)=&1+x+(1+y^2)\,x^2+ (1+2\, y^2+y^4+y^6)\,x^3\nonumber\\
 &\,\,\, + (1+3\,y^2+3\, y^4+ 3\, y^6+2\, y^8+ y^{10}+y^{12})\,x^4+\mathcal{O}(x^5)\,,\nonumber\\
\accentset{\circ}{\boldsymbol{\mathcal{Z}}}_{{}^{(0)}\mathcal{P}}^{(3),\textrm{ref}}(y, x)=&1+x+(1+y^2+y^4)\,x^2+ (1+2\, y^2+3\, y^4+2\, y^6+2\, y^8+y^{10}+y^{12})\,x^3 \nonumber\\
& \,\,\,+(1+3 y^2+6 y^4+7y^6+8y^8+7 y^{10}+7 y^{12}+5 y^{14} \nonumber\\
&\qquad \qquad \qquad \qquad +4 y^{16}+3 y^{18} +2 y^{20}+ y^{22}+y^{24})\,x^4+\mathcal{O}(x^5)\nonumber
\end{align}
as can be easily checked by hand.
Finally, $\accentset{\circ}{\boldsymbol{\mathcal{Z}}}_{{}^{(0)}\mathcal{P}}^{(m),\textrm{ref}}(y, x)$ has a closed-form formula\footnote{
Such a formula already appeared in \cite{Reineke2011DegenerateCH}, however, it is $\accentset{\circ}{\boldsymbol{\mathcal{Z}}}_{{}^{(0)}\mathcal{P}}^{(m),\textrm{ref}}(x)$, rather than $\boldsymbol{\mathcal{Z}}_{{}^{(0)}\mathcal{P}}^{(m),\textrm{ref}}(x)$, that coincides with the generating function of the Poincar\'e polynomials in \cite{Reineke2011DegenerateCH}.
}
\begin{equation}
\accentset{\circ}{\boldsymbol{\mathcal{Z}}}_{{}^{(0)}\mathcal{P}}^{(m),\textrm{ref}}(x)=    \frac{H_m(y,x)}{H_m(y,\tfrac{x}{y^2})}\quad \textrm{with} \quad  
H_m(y,x)\equiv \sum^{\infty}_{n=0} \frac{y^{2(m-1)\binom{n}{2}}}{(y^{-2},y^{-2})_n}\, x^n    \,.
\end{equation}
One can check that $\accentset{\circ}{\boldsymbol{\mathcal{Z}}}_{{}^{(0)}\mathcal{P}}^{(m),\textrm{ref}}(x)$ satisfies a self-consistency condition that is a refined version of  \eqref{eq:treeformula}: 
\begin{equation}
\accentset{\circ}{\boldsymbol{\mathcal{Z}}}_{{}^{(0)}\mathcal{P}}^{(m),\textrm{ref}}(y, x)  =
1+x \prod^{m-1}_{k=0} \accentset{\circ}{\boldsymbol{\mathcal{Z}}}_{{}^{(0)}\mathcal{P}}^{(m),\textrm{ref}}(y, y^{2k} x )   \,.
\end{equation}

\medskip

As in Sec.\ \ref{ssec:Kronecker}, another way to reproduce the (refined) framed generating series $\mathcal{Z}^{(m)}_{\textrm{NC}}(y,x)$ is by directly enumerating states that are obtained by applying $e_n$ on the vacuum, after removing dependent states using the quadratic relations. 
Using \eqref{eq:GSdef}, the ground state for the NC chamber satisfies $
e_{n\geq  1}|\varnothing\rangle =0$.
Assigning $y^{2n+1}x$ for each $e_n$  mode and shifting the degree for the level-$d$ states by $y^{\#}\rightarrow y^{\#+(m-1)d}$ then reproduces $\mathcal{Z}^{(m)}_{\textrm{NC}}(y,x)$, after the sign is taken care of by $x\rightarrow -x$.

\subsubsection{\texorpdfstring{$m=1$}{m=1}: Jordan quiver}
\label{sssec:ExampleJordanQuiver}

The case with $m=0$ was already considered in Sec.\ \ref{ssec:m0loop}.
When $m=1$, the quiver \eqref{eq:mloopQuiver} is the Jordan quiver and its quiver Yangian is given by setting $m=1$ and $h_i=h$ in \eqref{eq:mloopQYQuadratic}:
\begin{equation}\label{eq:QYJordanField}
\begin{aligned}
\psi(z)\, \psi(w)&=   \psi(w)\, \psi(z)  \;,\\
(\Delta-h)\,\psi(z) \, e(w)&\simeq (\Delta+h)\, e(w)\,\psi(z) \,, \\
(\Delta-h)\,e(z)\, e(w)&\sim (\Delta+h)\, e(w)\, e(z) \,,\\
(\Delta+h)\,\psi(z)\, f(w)&\simeq (\Delta-h)\, f(w)\, \psi(z)\,, \\
(\Delta+h)\,f(z)\, f(w)&\sim (\Delta-h)\, f(w)\, f(z) \,,\\
[e(z)\, , \, f(w)]&\sim -\frac{\psi(z)-\psi(w)}{z-w} \,,
\end{aligned}
\end{equation}
where we have also adopted the choice of statistic factors \eqref{eq:stmloop} for simplicity. 
The relations in terms of the modes are
\begin{equation}\label{eq:QYJordanMode}
\begin{aligned}
[\psi_{j}\,,\, \psi_{k}]  &= 0 \,,\\
[\psi_{j+1}\,,\, e_{k}] -[\psi_{j}\,,\, e_{k+1}] &= h \{\psi_{j}\, , \, e_{k}\} \,,\\
[e_{j+1}\,,\, e_{k}] -[e_{j}\,,\, e_{k+1}] &= h \{e_{j}\, , \, e_{k}\}\,,\\
[\psi_{j+1}\,,\, f_{k}] -[\psi_{j}\,,\, f_{k+1}] &=- h \{\psi_{j}\, , \, f_{k}\}\,,\\
[f_{j+1}\,,\, f_{k}] -[f_{j}\,,\, f_{k+1}] &=- h \{f_{j}\, , \, f_{k}\}\,,\\
[e_{j}\,,\, f_{k}]&=\psi_{j+k}\,.
\end{aligned}    
\end{equation}
Although looking complicated, \eqref{eq:QYJordanMode} is actually isomorphic to the UEA of the Heisenberg algebra, which is defined by 
\begin{equation}\label{eq:Heisenberg}
[\mathtt{a}\, , \, \mathtt{a}^{\dagger}]=1\,,
\end{equation}
as we will show in this subsubsection.

The hint comes from the unrefined character of its vacuum representation 
\begin{equation}\label{eq:Zm1}
\boldsymbol{\mathcal{Z}}_{{}^{(0)}\mathcal{P}}^{(1),\textrm{unref}}(y, x)=1+x+x^2+x^3+x^4+x^5+\mathcal{O}(x^6) =\frac{1}{1-x}  \,, \\
\end{equation}
which coincides with the vacuum character of the Heisenberg algebra \eqref{eq:Heisenberg}.
Let us first check \eqref{eq:Zm1} by explicitly writing down the vacuum representation of the quiver Yangian \eqref{eq:QYJordanMode}, which is given by  specializing the result in Sec.\ \ref{sssec:mloopVacuumRep} to $m=1$.
Since there is only one arrow $I=I^{0\rightarrow0}$, we see that at level-$d$, there is only one state:
\begin{equation}
\Pi_{d}=\{I^{(0)},I^{(0)}\cdot I,\dots,I^{(0)}\cdot \underbrace{I\cdot \dots \cdot I}_{d-1}\}\,,
\end{equation}
with charge function
\begin{equation}\label{eq:PiellJordan}
\Psi_{\Pi_{d}}(z)=\frac{z+h}{(z-(d-1)\, h)\, (z-d \, h)} \,.
\end{equation}
It has only one removable atom and one addable atom:
\begin{equation}
\textrm{Rem}(\Pi_{d})=\{I^{(0)}\cdot \underbrace{I \cdot \dots \cdot I}_{d-1}\} \qquad \textrm{and} \qquad \textrm{Add}(\Pi_{d})=\{I^{(0)}\cdot \underbrace{I \cdot \dots \cdot I}_{d}\} \,. 
\end{equation}
The single addable atom then gives rise to the single state at level-$(d+1)$.
Therefore, the vacuum representation gives the character \eqref{eq:Zm1}, which one recognizes as the vacuum character of the Heisenberg algebra \eqref{eq:Heisenberg}.
Now we will show explicitly that the quiver Yangian for the Jordan quiver is indeed isomorphic to the UEA of the Heisenberg algebra.

To derive the map between the \eqref{eq:QYJordanMode} and \eqref{eq:Heisenberg}, it is easier to first compare the vacuum representation constructed in Sec.\ \ref{sssec:ExampleJordanQuiver} and the highest weight representation of the Heisenberg algebra. 

The action of the Jordan quiver algebra on its  vacuum representation is 
\begin{equation}\label{eq:ActionJordanField}
\begin{aligned}
\psi(z)\,|d\rangle &= \frac{z+h}{(z-(d-1)\,h)(z-d\,h)}\, |d\rangle \,, \\
e(z)\,|d\rangle &= \frac{\sqrt{-(d+1)}}{z-d\,h}\, |d+1\rangle\,,\qquad 
f(z)\,|d\rangle = \frac{\sqrt{-d}}{z-(d-1)\,h}\, |d-1\rangle\,,\\
\end{aligned}
\end{equation}
where we use $|d\rangle$ to denote the unique state at level $d$, labeled by $\Pi_{d}$ in \eqref{eq:PiellJordan}.
Translating into the actions of the modes, we have
\begin{equation}\label{eq:ActionJordanMode}
\begin{aligned}
\psi_j\, |d\rangle &= h^j \, \left((d+1) \, d^j-d\, (d-1)^j \right)
 |d\rangle\,,\\
e_j \, |d\rangle &=h^j\, d^j\, \sqrt{-(d+1)} \, |d+1\rangle\,,\qquad f_j\, |d\rangle =h^j\, (d-1)^j\, \sqrt{-d} \,  |d-1\rangle\,.\\
\end{aligned}
\end{equation}

On the other hand, the highest weight representation of the Heisenberg algebra \eqref{eq:Heisenberg}, in  the orthonormal basis $| d\rrangle$
\begin{equation}
\llangle d | d'\rrangle=\delta_{d,d'}\,, \qquad \sum^{\infty}_{d=0} |d\rrangle \llangle d |=1\,,
\end{equation}
is given by 
\begin{equation}\label{eq:HeisenbergAction}
\mathtt{a}\, |d\rrangle= \sqrt{d} \, |d-1\rrangle \qquad \textrm{and}  \qquad   \mathtt{a}^{\dagger}\, |d\rrangle= \sqrt{d+1}\, |d+1\rrangle\,.
\end{equation}
The state $| d\rrangle$ is the eigenstate of the number operator
\begin{equation}
\mathtt{n} \equiv \mathtt{a}^{\dagger}\mathtt{a}\,, 
\qquad \qquad
\mathtt{n}\, |d\rrangle=   d\, |d\rrangle \,.
\end{equation}

Comparing the representation \eqref{eq:ActionJordanMode} and \eqref{eq:HeisenbergAction}, we see that the translation between the two algebras is given by
\begin{equation}\label{eq:mapsl2toJordanQY}
\begin{aligned}
&\psi_j= h^j \, \left((\mathtt{n}+1) \, \mathtt{n}^j-\mathtt{n}\, (\mathtt{n}-1)^j \right)\,,\qquad     
e_j= i\,  h^j\,  \mathtt{a}^{\dagger} \,  \mathtt{n}^j\,,  
\qquad  f_j= i\,  h^j\, \mathtt{n}^j \, \mathtt{a}\,,
\end{aligned}
\end{equation}
together with the map between the states in their vacuum representations:
\begin{equation}
|d\rangle = |d\rrangle\,.
\end{equation}
Finally, plugging the map \eqref{eq:mapsl2toJordanQY} into the relations \eqref{eq:QYJordanMode}, and using the Heisenberg algebra relation \eqref{eq:Heisenberg}, we confirm that the map indeed satisfies all the relations in \eqref{eq:QYJordanMode}.

\begin{figure}
\begin{center}
\includegraphics[width=3 cm]{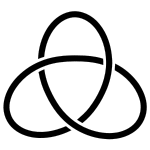}
\end{center}
\caption{The trefoil knot}
\label{fig:trefoil}
\end{figure}

\subsection{Example: trefoil knot}

In this section, we consider the trefoil knot.
The corresponding quiver has three vertices, and depending on which vertex we use as the framed vertex, we can have three different vacuum representations. 
For all of them, we find that the refinement prescription \eqref{eq:upsilonKQ} with $Q'_0=\{1,2\}\}$ together with \eqref{eq:UpsilonAtom} applies, namely, we can always find the $\Upsilon$-charge assignment on the arrows such that the refined vacuum characters reproduce the motivic DT invariants computed via \eqref{eq:GQfromC} and hence the LMOV invariants.

\subsubsection{Quiver from trefoil knot}

Via the knot-quiver correspondence, for the trefoil knot shown in Fig.\ \ref{fig:trefoil}, the quiver and its adjacency matrix are \cite{Kucharski:2017poe,Kucharski:2017ogk}:
\begin{equation}\label{eq:TrefoilQuiver}
Q_{\textrm{trefoil}}=\begin{array}{c}
\begin{tikzpicture}[scale=0.8]
%\node%[state]  
%[regular polygon, regular polygon sides=4, draw=blue!50, very thick, fill=blue!10] 
\node[vertex,minimum size=0.5mm,font=\footnotesize] (a0) at (0,0.5)  {$0$};
%\draw[fill=white] (a1) (0,0) circle (0.5) ;
%\node (f) at (0,3) {$\infty$};
\node[vertex,minimum size=0.5mm,font=\footnotesize] (a1) at (-2,-2)  {$1$};
\node[vertex,minimum size=0.5mm,font=\footnotesize] (a2) at (2,-2)  {$2$};
%\draw[thick, -{Latex[length=2mm, width=1.5mm]},every loop/.append style=-{Latex[length=2mm, width=1.5mm]}] (f) -- (a0); 
%\node[below] {$0$};
%\path[thick, ->] (f) -- (a1) [below] {$0$};
\path[thick, -{Latex[length=2mm, width=1.0mm]}
,every loop/.append style=-{Latex[length=2mm, width=1.0mm]}
] 
%(a1) edge [in=90, out=150, loop, thin, above left] node {$h_3$} ()
%(a1) edge [in=210, out=270, loop, thin, below left] node {$h_1$} ()
(a1) edge[->>] [out=150, in=210, loop, thin, left, ->>] node {} (a1)
(a1) edge[->>] [out=160, in=2000, loop, thin, left, ->>] node {} (a1)
(a2) edge[->>>] [out=350, in=10, loop, thin, right] node {} ()
(a2) edge[->>>] [out=340, in=20, loop, thin, right] node {} ()
(a2) edge[->>>] [out=330, in=30, loop, thin, right] node {} ()
(a0) edge  [thin, bend right]  node [left] {} (a1) 
(a1) edge   [thin]  node [right] {} (a0) 
(a1) edge[->>]   [thin,bend right]  node [right] {} (a2)
(a2) edge[->>]   [thin]  node [right] {} (a1) 
(a0) edge   [thin]  node [below] {} (a2) 
(a2) edge   [thin, bend right]  node [right] {} (a0) 
;
\end{tikzpicture}
\end{array}\,,
\qquad 
\boldsymbol{C}=\begin{pmatrix}
0 & 1 & 1 \\
1 & 2 & 2 \\
1  & 2  & 3
\end{pmatrix}
\,,
\end{equation}
and $W=0$.
Note that there are two arrows from vertex  $1$ to $2$ and two from $2$ back to $1$.
From the adjacency matrix in \eqref{eq:TrefoilQuiver},  one can compute the generating function of the motivic DT invariants via \eqref{eq:GQfromC}:

\vspace{-1em}
{\footnotesize
\begin{equation}\label{eq:OmegaTF}
\begin{aligned}
\boldsymbol{\Omega}(y,\boldsymbol{x})
=&\left(
-x_0-y^2 x_1+y^3 x_2 
\right)  
+\left(
y^4  x_1^2 + y^8 x_2^2 +y^2 x_0 x_1  - y^3 x_0 x_2 -(y^5 +y^7)x_1 x_2   
\right)  
 \\
& 
+\left(
-y^8 x_1^3 
+ (y^{11} + y^{13} + y^{17})x_2^3  
- (y^4 + y^6) x_0 x_1^2  
- (y^6 + y^8 + y^{10}) x_0 x_2^2 
\right.
\\
& \left.
 \quad \, \, 
 + (y^7 + 2 y^9 + y^{11} + y^{13})x_1^2 x_2 
- (y^8 + 2 y^{10} +2 y^{12} + y^{14} + y^{16}) x_1 x_2^2 \right.\\ 
&\qquad \qquad \qquad \qquad \qquad \qquad\qquad \qquad \qquad\quad\,  \left. 
+ (2 y^5 + 2 y^7 + y^9)x_0 x_1 x_2  \right)  
+\mathcal{O}(\boldsymbol{x}^4) \,.
\end{aligned}
\end{equation}
}\normalsize

From the quiver \eqref{eq:TrefoilQuiver}, one can immediately read off the bonding factors \eqref{eq:BondingFactorDef}; and plugging them into \eqref{eq:QuadraticFields} one has the quiver Yangian for  \eqref{eq:TrefoilQuiver}. 
We now check that there exists a refinement prescription for $\upsilon(\sqbox{$a$})$ together with a $\Upsilon$-charge assignment on all the arrows such that the refined vacuum character reproduces the NCDT generating functions of the motivic DT invariants.

\subsubsection{Representation for the trefoil knot quiver}

We first construct the vacuum-like representation using the procedure of Sec.\ \ref{sssec:RepresentationConstruction}. 
To do so, we first choose a vertex out of $0,1,2$ to frame. 
Different choices lead to different vacuum-like representations, but all reproduce the DT invariants $\Omega_{\boldsymbol{d}}(y)$. 
Let us first frame the vertex $0$ to explain the computation. 

In Appx.\ \ref{appsec:TFquiver}, we explicitly construct the states up to level $4$; here we show the final result for up to level $3$:
\begin{align}
\textrm{level }0:\qquad & \{\varnothing\}\,;\\    
\textrm{level }1:\qquad & \{I^{(0)}\}\,;\nonumber\\   
\textrm{level }2:\qquad & 
\{I^{(0)},\, I^{(0)}\cdot I^{0\rightarrow 1}\}\,,
\{I^{(0)},\, I^{(0)}\cdot I^{0\rightarrow 2}\}\,; \nonumber\\   
\textrm{level }3:\qquad & \{I^{(0)},\, I^{(0)}\cdot I^{0\rightarrow 1},\, I^{(0)}\cdot I^{0\rightarrow 1\rightarrow 0} \},\, 
\{I^{(0)},\, I^{(0)}\cdot I^{0\rightarrow 2},\, I^{(0)}\cdot I^{0\rightarrow 2\rightarrow 0} \}\,, \nonumber \\
&\{I^{(0)},\, I^{(0)}\cdot I^{0\rightarrow 1},\, I^{(0)}\cdot I^{0\rightarrow 1}\cdot I^{1\rightarrow 1}_{i=1,2} \},\, 
\{I^{(0)},\, I^{(0)}\cdot I^{0\rightarrow 2},\, I^{(0)}\cdot I^{0\rightarrow 2}\cdot I^{2\rightarrow 2}_{i=1,2,3} \}\,, \nonumber\\
&\{I^{(0)},\, I^{(0)}\cdot I^{0\rightarrow 1},\, I^{(0)}\cdot I^{0\rightarrow 1}\cdot I^{1\rightarrow 2}_{i=1,2} \}\,,
\{I^{(0)},\, I^{(0)}\cdot I^{0\rightarrow 2},\, I^{(0)}\cdot I^{0\rightarrow 2}\cdot I^{2\rightarrow 1}_{i=1,2} \}\,, \nonumber\\
& \{I^{(0)},\, I^{(0)}\cdot I^{0\rightarrow 1},\, I^{(0)}\cdot I^{0\rightarrow 2} \}\,; \nonumber
%\textrm{level }4:\qquad & \{I^{(0)},\, I^{(0)}\cdot I^{0\rightarrow 1},\, I^{(0)}\cdot I^{0\rightarrow 1\rightarrow 0},\, I^{(0)}\cdot I^{0\rightarrow 1\rightarrow 0\rightarrow 1}  \},\, 
%\{I^{(0)},\, I^{(0)}\cdot I^{0\rightarrow 2},\, I^{(0)}\cdot I^{0\rightarrow 2\rightarrow 0} \}\,, \\
%&\{I^{(0)},\, I^{(0)}\cdot I^{0\rightarrow 1},\, I^{(0)}\cdot I^{0\rightarrow 1}\cdot I^{1\rightarrow 1}_{i=1,2} \},\, 
%\{I^{(0)},\, I^{(0)}\cdot I^{0\rightarrow 2},\, I^{(0)}\cdot I^{0\rightarrow 2}\cdot I^{2\rightarrow 2}_{i=1,2,3} \}\,, \\
%&\{I^{(0)},\, I^{(0)}\cdot I^{0\rightarrow 1},\, I^{(0)}\cdot I^{0\rightarrow 1}\cdot I^{1\rightarrow 2}_{i=1,2} \}\,,
%\{I^{(0)},\, I^{(0)}\cdot I^{0\rightarrow 2},\, I^{(0)}\cdot I^{0\rightarrow 2}\cdot I^{2\rightarrow 1}_{i=1,2} \}\,,\\
%& \{I^{(0)},\, I^{(0)}\cdot I^{0\rightarrow 1},\, I^{(0)}\cdot I^{0\rightarrow 2} \}\,; \\
%\vdots &\\
\end{align}
where in $I^{1\rightarrow 1}_{i=1,2}$, $I^{2\rightarrow 2}_{i=1,2,3}$, $I^{1\rightarrow 2}_{i=1,2}$, and $I^{2\rightarrow 1}_{i=1,2}$ the subscripts distinguish the different arrows that share the same initial and final points, and we have used the shorthand $I^{a\rightarrow b\rightarrow c}=I^{a \rightarrow b} \cdot I^{b\rightarrow c}$ whenever there is no ambiguity, i.e.\ when there is only one single arrow from $a$ to $b$ and one arrow from $b$ to $c$.

This iterative result allows us to compute the unrefined character of ${}^{(0)}\mathcal{P}$ as the power series:

\vspace{-1em}
{\footnotesize
\begin{align}\label{eq:ZunrefTrefoil}
\boldsymbol{\mathcal{Z}}^{\textrm{unref}}_{\textrm{trefoil},0}(\boldsymbol{x})  
=
&1+x_0+\left(x_0x_1+x_0x_2\right)+\left(x_0^2x_1+x_0^2x_2+2\, x_0x_1^2+3\, x_0x_2^2+5\, x_0 x_1 x_2\right) \\
&+\left(5\, x_0^2 x_1^2 + 
7\, x_0^2 x_2^2+ 12 \, x_0^2 x_1 x_2 + 5\, x_0 x_1^3+ 12 \,x_0 x_2^3 + 21\, x_0 x_1^2 x_2  + 28\, x_0 x_1 x_2^2 \right)+\mathcal{O}(\boldsymbol{x}^5)  \,, \nonumber
\end{align}
}\normalsize
\noindent 
where the subscript in $\boldsymbol{\mathcal{Z}}^{\textrm{unref}}_{\textrm{trefoil},0}(\boldsymbol{x})$ indicates that the framed vertex is $0$.
From this iterative result, one can check with Mathematica that the unrefined  character  of ${}^{(0)}\mathcal{P}$ reproduces
\begin{equation}
\begin{aligned}
\boldsymbol{\mathcal{Z}}^{\textrm{unref}}_{\textrm{trefoil},0}(\boldsymbol{x})  &=\widehat{\textrm{Sgn}}\cdot\textrm{PE}\left[\sum_{\boldsymbol{d}} d_0 \, \Omega_{\boldsymbol{d}}(1) \boldsymbol{x}^{\boldsymbol{d}}\right]    \,,
\end{aligned}
\end{equation}
where 
\begin{equation}\label{eq:SgnTrefoil}
\widehat{\textrm{Sgn}}:\qquad x_0\rightarrow -x_0\,, x_1 \rightarrow - x_1    
\end{equation}
and $\Omega_{d}(1)$ are the numerical DT invariants computed from \eqref{eq:GQfromC} using the adjacency matrix $ \boldsymbol{C}$ in \eqref{eq:TrefoilQuiver} and setting $y=1$.

\subsubsection{Refined characters of vacuum-like representations}

Now we describe a refinement prescription such that  the refined characters of all the vacuum-like representations reproduce the motivic DT invariants computed from \eqref{eq:GQfromC} using the adjacency matrix $\boldsymbol{C}$ in \eqref{eq:TrefoilQuiver}.

Recall that for the $m$-loop quiver, the refinement prescription on $\upsilon(\sqbox{$0$})$ is given by \eqref{eq:upsilonmloop}.
Since the $m$-loop quivers can be considered as the building blocks for the symmetric quivers without potential \cite{Jankowski:2022qdp},  when there is more than one vertex, it is natural to generalize \eqref{eq:upsilonmloop} to
\begin{equation}\label{eq:upsilonKQansatz}
\upsilon(\sqbox{$a$})=\delta_{a\in Q_0'}\cdot\Upsilon(\sqbox{$a$})    \qquad  \textrm{with} \quad \Upsilon(\sqbox{$a$})= \sum_{I\in \mathfrak{p}^{\infty\rightarrow a}} \Upsilon(I) \,,
\end{equation}
where $\delta_a=1$ when $a$ is in the subset $Q_0'\subset Q_0$ to be determined, otherwise 0.
We will confirm the ansatz \eqref{eq:upsilonKQansatz} with explicit computations later.

Next, in the quiver \eqref{eq:TrefoilQuiver}, let us label the charges for the $i^{\textrm{th}}$ arrows from $a$ to $b$ by $\Upsilon_{i}^{a\rightarrow b}$;  
in this case we have $13$ parameters to determine:
\begin{equation}\label{eq:UpsilonTF}
\begin{aligned}
&\Upsilon_{i=1,2}^{1\rightarrow 1}\,, \quad \Upsilon_{i=1,2,3}^{2\rightarrow 2}\,, \quad \Upsilon^{0\rightarrow 1}\,, \quad \Upsilon^{1\rightarrow 0}\,, \quad \Upsilon^{0\rightarrow 2}\,, \quad  \Upsilon^{2\rightarrow 0}, \quad  \Upsilon_{i=1,2}^{1\rightarrow 2}\,, \quad \Upsilon_{i=1,2}^{2\rightarrow 1}\,.
\end{aligned}
\end{equation}
The goal is to fix these $13$ parameters, together with the subset $Q_0'$, in such a way that the resulting refined vacuum character $\boldsymbol{\mathcal{Z}}^{\textrm{ref}}_{\textrm{trefoil},\mathfrak{a}}(y,\boldsymbol{x})$, 
as computed from definition \eqref{eq:ZrefinedP} with prescription \eqref{eq:upsilonKQansatz}, agrees with  $\widehat{\textrm{Sgn}}\cdot \mathcal{Z}_{\textrm{NC},\mathfrak{a}}(y,\boldsymbol{x})$, as computed from \eqref{eq:OmegaTildeKQ} and \eqref{eq:GQfromC} and with $\widehat{\textrm{Sgn}}$ given by \eqref{eq:SgnTrefoil}:
\begin{equation}\label{eq:ZreftoZNCTFf}
\boldsymbol{\mathcal{Z}}^{\textrm{ref}}_{\textrm{trefoil},\mathfrak{a}}(y,\boldsymbol{x})
=\left(\sum_{\Pi\in {}^{(\mathfrak{a})}\mathcal{P} }y^{\upsilon(\Pi)}\, \boldsymbol{x}^{|\Pi|}\right)
=\widehat{\textrm{Sgn}}\cdot \mathcal{Z}_{\textrm{NC},\mathfrak{a}}(y,\boldsymbol{x})
=\widehat{\textrm{Sgn}}\cdot \textrm{PE}\left[\frac{\widehat{\boldsymbol{\Omega}}_{\mathfrak{a}}(y,\boldsymbol{x})}{y^2-1}\right] \,.
\end{equation}

In \eqref{eq:ZreftoZNCTFf}, the subscript $\mathfrak{a}$ denotes the choice of the framed vertex in the canonical framing; it enters ${}^{(\mathfrak{a})}\mathcal{P}$ as the color of the leading atom $\sqbox{$\mathfrak{a}$}$, and $\widehat{\boldsymbol{\Omega}}_{\mathfrak{a}}(y,\boldsymbol{x})$ via its definition \eqref{eq:OmegaTildeKQ} that depends on $\mathfrak{a}$ explicitly.   
Therefore, a priori the solution for the subset $Q_0'$ and the $13$ parameters in \eqref{eq:UpsilonTF} depends on the choice of $\mathfrak{a}$.

\medskip

Let us first consider the case with $\mathfrak{a}=0$.
We leave the details to Appx.\ \ref{appsec:TFquiver}.
The final result is that, when $\mathfrak{a}=0$, with the  prescription for step 3 of the refinement procedure (explained in Sec.\ \ref{ssec:Refining}) given by
\begin{equation}\label{eq:upsilonTF}
\begin{aligned}
&\upsilon(\sqbox{$0$})=  0\quad \textrm{and}\quad 
&\upsilon(\sqbox{$a$})=  \Upsilon(\sqbox{$a$})\equiv \sum_{I\in \mathfrak{p}^{\infty\rightarrow a}}    \Upsilon(I) \qquad a=1,2
\end{aligned}
\end{equation}
namely \eqref{eq:upsilonKQansatz} with $Q_0'=\{1,2\}$, we find a unique $\Upsilon$-charge assignment 
\begin{equation}\label{eq:TrefoilQuiverFrame0}
^{\sharp}Q_{\textrm{trefoil},0}=\begin{array}{c}
\begin{tikzpicture}[scale=0.8]
%\node%[state]  
%[regular polygon, regular polygon sides=4, draw=blue!50, very thick, fill=blue!10] 
\node[vertex,minimum size=0.5mm,font=\footnotesize] (a0) at (0,0.5)  {$0$};
%\draw[fill=white] (a1) (0,0) circle (0.5) ;
\node (f) at (-3,0.5) {$\infty$};
\node[vertex,minimum size=0.5mm,font=\footnotesize] (a1) at (-2,-2)  {$1$};
\node[vertex,minimum size=0.5mm,font=\footnotesize] (a2) at (2,-2)  {$2$};
\draw[thick, -{Latex[length=2mm, width=1.0mm]},every loop/.append style=-{Latex[length=2mm, width=1.0mm]}, above]  (f) -- node[above]{$0$} (a0); 
%\node[below] {$0$};
%\path[thick, ->] (f) -- (a1) [below] {$0$};
\path[thick, -{Latex[length=2mm, width=1.0mm]}
,every loop/.append style=-{Latex[length=2mm, width=1.0mm]}
] 
%(a1) edge [in=90, out=150, loop, thin, above left] node {$h_3$} ()
%(a1) edge [in=210, out=270, loop, thin, below left] node {$h_1$} ()
(a1) edge[->>] [out=150, in=210, loop, thin, left, ->>,font=\footnotesize] node {$0,2$} (a1)
(a1) edge[->>] [out=160, in=2000, loop, thin, left, ->>,font=\footnotesize] node {} (a1)
(a2) edge[->>>] [out=350, in=10, loop, thin, right,font=\footnotesize] node {} ()
(a2) edge[->>>] [out=340, in=20, loop, thin, right,font=\footnotesize] node {} ()
(a2) edge[->>>] [out=330, in=30, loop, thin, right,font=\footnotesize] node {$0,2,4$} ()
(a0) edge  [thin, bend right,font=\footnotesize]  node [left] {$2$} (a1) 
(a1) edge   [thin,font=\footnotesize]  node [right] {$2$} (a0) 
(a1) edge[->>]   [thin,bend right,font=\footnotesize]  node [below] {$1,3$} (a2)
(a2) edge[->>]   [thin,font=\footnotesize]  node [above] {$1,3$} (a1) 
(a0) edge   [thin,font=\footnotesize]  node [below] {$3$} (a2) 
(a2) edge   [thin, bend right,font=\footnotesize]  node [right] {$3$} (a0) 
;
\end{tikzpicture}
\end{array}
\end{equation}
More explicitly, in Appx.\ \ref{appsec:TFquiver} we determine the prescription \eqref{eq:upsilonTF} and the 
$\Upsilon$-charge assignment 
\eqref{eq:TrefoilQuiverFrame0} by  matching $\boldsymbol{\mathcal{Z}}^{\textrm{ref}}_{\textrm{trefoil},0}(y, \boldsymbol{x})$, as computed from definition \eqref{eq:ZrefinedP} with prescription \eqref{eq:upsilonTF}, to  $\widehat{\textrm{Sgn}}\cdot \mathcal{Z}_{\textrm{NC},0}(y,\boldsymbol{x})$, as computed from \eqref{eq:OmegaTildeKQ} and \eqref{eq:GQfromC},  up to level $4$:

\vspace{-1em}
{\footnotesize
\begin{align}\label{eq:ZrefTrefoilF0}
&\boldsymbol{\mathcal{Z}}^{\textrm{ref}}_{\textrm{trefoil},0}(y, \boldsymbol{x})  
=\widehat{\textrm{Sgn}}\cdot \mathcal{Z}_{\textrm{NC},0}(y,\boldsymbol{x}) \\
&=1+x_0+\left(y^2 x_0x_1+y^3 x_0x_2 \right) \nonumber\\
&\quad +\left(y^2x_0^2x_1+y^3x_0^2x_2+(y^4+y^6)\, x_0x^2_1+(y^6+y^8+y^{10})\, x_0x_2^2+(2y^5+2y^7+y^9)\, x_0 x_1 x_2\right) \nonumber\\
&\quad +\left(
(2 y^4 + 2 y^6 + y^8)\, x_0^2 x_1^2 + 
(2 y^6 + 2 y^8 + 2 y^{10} + y^{12}) \, x_0^2 x_2^2 + (4 y^5 + 4 y^7 + 3 y^9 + y^{11}) \, x_0^2 x_1 x_2 
\right.\nonumber\\
 &\quad \qquad  + (y^6+2 y^8+y^{10}+y^{12})\, x_0 x_1^3+ (y^9+2 y^{11}+3y^{13}+2y^{15}+2y^{17}+y^{19}+y^{21}) \,x_0 x_2^3 \nonumber \\
 &\quad \qquad \qquad \qquad \qquad \qquad \qquad
 + (3 y^7 + 6 y^9 + 5 y^{11} + 4 y^{13} + 2 y^{15} + y^{17})\, x_0 x_1^2 x_2\nonumber \\
 &\quad \qquad \left.
 \qquad \qquad \qquad \qquad \qquad + (3 y^8 + 6 y^{10} + 7 y^{12} + 5 y^{14} + 4 y^{16} + 2 y^{18} + y^{20})\, x_0 x_1 x_2^2 \right)
 +\mathcal{O}(\boldsymbol{x}^5)\nonumber\,.
\end{align}
}\normalsize
\noindent
One then uses Mathematica to check that this assignment reproduces also the higher order terms in the expansion of $\widehat{\textrm{Sgn}}\cdot \mathcal{Z}_{\textrm{NC},0}(y,\boldsymbol{x})$.
Note that since the refined character contains an infinite number of terms, the fact that the charges fixed at low orders allow the identity \eqref{eq:ZreftoZNCTFf} is non-trivial.

\medskip

As a consistency check, we have also done this computation for the other two canonical framings $\mathfrak{a}=1,2$.
Let us first consider how the framing choice $\mathfrak{a}$ enters the computation.
The partition function $ \mathcal{Z}_{\textrm{NC},\mathfrak{a}}(y,\boldsymbol{x})$ computed from \eqref{eq:OmegaTildeKQ} starts with 
\begin{equation}
\mathcal{Z}_{\textrm{NC},\mathfrak{a}}(y,\boldsymbol{x}) = 1+\Omega_{\boldsymbol{f}_{\mathfrak{a}}}(y)+\cdots \,,
\end{equation}
where $\boldsymbol{f}_{\mathfrak{a}}$ is the framing vector $\textrm{e}_{\mathfrak{a}}$. 
In this case, there are three possible framing vectors $\boldsymbol{f}_0=(1,0,0)$, $\boldsymbol{f}_1=(0,1,0)$, $\boldsymbol{f}_2=(0,0,1)$, corresponding to $\mathfrak{a}=0,1,2$, respectively.
From  the leading terms in \eqref{eq:OmegaTF}:
\begin{equation}
\boldsymbol{\Omega}(y,\boldsymbol{x})=-x_0-y^2 x_1+y^3 x_2 +\mathcal{O}(\boldsymbol{x}^2)\,,
\end{equation}
 we have 
\begin{equation}\label{eq:ZNCleadingTF}
\mathcal{Z}_{\textrm{NC},\mathfrak{a}}(y,\boldsymbol{x})=
\begin{cases}
\begin{aligned}
&1-x_0+\mathcal{O}(\boldsymbol{x}^2) \quad\qquad \textrm{for} \quad\mathfrak{a}=0\,,\\
&1-y^2x_1+\mathcal{O}(\boldsymbol{x}^2)  \qquad  \textrm{for} \quad\mathfrak{a}=1 \,,\\
&1+y^3x_2+\mathcal{O}(\boldsymbol{x}^2)  \qquad  \textrm{for} \quad\mathfrak{a}=2 \,,\\
\end{aligned}  
\end{cases}  
\end{equation}
which can be collectively written as
\begin{equation}
\widehat{\textrm{Sgn}}\cdot \mathcal{Z}_{\textrm{NC},\mathfrak{a}}(y,\boldsymbol{x})=1+ y^{m_{\mathfrak{a}}} x_{\mathfrak{a}}+\cdots \,,   
\end{equation}
where $\widehat{\textrm{Sgn}}$ was defined in \eqref{eq:SgnTrefoil},  $x_{\mathfrak{a}}$ is the fugacity for the framed vertex $\mathfrak{a}$, and $m_{\mathfrak{a}}$ is the number of self-loops at vertex $\mathfrak{a}$.
Therefore, the $\Upsilon$-charge for the arrow $I^{\infty\rightarrow \mathfrak{a}}$ from the framing node $\infty$ to the framed node $ \mathfrak{a}$ is
\begin{equation}
\Upsilon^{\infty\rightarrow \mathfrak{a}}\equiv \Upsilon(I^{\infty\rightarrow \mathfrak{a}}) = m_{\mathfrak{a}}   \,.
\end{equation}

Since the arrow $I^{\infty \rightarrow \mathfrak{a}}$ is the first arrow in all atoms, it is convenient to define a shift in the variable $x_{\mathfrak{a}}$:
\begin{equation}
\hat{\textrm{S}}_{\mathfrak{a}}: \qquad x_{\mathfrak{a}} \rightarrow y^{-m_{\mathfrak{a}}} \, x_{\mathfrak{a}} \,,   
\end{equation}
under which 
\begin{equation}
\accentset{\circ}{\boldsymbol{\mathcal{Z}}}^{\textrm{ref}}_{\textrm{trefoil},\mathfrak{a}}(y, \boldsymbol{x})  
\equiv \hat{\textrm{S}}_{\mathfrak{a}}\cdot \boldsymbol{\mathcal{Z}}^{\textrm{ref}}_{\textrm{trefoil},\mathfrak{a}}(y, \boldsymbol{x})  
=\widehat{\textrm{S}}_{\mathfrak{a}}\cdot     \widehat{\textrm{Sgn}}\cdot \mathcal{Z}_{\textrm{NC},\mathfrak{a}}(y,\boldsymbol{x}) =1+x_{\mathfrak{a}}+\cdots \,.
\end{equation}
Namely, the only difference between $\accentset{\circ}{\boldsymbol{\mathcal{Z}}}^{\textrm{ref}}_{\textrm{trefoil},\mathfrak{a}}(y, \boldsymbol{x})$ and  $\boldsymbol{\mathcal{Z}}^{\textrm{ref}}_{\textrm{trefoil},\mathfrak{a}}(y, \boldsymbol{x})$ is that  for the former $I^{\infty\rightarrow \mathfrak{a}}=0$ and  for the latter $I^{\infty\rightarrow \mathfrak{a}}=m_{\mathfrak{a}}$. 
Then in practice, one can first try to match $\accentset{\circ}{\boldsymbol{\mathcal{Z}}}^{\textrm{ref}}_{\textrm{trefoil},\mathfrak{a}}(y, \boldsymbol{x})  $ with $\widehat{\textrm{S}}_{\mathfrak{a}}\cdot     \widehat{\textrm{Sgn}}\cdot \mathcal{Z}_{\textrm{NC},\mathfrak{a}}(y,\boldsymbol{x})$ by setting $\Upsilon^{\infty\rightarrow \mathfrak{a}}=0$ and determining the subset $Q_0'$ in the prescription \eqref{eq:upsilonKQansatz} together with the charges in \eqref{eq:UpsilonTF}. 
The final solution then has the same solution for $Q_0'$ and all the charges in \eqref{eq:UpsilonTF} together with $I^{\infty\rightarrow \mathfrak{a}}=m_{\mathfrak{a}}$.

\medskip

We omit the details and only quote the final result. 
For the framed vertex  $\mathfrak{a}=1$,  we find that the procedure above fixes the prescription \eqref{eq:upsilonTF} and the $\Upsilon$-charge assignment for the arrows
\begin{equation}\label{eq:TrefoilQuiverFrame1}
^{\sharp}Q_{\textrm{trefoil},1}=\begin{array}{c}
\begin{tikzpicture}[scale=0.8]
%\node%[state]  
%[regular polygon, regular polygon sides=4, draw=blue!50, very thick, fill=blue!10] 
\node[vertex,minimum size=0.5mm,font=\footnotesize] (a0) at (0,0.5)  {$0$};
%\draw[fill=white] (a1) (0,0) circle (0.5) ;
\node (f) at (-2,-4) {$\infty$};
\node[vertex,minimum size=0.5mm,font=\footnotesize] (a1) at (-2,-2)  {$1$};
\node[vertex,minimum size=0.5mm,font=\footnotesize] (a2) at (2,-2)  {$2$};
\draw[thick, -{Latex[length=2mm, width=1.5mm]},every loop/.append style=-{Latex[length=2mm, width=1.5mm]}] (f) -- node[left]{$2$}(a1); 
%\node[below] {$0$};
%\path[thick, ->] (f) -- (a1) [below] {$0$};
\path[thick, -{Latex[length=2mm, width=1.0mm]}
,every loop/.append style=-{Latex[length=2mm, width=1.0mm]}
] 
%(a1) edge [in=90, out=150, loop, thin, above left] node {$h_3$} ()
%(a1) edge [in=210, out=270, loop, thin, below left] node {$h_1$} ()
(a1) edge[->>] [out=150, in=210, loop, thin, left, ->>,font=\footnotesize] node {$0,2$} (a1)
(a1) edge[->>] [out=160, in=2000, loop, thin, left, ->>,font=\footnotesize] node {} (a1)
(a2) edge[->>>] [out=350, in=10, loop, thin, right,font=\footnotesize] node {} ()
(a2) edge[->>>] [out=340, in=20, loop, thin, right,font=\footnotesize] node {} ()
(a2) edge[->>>] [out=330, in=30, loop, thin, right,font=\footnotesize] node {$0,2,4$} ()
(a0) edge  [thin, bend right,font=\footnotesize]  node [left] {$4-\alpha$} (a1) 
(a1) edge   [thin,font=\footnotesize]  node [right] {$\alpha$} (a0) 
(a1) edge[->>]   [thin,bend right,font=\footnotesize]  node [below] {$3,5$} (a2)
(a2) edge[->>]   [thin,font=\footnotesize]  node [above] {$-1,1$} (a1) 
(a0) edge   [thin,font=\footnotesize]  node [left] {$7-\alpha$} (a2) 
(a2) edge   [thin, bend right,font=\footnotesize]  node [right] {$\alpha-1$} (a0) 
;
\end{tikzpicture}
\end{array}
\end{equation}
where $\alpha=\Upsilon(I^{1\rightarrow 0})$.
Similar to the case with $\mathfrak{a}=0$, the 
$\Upsilon$-charge assignment 
\eqref{eq:TrefoilQuiverFrame1} is determined by matching $\accentset{\circ}{\boldsymbol{\mathcal{Z}}}^{\textrm{ref}}_{\textrm{trefoil},1}(y, \boldsymbol{x})  $ with $\widehat{\textrm{S}}_{\texttt{1}}\cdot     \widehat{\textrm{Sgn}}\cdot \mathcal{Z}_{\textrm{NC},1}(y,\boldsymbol{x})$ up to level $4$, and for which we give the first three levels here 

\vspace{-1em}
{\footnotesize
\begin{align}\label{eq:ZrefTrefoilF1}
\accentset{\circ}{\boldsymbol{\mathcal{Z}}}^{\textrm{ref}}_{\textrm{trefoil},1}(y, \boldsymbol{x})
&=
1+x_1
+\left( x_0x_1 +(1+y^2)x_1^2
+ (y^3+y^5)x_1x_2
\right)\\
&\, \,\, +\left(
(1 + 2 y^2 + y^4 + y^6)x_1^3 
+(2 + 2 y^2 + y^4) x_0 x_1^2 
+ (2 y^3 + 2 y^5 + y^7) x_0 x_1 x_2 
\right.\nonumber\\
&\, \quad \,\, \left.
+(2 y^3 + 4 y^5 + 3 y^7 + 2 y^9 + y^{11})x_1^2 x_2  
+(y^6 + 2 y^8 + 2 y^{10} + y^{12} + y^{14})  x_1 x_2^2 \right)
+\mathcal{O}(\boldsymbol{x}^4)\nonumber \,.
\end{align}
}\normalsize
\noindent
One can then use Mathematica to check that the assignment \eqref{eq:TrefoilQuiverFrame1} reproduces also the  higher order terms in the expansion of $\widehat{\textrm{S}}_{\texttt{1}}\cdot     \widehat{\textrm{Sgn}}\cdot \mathcal{Z}_{\textrm{NC},1}(y,\boldsymbol{x})$.
Note that since the atom of color $0$ doesn't contribute to the $\Upsilon$-charge, and in the expansion \eqref{eq:ZrefTrefoilF1},  $\Upsilon^{0\rightarrow 1,2}$ and $\Upsilon^{1,2\rightarrow 0}$ always appear in the pairs 
\begin{equation}
\Upsilon^{1\rightarrow 0}+\Upsilon^{0\rightarrow 1} \,, \quad 
\Upsilon^{1\rightarrow 0}+\Upsilon^{0\rightarrow 2} \,, \quad 
\Upsilon^{2\rightarrow 0}+\Upsilon^{0\rightarrow 2} \,, \quad
\Upsilon^{2\rightarrow 0}+\Upsilon^{0\rightarrow 1}\,, 
\end{equation} 
there is one free parameter in the assignment \eqref{eq:TrefoilQuiverFrame1}.

\medskip

Similarly, for the framed vertex $\mathfrak{a}=2$, the procedure above fixes the prescription \eqref{eq:upsilonTF} and the $\Upsilon$-charge assignment for the arrows
\begin{equation}\label{eq:TrefoilQuiverFrame2}
^{\sharp}Q_{\textrm{trefoil},2}=\begin{array}{c}
\begin{tikzpicture}[scale=0.8]
%\node%[state]  
%[regular polygon, regular polygon sides=4, draw=blue!50, very thick, fill=blue!10] 
\node[vertex,minimum size=0.5mm,font=\footnotesize] (a0) at (0,0.5)  {$0$};
%\draw[fill=white] (a1) (0,0) circle (0.5) ;
\node (f) at (2,-4) {$\infty$};
\node[vertex,minimum size=0.5mm,font=\footnotesize] (a1) at (-2,-2)  {$1$};
\node[vertex,minimum size=0.5mm,font=\footnotesize] (a2) at (2,-2)  {$2$};
\draw[thick, -{Latex[length=2mm, width=1.0mm]},every loop/.append style=-{Latex[length=2mm, width=1.0mm]}] (f) -- node[right]{$3$}(a2); 
%\node[below] {$0$};
%\path[thick, ->] (f) -- (a1) [below] {$0$};
\path[thick, -{Latex[length=2mm, width=1.0mm]}
,every loop/.append style=-{Latex[length=2mm, width=1.0mm]}
] 
%(a1) edge [in=90, out=150, loop, thin, above left] node {$h_3$} ()
%(a1) edge [in=210, out=270, loop, thin, below left] node {$h_1$} ()
(a1) edge[->>] [out=150, in=210, loop, thin, left, ->>,font=\footnotesize] node {$0,2$} (a1)
(a1) edge[->>] [out=160, in=2000, loop, thin, left, ->>,font=\footnotesize] node {} (a1)
(a2) edge[->>>] [out=350, in=10, loop, thin, right,font=\footnotesize] node {} ()
(a2) edge[->>>] [out=340, in=20, loop, thin, right,font=\footnotesize] node {} ()
(a2) edge[->>>] [out=330, in=30, loop, thin, right,font=\footnotesize] node {$0,2,4$} ()
(a0) edge  [thin, bend right,font=\footnotesize]  node [left] {$4-\alpha$} (a1) 
(a1) edge   [thin,font=\footnotesize]  node [right] {$\alpha$} (a0) 
(a1) edge[->>]   [thin,bend right,font=\footnotesize]  node [below] {$0,2$} (a2)
(a2) edge[->>]   [thin,font=\footnotesize]  node [above] {$2,4$} (a1) 
(a0) edge   [thin,font=\footnotesize]  node [left] {$4-\alpha$} (a2) 
(a2) edge   [thin, bend right,font=\footnotesize]  node [right] {$2+\alpha$} (a0) 
;
\end{tikzpicture}
\end{array}
\end{equation}
where $\alpha=\Upsilon(I^{1\rightarrow 0})$.
Again, the 
$\Upsilon$-charge assignment 
\eqref{eq:TrefoilQuiverFrame2} is determined by matching $\accentset{\circ}{\boldsymbol{\mathcal{Z}}}^{\textrm{ref}}_{\textrm{trefoil},2}(y, \boldsymbol{x})  $ with $\widehat{\textrm{S}}_{\texttt{2}}\cdot     \widehat{\textrm{Sgn}}\cdot \mathcal{Z}_{\textrm{NC},2}(y,\boldsymbol{x})$ up to level $4$, and for which we give the first three levels here 

\vspace{-1em}
{\footnotesize
\begin{flalign}\label{eq:ZrefTrefoilF2}
\accentset{\circ}{\boldsymbol{\mathcal{Z}}}^{\textrm{ref}}_{\textrm{trefoil},2}(y, \boldsymbol{x})
=&
1+x_2
+\left(  (1 + y^2 + y^4)x_2^2 
+ x_0 x_2 + (y^2 + y^4) x_1 x_2 
\right)\\
&+\left(  
 (1 + 2 y^2 + 3 y^4 + 2 y^6 + 2 y^8 + y^{10} + y^{12})x_2^3 
+(2 + 2 y^2 + 2 y^4 + y^6) x_0 x_2^2 \right. \nonumber\\
&  \quad + (2 y^2 + 4 y^4 + 4 y^6 + 3 y^8 + 2 y^{10} + y^{12})  x_1 x_2^2
+  (y^4 + 2 y^6 + y^8 + y^{10})x_1^2 x_2 \nonumber\\
&\left.\qquad\qquad\qquad \qquad\qquad\qquad\qquad\qquad\qquad\qquad +(2 y^2 + 2 y^4 + y^6) x_0 x_1 x_2  \right)
+\mathcal{O}(\boldsymbol{x}^4)\,. \nonumber
\end{flalign}
}\normalsize
\noindent
One can then use Mathematica to check that the assignment \eqref{eq:TrefoilQuiverFrame2} also reproduces the higher order terms in the  expansion of $\widehat{\textrm{S}}_{\texttt{2}}\cdot     \widehat{\textrm{Sgn}}\cdot \mathcal{Z}_{\textrm{NC},2}(y,\boldsymbol{x})$.

Finally, we observe that in all three cases, the charges satisfy
\begin{equation}
\begin{aligned}
&\Upsilon^{0\rightarrow 1}+\Upsilon^{1\rightarrow 0}=4\,, \quad \Upsilon^{0\rightarrow 2}+\Upsilon^{2\rightarrow 0}=6\,,\quad\sum^{2}_{j=1} (\Upsilon_j^{1\rightarrow2}+\Upsilon_j^{2\rightarrow 1})=8 \,.
\end{aligned}
\end{equation}

\section{Summary and discussion}
\label{sec:Summary}

In this work, we have generalized the study of quiver Yangians from the toric CY$_3$ quivers to general quivers. 
The definition of the quiver Yangian is formally unchanged.
However, unlike the toric CY$_3$ quivers, for which the representations were known to be given by 3D colored crystals, the representations of the quiver Yangians for general quivers are unknown.

In this paper, we first described a procedure for constructing the vacuum representation for an arbitrary quiver Yangian, and then showed that they indeed form a representation of the quiver Yangian. 
We then explained how to construct more general non-vacuum representations from these vacuum representations. 
(Note that for both the vacuum-like representations and non-vacuum representations, we focus on those representations of which for each state,  the constraints $\partial_{I} W=0$ (if $W$ is present) always eliminate all but one path with the same weight.)
These representations allow us to compute the characters of the quiver Yangian straightforwardly.

The construction above applies to all quivers. 
We then focused on two families of quivers that are particularly fascinating. 

\medskip

For the BPS quivers of 4D $\mathcal{N}=2$ theories, the quiver Yangians are the BPS algebras of these systems. 
Besides the toric CY$_3$ quivers, we studied examples such as the $m$-Kronecker quivers and affine ADE quivers and computed the vacuum characters of these quiver Yangians, which map to the BPS partition functions and generating functions of the framed DT invariants (in the NCDT chamber) of these systems. 
We also studied the refinement of the vacuum characters in order to reproduce the refined BPS partition functions of the framed refined DT invariants. 
For the two-acyclic quivers of the 4D $\mathcal{N}=2$ theories, we can refine the characters using the physical spin-grading of the $e^{(a)}_n$ generators; and for several simple examples the results can also be reproduced by directly computing in terms of ideal posets, using rather simple prescriptions.
For the Calabi-Yau three quivers, while the refinement choice is not unique, we gave one simple and universal refinement procedure and also discussed various other choices.

\medskip

For the quivers that come from the knot-quiver correspondence, we conjecture that the quiver Yangians are the BPS algebras for the underlying 3D $\mathcal{N}=2$ theory of \cite{Ekholm:2018eee}.
As examples of this type of quivers, we studied the $m$-loop quivers (which can be viewed as the building block of these knot quivers) and the trefoil knot quiver, and found a refinement prescription with which the refined vacuum character of the corresponding quiver Yangian reproduces the motivic DT invariants and hence the knot invariants by the knot-quiver correspondence.\footnote{We observe that this refinement is the same as the one used for the simple examples of two-acyclic BPS quivers but slightly different from the one used Calabi-Yau-three quivers.}
This provides strong evidence that the quiver Yangians are also the BPS algebras for the system that gives rise to the knot-quiver correspondence.
\medskip

Finally, it is straightforward to generalize these quiver Yangians for arbitrary quivers to the trigonometric, elliptic, and generalized cohomological versions, following \cite{Galakhov:2021omc}, see Appx.\ \ref{appsec:TrigElliptic}.

\bigskip

Let us list some interesting directions for future research.
For the two-acyclic BPS quivers of the 4D $\mathcal{N}=2$  theories:
\begin{itemize}
\item It would be nice to explore more general representations (corresponding to line defects of the theory) for more complicated examples, and compute their refined characters using methods of this paper. The presence of physical spin-grading for $e^{(a)}_n$ generators would help us to better understand the refinement prescription in terms of the ideal posets. 
See also related discussion (in terms of shuffle algebras) in \cite{Gaiotto:2024fso}.
\item We would also like to extend the grading to the $\psi$ and $f$ generators of the quiver Yangian, which might play an important role in understanding the Koszul dual of the algebra.
\item It would also be good to explore more the application of the quiver Yangians and their representations for the Class $\mathcal{S}$ theories, generalizing the results of \cite{Li:2020rij,Galakhov:2021xum} for toric CY$_3$ quivers.
\end{itemize}
For the Calabi-Yau-three quivers:
\begin{itemize}
\item We would like to better understand the refinement prescription we have used.
First of all, recall that the refinement is not unique, see the discussion in Sec.\ \ref{sec:BPSInvariants}.
In this paper, we have mainly discussed one refinement procedure that is particularly simple and universal, and in particular, respects the Poincar\'e symmetry.
It would be useful to study possible refinements systematically.
\item Once we choose a refinement prescription, the refined characters of the quiver Yangian offers a straightforward method to compute the framed refined DT invariants for arbitrary BPS quivers in the NCDT chamber. 
On the other hand, one can use the attractor tree formula of \cite{Mozgovoy:2020has,Mozgovoy:2021iwz} or the flow tree formula of \cite{Arguz:2021zpx} to compute the unframed refined DT invariants at the attractor points for an arbitrary BPS quiver. 
Since their results generically capture cohomologies with compact support (on a non-compact variety), and hence break Poincar\'e symmetry, we would like to understand how to compare to their results and whether it is possible to find a prescription to directly match their results.

\end{itemize}
For the quivers from the knot-quiver correspondence: 
\begin{itemize}
\item Similar to previous classes, we would like to understand better the refinement prescription from first principles, e.g.\ by matching with the results obtained by considering products of $e^{(a)}_n$'s acting on the ground states. 

\item In the quivers from the knot-quiver correspondence, there are also ``generalized" quivers that involve ``negative" arrows --- accordingly the adjacency matrix $\mathcal{C}$ contains negative entries \cite{Jankowski:2022qdp}. 
We didn't discuss this type of generalized quiver in this paper. 
One may hope that the definition of the quiver Yangian can be slightly modified in order to incorporate also them. 

\item It would be nice to use the quiver Yangian to provide some algebraic understanding of the quiver diagonalization of \cite{Ekholm:2019lmb,Jankowski:2022qdp}.

\item  The knot-quiver correspondence was explained in terms of topological strings and 3D $\mathcal{N}=2$ gauge theory in \cite{Ekholm:2018eee}.
It is natural to expect that the quiver Yangians are the BPS quivers of these theories, generalizing the statement from the 4D $\mathcal{N}=2$ gauge theories.
The fact that, for the examples that we checked, the refined vacuum characters of the quiver Yangians reproduce precisely the motive DT invariants of the knot quiver and the knot invariants provides strong support for this physical explanation of the knot-quiver correspondence. 
It would be great to have a more physical derivation of the quiver Yangians for the knot quivers, which might help elucidate the nature of the knot-quiver correspondence itself. 
\end{itemize}
Finally, one could try to apply the quiver Yangians to study quiver systems other than those from the BPS sectors of the 4D $\mathcal{N}=2$ theories or those from the knot-quiver correspondence. 
We hope to report on these in the future. 

\bigskip

\section*{Acknowledgements}
We thank Tiantai Chen, Matthias Gaberdiel, Pietro Longhi, Boris Pioline, Mauricio Romo, and Yegor Zenkevich for helpful discussions.
This work is supported by NSFC No.\ 11875064, No.\ 12275334, No.\ 11947302, and the Max-Planck Partnergruppen fund.
We are also grateful for the hospitality of ETH Zurich, the Kavli Institute for Theoretical Physics, and Perimeter Institute, where part of this work was done.

\appendix

\section{Constructing non-vacuum representations from vacuum-like representations}
\label{appsec:NonVacuumReps}

In this appendix, we explain in more detail the two methods of constructing non-vacuum representations ${}^{\sharp}\mathcal{P}_{(Q,W)}$ starting from the vacuum-like representations ${}^{(\mathfrak{a})}\mathcal{P}_{(Q,W)}$. 

\subsection{Non-vacuum representation as superposition of positive and negative vacuum-like representations}
\label{appssec:Method1}

For this method, we view the non-vacuum representation  ${}^{\sharp}\mathcal{P}_{(Q,W)}$ as a superposition of  ``positive" and ``negative"  vacuum-like representations ${}^{(\mathfrak{a})}\mathcal{P}_{(Q,W)}$.
The procedure of determining the ground state charge function ${}^{\sharp}\psi^{(a)}_0(z)$ takes four steps.\footnote{
These four steps are in complete parallel to those in \cite[Sec.\ 3]{Galakhov:2021xum} for the toric 
CY$_3$ quivers.
}
\begin{enumerate}
\item Choose a set of starters $\{\sqbox{$\mathfrak{b}$}^{\textrm{starter}}\}$. 
A ``starter" is defined as the starting point from which we can grow a (shifted) vacuum-like representation ${}^{(\mathfrak{b})}\mathcal{P}_{(Q,W)}$. 
It is chosen among the atoms of a vacuum-like representation ${}^{(\mathfrak{a})}\mathcal{P}_{(Q,W)}$ in which ${}^{\sharp}\mathcal{P}_{(Q,W)}$ can be embedded. 
In this computation, once we have chosen this $\mathfrak{a}$, let us relabel it as $\mathfrak{a}=0$, and call ${}^{(0)}\mathcal{P}_{(Q,W)}$ the vacuum representation to distinguish it from the vacuum-like representations led by starters/stoppers/pausers.
A starter of color $\mathfrak{b}\in Q_{0}$ is
\begin{equation}\label{eq:starterDef}
\sqbox{$\mathfrak{b}$}^{\textrm{starter}} 
=I^{(0)}\cdot\mathfrak{p}^{0\rightarrow\mathfrak{b}}   
\end{equation}
for some path $\mathfrak{p}^{0\rightarrow\mathfrak{b}}\in J^{(0)}_{(Q,W)}$.
It gives rise to a ``shifted"
vacuum-like representation ${}^{(\mathfrak{b})}\mathcal{P}_{(Q,W)}$, namely a representation whose ground state charge function is 
\begin{equation}
{}^{\sharp}\psi^{(a)}_0(z)
=\left(\frac{1}{z-h(\sqbox{$\mathfrak{b}$}^{\textrm{starter}}) } \right)^{\delta_{a,\mathfrak{b}}} \,.
\end{equation}
One can start from this ${}^{\sharp}\psi^{(a)}_0(z)$ and apply the procedure of Sec.\ \ref{sssec:RepresentationConstruction} and construct the corresponding (positive) representation ${}^{(\mathfrak{b})}\mathcal{P}_{(Q,W)}$, which by definition can be embedded into the original vacuum representation ${}^{(0)}\mathcal{P}_{(Q,W)}$.\footnote{
This is the generalization of the statement in the toric CY$_3$ case that ${}^{\sharp}\mathcal{C}$ can be viewed as a subcrystal of the canonical crystal $\mathcal{C}_0$ \cite{Galakhov:2021xum}.
} 
Let us consider a set of such starters $\{\sqbox{$\mathfrak{b}$}^{\textrm{starter}}\}$ and the corresponding positive $\{{}^{(\mathfrak{b})}\mathcal{P}_{(Q,W)}\}$.

\item As we ``grow" these positive posets level by level, at some point, two different  ${}^{(\mathfrak{b}_1)}\mathcal{P}_{(Q,W)}$ and  ${}^{(\mathfrak{b}_2)}\mathcal{P}_{(Q,W)}$ may ``meet" by having a common addable atom $\sqbox{$\mathfrak{c}$}=\mathfrak{p}^{0\rightarrow\mathfrak{c}}\in J^{(0)}_{(Q,W)}$, with $\mathfrak{c}\in Q_0$.\footnote{
Note that in general $p^{(\mathfrak{c})}$ appears at different levels of ${}^{(\mathfrak{b}_1)}\mathcal{P}_{(Q,W)}$ and  ${}^{(\mathfrak{b}_2)}\mathcal{P}_{(Q,W)}$.
}
Therefore, starting from $\sqbox{$\mathfrak{c}$}$, the set of addable atoms from ${}^{(\mathfrak{b}_1)}\mathcal{P}_{(Q,W)}$ and  ${}^{(\mathfrak{b}_2)}\mathcal{P}_{(Q,W)}$  will be identical. 
To avoid double counting, we need to introduce a negative ${}^{(\mathfrak{c})}\mathcal{P}_{(Q,W)}$.
Since such a negative poset slows down the growth of the joint poset, we call its leading atom $\sqbox{$\mathfrak{c}$}$ ``pauser" (borrowing the terminology of \cite{Galakhov:2021xum}) and moreover attach a ``-" sign to indicate that it corresponds to a negative poset:
\begin{equation}\label{eq:pauser-Def}
\sqbox{$\mathfrak{c}$}^{\textrm{pauser-}}=I^{(0)}\cdot\mathfrak{p}^{0\rightarrow\mathfrak{c}}   \,.
\end{equation}
Any pair of two positive posets may result in such a ``$\sqbox{$\mathfrak{c}$}^{\textrm{pauser-}}$" and a corresponding ${}^{(\mathfrak{c})}\mathcal{P}_{(Q,W)}$.

\item Since two negative posets ${}^{(\mathfrak{c}_1)}\mathcal{P}_{(Q,W)}$ and  ${}^{(\mathfrak{c}_2)}\mathcal{P}_{(Q,W)}$ can also overlap, say at 
\begin{equation}\label{eq:pauser+Def}
\sqbox{$\mathfrak{d}$}^{\textrm{pauser+}}=I^{(0)}\cdot\mathfrak{p}^{0\rightarrow\mathfrak{d}}   \,,
\end{equation}
one then needs to introduce an additional positive poset ${}^{(\mathfrak{d})}\mathcal{P}_{(Q,W)}$ to cancel each such negative overlap. 
After obtaining all such ${}^{(\mathfrak{d})}\mathcal{P}_{(Q,W)}$, one then goes back to step 2 to check whether these new positive posets overlap. 
Using the inclusion-exclusion principle, it takes a finite number of such steps to finish the iteration. 
\item Finally, one can choose to place negative posets ``by hand" to truncate the poset, led by the atom called ``stopper" because the growth of the poset stops there:
\begin{equation}\label{eq:stopperDef}
\sqbox{$\mathfrak{e}$}^{\textrm{stopper}}=I^{(0)}\cdot\mathfrak{p}^{0\rightarrow\mathfrak{e}}   \,.
\end{equation}
This final step is optional.
\end{enumerate}
Once we have determined the set of starters, pausers$\pm$, and the stoppers, the final representation  is defined by
\begin{equation}\label{eq:varphi0M1}
{}^{\sharp}\psi_{0}^{(a)}(z)=\frac{\prod_{\sqbox{$a$}\in \textrm{pauser-}}(z-h(\sqbox{$a$}))\prod_{\sqbox{$a$}\in \textrm{stopper}}(z-h(\sqbox{$a$}))}{\prod_{\sqbox{$a$}\in \textrm{starter}}(z-h(\sqbox{$a$}))\cdot \prod_{\sqbox{$a$}\in \textrm{pauser+}}(z-h(\sqbox{$a$}))}    \,.
\end{equation}
This is the ground state charge function defined in \eqref{eq:varphi0Def}, whose poles and zeros are related to \eqref{eq:varphi0M1} by 
\begin{equation}
\begin{aligned}
\{\mathtt{p}^{(a)} \}&=\{h(\sqbox{$a$}^{\textrm{starter}})\} \cup   \{h(\sqbox{$a$}^{\textrm{pauser+}}) \}  
\quad \textrm{and} \quad
\{\mathtt{z}^{(a)} \}=\{h(\sqbox{$a$}^{\textrm{pauser-}})\} \cup   \{h(\sqbox{$a$}^{\textrm{stopper}}) \} \,,
\end{aligned}
\end{equation}
together with
\begin{equation}
\mathfrak{s}^{(a)}_{+}\equiv |\{\mathtt{p}^{(a)} \}|    \qquad \textrm{and} \qquad \mathfrak{s}^{(a)}_{-}\equiv |\{\mathtt{z}^{(a)} \}| \,.
\end{equation}
One can then use the procedure of Sec.\ \ref{sssec:RepresentationConstruction} to determine all the states in ${}^{\sharp}\mathcal{P}$ starting from ${}^{\sharp}\psi_{0}^{(a)}(z)$.

\subsection{Non-vacuum representation as sub-compound of vacuum-like representation}
\label{appssec:Method2}

In this method, we view ${}^{\sharp}\mathcal{P}$ as a  ``sub-compound" of ${}^{(\mathfrak{a})}\mathcal{P}$ for some $\mathfrak{a}$ and convert some excited state of ${}^{(\mathfrak{a})}\mathcal{P}$ into the ground state of ${}^{\sharp}\mathcal{P}$.
For this method to work, $\mathfrak{a}$ needs to be chosen such that ${}^{\sharp}\mathcal{P}$ can indeed be embedded in ${}^{(\mathfrak{a})}\mathcal{P}$.
In this computation, once $\mathfrak{a}$ is chosen, let us relabel it as $\mathfrak{a}=0$ and call the vacuum-like representation ${}^{(0)}\mathcal{P}$ the vacuum representation, since there is no risk of confusion.

\begin{enumerate}
\item The first step is optional. 
One can choose to truncate the vacuum representation ${}^{(0)}\mathcal{P}$ by placing some stoppers, similar to the $4^\textrm{th}$ step in Method 2.\footnote{
Note that this truncation step happens last in Method 1 but first in Method 2.
}
\item Consider an excited state ${}^{(0)}\Pi_{\sharp}$ of the (truncated) vacuum representation ${}^{(0)}\mathcal{P}$, and compute its charge function $\Psi^{(a)}_{{}^{(0)}\Pi_{\sharp}}(u)$ 
\begin{equation}\label{eq:PsiPi1}
\Psi^{(a)}_{{}^{(0)}\Pi_{\sharp}}(z)
=\prod_{\sqbox{$a$}\in \textrm{stopper}}(z-h(\sqbox{$a$}))\cdot(\frac{1}{z})^{\delta_{a,0}}   \cdot \prod_{b\in Q_0}\prod_{\sqbox{$b$}\in {}^{(0)}\Pi_{\sharp}} \varphi^{a\Leftarrow b}(z-h(\sqbox{$b$}) \,.
\end{equation}

\item When viewed as an excited state of the (truncated) vacuum representation, ${}^{(0)}\Pi_{\sharp}$ has a non-empty set of removable atoms Rem$(\Pi_{\sharp})$, defined in \eqref{eq:RemPi}, corresponding to the set of removing poles in $\Psi^{(a)}_{{}^{(0)}\Pi_{\sharp}}(u)$. 
Now we view ${}^{(0)}\Pi_{\sharp}$ as the ground state of a new, non-vacuum, representation, namely we declare
\begin{equation}\label{eq:Step3Method2}
{}^{(0)}\Pi_{\sharp}
\quad  \longrightarrow  \quad {}^{\sharp}\Pi_{\textrm{g.s.}}=\{\varnothing\}\,.
\end{equation}
Therefore, by definition \eqref{eq:RemPi}, the set of removable atoms from the ground state ${}^{\sharp}\Pi_{\textrm{g.s.}}$ is now empty
\begin{equation}
\textrm{Rem}( {}^{\sharp}\Pi_{\textrm{g.s.}})=\{\varnothing\}\,.
\end{equation}
From now on, a state in the non-vacuum representation ${}^{\sharp}\mathcal{P}$ is denoted as ${}^{\sharp}\Pi$, to distinguish it from a state ${}^{(0)}\Pi$ in the vacuum representation ${}^{(0)}\mathcal{P}$.

\item Since the removing poles in $\Psi^{(a)}_{{}^{(0)}\Pi_{\sharp}}(z)$ do not correspond to any removable atoms for the ground state of the new non-vacuum representation, the new ground state charge function should be defined to be
\begin{equation}\label{eq:varphi0M22}
{}^{\sharp}\psi_{0}^{(a)}(z) \equiv \Psi^{(a)}_{{}^{(0)}\Pi_{\sharp}}(z)\cdot \prod_{\sqbox{$a$}\in \textrm{Rem}({}^{(0)}\Pi_{\sharp})}(z-h(\sqbox{$a$}))
\end{equation}
in order not to have any of these old removing poles.\footnote{ 
One can also choose to define ${}^{\sharp}\psi_{0}^{(a)}(z) \equiv \Psi^{(a)}_{{}^{(0)}\Pi_{\sharp}}(z)$, and pay the price of having spurious (removing) poles.}
\end{enumerate}
In both cases, the final results \eqref{eq:varphi0M22} agree with \eqref{eq:varphi0M1} obtained in Method 1, and is the ground state charge function \eqref{eq:varphi0Def} of a new non-vacuum representation ${}^{\sharp}\mathcal{P}$; one can then use the procedure of Sec.\ \ref{sssec:RepresentationConstruction} to determine all the states in ${}^{\sharp}\mathcal{P}$ starting from ${}^{\sharp}\psi_{0}^{(a)}(z)$.
 Comparing the final expressions \eqref{eq:varphi0M1} with   \eqref{eq:varphi0M22}, we see that the addable atoms of ${}^{(0)}\Pi_{\sharp}$ are the starters in Method 1.

\subsection{Comparison of two methods}
\label{appssec:Comparison2methods}

Finally let us discuss the relative advantage of the two methods. 
Although in simple examples the second method is more straightforward, very often we encounter the situation where the ground state of ${}^{\sharp}\mathcal{P}$ corresponds to some excited state ${}^{(0)}\Pi_{\sharp}$ (of ${}^{(0)}\mathcal{P}$) with an infinite number of atoms.
The infinite number of atoms then requires a regularization procedure to compute ${}^{\sharp}\psi^{(a)}_0(z)$, which, although straightforward, is a bit lengthy. 
For such cases, Method 1 is more suitable. 

\medskip

For example, this situation arises when we consider open BPS invariants, namely when there are D2-branes wrapping holomorphic disks and ending on D4-branes wrapping Lagrangian 3-cycles.
The simplest such example arises for the affine Yangian of $\mathfrak{gl}_1$ when we consider representations that correspond to plane partitions with non-trivial asymptotics. 
We have shown that Method 1 \cite[Sec.\ 6.1]{Galakhov:2021xum} is more efficient than Method 2 (see \cite[eq.\ (3.31)]{Gaberdiel:2017hcn} or for more details see \cite[eq.\ (2.32)]{Gaberdiel:2018nbs} and \cite[Sec.\ B.1]{Galakhov:2021xum}).

Another example where this situation occurs is when we consider a wall-crossing chamber away from the Non-commutative Donaldson-Thomas (NCDT) chamber. 
The NCDT chamber corresponds to a vacuum-like representation, whereas all the other chambers correspond to non-vacuum representations. 
For the case of the resolved conifold, one can compare the computation of ${}^{\sharp}\psi_{0}^{(a)}(z)$ using Method 1 (see \cite[Sec.\ 5.1]{Galakhov:2021xum}) and the one using Method 2 (see \cite[Sec.\ B.2]{Galakhov:2021xum}) and see that Method 2 is much more efficient. 

\medskip

As far as we are aware, the only situation in which Method 1 is awkward and one has to resort to Method 2 is when we consider the conjugate representation, for example, the so-called high-wall representations for the affine Yangian of $\mathfrak{gl}_1$ 
(see \cite[eq (3.16)]{Gaberdiel:2017hcn} or for more details \cite[Sec.\ 3]{Gaberdiel:2018nbs}).
It would be interesting to develop Method 1 also for this class of representations. 

\section{Adding Rule and Rule of Simple Pole for \texorpdfstring{$n=2,3$}{n23}}
\label{appsec:AddingRulen=23}

In this appendix, we illustrate the proofs of the Adding Rule and the Rule of Simple Pole with the examples of $n=2$ and $n=3$.
We remind that it can happen that a different atom gives rise to different simple pole but with the same value, in which case the pole need to be resolved, see Appendix \ref{appsec:poleresolution}.

\subsection{\texorpdfstring{$n=2$}{n=2}}
\label{appssec:AddingRulen=2}

This situation corresponds to the graph \eqref{eq:pathequivalenceFig} with $i=1$, which we redraw here, with $c_1=c$
\begin{equation}\label{eq:pathequivalenceFign2}
\begin{array}{c}
\begin{tikzpicture}[scale=1]
%\node%[state]  
%[regular polygon, regular polygon sides=4, draw=blue!50, very thick, fill=blue!10] (a1) at (0,0)  {$1$};
\vertex (c) at (0, 1.5)  {$c$};
\vertex (b1) at (-2,0)  {$b_1$};
\vertex (b2) at (2,0)  {$b_2$};
\vertex (a) at (0,-1)  {$a$};
%\draw[fill=white] (a1) (0,0) circle (0.5) ;
%\node (f) at (-2,0) {$\infty$};
%\draw[thick, ->] (f) -- (a1);
%\path[thick, -{Latex[length=2mm, width=1.5mm]},every loop/.append style=-{Latex[length=2mm, width=1.5mm]}] 
\draw[->,very thick] (c) --  (b1);
\draw[->,very thick] (c) --  (b2);
\draw[->] (b1) --  (a) ;
\draw[->] (b2) --  (a) ;
\draw[->] (a) --  (c) 
;
\end{tikzpicture}
\end{array} 
\end{equation}

Consider a state $\Pi$ to which we want to add the atom $\sqbox{$a$}$.
A priori, either or both paths in \eqref{eq:pathequivalenceFign2} have to be in $\Pi$ in order for us to add the atom $\sqbox{$a$}$.
For either paths, the atom $\sqbox{$c$}$ has to be in $\Pi$, and since there is an arrow $I^{a\rightarrow c}\in Q_1$, it contributes to $\Psi^{(a)}_{\Pi}(u)$ a factor
\begin{equation}\label{eq:zerofromcn=2}
\Psi^{(a)}_{\Pi}(u) \ni \varphi^{a\Leftarrow c}(u- h(\sqbox{$c$}))     \ni (u +h(I^{a\rightarrow c}) -h(\sqbox{$c$}) ) \,.
\end{equation}
The atom $\sqboxs{$b_i$}$, with $i=1,2$, if in $\Pi$, would contribute a factor
\begin{equation}\label{eq:polefrombin=2}
\Psi^{(a)}_{\Pi}(u) \ni \varphi^{a\Leftarrow b_i}(u- h(\sqboxs{$b_i$}) )     \ni \frac{1}{u-h(I^{b_i\rightarrow a})-h(\sqboxs{$b_i$})  } \qquad i=1,2\,.
\end{equation}
Using \eqref{eq:pathequivalenceFign2}, we see that we need both $\sqboxs{$b_1$}$ and $\sqboxs{$b_2$}$ in $\Pi$ in order to cancel the zero from $\sqbox{$c$}$ in \eqref{eq:zerofromcn=2}.
This thus proves the Adding Rule for $n=2$.
It also shows that the atom $\sqbox{$a$}$ to be added corresponds to a simple pole in  $\Psi^{(a)}_{\Pi}(u)$.

\subsection{\texorpdfstring{$n=3$}{n=3}}

This is the situation \eqref{eq:2ctobtoa} with $n=3$.
We need to combine two sub-graphs, \eqref{eq:pathequivalenceFig} with $i=1,2$, to form the full graph. 
Depending on whether the colors of  $\sqboxs{$c_1$}$ and $\sqboxs{$c_2$}$ are the same, there are two possible scenarios:
\begin{equation}\label{eq:pathequivalenceFign3}
\begin{array}{cc}
\begin{tikzpicture}[scale=0.7]
%\node%[state]  
%[regular polygon, regular polygon sides=4, draw=blue!50, very thick, fill=blue!10] (a1) at (0,0)  {$1$};
\vertex (c) at (0,0)  {$c$};
\vertex (b1) at (-3,-2)  {$b_1$};
\vertex (b2) at (0,-2)  {$b_2$};
\vertex (b3) at (3,-2)  {$b_3$};
\vertex (a) at (0,-4)  {$a$};
%\draw[fill=white] (a1) (0,0) circle (0.5) ;
%\node (f) at (-2,0) {$\infty$};
%\draw[thick, ->] (f) -- (a1);
%\path[thick, -{Latex[length=2mm, width=1.5mm]},every loop/.append style=-{Latex[length=2mm, width=1.5mm]}] 
\draw[->,very thick] (c) --  (b1); %{$I$} 
\draw[->,very thick] (c) --  (b2) ;
\draw[->,very thick] (c) --  (b3) ;
\draw[->] (b1) --  (a) ;
\draw[->] (b2) --  (a) ;
\draw[->] (b3) --  (a) ;
\path[-{Latex[length=2mm, width=1.5mm]},every loop/.append style=-{Latex[length=2mm, width=1.5mm]}] 
(a) edge   [thin, bend left]  node [left] {$I_1$} (c) 
(a) edge   [thin, bend right]  node [right] {$I_2$} (c)
%\draw[->] (a) --  (c) 
;
\end{tikzpicture} \qquad \qquad 
\begin{tikzpicture}[scale=0.7]
%\node%[state]  
%[regular polygon, regular polygon sides=4, draw=blue!50, very thick, fill=blue!10] (a1) at (0,0)  {$1$};
\vertex (c1) at (-2,1)  {$c_1$};
\vertex (c2) at (2,1)  {$c_2$};
\vertex (b1) at (-3,-1)  {$b_1$};
\vertex (b2) at (0,-1)  {$b_{2}$};
\vertex (b3) at (3,-1)  {$b_{3}$};
\vertex (a) at (0,-3)  {$a$};
%\draw[fill=white] (a1) (0,0) circle (0.5) ;
%\node (f) at (-2,0) {$\infty$};
%\draw[thick, ->] (f) -- (a1);
%\path[thick, -{Latex[length=2mm, width=1.5mm]},every loop/.append style=-{Latex[length=2mm, width=1.5mm]}] 
\draw[->,very thick] (c1) --  (b1); %{$I$} 
\draw[->,very thick] (c1) --  (b2) ;
\draw[->,very thick] (c2) --  (b2) ;
\draw[->,very thick] (c2) --  (b3) ;
\draw[->] (b1) --  (a) ;
\draw[->] (b2) --  (a) ;
\draw[->] (b3) --  (a) ;
\draw[->] (a) --  (c1) ;
\draw[->] (a) --  (c2) ;
;
\end{tikzpicture}
\end{array} 
\end{equation}
In the first scenario, $\sqbox{$c$}$ has to be in $\Pi$ since it lies on all the three paths. 
And since there are two different arrows $I^{a\rightarrow c}_1, I^{a\rightarrow c}_2\in Q_1$, the atom $\sqbox{$c$}$ 
contributes a factor to $\Psi^{(a)}_{\Pi}(u)$:
\begin{equation}\label{eq:zerofromc2}
\Psi^{(a)}_{\Pi}(u) \ni \varphi^{a\Leftarrow c}(u- h(\sqbox{$c$}))     \ni (u +h(I^{a\rightarrow c}_1) -h(\sqbox{$c$})))  (u +h(I^{a\rightarrow c}_2) -h(\sqbox{$c$})) \,.
\end{equation}
Each $\sqboxs{$b_i$}$, with $i=1,2,3$, if in $\Pi$, will contribute a factor as in \eqref{eq:polefrombi}. 
So we need to have all three of them in $\Pi$ in order to cancel the two zeros in \eqref{eq:zerofromc2}, and hence the Adding Rule and the Rule of Simple Pole hold.

In the second scenario, we first check that when all three $\sqbox{$b$}$'s and both $\sqbox{$c$}$'s are present, there is precisely a simple pole at $\sqbox{$a$}$.
Then if we want to remove any of the $\sqbox{$b$}$'s, the degree of the pole would change by $-1$. 
Removing one of the $\sqbox{$c$}$'s on top (one cannot remove both), say $\sqboxs{$c_1$}$, doesn't help because that would necessitate also removing yet another $\sqbox{$b$}$ --- the net change of degree is at least $-1$, and thus the pole would disappear.
To summarize, in both scenarios, the Adding Rule and the Rule of Simple Pole apply.

\section{Treating higher-order poles}
\label{appsec:poleresolution}

In the main text, we have mainly dealt with the situation where the adding/removing poles for the states in the representation are simple.
For a general representation of a general quiver, there can exist higher order poles in the charge functions. In this appendix, we discuss the various scenarios when this happens and how to treat these higher order poles. 

There are the following scenarios with higher order poles:
\begin{enumerate}
\item Simple poles collide when the weights of the arrows are fine-tuned. For example, in the three-loop quiver \eqref{fig:C3quiver} for the $\mathbb{C}^3$ case, suppose we set $h_1=h_2$, then the two adding poles at $z=h_1$ and $z=h_2$ of the one-box state (with charge function $\frac{1}{z}\prod^3_{I=1}\frac{(z+h_I)}{(z-h_I)}$) would coincide, leading to a double pole at $z=h_1=h_2$. 
Since this scenario is somewhat artificially degenerate, in this paper we will not consider it, namely we will only examine the cases where the higher-order poles arise even with generic $\{h_I\}$.
\item Two or more (non-equivalent) paths share the same adding/removing pole even with generic weights $\{h_I\}$. There are two subclasses:
\begin{enumerate}
\item The quiver has no potential and hence no path equivalence, and its representation contains states with two or more paths whose weight coincide even for generic value of $\{h_I\}$.
One prime example is the generic  representations for the $m$-loop quiver with $m\geq 2$.
The reason for the colliding of the poles corresponding to different paths (which are never equivalent in this class) is that the numerical value of the pole is insensitive to the order of the arrows in a path.
Therefore we can resolve these higher order poles by introducing a non-commutative addition and subtraction of weights, characterizing the ordering of arrows in the paths.
\item The quiver has a potential but for some representations, the potential is not strong enough to eliminate all but one path, resulting in higher order poles. 
One prime example is the triple quiver of the affine D$_4$. 
In general, if these higher order poles are pure, one can modify the action of the $e$ and $f$ generators.
Otherwise, one should only consider those representations for which in all states, the potential can eliminate all but one path. 
(These turn out to be also the interesting physical cases.)
\end{enumerate}

\end{enumerate}
We will now discuss these two cases.

\subsection{Resolution of higher order poles}
\label{appssec:resolving}

In this subsection, we consider the case 2-(a), where an atom corresponds to a unique path, which happens for quivers without potential, and the higher-order poles arise when different paths share the same weight.  
We will explain that in this case, one can resolve the higher-order poles by introducing a non-commutative addition and subtraction of weights.

Let's first discuss the general problem of higher-order poles.
Recall that although each atom corresponds to a 1st-order pole in the charge function, for a general quiver it can happen that $n\geq 2$ different atoms, $\sqbox{$a_i$}$ with $i=1,2,\dots, n$, give rise to poles with the same value, contributing a $n^{\textrm{th}}$-order pole to the charge function:
\begin{equation}\label{eq:Psinthorder}
\Psi_{\Pi}(u)=\frac{f(u)}{(u-u^{*})^n}\,, \qquad u^*=h(\sqbox{$a_1$})=\dots =h(\sqbox{$a_n$})\,.
\end{equation}
Since an $n^{\textrm{th}}$-order pole corresponds to $n$ different atoms and in this case $n$ different paths, the proper way of dealing with it is to resolve it into $n$ simple poles. 
We now explain this resolution procedure and show that it does not affect the algebra, in particular, does not introduce new parameters to the algebra. 

The charge function \eqref{eq:Psinthorder} can be resolved into
\begin{equation}\label{eq:PsinthorderResolution}
\begin{aligned}
\Psi_{\Pi}(u)=\frac{f(u)}{(u-u^{*})^n}\rightarrow \tilde{\Psi}_{\Pi}(u)&= f(u)\prod^{n}_{i=1}\frac{1}{u-(u^{*}+\epsilon_i)}\\	&=f(u)\sum^{n}_{i=1}\frac{\delta_i}{u-(u^{*}+\epsilon_i)}\,,	
	\end{aligned}
\end{equation}
where 
\begin{equation}
	\delta_i\equiv (-1)^{n-1}\prod^{n}_{j\neq i}(\epsilon_j-\epsilon_i)^{-1}\,.
\end{equation}
The residues in \eqref{eq:constraintfromef} and the subsequent formulae then become residues for simple poles again, and the derivation in Sec \ref{ssec:Bootstrap} goes through as in the toric Calabi-Yau-three case when there is never any accidental higher-order pole, and gives the action on the representation \eqref{eq:ActiononRep}.

Now there are two ways to treat the parameters $\epsilon_i$ that arise in this resolution process.
One might apply the following procedure 
\begin{equation}\label{eq:ResDef}
	\textrm{Res}_{u=u^*+\epsilon_i}\tilde{\Psi}_{\Pi}(u)\rightarrow \left(\frac{1}{\delta_i}\textrm{Res}_{u=u^*+\epsilon_i}\tilde{\Psi}_{\Pi}(u)\right)|_{\textrm{all } \epsilon_k\rightarrow 0}
\end{equation}
to eliminate all $\epsilon_i$ after each resolution.
One drawback of this approach is that the $n$ atoms would have the same simple pole in the action of $e(z)$, resulting in null states in the representation, which requires further quotienting them out.

One can also regard the parameter $\epsilon_i$ as parameters that characterize the representation, as long as only a finite number of $\epsilon_i$ are needed to resolve all higher-order poles for the entire representation. 
This is the case for the $m$-loop quiver, and hence for those symmetric quivers that appear in the knot-quiver correspondence. 
We will now illustrate this approach for the simplest case: the quiver with one vertex and $m$ loops. 
The method can be straightforwardly generalized to more complicated cases.

\subsubsection{Example: $m$-loop quivers}

For the $m$-loop quivers (with the arrows labeled by $I_{i}$, $i=1,2,\dots,m$), the higher-order poles can appear because two different paths $I_i\cdot I_{j }\neq I_{j}\cdot I_i$ share the same weight $h_i+h_j$, which results in multiple atoms sharing the same numerical value for their weights. 
However, the ordering of these arrows matters in the description of the atoms as paths in the $m$-dimensional lattice. 
Therefore, one simple way to resolve these higher-order poles is just to introduce a "non-commutative" addition (and subtraction) of the weights $\{h_{i}\}$, as a way to reflect the ordering of the arrows in the path. 

We define the non-commutative addition $\dot{+}$ of $h_i$ as
\begin{equation}
	h_{i}\dot{+}h_{j}\equiv h_i+h_j+\epsilon_{ij}\,,
\end{equation}
where $\epsilon_{ij}+\epsilon_{ji}=0$, namely we are introducing $m(m-1)/2$ parameters in order to describe the representations of the algebra. (As we will show momentarily, these parameters do not affect the algebra itself.)
The binary operation $\dot{+}$ satisfies $a\dot{+}0=0\dot{+}a=a$ and the associativity
\begin{equation}
	h_{i}\dot{+}h_{j}\dot{+}h_k= (h_{i}\dot{+}h_{j})\dot{+}h_k=h_{i}\dot{+}(h_{j}\dot{+}h_k)= h_i+h_j+h_k+\epsilon_{ij}+\epsilon_{ik}+\epsilon_{jk}\,.	
\end{equation}
The non-commutative subtraction $\dot{-}$ is then just the inverse of $\dot{+}$: 
\begin{equation}
	a\dot{+}b\dot{-}b
	=a\dot{-}b\dot{+}b
	=(a\dot{+}b)\dot{-}b =a\dot{+}(b\dot{-}b)=a\,,
\end{equation}
and satisfies 
\begin{equation}
	a\dot{-}a=0 
	\quad \textrm{and}
	\quad a\dot{-}(b\dot{+}c)=a\dot{-}b\dot{-}c\,.
\end{equation}

With the introduction of the non-commutative $\dot{+}$ and $\dot{-}$, the definition for the weight of a path needs to be modified accordingly:
\begin{equation}
	h(\sqbox{$0$})
	\rightarrow\dot{h}(\sqbox{$0$})
	=\dot{\Sigma}_{I\in\mathfrak{p}(\sqbox{$0$})}h_I
	=h_{i_{1}}\dot{+}h_{i_{2}}\dot{+}\dots=h(\sqbox{$0$})+\sum_{i_n<i_m}\epsilon_{i_ni_m}\,,
\end{equation}
where the path is given by $I_{i_1}I_{i_2}\cdots$.
Similarly, the charge function is now
\begin{equation}
	\Psi_{\Kappa}(u)=\prod_{\sqbox{$0$}\in \Kappa}\varphi^{0\Leftarrow0}(u\dot{-}\dot{h}(\sqbox{$0$}))
	\quad \textrm{with} \quad
	\varphi(u)=\prod^{m}_{i=1}\frac{u\dot{+}h_I}{u\dot{-}h_J}\,,
\end{equation}
where
\begin{equation}
	\varphi^{0\Leftarrow0}(u\dot{-}\dot{h}(\sqbox{$0$}))
	=\prod^{m}_{i=1}\frac{u\dot{-}\dot{h}(\sqbox{$0$})\dot{+}h_I}{u\dot{-}\dot{h}(\sqbox{$0$})\dot{-}h_J} 
	=\prod^{m}_{i=1}\frac{u\dot{-}(\dot{h}(\sqbox{$0$})\dot{-}h_I)}{u\dot{-}(\dot{h}(\sqbox{$0$})\dot{+}h_J)} 
	\rightarrow \prod^{m}_{i=1}\frac{u-(\dot{h}(\sqbox{$0$})\dot{-}h_I)}{u-(\dot{h}(\sqbox{$0$})\dot{+}h_J)} \,.
\end{equation}
In the last step, we have specified $u\dot{-} a=u-a$ since the goal of using the non-commutative addition and subtraction to keep track of the ordering in the paths has been achieved and the quantities inside the brackets now directly correspond to the adding/removing poles and canceling zeros.

After the resolution of the higher-order poles this way, one can check that the action \eqref{eq:ActiononRep} respects the algebra. 
In particular, the algebra remains unchanged, i.e.\ the new $\epsilon_{ij}$ parameters do not enter the algebraic relations.
This method of resolution applies to all cases where different paths can accidentally share the same numerical values for their weights, in particular, to all the quivers from the knot-quiver correspondence.\footnote{It would be interesting to see whether this non-commutative addition/subtraction  is related to the skein-valued knot-quiver correspondence in \cite{Ekholm:2024ceb}.}

\subsection{Condition on framing}
\label{appssec:onframing}
In this subsection we consider the 2-(b) case, where the higher-order poles arise because for some representation, the potential is not strong enough to eliminate all but one path.

Recall that in studying the representation of the quiver Yangian, we have focused on those framings $\sharp$ such that in the resulting representation, for each state of the representation,  the constraints $\partial_{I} W=0$ (if $W$ is present) always eliminate all but one path with the same weight.
This is crucial in order to define the ``atom" as a path up to path equivalence, see \eqref{eq:AtomDef} and to further define a meaningful charge function and the action of the quiver Yangian on the states. 
If one ignores this condition and apply the definition of the charge function and the action, one would encounter higher-order poles. 

First of all, if the higher order pole happens to be pure:
\begin{equation}
\begin{aligned}
\Psi^{(a)}_{\Pi}(u)=\frac{f(u)}{(u-h^{*})^n}&=\frac{f(h^*)}{(u-h^{*})^n}\,,
\end{aligned}
\end{equation}
namely if
\begin{equation}
f'(h^{*})=\dots = f^{(n-1)}(h^{*})=0\,,
\end{equation}
then we can simply modify the action of the $e$ and $f$ (on these states) to be
\begin{equation}\label{eq:modifideefaction}
\begin{aligned}
e^{(a)}(z)|\Pi\rangle &= \sum^{n}_{m=1}\frac{E}{(z-h^{*})^m}|\Pi+\sqbox{$a$}\rangle_m\,, \\
f^{(a)}(z)|\Pi+\sqbox{$a$}\rangle_m&=\frac{F}{(z-h^{*})^{n-m+1}}|\Pi\rangle
\end{aligned}
\end{equation}
where the $n$ states $|\Pi+\sqbox{$a$}\rangle_m$ with $m=1,2,\dots,n$ span the vector space of the relevant paths after imposing the constraint $\partial W=0$, and
\begin{equation} E\sim F\sim(\textrm{lim}_{u\rightarrow h^*}((u-h^*)^n \Psi^{(a)}_{\Pi}(u)))^{\frac{1}{2}}=f(h^*)^{\frac{1}{2}}
\end{equation}
where $\sim$ means up to the phase factor that is to be fixed in the same way as in the simple pole case.

However, if the poles are not pure, one needs to choose the framing such that within the corresponding representation, for each state of the representation, the constraints $\partial_{I} W=0$ (if $W$ is present) always eliminate all but one path with the same weight.  
It is easiest to illustrate this using an example: the triple quiver of affine $D_4$.\footnote{We thank Yegor Zenkevich for suggesting this example.}

%\subsection{Condition of framings}
%\label{appssec:framingconstraint}

\subsubsection{Example: triple quiver of affine $D_4$}
\label{appsssec:PosetD4}

Consider the triple quiver of the affine D$_4$ Dynkin diagram:
\begin{equation}\label{eq:AffineD4}
D_{4}:\qquad
\begin{array}{c}
\begin{tikzpicture}[
block/.style={circle, draw, minimum size={width("$\tiny{n-2}$")+0pt},
font=\small,scaling=0.9}]
\node[vertex,minimum size=0.5mm,font=\scriptsize] (a0) at (-2,2) {0};
\node[vertex,minimum size=0.5mm,font=\scriptsize] (a1) at (-4,0) {1};
\node[vertex,minimum size=0.5mm,font=\scriptsize] (a2) at (-2,0) {2};
\node[vertex,minimum size=0.5mm,font=\scriptsize] (a4) at (0,0) {4};
\node[vertex,minimum size=0.5mm,font=\scriptsize] (a3) at (-2,-2) {3};
\path[-{Latex[length=2mm, width=1.0mm]},every loop/.append style=-{Latex[length=2mm, width=1.0mm]}] 
(a0) edge [in=60, out=120, loop, thin, above,font=\scriptsize] node {$h_3$} ()
(a1) edge [in=150, out=210, loop, thin, left,font=\scriptsize] node {$h_3$} ()
(a4) edge [in=330, out=30, loop, thin, right,font=\scriptsize] node {$h_3$} ()
(a2) edge [in=265, out=220, loop, thin, left,font=\scriptsize] node {$h_3$} ()	
(a3) edge [in=300, out=240, loop, thin, above,font=\scriptsize] node {$h_3$} ()
(a0) edge   [thin, bend left,font=\scriptsize]  node [right] {$h_1$} (a2) 
(a2) edge   [thin, bend left,font=\scriptsize]  node [left]{$h_2$} (a0)
(a2) edge   [thin,font=\scriptsize]  node [above] {$h_1$} (a1) 
(a1) edge   [thin, bend right,font=\scriptsize]  node [below]{$h_2$} (a2)
(a2) edge   [thin,font=\scriptsize]  node [above] {$h_1$} (a4) 
(a4) edge   [thin, bend left,font=\scriptsize]  node [below]{$h_2$} (a2)
%			%
(a2) edge   [thin, bend left,font=\scriptsize]  node [right] {$h_2$} (a3) 
(a3) edge   [thin, 
%bend left,
font=\scriptsize]  node [left]{$h_1$} (a2);
\end{tikzpicture}
\end{array}
\end{equation}
Let's consider the vacuum representation given by the framing with $\infty \rightarrow 0$.
The first state at which the a non-simple adding pole arise is the following state at level $5$:
\begin{equation}\label{eq:D4leve5state}
\{I^{\infty \rightarrow 0}
\,,\,
I^{\infty \rightarrow 0 \rightarrow 2}\,,\, 
I^{\infty \rightarrow 0 \rightarrow 2 \rightarrow 1}\,,\, 
I^{\infty \rightarrow 0 \rightarrow 2 \rightarrow 3}\,,\, 
I^{\infty \rightarrow 0 \rightarrow 2 \rightarrow 4} \}
\end{equation}
with charge function
\begin{equation}\label{eq:chargefunctionD4}
\Psi^{(2)}(u)=\frac{(u-h_1)^2}{(u-(2h_1+h_2))^2}
\end{equation}
where the double pole at $2h_1+h_2$ is due to the fact that at the next level, the three paths
\begin{equation}\label{eq:D4threepaths}
\{I^{\infty \rightarrow 0 \rightarrow 2 \rightarrow 1\rightarrow 2}\,,\,
I^{\infty \rightarrow 0 \rightarrow 2 \rightarrow 3\rightarrow 2}\,,\, I^{\infty \rightarrow 0 \rightarrow 2 \rightarrow 4\rightarrow 2} \}
\end{equation}
are subject to only one linear constraint, from the terms in the potential:
\begin{equation}
W\supset I^{2\rightarrow 2} I^{2\rightarrow 1}I^{1\rightarrow 2} 
+I^{2\rightarrow 2} I^{2\rightarrow 3}I^{3\rightarrow 2}
-I^{2\rightarrow 2} I^{2\rightarrow 4}I^{4\rightarrow 2}
\end{equation}
Note that this is due to the 4-valence of the Dynkin diagram at  node $2$, hence within the affine ADE series, this problem for the vacuum representation only arises for the affine D$_4$ case.
For all other cases, the three paths in \eqref{eq:D4threepaths} would become only two, and after the constraint is imposed, one can interpret the one independent path as an atom.

Note that this double pole is not pure, therefore one cannot just modify the action of $e$ and $f$ as in \eqref{eq:modifideefaction}.
Instead, one should only consider smaller representations that do not suffer from this problem of not enough constraints. 
For example, one can add a framing arrow $4\rightarrow \infty$ with weight $-2h_1$, canceling the adding pole for creating the path $I^{\infty \rightarrow 0 \rightarrow 2 \rightarrow 4}$ in \eqref{eq:D4leve5state}.
As a result, the path  $I^{\infty \rightarrow 0 \rightarrow 2 \rightarrow 4\rightarrow 2}$ would also disappear from  \eqref{eq:D4threepaths}, and the remaining two paths, after the loop constraint is imposed, produce only one independent path and hence can be interpreted as an atom again. 
Similarly, one can also add a framing arrow $2\rightarrow \infty$ with weight $-(2h_1+h_2)$, canceling one of the poles in \eqref{eq:chargefunctionD4}; the effect of this is to add a ``negative" path that cancels one of the three paths in \eqref{eq:D4threepaths}, resulting in only one independent path after the constraint.\footnote{Note that this negative path also automatically introduces more negative paths at the higher level and they cancel more poles at higher levels.}
Interestingly, we have seen in a few physical setups with potential higher order poles that the physically relevant representations always have the framing such that higher order poles do not arise, and we hope to report this in a future work.

\section{Examples of poset constructions}
\label{appsec:ExamplePosetConstruction}

\subsection{Toric \texorpdfstring{CY$_3$}{CY3}  quivers}
\label{appssec:toricCY3ex}

In this subsection, we reexamine quivers with potential that come from toric CY$_3$'s, whose corresponding quiver Yangians were defined in \cite{Li:2020rij}.
We will confirm that the poset-generating procedure of Sec.\ \ref{sssec:RepresentationConstruction} indeed reproduces the 3D colored crystals in \cite{Ooguri:2008yb}.

For IIA string theory on toric CY$_3$, the BPS sector can be described by an $\mathcal{N}=4$ quiver quantum mechanics, with quiver $Q$ and superpotential $W$ \cite{Denef:2002ru}. 
The superpotential $W$ has the property that each $I$ appears twice, once in a term with ``$+$" sign and once in a term with ``$-$" sign \cite{Feng:2000mi}.
This property allows one to translate the quiver and superpotential pair into a periodic quiver \cite{Franco:2005rj,Franco:2005sm}
\begin{equation}\label{eq:periodicQDef}
    \tilde{Q}=(Q_0,Q_1,s,t, Q_2)\,,
\end{equation}
which can be drawn on $\mathbb{R}^2$ \textit{periodically}, with the properties that (1) along each face, the arrows point along the same direction, and $Q_2$ denotes the set of faces in the fundamental domain of the periodic quiver $\tilde{Q}$; and (2) each face $f\in Q_{2}$ in the fundamental domain corresponds to a term in the superpotential $W$, with the loop of the arrows around the face corresponding to the monomial and the orientation of the loop corresponding to the sign of the term.
Therefore in this case, the constraint \eqref{eq:LoopConstraints} that comes from each term of $W$ can be associated to each loop in the periodic quiver (named loop constraint in \cite{Li:2020rij}):
\begin{equation}\label{eq:LoopConstraintsToricCY3}
    \textrm{loop constraint:} \qquad \sum_{I \in L} h_I=0 \qquad \forall \, L\in Q_2 \,.
\end{equation}
This reformulation of $(Q,W)$ as a 2D periodic quiver is the first step in translating the poset into a 3D crystal. 

Let us first consider the canonically framed quiver defined in \eqref{eq:CanonicalQW}. 
For the $(Q,W)$ that comes from a toric CY$_3$, each path $\mathfrak{p}^{0\rightarrow a}$ can be viewed as an atom with color $a$, denoted by $\sqbox{$a$}$, in a 3D crystal uplift from the periodic quiver $\tilde{Q}$ defined in \eqref{eq:periodicQDef}:
\begin{equation}
\mathfrak{p}^{0\rightarrow a}_{d}  \quad \longleftrightarrow \quad \sqbox{$a$}_{(h,d)}\,.
\end{equation}
Here $d$ is the level of the state $\Pi_{d}$, namely the number of steps to reach $\Pi_{d}$ in the poset-generating procedure of Sec.\ \ref{sssec:RepresentationConstruction}, and it also gives the number of arrows (including the first zero-length one $I^{(0)}_0$) in $\mathfrak{p}^{0\rightarrow a}$.
The atom $\sqbox{$a$}$ is specified by three pieces of information: (1) the color $a\in Q_0$; (2) the coordinate function $h(\sqbox{$a$})$, defined as
\begin{equation}
h(\sqbox{$a$}) \equiv \sum_{I\in \textrm{path}[\mathfrak{o}\rightarrow \sqbox{$a$}]} h_I     \,,
\end{equation}
which corresponds to the equivariant weight of the atom $\sqbox{$a$}$; and finally (3) the depth $n$, to be defined momentarily.
As before, this 3D crystal structure can be explained level by level.

\medskip

\noindent \textbf{Level 0}. 
At level 0, there is only one element $\Pi=\{\varnothing\}$, which has no pole.
Hence there is no path and no atom corresponding to it.

\medskip

\noindent \textbf{Level 1}. 
At level 1, since we are considering a vacuum module, namely the canonically framed $(^{0}Q,^{0}W)$ where there is only one arrow from the framing vertex $\infty$ to the framed vertex $a=0$, the pole in $\Pi_{1}$ is $(0)^{(0)}$, corresponding to the zero-length path $ \mathfrak{p}^{0}=I^{(0)}_0$.
In the periodic quiver $\tilde{Q}$, we can view this zero-length path as starting from a vertex $a=0$ in $\tilde{Q}$ and ending right there, which in turn can be viewed as an ``atom" of color $0$, at this vertex $a=0$. (There are an infinite number of such vertices since we considering the periodic quiver, and we can choose an arbitrary one and fix it as the origin, thus breaking periodicity.) 
This will be the \textit{leading atom} of the crystal. 

\medskip

\noindent \textbf{Level 2}. 
Recall from the discussion around \eqref{eq:StatesLevel2}  that there are in total \linebreak
$\sum_{a\in Q_0}|0\rightarrow a|$ poles at level 2, each corresponding to an arrow $J$ from $a=0$ to $t(J)$, namely a length-one path $\mathfrak{p}^{0\rightarrow t(J)}=
I^{(0)}_0\cdot J$.
Each such path goes from the vertex $a=0$ at the origin of the periodic quiver $\tilde{Q}$ to an adjacent vertex (with color $t(J)$), and can be interpreted as an atom of color $t(J)$, sitting at the corresponding vertex of color $t(J)$.

\medskip

\noindent \textbf{Level 3}. Similarly, a pole at level 3 corresponds to a path consisting of $2$ arrows, and in turn to an atom that sits adjacent to the atom at level 2. 

\medskip

\noindent \textbf{Level 4.} A pole at level 4 corresponds to a path consisting of $3$ arrows. 
Depending on the quiver, starting from this level, it is possible to have a closed loop in the periodic quiver. 
The depth $n$ of an atom $\sqbox{$a$}$ is the number of closed loops in its corresponding path:
\begin{equation}
\textrm{depth } n \equiv \# \textrm{ of loops in }  \mathfrak{p}^{0\rightarrow a} \,.
\end{equation}

To summarize, we have the dictionary for the toric CY$_3$ quivers:
\begin{equation}
\begin{aligned}
\textrm{poset } \mathcal{P}
\quad &\longleftrightarrow \quad 
\textrm{full crystal }\mathcal{C}
\\
\textrm{poset element } \Pi \in  \mathcal{P}
\quad &\longleftrightarrow \quad
\textrm{crystal state }\Kappa \in \mathcal{C} 
\\
\textrm{path } p^{(a)}\in \Pi 
\quad &\longleftrightarrow \quad 
\textrm{atom } \sqbox{$a$}\in\Kappa
\,.
\end{aligned}
\end{equation}
Finally, the loop constraint corresponds to the global symmetry of the theory. 
It is also possible to impose the vertex constraints
\begin{equation}
\textrm{vertex constraint:} \qquad \sum_{I \in \{a\rightarrow * \}} h_I =\sum_{J \in \{ * \rightarrow a \}} h_J \qquad \forall a \, \in Q_0\,,
\end{equation}
which corresponds to the gauge symmetries of the theory.
After imposing both the loop and vertex constraints, the number of parameters is \cite{Li:2020rij}
\begin{equation}
|Q_0|+|Q_2|-|Q_1|=2    \,,
\end{equation}
which corresponds to the two equivariant parameters of a toric CY$_3$. 

\subsection{Poset construction for \texorpdfstring{$\mathbb{C}^2/\mathbb{Z}_{n+1}\times \mathbb{C}$}{C2/Zn+1 C} }
\label{appssec:PosetAn}

In this subsection, we construct the vacuum representation for the affine $A$-type triple quiver \eqref{eq:AffineAnQuiver}.
The weight assignment of the arrows are
\begin{equation}
h(I^{a\rightarrow a+1})=h_1\,, \qquad h(I^{a\rightarrow a-1})=h_2\,, \qquad h(I^{a\rightarrow a})=h_3\,,
\end{equation}
where $a\in \mathbb{Z}/(n+1)\mathbb{Z}$ and $h_1+h_2+h_3=0$.
As we will see, this precisely reproduces the colored plane partition configurations that describe the 1/2 BPS states of type IIA string on the toric CY$_3$ $(\mathbb{C}^2)/\mathbb{Z}_{n+1}\times \mathbb{C}$. 

\noindent \textbf{Level 0.} There is only one state at level 0: the ground state $\Pi_{0}=\{\varnothing\}$, which is the empty set. 
By definition \eqref{eq:ChargeFunctionDef}, the charge function of $\Pi_{0}$, is just the vacuum charge function, which can be read off from the framed quiver  by definition \eqref{eq:varphi0Def} to be
\begin{equation}
\Psi^{(0)}_{\Pi_0}(z)=   \psi_{0}(z)=\frac{1}{z}\,, \qquad
\Psi^{(a)}_{\Pi_0}(z)=   1\,, \,\,\, \textrm{for} \,\, a=1,2,\dots,n \,,\\
\end{equation}
which has only one pole at $z=0$ for color $0$.
Together with the fact that the state $\Pi_{0}$ is the empty set,  by definition \eqref{eq:RemPi} and \eqref{eq:AddPi} (or equivalently from \eqref{eq:RemVacuum} and \eqref{eq:AddVacuumAtom}), we have
\begin{equation}
\textrm{Rem}(\Pi_{0})=\{\varnothing\} \qquad \textrm{and} \qquad \textrm{Add}(\Pi_{0})=\{\mathfrak{p}_0\} \,,
\end{equation}
where we have used the shorthand notation introduced in \eqref{eq:ADEshorthand}.

\noindent \textbf{Level 1.} From the single atom in $\textrm{Add}(\Pi_0)$, we obtain one state at level-1:
\begin{equation}
    \Pi_1=\{\mathfrak{p}_0\}\,,
\end{equation}
by Step-4 of the procedure in Sec.\ \ref{sssec:RepresentationConstruction}, or directly from \eqref{eq:StatesLevel1}.
Its charge functions are
\begin{equation}
\begin{aligned}
&\Psi^{(0)}_{\Pi_1}(z)=\frac{1}{z}\cdot\frac{z+h_3}{z-h_3}\,, \qquad \Psi^{(1)}_{\Pi_1}(z)=\frac{z+h_2}{z-h_1}\,, \qquad\\
&\Psi^{(n)}_{\Pi_1}(z)=\frac{z+h_1}{z-h_2} \,, \qquad \quad \,\,
\Psi^{(a)}_{\Pi_1}(z)=   1 \,\,\, \textrm{for} \,\, a=2,\dots,n-1\,,
\end{aligned}
\end{equation}
by definition \eqref{eq:ChargeFunctionDef}.
Then by definition \eqref{eq:RemPi} and \eqref{eq:AddPi}, we have
\begin{equation}
\textrm{Rem}(\Pi_{1})=\{
\mathfrak{p}_0
\} \qquad \textrm{and} \qquad 
\textrm{Add}(\Pi_{1})=\{
\mathfrak{p}_{0}^{0\rightarrow 1},\, 
\mathfrak{p}_{0}^{0\rightarrow n},\, 
\mathfrak{p}_{0}^{0\rightarrow 0}
\}\,.
\end{equation}

\noindent \textbf{Level 2.} From the three atoms in $\textrm{Add}(\Pi_1)$, we obtain three states at level 2 
\begin{equation}
\Pi_{2,1}= \{
\mathfrak{p}_{0},\,
\mathfrak{p}_{0}^{0\rightarrow 1}
\} \,,\qquad 
\Pi_{2,2}= \{
\mathfrak{p}_{0},\,
\mathfrak{p}_{0}^{0\rightarrow n}
\} \,,\qquad
\Pi_{2,3}=\{
\mathfrak{p}_{0},\,
\mathfrak{p}_{0}^{0\rightarrow 0}
\}\,.
\end{equation}
The charge functions of $\Pi_{2,1}$ are
\begin{equation} 
\begin{aligned}
&\Psi^{(0)}_{\Pi_{2,1}}(z)=\frac{1}{z-h_3}\,,  \qquad \Psi^{(1)}_{\Pi_{2,1}}(z)=\frac{z+h_3-h_1}{z-h_1}   \,, \qquad \Psi^{(2)}_{\Pi_{2,1}}(z)=\frac{z+h_2-h_1}{z-2h_1}\,,  \\
&\Psi^{(n)}_{\Pi_{2,1}}(z)=\frac{z+h_1}{z-h_2}\,,
\qquad \Psi^{(a)}_{\Pi_{2,1}}(z)=   1 \,\,\, \textrm{for} \,\, a=3,\dots, n-1\,,
\end{aligned}
\end{equation}
which gives the set of removable and addable atoms as
\begin{equation}
\begin{aligned}
\textrm{Rem}(\Pi_{2,1})=\{
\mathfrak{p}_{0}^{0\rightarrow 1}
\} \qquad \textrm{and} \qquad
\textrm{Add}(\Pi_{2,1})= \{
\mathfrak{p}_{0}^{0\rightarrow 1\rightarrow 2},\, 
\mathfrak{p}_{0}^{0\rightarrow n} ,\,
\mathfrak{p}_{0}^{0\rightarrow 0}
\}\,,
\end{aligned}
\end{equation}
among which the addable atoms give rise to three states at level $3$:
\begin{equation}\label{eq:AnNewStates21}
\begin{aligned}
&\{
\mathfrak{p}_{0},\,
\mathfrak{p}_{0}^{0\rightarrow 1},\, \mathfrak{p}_{0}^{0\rightarrow 1\rightarrow 2}
\}\,, \qquad 
\{
\mathfrak{p}_{0} ,\,
\mathfrak{p}_{0}^{0\rightarrow 1} ,\, 
\mathfrak{p}_{0}^{0\rightarrow n}
\}\,,\qquad 
\{
\mathfrak{p}_{0} ,\,
\mathfrak{p}_{0}^{0\rightarrow 1} ,\, 
\mathfrak{p}_{0}^{0\rightarrow 0}
\}\,.
\end{aligned}
\end{equation}
The charge functions of $\Pi_{2,2}$ are
\begin{equation}
\begin{aligned}
&\Psi^{(0)}_{\Pi_{2,2}}(z)=\frac{1}{z-h_3}\,, \qquad \qquad \,\,\, \Psi^{(1)}_{\Pi_{2,2}}(z)=\frac{z+h_2}{z-h_1}   \\
&\Psi^{(n-1)}_{\Pi_{2,2}}(z)=\frac{z+h_1-h_2}{z-2h_2}\,,  \qquad 
\Psi^{(n)}_{\Pi_{2,2}}(z)=\frac{z+h_3-h_2}{z-h_2}\,,\\
&\Psi^{(a)}_{\Pi_{2,2}}(z)=   1 \,\,\, \textrm{for} \,\, a=2,\dots, n-2\,,
\end{aligned}
\end{equation}
which give the set of removable and addable atoms as
\begin{equation}
\begin{aligned}
\textrm{Rem}(\Pi_{2,2})=\{
\mathfrak{p}_{0}^{0\rightarrow n}
\} \qquad \textrm{and} \qquad 
\textrm{Add}(\Pi_{2,2})= \{
\mathfrak{p}_{0}^{0\rightarrow 1} ,\,
\mathfrak{p}_{0}^{0\rightarrow n \rightarrow n-1} ,\, 
\mathfrak{p}_{0}^{0\rightarrow 0} 
\}\,,
\end{aligned}
\end{equation}
among which the addable atoms give rise to three states at level $3$:
\begin{equation}\label{eq:AnNewStates22}
\begin{aligned}
&\{
\mathfrak{p}_{0} ,\,
\mathfrak{p}_{0}^{0\rightarrow n} ,\, \mathfrak{p}_{0}^{0\rightarrow 1}
\}\,,\qquad 
\{
\mathfrak{p}_{0} ,\,
\mathfrak{p}_{0}^{0\rightarrow n} ,\, \mathfrak{p}_{0}^{0\rightarrow n\rightarrow n-1}
\}\,,  \qquad 
\{
\mathfrak{p}_{0} ,\,
\mathfrak{p}_{0}^{0\rightarrow n} ,\, \mathfrak{p}_{0}^{0\rightarrow 0}
\}\,.
\end{aligned}
\end{equation}
The charge functions of $\Pi_{2,3}$ are
\begin{equation} 
\begin{aligned}
&\Psi^{(0)}_{\Pi_{2,3}}(z)=\frac{z+h_3}{(z-h_3)(z-2h_3)}\,, \qquad  \Psi^{(1)}_{\Pi_{2,3}}(z)=\frac{z+h_2-h_3}{z-h_1}\,, \\
&  \Psi^{(n)}_{\Pi_{2,3}}(z)=\frac{z+h_1-h_3}{z-h_2} \,, \qquad
\Psi^{(a)}_{\Pi_{2,3}}(z)=   1 \,\,\, \textrm{for} \,\, a=2,\dots,n-1\,,
\end{aligned}
\end{equation}
which gives the set of removable and addable atoms as
\begin{equation}
\begin{aligned}
\textrm{Rem}(\Pi_{2,3})=\{
\mathfrak{p}_{0}^{0\rightarrow 0}
\} \qquad \textrm{and} \qquad 
\textrm{Add}(\Pi_{2,3})= \{
\mathfrak{p}_{0}^{0\rightarrow 1} ,\, 
\mathfrak{p}_{0}^{0\rightarrow n},\,
\mathfrak{p}_{0}^{0\rightarrow 0\rightarrow 0}
\}\,,
\end{aligned}
\end{equation}
among which the addable atoms give rise to three states at level $3$:
\begin{equation}\label{eq:AnNewStates23}
\begin{aligned}
&\{
\mathfrak{p}_{0},\,
\mathfrak{p}_{0}^{0\rightarrow 0} ,\, \mathfrak{p}_{0}^{0\rightarrow 1}
\} \,, \qquad 
\{
\mathfrak{p}_{0},\,
\mathfrak{p}_{0}^{0\rightarrow 0},\, \mathfrak{p}_{0}^{0\rightarrow n}
\}\,,\qquad
\{
\mathfrak{p}_{0},\,
\mathfrak{p}_{0}^{0\rightarrow 0} ,\, \mathfrak{p}_{0}^{0\rightarrow 0\rightarrow 0}
\}\,.
\end{aligned}
\end{equation}
Finally, among the nine states in \eqref{eq:AnNewStates21}, \eqref{eq:AnNewStates22} and \eqref{eq:AnNewStates23}, there are only $6$ distinct ones, namely, there are $6$ new states at level 3:
\begin{equation}
\begin{aligned}
&\Pi_{3,1}=\{
\mathfrak{p}_{0},\,
\mathfrak{p}_{0}^{0\rightarrow 1} ,\, \mathfrak{p}_{0}^{0\rightarrow 1 \rightarrow 2}
\}\,,\quad \,\,\,\,
\Pi_{3,2}=\{
\mathfrak{p}_{0} ,\, 
\mathfrak{p}_{0}^{0\rightarrow 1},\, \mathfrak{p}_{0}^{0\rightarrow n}
\}\,,\,\,\,
\Pi_{3,3}=\{
\mathfrak{p}_{0},\,
\mathfrak{p}_{0}^{0\rightarrow 1} ,\, \mathfrak{p}_{0}^{0\rightarrow 0}
\}\,,\\%\quad
&\Pi_{3,4}=\{
\mathfrak{p}_{0},\,
\mathfrak{p}_{0}^{0\rightarrow n} ,\, \mathfrak{p}_{0}^{0\rightarrow n\rightarrow n-1}
\}\,,\,\,\, 
\Pi_{3,5}=\{
\mathfrak{p}_{0} ,\,
\mathfrak{p}_{0}^{0\rightarrow n} ,\, \mathfrak{p}_{0}^{0\rightarrow 0}
\}\,,\,\,\, 
\Pi_{3,6}=\{
\mathfrak{p}_{0} ,\,
\mathfrak{p}_{0}^{0\rightarrow 0} ,\, \mathfrak{p}_{0}^{0\rightarrow 0\rightarrow 0}
\}\,.
\end{aligned}
\end{equation}

One can repeat this procedure and construct the entire poset. 
We show the Hasse diagram of the poset up to level 3:
\begin{equation}
\begin{array}{c}
\begin{tikzpicture}
\node (a0) at (0,0)  {$\{\varnothing\}$};
%%%%%%%%%%%%%%%%%%%
\node (a11) at (0,-1)  {$\{
\mathfrak{p}_{0}
\}$};
%%%%%%%%%%%%%%%%%%%
\node (a21) at (-4,-2)  {$\{
\mathfrak{p}_{0},\,
\mathfrak{p}_{0}^{0\rightarrow 1}
\}$};
\node (a22) at (0,-2)  {$\{
\mathfrak{p}_{0},\,
\mathfrak{p}_{0}^{0\rightarrow n}
\}$};
\node (a23) at (4,-2)  {$\{
\mathfrak{p}_{0},\,
\mathfrak{p}_{0}^{0\rightarrow 0}
\}$};
%%%%%%%%%%%%%%%%%%%
\node (a31) at (-6.15,-3)  {$\Pi_{3,1}$};
\node (a312) at (-3.65,-3)  {$\Pi_{3,2}$};
\node (a313) at (-1.15,-3)  {$\Pi_{3,3}$};
\node (a32) at (1.15,-3)  {$\Pi_{3,4}$};
\node (a323) at (3.65,-3)  {$\Pi_{3,5}$};
\node (a33) at (6.15,-3)  {$\Pi_{3,6}$};

%%%%%%%%%%%%%%%%%%%
%\node (dots1) at (-5,-4)  {$\vdots$};
%\node (dots1) at (-3,-4)  {$\vdots$};
%\node (dots1) at (-1,-4)  {$\vdots$};
%\node (dots1) at (1,-4)  {$\vdots$};
%\node (dots1) at (3,-4)  {$\vdots$};
%\node (dots1) at (5,-4)  {$\vdots$};
%%%%%%%%%%%%%%%%%%%
\draw[->] (a0) -- (a11);
%%%%%%%%%%%%%%%%%%%
\draw[->] (a11) -- (a21);
\draw[->] (a11) -- (a22);
\draw[->] (a11) -- (a23);
%%%%%%%%%%%%%%%%%%%
\draw[->] (a21) -- (a31);
\draw[->] (a21) -- (a312);
\draw[->] (a21) -- (a313);
\draw[->] (a22) -- (a32);
\draw[->] (a22) -- (a312);
\draw[->] (a22) -- (a323);
\draw[->] (a23) -- (a33);
\draw[->] (a23) -- (a313);
\draw[->] (a23) -- (a323);
\end{tikzpicture}
\end{array}
\end{equation}

\subsection{Poset construction for quiver Yangian from the affine \texorpdfstring{$E_{6,7,8}$}{E678} quivers}
\label{appssec:PosetE678}

Finally, let us construct the vacuum representation for the quiver Yangian from the affine $E_{6,7,8}$ quiver \eqref{eq:AffineE6Quiver}-\eqref{eq:AffineE8Quiver}.
Since the corresponding CY$_3$'s $(\mathbb{C}^2)/\Gamma\times \mathbb{C}$ with $\Gamma$ being the binary tetrahedral, octahedral, and icosahedral groups (of order 24, 48 and 120, respectively) are not toric, this set of states does not have a crystal description.
The weight assignment of the arrows can be read off from \eqref{eq:AffineE6Quiver}, \eqref{eq:AffineE7Quiver}, and \eqref{eq:AffineE8Quiver}.

Let us first construct the vacuum representation of the affine $E_6$ quiver \eqref{eq:AffineE6Quiver} using the procedure of Sec.\ \ref{sssec:RepresentationConstruction}.

\medskip

\noindent \textbf{Level 0.} There is only one state at level 0: the ground state $\Pi_{0}=\{\varnothing\}$, which is the empty set. 
By definition \eqref{eq:ChargeFunctionDef}, the charge function of $\Pi_{0}$, is just the vacuum charge function, which can be read off from the framed quiver  by definition \eqref{eq:varphi0Def} to be
\begin{equation}
\begin{aligned}
\Psi^{(0)}_{\Pi_0}(z)=   \psi_{0}(z)=\frac{1}{z}\,, \qquad
\Psi^{(a)}_{\Pi_0}(z)=   1  \quad \textrm{with} \quad a=1,2,\dots,6\,,
   \end{aligned}
\end{equation}
which has only one pole at $z=0$ for color $0$.
Together with the fact that the state $\Pi_{0}$ is the empty set, by definition \eqref{eq:RemPi} and \eqref{eq:AddPi} (or equivalently from \eqref{eq:RemVacuum} and \eqref{eq:AddVacuumAtom}), we have
\begin{equation}
\textrm{Rem}(\Pi_{0})=\{ \varnothing \} \qquad \textrm{and} \qquad 
\textrm{Add}(\Pi_{0})=\{ \mathfrak{p}_0 \} \,.
\end{equation}

\noindent \textbf{Level 1.} From the single atom in $\textrm{Add}(\Pi_0)$, we obtain one state at level-1:
\begin{equation}
\Pi_1=\{
 \mathfrak{p}_0 
\}\,,
\end{equation}
by Step-4 of the procedure in Sec.\ \ref{sssec:RepresentationConstruction}, or directly from \eqref{eq:StatesLevel1}.
Its charge functions are
\begin{equation}
\begin{aligned}
&\Psi^{(0)}_{\Pi_1}(z)=\frac{1}{z}\cdot\frac{z+h_3}{z-h_3}\,, \qquad \qquad \Psi^{(6)}_{\Pi_1}(z)=\frac{z+h_2}{z-h_1}  \\
&\Psi^{(a)}_{\Pi_1}(z)=   1 \qquad \qquad \textrm{for} \quad a=1,2,\dots,5\\
\end{aligned}
\end{equation}
by definition \eqref{eq:ChargeFunctionDef}.
Then by definition \eqref{eq:RemPi} and \eqref{eq:AddPi}, we have
\begin{equation}
\textrm{Rem}(\Pi_{1})=\{
 \mathfrak{p}_0 
\} \qquad \textrm{and} \qquad 
\textrm{Add}(\Pi_{1})=\{
 \mathfrak{p}_0^{0\rightarrow 6} ,\,
 \mathfrak{p}_0^{0\rightarrow 0}
\} \,.
\end{equation}

\noindent \textbf{Level 2.} From the two atoms in $\textrm{Add}(\Pi_1)$, we obtain two states at level 2 
\begin{equation}
\Pi_{2,1}= \{
\mathfrak{p}_0  ,\,
\mathfrak{p}_0^{0\rightarrow 6} 
\} \qquad \textrm{and} \qquad 
\Pi_{2,2}=\{
\mathfrak{p}_0 ,\,
\mathfrak{p}_0^{0\rightarrow 0} 
\}  \,.
\end{equation}
The charge functions of $\Pi_{2,1}$ are
\begin{equation} 
\begin{aligned}
&\Psi^{(0)}_{\Pi_{2,1}}(z)=\frac{1}{z-h_3}\,,  \qquad   \Psi^{(3)}_{\Pi_{2,1}}(z)=\frac{z+h_2-h_1}{z-2h_1} \,, \qquad\Psi^{(6)}_{\Pi_{2,1}}(z)=\frac{z+h_3-h_1}{z-h_1}  \\
&\Psi^{(a)}_{\Pi_{2,1}}(z)=   1 \qquad \qquad \textrm{for} \quad a=1,2,4,5\,,
\end{aligned}
\end{equation}
which gives the set of removable and addable atoms as
\begin{equation}
\begin{aligned}
\textrm{Rem}(\Pi_{2,1})=\{
\mathfrak{p}_0^{0\rightarrow 6}
\}  \qquad \textrm{and} \qquad 
\textrm{Add}(\Pi_{2,1})= \{
\mathfrak{p}_0^{0\rightarrow 6\rightarrow 3} ,\, 
\mathfrak{p}_0^{0\rightarrow 0}
\}\,,
\end{aligned}
\end{equation}
among which the addable atoms give rise to two states at level $3$:
\begin{equation}\label{eq:E6NewStates21}
\begin{aligned}
&\{
\mathfrak{p}_0,\,
\mathfrak{p}_0^{0\rightarrow 6} ,\, 
\mathfrak{p}_0^{0\rightarrow 6\rightarrow 3} 
\}  \qquad \textrm{and} \qquad
\{
\mathfrak{p}_0,\,
\mathfrak{p}_0^{0\rightarrow 6},\, 
\mathfrak{p}_0^{0\rightarrow 0}
\}\,.
\end{aligned}
\end{equation}
The charge functions of $\Pi_{2,2}$ are
\begin{equation} 
\begin{aligned}
&\Psi^{(0)}_{\Pi_{2,1}}(z)=\frac{z+h_3}{(z-h_3)(z-2h_3)}\,, \qquad \qquad 
\Psi^{(6)}_{\Pi_{2,2}}(z)=\frac{z+h_2-h_3}{z-h_1}  \\
&\Psi^{(a)}_{\Pi_{2,2}}(z)=   1 \qquad \qquad \textrm{for} \quad a=1,2,\dots,5\,,
\end{aligned}
\end{equation}
which gives the set of removable and addable atoms as
\begin{equation}
\begin{aligned}
\textrm{Rem}(\Pi_{2,2})=\{
\mathfrak{p}_0^{0\rightarrow 0}
\} \qquad \textrm{and}\qquad  
\textrm{Add}(\Pi_{2,2})= \{
\mathfrak{p}_0^{0\rightarrow 6} ,\, 
\mathfrak{p}_0^{0\rightarrow 0\rightarrow 0} 
\}\,,
\end{aligned}
\end{equation}
among which the addable atoms give rise to two states at level $3$:
\begin{equation}\label{eq:E6NewStates22}
\begin{aligned}
&\{
\mathfrak{p}_0,\,
\mathfrak{p}_0^{0\rightarrow 0} ,\, 
\mathfrak{p}_0^{0\rightarrow 6}
\}  \qquad \textrm{and} \qquad
\{
\mathfrak{p}_0,\,
\mathfrak{p}_0^{0\rightarrow 0},\, 
\mathfrak{p}_0^{0\rightarrow 0\rightarrow 0} 
\}\,.
\end{aligned}
\end{equation}
Finally, among the four states in \eqref{eq:E6NewStates21} and \eqref{eq:E6NewStates22}, there are only $3$ distinct ones, namely, there are $3$ new states at level 3:
\begin{equation}
\begin{aligned}
\Pi_{3,1}=\{
\mathfrak{p}_0,\,
\mathfrak{p}_0^{0\rightarrow 6} ,\, 
\mathfrak{p}_0^{0\rightarrow 6\rightarrow 3}
\}\,,\quad 
\Pi_{3,2}=\{
\mathfrak{p}_0,\,
\mathfrak{p}_0^{0\rightarrow 6},\, 
\mathfrak{p}_0^{0\rightarrow 0}
\}\,, \quad
\Pi_{3,3}=\{
\mathfrak{p}_0,\,
\mathfrak{p}_0^{0\rightarrow 0},\, 
\mathfrak{p}_0^{0\rightarrow 0\rightarrow 0}
\}\,.
\end{aligned}
\end{equation}

One can repeat this procedure and construct the entire representation. 
We show the Hasse diagram of the poset up to level 3:
\begin{equation}
\begin{array}{c}
\begin{tikzpicture}[scale=0.9]
\node (a0) at (0,0)  {$\{\varnothing\}$};
%%%%%%%%%%%%%%%%%%%
\node (a11) at (0,-1)  {$\{ \mathfrak{p}_0
\}$};
%%%%%%%%%%%%%%%%%%%
\node (a21) at (-2,-2)  {$\{
\mathfrak{p}_0 ,\,
\mathfrak{p}_0^{0\rightarrow 6} 
\}$};
\node (a22) at (2,-2)  {$\{
\mathfrak{p}_0 ,\,
\mathfrak{p}_0^{0\rightarrow 0} 
\}$};
%%%%%%%%%%%%%%%%%%%
\node (a31) at (-6,-3)  {$\{
\mathfrak{p}_0,\,
\mathfrak{p}_0^{0\rightarrow 6} ,\, 
\mathfrak{p}_0^{0\rightarrow 6\rightarrow 3}
\}$};
\node (a32) at (0,-3)  {$\{
\mathfrak{p}_0,\,
\mathfrak{p}_0^{0\rightarrow 6},\, 
\mathfrak{p}_0^{0\rightarrow 0}
\}$};
\node (a33) at (6,-3)  {$\{
\mathfrak{p}_0,\,
\mathfrak{p}_0^{0\rightarrow 0},\, 
\mathfrak{p}_0^{0\rightarrow 0\rightarrow 0}
\}$};
%%%%%%%%%%%%%%%%%%%
%\node (dots1) at (-5,-4)  {$\vdots$};
%\node (dots2) at (0,-4)  {$\vdots$};
%\node (dots3) at (5,-4)  {$\vdots$};
%%%%%%%%%%%%%%%%%%%
\draw[->] (a0) -- (a11);
%%%%%%%%%%%%%%%%%%%
\draw[->] (a11) -- (a21);
\draw[->] (a11) -- (a22);
%%%%%%%%%%%%%%%%%%%
\draw[->] (a21) -- (a31);
\draw[->] (a21) -- (a32);
\draw[->] (a22) -- (a32);
\draw[->] (a22) -- (a33);
\end{tikzpicture}
\end{array}
\end{equation}
The vacuum character is then
\begin{equation}\label{eq:ZE6expansion}
\begin{aligned}
Z^{E_6}(\boldsymbol{x})=&1+x_0+(x_0^2+x_0x_6)+(x_0^3+x^2_0 x_6+x_0 x_6 x_3)\\
&+(x_0^4+x_0^3 x_6 + x_0^2x_6^2
   +x_0^2 x_6 x_3 +  
    x_0  x_6 x_3 x_2 +  
   x_0 x_6 x_3x_4)+\mathcal{O}(\boldsymbol{x}^5) \,.
%   +x_0^5+x_6 x_0^4+x_6^2
%   x_0^3+x_3 x_6 x_0^3+x_3
%   x_6^2 x_0^2+x_2 x_3 x_6
%   x_0^2+x_3 x_4 x_6 x_0^2+x_1
%   x_2 x_3 x_6 x_0+x_2 x_3 x_4
%   x_6 x_0+x_3 x_4 x_5 x_6
%   x_0,x_0^6+x_6 x_0^5+x_6^2
%   x_0^4+x_3 x_6 x_0^4+x_6^3
%   x_0^3+x_3 x_6^2 x_0^3+x_2
%   x_3 x_6 x_0^3+x_3 x_4 x_6
%   x_0^3+x_3^2 x_6^2 x_0^2+x_2
%   x_3 x_6^2 x_0^2+x_3 x_4
%   x_6^2 x_0^2+x_1 x_2 x_3 x_6
%   x_0^2+x_2 x_3 x_4 x_6
%   x_0^2+x_3 x_4 x_5 x_6
%   x_0^2+x_2 x_3^2 x_4 x_6
%   x_0+x_1 x_2 x_3 x_4 x_6
%   x_0+x_2 x_3 x_4 x_5 x_6
%   x_0,x_0\right\}
\end{aligned}
\end{equation}

The vacuum representations for the $E_7$ quiver \eqref{eq:AffineE7Quiver} and for the $E_8$ quiver \eqref{eq:AffineE7Quiver} can be constructed in the same manner. 
Here we omit the intermediate details, and only show the Hasse diagram for the $E_7$ poset up to level 3:
\begin{equation}
\begin{array}{c}
\begin{tikzpicture}[scale=0.9]
\node (a0) at (0,0)  {$\{\varnothing\}$};
%%%%%%%%%%%%%%%%%%%
\node (a11) at (0,-1)  {$\{ \mathfrak{p}_0
\}$};
%%%%%%%%%%%%%%%%%%%
\node (a21) at (-2,-2)  {$\{
\mathfrak{p}_0 ,\,
\mathfrak{p}_0^{0\rightarrow 1} 
\}$};
\node (a22) at (2,-2)  {$\{
\mathfrak{p}_0 ,\,
\mathfrak{p}_0^{0\rightarrow 0} 
\}$};
%%%%%%%%%%%%%%%%%%%
\node (a31) at (-6,-3)  {$\{
\mathfrak{p}_0,\,
\mathfrak{p}_0^{0\rightarrow 1} ,\, 
\mathfrak{p}_0^{0\rightarrow 1\rightarrow 2}
\}$};
\node (a32) at (0,-3)  {$\{
\mathfrak{p}_0,\,
\mathfrak{p}_0^{0\rightarrow 1},\, 
\mathfrak{p}_0^{0\rightarrow 0}
\}$};
\node (a33) at (6,-3)  {$\{
\mathfrak{p}_0,\,
\mathfrak{p}_0^{0\rightarrow 0},\, 
\mathfrak{p}_0^{0\rightarrow 0\rightarrow 0}
\}$};
%%%%%%%%%%%%%%%%%%%
%\node (dots1) at (-5,-4)  {$\vdots$};
%\node (dots2) at (0,-4)  {$\vdots$};
%\node (dots3) at (5,-4)  {$\vdots$};
%%%%%%%%%%%%%%%%%%%
\draw[->] (a0) -- (a11);
%%%%%%%%%%%%%%%%%%%
\draw[->] (a11) -- (a21);
\draw[->] (a11) -- (a22);
%%%%%%%%%%%%%%%%%%%
\draw[->] (a21) -- (a31);
\draw[->] (a21) -- (a32);
\draw[->] (a22) -- (a32);
\draw[->] (a22) -- (a33);
\end{tikzpicture}
\end{array}
\end{equation}
with the vacuum character
\begin{equation}\label{eq:ZE7expansion}
\begin{aligned}
Z^{E_7}(\boldsymbol{x})=&1+ x_0+ (x_0^2+x_0x_1)+  (x_0^3+x_0^2x_1 +x_0 x_1 x_2)\\
&+(x_0^4+ x_0^3x_1+x_0^2 x_1^2+ x_0^2x_1 x_2+x_0x_1 x_2 x_3 )+\mathcal{O}(\boldsymbol{x}^5) \,.
	\end{aligned}
\end{equation}
%\begin{equation}\label{eq:ZE7expansion}
%\begin{aligned}
%Z^{E_7}(\boldsymbol{x})=&1+x_0+(x_0^2+x_0x_7)+(x_0^3+x^2_0 x_7+x_0 x_7 x_3)\\
%&+(x_0^4+x_0^3 x_7 + x_0^2x_7^2
%   +x_0^2 x_7 x_3 +  
%    x_0  x_7 x_3 x_2 +  
%   x_0 x_7 x_3x_4)+\mathcal{O}(\boldsymbol{x}^5) \,.
%%   +x_0^5+x_6 x_0^4+x_6^2
%%   x_0^3+x_3 x_6 x_0^3+x_3
%%   x_6^2 x_0^2+x_2 x_3 x_6
%%   x_0^2+x_3 x_4 x_6 x_0^2+x_1
%%   x_2 x_3 x_6 x_0+x_2 x_3 x_4
%%   x_6 x_0+x_3 x_4 x_5 x_6
%%   x_0,x_0^6+x_6 x_0^5+x_6^2
%%   x_0^4+x_3 x_6 x_0^4+x_6^3
%%   x_0^3+x_3 x_6^2 x_0^3+x_2
%%   x_3 x_6 x_0^3+x_3 x_4 x_6
%%   x_0^3+x_3^2 x_6^2 x_0^2+x_2
%%   x_3 x_6^2 x_0^2+x_3 x_4
%%   x_6^2 x_0^2+x_1 x_2 x_3 x_6
%%   x_0^2+x_2 x_3 x_4 x_6
%%   x_0^2+x_3 x_4 x_5 x_6
%%   x_0^2+x_2 x_3^2 x_4 x_6
%%   x_0+x_1 x_2 x_3 x_4 x_6
%%   x_0+x_2 x_3 x_4 x_5 x_6
%%   x_0,x_0\right\}
%\end{aligned}
%\end{equation}
And similarly, for the $E_8$ quiver
\begin{equation}
\begin{array}{c}
\begin{tikzpicture}[scale=0.9]
\node (a0) at (0,0)  {$\{\varnothing\}$};
%%%%%%%%%%%%%%%%%%%
\node (a11) at (0,-1)  {$\{ \mathfrak{p}_0
\}$};
%%%%%%%%%%%%%%%%%%%
\node (a21) at (-2,-2)  {$\{
\mathfrak{p}_0 ,\,
\mathfrak{p}_0^{0\rightarrow 7} 
\}$};
\node (a22) at (2,-2)  {$\{
\mathfrak{p}_0 ,\,
\mathfrak{p}_0^{0\rightarrow 0} 
\}$};
%%%%%%%%%%%%%%%%%%%
\node (a31) at (-6,-3)  {$\{
\mathfrak{p}_0,\,
\mathfrak{p}_0^{0\rightarrow 7} ,\, 
\mathfrak{p}_0^{0\rightarrow 7\rightarrow 6}
\}$};
\node (a32) at (0,-3)  {$\{
\mathfrak{p}_0,\,
\mathfrak{p}_0^{0\rightarrow 7},\, 
\mathfrak{p}_0^{0\rightarrow 0}
\}$};
\node (a33) at (6,-3)  {$\{
\mathfrak{p}_0,\,
\mathfrak{p}_0^{0\rightarrow 0},\, 
\mathfrak{p}_0^{0\rightarrow 0\rightarrow 0}
\}$};
%%%%%%%%%%%%%%%%%%%
%\node (dots1) at (-5,-4)  {$\vdots$};
%\node (dots2) at (0,-4)  {$\vdots$};
%\node (dots3) at (5,-4)  {$\vdots$};
%%%%%%%%%%%%%%%%%%%
\draw[->] (a0) -- (a11);
%%%%%%%%%%%%%%%%%%%
\draw[->] (a11) -- (a21);
\draw[->] (a11) -- (a22);
%%%%%%%%%%%%%%%%%%%
\draw[->] (a21) -- (a31);
\draw[->] (a21) -- (a32);
\draw[->] (a22) -- (a32);
\draw[->] (a22) -- (a33);
\end{tikzpicture}
\end{array}
\end{equation}
with the vacuum character
\begin{equation}\label{eq:ZE8expansion}
\begin{aligned}
Z^{E_8}(\boldsymbol{x})=&1+x_0+(x_0^2+x_0x_7)+(x_0^3+x^2_0 x_7+x_0 x_7 x_6)\\
&+(x_0^4+x_0^3 x_7 + x_0^2x_7^2
   +x_0^2 x_7 x_6 +  
    x_0  x_7 x_6 x_5 )+\mathcal{O}(\boldsymbol{x}^5) \,.
%   +x_0^5+x_6 x_0^4+x_6^2
%   x_0^3+x_3 x_6 x_0^3+x_3
%   x_6^2 x_0^2+x_2 x_3 x_6
%   x_0^2+x_3 x_4 x_6 x_0^2+x_1
%   x_2 x_3 x_6 x_0+x_2 x_3 x_4
%   x_6 x_0+x_3 x_4 x_5 x_6
%   x_0,x_0^6+x_6 x_0^5+x_6^2
%   x_0^4+x_3 x_6 x_0^4+x_6^3
%   x_0^3+x_3 x_6^2 x_0^3+x_2
%   x_3 x_6 x_0^3+x_3 x_4 x_6
%   x_0^3+x_3^2 x_6^2 x_0^2+x_2
%   x_3 x_6^2 x_0^2+x_3 x_4
%   x_6^2 x_0^2+x_1 x_2 x_3 x_6
%   x_0^2+x_2 x_3 x_4 x_6
%   x_0^2+x_3 x_4 x_5 x_6
%   x_0^2+x_2 x_3^2 x_4 x_6
%   x_0+x_1 x_2 x_3 q_4 q_6
%   q_0+q_2 q_3 q_4 q_5 q_6
%   q_0,q_0\right\}
\end{aligned}
\end{equation}
We have checked \eqref{eq:ZE6expansion}, \eqref{eq:ZE7expansion}, and \eqref{eq:ZE8expansion} to higher orders and confirmed that they obey the closed-form formula \eqref{eq:VacuumCharacterAffineADEProduct} with the Kac labels given in the second line of \eqref{eq:KacLabelADE}.
We discuss the refinement of these vacuum characters in Sec.\ \ref{ssec:RefinementADE}.

\section{Refining character for trefoil knot quiver}
\label{appsec:TFquiver}

In this appendix, we explain in more detail the procedure of constructing the vacuum representation and determining the refinement prescription for the quiver from the trefoil knot, shown in Figure \ref{fig:trefoil}. 

\medskip

The quiver is given in \eqref{eq:TrefoilQuiver}.
Let us first consider the framing $\infty\rightarrow 0$, with the framed quiver given in \eqref{eq:TrefoilQuiverFrame0}.
Using the procedure of Sec.\ \ref{sssec:RepresentationConstruction}, we can construct the vacuum representation $\mathcal{P}_{0}$ (with framed vertex $0$) to arbitrarily high level.
A direct counting of states reproduces correctly the unrefined limit $y\rightarrow 1$ of $\widehat{\textrm{Sgn}}\cdot \mathcal{Z}_{\textrm{NC}}(y,\boldsymbol{x})$, with the first few levels shown in \eqref{eq:ZunrefTrefoil}.

To refine the vacuum character, we need to determine the subset $Q_0'$ in the prescription \eqref{eq:upsilonKQansatz} together with the $13$ $\Upsilon$-charges in \eqref{eq:UpsilonTF}, by comparing  the refined vacuum character $\boldsymbol{\mathcal{Z}}^{\textrm{ref}}_{\textrm{trefoil},0}(y, \boldsymbol{x})$, computed from definition \eqref{eq:ZrefinedP} with prescription \eqref{eq:upsilonKQansatz}, with the expansion of $\widehat{\textrm{Sgn}}\cdot \mathcal{Z}_{\textrm{NC},0}(y,\boldsymbol{x})$, with the first few levels shown in \eqref{eq:ZrefTrefoilF0}.
 
\medskip

\noindent \textbf{Ground state (level 0).} There is one state at level 0: the vacuum $\{\varnothing\}$, which  corresponds to the leading term $1$ in $\boldsymbol{\mathcal{Z}}^{\textrm{ref}}_{\textrm{trefoil},0}(y, \boldsymbol{x})  
=\widehat{\textrm{Sgn}}\cdot \mathcal{Z}_{\textrm{NC},0}(y,\boldsymbol{x})$. 

\smallskip

\noindent \textbf{Level 1.} Since the framing is $\infty\rightarrow 0$, there is only state at level 1:  $\{I^{(0)}\}$, which corresponds to the term 
$x_0$ in $\widehat{\textrm{Sgn}}\cdot \mathcal{Z}_{\textrm{NC},0}(y,\boldsymbol{x})$. 
This fixes 
\begin{equation}
\Upsilon^{\infty\rightarrow 0}\equiv \Upsilon(I^{\infty \rightarrow 0})=0    \,.
\end{equation}

\noindent \textbf{Level 2.} There are two states at this level: 
\begin{equation}
\{I^{(0)},\, I^{(0)}\cdot I^{0\rightarrow 1}\} \qquad \textrm{and} \qquad \{I^{(0)},\, I^{(0)}\cdot I^{0\rightarrow 2}\}\,,
\end{equation}
which gives 
\begin{equation}
y^{\Upsilon^{0\rightarrow 1}}   x_0 x_1 +   y^{\Upsilon^{0\rightarrow 2}}   x_0 x_2
\end{equation}
in $\boldsymbol{\mathcal{Z}}^{\textrm{ref}}_{\textrm{trefoil},0}(y, \boldsymbol{x})$.
Comparing with the corresponding term in \eqref{eq:ZrefTrefoilF0}, $y^2 x_0 x_1 + y^3 x_0 x_2  $, we have 
 \begin{equation}
\Upsilon^{0\rightarrow 1}=2    
\qquad \textrm{and} \qquad  \Upsilon^{0\rightarrow2}=3  \,.
 \end{equation}

\smallskip
 
\noindent \textbf{Level 3.}  There are $12$ states at this level.
\begin{itemize}
\item The two states 
\begin{equation}
\{I^{(0)},\, I^{(0)}\cdot I^{0\rightarrow 1},\, I^{(0)}\cdot I^{0\rightarrow 1\rightarrow 0} \}
\qquad \textrm{and} \qquad  
\{I^{(0)},\, I^{(0)}\cdot I^{0\rightarrow 2},\, I^{(0)}\cdot I^{0\rightarrow 2\rightarrow 0} \}
\end{equation} 
should correspond to the terms
\begin{equation}
y^2 x_0^2 x_1    
\qquad \textrm{and} \qquad  
y^3 x_0^2 x_2 
\end{equation}
in $\boldsymbol{\mathcal{Z}}^{\textrm{ref}}_{\textrm{trefoil},0}(y, \boldsymbol{x})$, respectively. 
 There are two ways to solve this: 
\begin{itemize}
\item $\upsilon(\sqbox{$0$})\neq 0$, together with $\Upsilon^{1\rightarrow 0}=-2 $ and $\Upsilon^{2\rightarrow 0}=-3$.
\item Or $\upsilon(\sqbox{$0$})=0$ irrespective of the values of $\Upsilon^{1\rightarrow 0}$ and $\Upsilon^{2\rightarrow 0}$.
\end{itemize}
\item The two states
\begin{equation}
\{I^{(0)},\, I^{(0)}\cdot I^{0\rightarrow 1},\, I^{(0)}\cdot I^{0\rightarrow 1}\cdot I^{1\rightarrow 1}_{i=1,2} \}
\end{equation}
give 
\begin{equation}
\left(\sum^2_{i=1}y^{4+\Upsilon_i^{1\rightarrow 1}}\right)   x_0 x^2_1 
\end{equation}
in $\boldsymbol{\mathcal{Z}}^{\textrm{ref}}_{\textrm{trefoil},0}(y, \boldsymbol{x})$.
Comparing with the corresponding term in \eqref{eq:ZrefTrefoilF0}, $(y^4+y^6) x_0 x_1^2 $, we have  
\begin{equation}
\Upsilon_{i}^{1\rightarrow 1}=2(i-1) \,,   \qquad i=1,2\,.
\end{equation}
\item Similarly,  the three states
\begin{equation}
\{I^{(0)},\, I^{(0)}\cdot I^{0\rightarrow 2},\, I^{(0)}\cdot I^{0\rightarrow 2}\cdot I^{2\rightarrow 2}_{i=1,2,3} \}
\end{equation}
give 
\begin{equation}
\left(\sum^3_{i=i}y^{6+\Upsilon_i^{2\rightarrow 2}}\right)x_0 x^2_2   
\end{equation}
in $\boldsymbol{\mathcal{Z}}^{\textrm{ref}}_{\textrm{trefoil},0}(y, \boldsymbol{x})$.
Comparing with the corresponding term in \eqref{eq:ZrefTrefoilF0}, $(y^6+y^8+y^{10}) x_0 x^2_2$, we have  
\begin{equation}
\Upsilon_{i}^{2\rightarrow 2}=2(i-1) \,,   \qquad i=1,2,3\,.
 \end{equation}
 \item The $5$ states
 \begin{equation}
 \begin{aligned}
&\{I^{(0)},\, I^{(0)}\cdot I^{0\rightarrow 1},\, I^{(0)}\cdot I^{0\rightarrow 1}\cdot I^{1\rightarrow 2}_{i=1,2} \}\,,
\{I^{(0)},\, I^{(0)}\cdot I^{0\rightarrow 2},\, I^{(0)}\cdot I^{0\rightarrow 2}\cdot I^{2\rightarrow 1}_{i=1,2} \}\,,\\
&\{I^{(0)},\, I^{(0)}\cdot I^{0\rightarrow 1},\, I^{(0)}\cdot I^{0\rightarrow 2} \} 
\end{aligned}
\end{equation}
give
\begin{equation}
\left(y^5+\sum^2_{i=1}y^{4+\Upsilon_i^{1\rightarrow 2}}
+\sum^2_{i=1} y^{6+\Upsilon_i^{2\rightarrow 1}}\right) x_0 x_1 x_2
\end{equation}
in $\boldsymbol{\mathcal{Z}}^{\textrm{ref}}_{\textrm{trefoil},0}(y, \boldsymbol{x})$, which should correspond to the term $(2y^5+2 y^7+y^9)x_0x_1x_2$ in \eqref{eq:ZrefTrefoilF0}.
This gives 
\begin{equation}\label{eq:TFlevel3x0x1x2}
 \sum^2_{i=1}y^{\Upsilon^{1\rightarrow 2}_i}
+y^2\sum^2_{i=1}y^{\Upsilon^{2\rightarrow 1}_i}
  = y + 2 y^3 +  y^5 \,.
\end{equation}
At this level, we don't have enough information to solve this equation.
\end{itemize}

\smallskip

\noindent \textbf{Level 4} 
\begin{itemize}
\item The $5$ states 
\begin{equation}
\begin{aligned}
&\{
I^{(0)},\, 
I^{(0)}\cdot I^{0\rightarrow 1},\,
I^{(0)}\cdot I^{0\rightarrow 1\rightarrow 0 },\,
I^{(0)}\cdot I^{0\rightarrow 1\rightarrow 0 \rightarrow 1}
\}\,,\\
&\{
I^{(0)},\, 
I^{(0)}\cdot I^{0\rightarrow 1},\, 
I^{(0)}\cdot I^{0\rightarrow 1}\cdot I^{1\rightarrow 1}_{i=1,2} ,\,
I^{(0)}\cdot I^{0\rightarrow 1}\cdot I^{1\rightarrow 1}_{i=1,2} \cdot I^{1\rightarrow 0} 
\}\,,\\
& \{
I^{(0)},\, 
I^{(0)}\cdot I^{0\rightarrow 1},\, 
I^{(0)}\cdot I^{0\rightarrow 1\rightarrow 0} ,\,
I^{(0)}\cdot I^{0\rightarrow 1} \cdot I^{1\rightarrow 1}_{i=1,2} 
\}        
\end{aligned}     
\end{equation}
should correspond to the term $(2y^4+2y^6+y^{8})x^2_0x^2_1$ in \eqref{eq:ZrefTrefoilF0}.
A priori, there are two ways to solve this.
\begin{itemize}
\item $\upsilon(\sqbox{$0$})\neq 0$ and
\begin{equation}
y^{4+\Upsilon^{1\rightarrow 0}}
+2 y^{6+\Upsilon^{1\rightarrow 0}}
+y^{8+\Upsilon^{1\rightarrow 0}}
+y^{10+\Upsilon^{1\rightarrow 0}}
 =2y^4+2y^6+y^{8}\,,
\end{equation}
which allows no solution for $\Upsilon^{1\rightarrow 0}$.
\item Or $\upsilon(\sqbox{$0$})=0$ and 
\begin{equation}
2 y^{4}
+2y^{6}+y^{6+\Upsilon^{1\rightarrow 0}}
=2y^4+2y^6+y^{8}\,,
\end{equation}
which gives $\Upsilon_1^{1\rightarrow 0}=2$.

\end{itemize}
Therefore, we have
\begin{equation}
\upsilon(\sqbox{$0$})=0
\qquad \textrm{and} \qquad    \Upsilon_1^{1\rightarrow 0}=2\,.    
\end{equation}
\item Since we have already fixed $\upsilon(\sqbox{$0$})=0$, the $7$ states 
\begin{equation}
\begin{aligned}
&\{
I^{(0)},\, 
I^{(0)}\cdot I^{0\rightarrow 2},\,
I^{(0)}\cdot I^{0\rightarrow 2\rightarrow 0 },\,
I^{(0)}\cdot I^{0\rightarrow 2\rightarrow 0 \rightarrow 2}
\}\,,\\
&\{
I^{(0)},\, 
I^{(0)}\cdot I^{0\rightarrow 2},\, 
I^{(0)}\cdot I^{0\rightarrow 2}\cdot I^{2\rightarrow 2}_{i=1,2,3} ,\,
I^{(0)}\cdot I^{0\rightarrow 2}\cdot I^{2\rightarrow 2}_{i=1,2,3} \cdot I^{2\rightarrow 0} 
\}\,,\\
& \{
I^{(0)},\, 
I^{(0)}\cdot I^{0\rightarrow 2},\, 
I^{(0)}\cdot I^{0\rightarrow 2\rightarrow 0} ,\,
I^{(0)}\cdot I^{0\rightarrow 2} \cdot I^{2\rightarrow 2}_{i=1,2,3} 
\}        
\end{aligned}     
\end{equation}
give
\begin{equation}
(y^{9+\Upsilon^{2\rightarrow 0}}+ 2y^{6}+2y^8+2 y^{10})x_0^2 x_2^2
\end{equation}
in $\boldsymbol{\mathcal{Z}}^{\textrm{ref}}_{\textrm{trefoil},0}(y, \boldsymbol{x})$.
Comparing with the corresponding term in \eqref{eq:ZrefTrefoilF0}, $(2y^6+2y^8+2y^{10}+y^{12}) x^2_0 x^2_2$, we have  
\begin{equation}
\Upsilon^{2\rightarrow 0}=3\,.    
\end{equation}

\item The $12$ states
\begin{equation}
\begin{aligned}
&\{
I^{(0)},\, 
I^{(0)}\cdot I^{0\rightarrow 1},\,
I^{(0)}\cdot I^{0\rightarrow 1}\cdot I^{1\rightarrow 2}_{j=1,2} ,\,
I^{(0)}\cdot I^{0\rightarrow 1}\cdot I^{1\rightarrow 2}_{j=1,2} \cdot I^{2\rightarrow 0}
\}\,,\\
&\{
I^{(0)},\, 
I^{(0)}\cdot I^{0\rightarrow 2},\,
I^{(0)}\cdot I^{0\rightarrow 2}\cdot I^{2\rightarrow 1}_{k=1,2} ,\,
I^{(0)}\cdot I^{0\rightarrow 2}\cdot I^{2\rightarrow 1}_{k=1,2} \cdot I^{1\rightarrow 0}
\}\,,\\
&\{
I^{(0)},\, 
I^{(0)}\cdot I^{0\rightarrow 1},\,
I^{(0)}\cdot I^{0\rightarrow 1\rightarrow 0} ,\,
I^{(0)}\cdot I^{0\rightarrow 1\rightarrow 0\rightarrow 2}
\}\,,\\
&\{
I^{(0)},\, 
I^{(0)}\cdot I^{0\rightarrow 2},\,
I^{(0)}\cdot I^{0\rightarrow 2\rightarrow 0} ,\,
I^{(0)}\cdot I^{0\rightarrow 2\rightarrow 0\rightarrow 1}
\}\,,\\
&\{
I^{(0)},\, 
I^{(0)}\cdot I^{0\rightarrow 1},\,
I^{(0)}\cdot I^{0\rightarrow 1\rightarrow 0} ,\,
I^{(0)}\cdot I^{0\rightarrow 1}\cdot I^{1\rightarrow 2}_{j=1,2} 
\}\,,\\  
&\{
I^{(0)},\, 
I^{(0)}\cdot I^{0\rightarrow 2},\,
I^{(0)}\cdot I^{0\rightarrow 2\rightarrow 0} ,\,
I^{(0)}\cdot I^{0\rightarrow 2}\cdot I^{2\rightarrow 1}_{k=1,2} 
\}\,,\\  
&\{
I^{(0)},\, 
I^{(0)}\cdot I^{0\rightarrow 1},\,
I^{(0)}\cdot I^{0\rightarrow 1\rightarrow 0} ,\,
I^{(0)}\cdot I^{0\rightarrow 2}
\}\,,\\  
&\{
I^{(0)},\, 
I^{(0)}\cdot I^{0\rightarrow 1},\,
I^{(0)}\cdot I^{0\rightarrow 2} ,\,
I^{(0)}\cdot I^{0\rightarrow 2\rightarrow 0}
\}\,,\\  
\end{aligned}     
\end{equation}
give

\vspace{-1em}
{\footnotesize
\begin{flalign}
&\left(
\sum^2_{j=1}y^{4+\Upsilon^{1\rightarrow 2}_j}
+\sum^2_{k=1}y^{6+\Upsilon^{2\rightarrow 1}_k}
+y^9+y^{11} +\sum^2_{j=1}y^{4+\Upsilon^{1\rightarrow 2}_j}
+\sum^{2}_{k=1}y^{6+\Upsilon^{2\rightarrow 1}_{k}}+2y^5
\right)    x^2_0x_1 x_2
\end{flalign}
}\normalsize
\noindent 
in $\boldsymbol{\mathcal{Z}}^{\textrm{ref}}_{\textrm{trefoil},0}(y, \boldsymbol{x})$, which should correspond to the term $(4 y^5 + 4 y^7 + 3 y^9 + y^{11}) x^2_0x_1 x_2$ in \eqref{eq:ZrefTrefoilF0}, which reproduces \eqref{eq:TFlevel3x0x1x2}.

\item The $21$ states
\begin{equation}
\begin{aligned}
&\{
I^{(0)},\, 
I^{(0)}\cdot I^{0\rightarrow 1},\,
I^{(0)}\cdot I^{0\rightarrow 1}\cdot I^{1\rightarrow 1}_{i=1,2} ,\,
I^{(0)}\cdot I^{0\rightarrow 1}\cdot I^{1\rightarrow 1}_{i=1,2} \cdot I^{1\rightarrow 2}_{j=1,2}
\}\,,\\
&\{
I^{(0)},\, 
I^{(0)}\cdot I^{0\rightarrow 1},\,
I^{(0)}\cdot I^{0\rightarrow 1}\cdot I^{1\rightarrow 2}_{j=1,2} ,\,
I^{(0)}\cdot I^{0\rightarrow 1}\cdot I^{1\rightarrow 2}_{j=1,2} \cdot I^{2\rightarrow 1}_{k=1,2}
\}\,,\\  
&\{
I^{(0)},\, 
I^{(0)}\cdot I^{0\rightarrow 2},\,
I^{(0)}\cdot I^{0\rightarrow 2}\cdot I^{2\rightarrow 1}_{k=1,2} ,\,
I^{(0)}\cdot I^{0\rightarrow 2}\cdot I^{2\rightarrow 1}_{k=1,2} \cdot I^{1\rightarrow 1}_{i=1,2}
\}\,,\\ 
&\{
I^{(0)},\, 
I^{(0)}\cdot I^{0\rightarrow 1},\,
I^{(0)}\cdot I^{0\rightarrow 1}\cdot I^{1\rightarrow 1}_{i=1,2} ,\,
I^{(0)}\cdot I^{0\rightarrow 1}\cdot I^{1\rightarrow 2}_{j=1,2} 
\}\,,\\  
&\{
I^{(0)},\, 
I^{(0)}\cdot I^{0\rightarrow 2},\,
I^{(0)}\cdot I^{0\rightarrow 2}\cdot I^{2\rightarrow 1}_{1} ,\,
I^{(0)}\cdot I^{0\rightarrow 2}\cdot I^{2\rightarrow 1}_{2} 
\}\,,\\  
&\{
I^{(0)},\, 
I^{(0)}\cdot I^{0\rightarrow 1},\,
I^{(0)}\cdot I^{0\rightarrow 1}\cdot I^{1\rightarrow 1}_{i=1,2} ,\,
I^{(0)}\cdot I^{0\rightarrow 2}
\}\,,\\  
&\{
I^{(0)},\, 
I^{(0)}\cdot I^{0\rightarrow 1},\,
I^{(0)}\cdot I^{0\rightarrow 2} ,\,
I^{(0)}\cdot I^{0\rightarrow 2}\cdot I^{2\rightarrow1}_{k=1,2}
\}\,,\\  
\end{aligned}     
\end{equation}
give
\begin{equation}
\begin{aligned}
&\left(
\sum^2_{j=1}(y^{6+\Upsilon^{1\rightarrow 2}_j}+y^{10+\Upsilon^{1\rightarrow 2}_j})  +\sum^2_{j=1} \sum^{2}_{k=1}  y^{6+2\Upsilon^{1\rightarrow 2}_j+\Upsilon^{2\rightarrow 1}_{k}}
+\sum^2_{k=1}(y^{9+2\Upsilon^{2\rightarrow 1}_k}+y^{11+2\Upsilon^{2\rightarrow 1}_k}) \right. \\
&\left. +\sum^2_{j=1}(y^{6+\Upsilon^{1\rightarrow 2}_j}+y^{8+\Upsilon^{1\rightarrow 2}_j}) 
+ y^{9+\Upsilon^{2\rightarrow 1}_1+\Upsilon^{2\rightarrow 1}_2}+(y^7+y^9)
+\sum^{2}_{k=1}y^{8+\Upsilon^{2\rightarrow 1}_{k}}
\right)    x_0x^2_1 x_2
\end{aligned}
\end{equation}
in $\boldsymbol{\mathcal{Z}}^{\textrm{ref}}_{\textrm{trefoil},0}(y, \boldsymbol{x})$, which should correspond to the term 
\begin{equation}
(3 y^7+6 y^9+5 y^{11}+4 y^{13}+2 y^{15}+y^{17}) x_0x^2_1 x_2    
\end{equation}
in \eqref{eq:ZrefTrefoilF0}.
Denoting $A_{j}=y^{\Upsilon^{1\rightarrow 2}_j}$ and $B_{k}=y^{\Upsilon^{2\rightarrow 1}_k}$, we have
\begin{flalign}\label{eq:A12B12eq2}
 & (2+ y^2+y^{4})  (A_1+A_2) +(A_1^2+A_2^2)(B_1+B_2)
+(y^3+y^5)(B_1^2+B_2^2) \\
&\qquad \qquad \qquad + y^{3}B_1B_2
+y^2(B_1+B_2)=    2 y+5 y^3+5 y^{5}+4 y^{7}+2 y^{9}+y^{11} \,. \nonumber
\end{flalign}

\item The $28$ states
\begin{equation}
\begin{aligned}
&\{
I^{(0)},\, 
I^{(0)}\cdot I^{0\rightarrow 2},\,
I^{(0)}\cdot I^{0\rightarrow 2}\cdot I^{2\rightarrow 2}_{i=1,2,3} ,\,
I^{(0)}\cdot I^{0\rightarrow 2}\cdot I^{2\rightarrow 2}_{i=1,2,3} \cdot I^{2\rightarrow 1}_{k=1,2}
\}\,,\\
&\{
I^{(0)},\, 
I^{(0)}\cdot I^{0\rightarrow 2},\,
I^{(0)}\cdot I^{0\rightarrow 2}\cdot I^{2\rightarrow 1}_{k=1,2} ,\,
I^{(0)}\cdot I^{0\rightarrow 2}\cdot I^{2\rightarrow 1}_{k=1,2} \cdot I^{1\rightarrow 2}_{j=1,2}
\}\,,\\  
&\{
I^{(0)},\, 
I^{(0)}\cdot I^{0\rightarrow 1},\,
I^{(0)}\cdot I^{0\rightarrow 1}\cdot I^{1\rightarrow 2}_{j=1,2} ,\,
I^{(0)}\cdot I^{0\rightarrow 1}\cdot I^{1\rightarrow 2}_{j=1,2} \cdot I^{2\rightarrow 2}_{i=1,2,3}
\}\,,\\ 
&\{
I^{(0)},\, 
I^{(0)}\cdot I^{0\rightarrow 2},\,
I^{(0)}\cdot I^{0\rightarrow 2}\cdot I^{2\rightarrow 2}_{i=1,2,3} ,\,
I^{(0)}\cdot I^{0\rightarrow 2}\cdot I^{2\rightarrow 1}_{k=1,2} 
\}\,,\\  
&\{
I^{(0)},\, 
I^{(0)}\cdot I^{0\rightarrow 1},\,
I^{(0)}\cdot I^{0\rightarrow 1}\cdot I^{1\rightarrow 2}_{1} ,\,
I^{(0)}\cdot I^{0\rightarrow 1}\cdot I^{1\rightarrow 2}_{2} 
\}\,,\\  
&\{
I^{(0)},\, 
I^{(0)}\cdot I^{0\rightarrow 2},\,
I^{(0)}\cdot I^{0\rightarrow 2}\cdot I^{2\rightarrow 2}_{i=1,2,3} ,\,
I^{(0)}\cdot I^{0\rightarrow 1}
\}\,,\\  
&\{
I^{(0)},\, 
I^{(0)}\cdot I^{0\rightarrow 2},\,
I^{(0)}\cdot I^{0\rightarrow 1} ,\,
I^{(0)}\cdot I^{0\rightarrow 1}\cdot I^{1\rightarrow2}_{j=1,2}
\}\,,\\  
\end{aligned}     
\end{equation}
give
\begin{align}
&\left(
\sum^2_{k=1}(y^{9+\Upsilon^{2\rightarrow 1}_k}+y^{13+\Upsilon^{2\rightarrow 1}_k}+y^{17+\Upsilon^{2\rightarrow 1}_k})  +\sum^2_{j=1} \sum^{2}_{k=1}  y^{9+2\Upsilon^{2\rightarrow 1}_k+\Upsilon^{1\rightarrow 2}_{j}} 
\right.
\nonumber\\
& 
+\sum^2_{j=1}(y^{6+2\Upsilon^{1\rightarrow 2}_j}+y^{8+2\Upsilon^{1\rightarrow 2}_j}+y^{10+2\Upsilon^{1\rightarrow 2}_j})  +\sum^2_{k=1}(y^{9+\Upsilon^{2\rightarrow 1}_k}+y^{11+\Upsilon^{2\rightarrow 1}_k}+y^{13+\Upsilon^{2\rightarrow 1}_k})\nonumber \\
&\left. 
+ y^{6+\Upsilon^{1\rightarrow 2}_1+\Upsilon^{1\rightarrow 2}_2}+(y^8+y^{10}+y^{12})
+\sum^{2}_{j=1}y^{7+\Upsilon^{1\rightarrow 2}_{j}}
\right)    x_0x_1 x^2_2
\end{align}
in $\boldsymbol{\mathcal{Z}}^{\textrm{ref}}_{\textrm{trefoil},0}(y, \boldsymbol{x})$, which should correspond to the term 
\begin{equation}
(3 y^8 + 6 y^{10} + 7 y^{12} + 5 y^{14} + 4 y^{16} + 2 y^{18} + y^{20}) x_0x_1 x^2_2    
\end{equation}
in \eqref{eq:ZrefTrefoilF0}.
We have
\begin{flalign}\label{eq:A12B12eq3}
&
(2y^3+y^5+2y^7+y^{11})(B_1+B_2)
+y^3(B_1^2+B_2^2)(A_1+A_2)
(1+y^2+y^4)(A_1^2+A_2^2)\nonumber \\
&
\,\, + A_1A_2
+y(A_1+A_2)=2 y^2 + 5 y^{4} + 6 y^{6} + 5 y^{8} + 4 y^{10} + 2 y^{12} + y^{14} \,.
\end{flalign}
The three equations \eqref{eq:TFlevel3x0x1x2}, \eqref{eq:A12B12eq2}, and \eqref{eq:A12B12eq3} are enough to solve for $\Upsilon^{1\rightarrow 2}_{j}$ and $\Upsilon^{2\rightarrow 1}_{k}$ and give:
\begin{equation}
\Upsilon^{1\rightarrow 2}_{1}=1\,, \qquad \Upsilon^{1\rightarrow 2}_{2}=3\,, \qquad   
\Upsilon^{2\rightarrow 1}_{1}=1\,, \qquad \Upsilon^{2\rightarrow 1}_{2}=3\,. 
\end{equation}
Thus we have determined the $\Upsilon$-charges of all 13 arrows in \eqref{eq:UpsilonTF}. 
Let us continue this computation for the remaining states at level 4 to check the consistency of the assignment of these $\Upsilon$-charges.
\item The $5$ states 
\begin{equation}
\begin{aligned}
&\{
I^{(0)},\, 
I^{(0)}\cdot I^{0\rightarrow 1},\,
I^{(0)}\cdot I^{0\rightarrow 1}\cdot I^{1\rightarrow 1}_{i=1,2} ,\,
I^{(0)}\cdot I^{0\rightarrow 1}\cdot I^{1\rightarrow 1}_{i=1,2} \cdot I^{1\rightarrow 1}_{j=1,2}
\}\,,\\ 
&\{
I^{(0)},\, 
I^{(0)}\cdot I^{0\rightarrow 1},\,
I^{(0)}\cdot I^{0\rightarrow 1}\cdot I^{1\rightarrow 1}_{1} ,\,
I^{(0)}\cdot I^{0\rightarrow 1}\cdot I^{1\rightarrow 1}_{2} 
\}\,,\\    
\end{aligned}    
\end{equation}
give the term
$y^6\,(1 + 2 y^2 + y^4 + y^6) \,x_0x_1^3$   
in \eqref{eq:ZrefTrefoilF0}.
\item The $12$ states 
\begin{equation}
\begin{aligned}
&\{
I^{(0)},\, 
I^{(0)}\cdot I^{0\rightarrow 2},\,
I^{(0)}\cdot I^{0\rightarrow 2}\cdot I^{2\rightarrow 2}_{i=1,2,3} ,\,
I^{(0)}\cdot I^{0\rightarrow 2}\cdot I^{2\rightarrow 2}_{i=1,2,3} \cdot I^{2\rightarrow 2}_{j=1,2,3}
\}\,,\\ 
&\{
I^{(0)},\, 
I^{(0)}\cdot I^{0\rightarrow 2},\,
I^{(0)}\cdot I^{0\rightarrow 2}\cdot I^{2\rightarrow 2}_{1} ,\,
I^{(0)}\cdot I^{0\rightarrow 2}\cdot I^{2\rightarrow 2}_{2} 
\}\,,\\ 
&\{
I^{(0)},\, 
I^{(0)}\cdot I^{0\rightarrow 2},\,
I^{(0)}\cdot I^{0\rightarrow 2}\cdot I^{2\rightarrow 2}_{1} ,\,
I^{(0)}\cdot I^{0\rightarrow 2}\cdot I^{2\rightarrow 2}_{3} 
\}\,,\\    
&\{
I^{(0)},\, 
I^{(0)}\cdot I^{0\rightarrow 2},\,
I^{(0)}\cdot I^{0\rightarrow 2}\cdot I^{2\rightarrow 2}_{2} ,\,
I^{(0)}\cdot I^{0\rightarrow 2}\cdot I^{2\rightarrow 2}_{3} 
\}\,,\\    
\end{aligned}    
\end{equation}
give the term $y^9\,(1 + 2 y^2 + 3 y^4 + 2 y^6 + 2 y^8 + y^{10} + y^{12}) \,x_0x_2^3$   in \eqref{eq:ZrefTrefoilF0}.
\end{itemize}
Thus the $\Upsilon$-charge assignment for the 13 arrows in \eqref{eq:UpsilonTF}, summarized in \eqref{eq:TrefoilQuiverFrame0}, ensures $\boldsymbol{\mathcal{Z}}^{\textrm{ref}}_{\textrm{trefoil},0}(y, \boldsymbol{x})  
=\widehat{\textrm{Sgn}}\cdot \mathcal{Z}_{\textrm{NC},0}(y,\boldsymbol{x})$ up to level 4. 
We have also checked this agreement to higher levels using Mathematica.

\section{Generalization to trigonometric, elliptic, and generalized cohomologies}
\label{appsec:TrigElliptic}

The quiver Yangians lie at the bottom of the rational/trigonometric/elliptic hierarchy. 
In this appendix, we generalize the (shifted) quiver Yangian for arbitrary quivers to their trigonometric, 
 elliptic, and generalized cohomology versions.
For the first two cases,  the resulting algebras are called toroidal quiver algebras and elliptic quiver algebras, respectively. 
For the last case, the corresponding algebra can be defined in terms of the inverse logarithm of the formal group law of the generalized cohomology theory, and we shall call it GCT quiver algebra.  
The quiver Yangian/toroidal quiver algebra/elliptic quiver algebra/GCT algebra are associated to the  (equivariant) ordinary cohomology/K-theory/elliptic cohomology/generalized cohomology of the moduli space of the quiver gauge theory.

Unlike quiver Yangians, these new algebras have a central extension. 
The posets discussed in the main text of the paper are their representations only when the central charge is turned off.

This appendix is a straightforward generalization of \cite{Galakhov:2021vbo}, which covers toric CY$_3$'s.
Indeed, most equations look almost identical to the corresponding ones in  \cite{Galakhov:2021vbo}, except for the fact that the statistical factors are more general here; and we claim that they are valid for any quiver, not just the toric CY$_3$ quivers.

\subsection{Toroidal and elliptic quiver algebras and beyond}

For an arbitrary quiver, we can write down a set of quadratic relations that encompass both the quiver Yangian \eqref{eq:QuadraticFields}, which is the rational case, and its trigonometric, elliptic, 
and generalized cohomology versions:
\begin{tcolorbox}[ams align]\label{eq:QuadraticFieldsTE}
\myk^{(a)}_{\epsilon}(z)\, \myk^{(b)}_{\epsilon}(w) &\simeq C^{\epsilon\,\extra\,\chi_{ab}}\,\myk^{(b)}_{\epsilon}(w)\, \myk^{(a)}_{\epsilon}(z) \qquad \epsilon = \pm \,, \\
\myk^{(a)}_{+}(z)\, \myk^{(b)}_{-}(w) &\simeq \frac{ \myphi^{a\Leftarrow b}\left(z+\frac{c}{2},w-\frac{c}{2}\right) }{\myphi^{a\Leftarrow b}\left(z-\frac{c}{2},w+\frac{c}{2}\right)}\, \myk^{(b)}_{-}(w)\, \myk^{(a)}_{+}(z)\,, \nonumber\\
\myk^{(a)}_{\pm}(z)\,\mye^{(b)}(w) &\simeq \myphi^{a\Leftarrow b}\left(z\pm \frac{c}{2},w\right) \,\mye^{(b)}(w)\, \myk^{(a)}_{\pm}(z)\,, \nonumber\\
\myk^{(a)}_{\pm}(z)\,\myf^{(b)}(w) &\simeq \myphi^{a\Leftarrow b}\left(z\mp \frac{c}{2},w\right)^{-1}\, \myf^{(b)}(w)\, \myk^{(a)}_{\pm}(z)\,, \nonumber\\
\mye^{(a)}(z)\mye^{(b)}(w) &\simeq e^{\pi i s_{ab}}\,\myphi^{a\Leftarrow b}(z,w) \mye^{(b)}(w) \mye^{(a)}(z)\,, \nonumber\\
\myf^{(a)}(z)\myf^{(b)}(w) &\simeq e^{-\pi i s_{ab}}\, \myphi^{a\Leftarrow b}(z,w)^{-1} \myf^{(b)}(w) \myf^{(a)}(z)\,, \nonumber\\
\left[\mye^{(a)}(z)\,,\myf^{(b)}(w)\right\}&\simeq -\delta_{a,b} \left(\myp(\Delta-c) \myk^{(a)}_{+}\left(z-\frac{c}{2}\right)- \myp(\Delta+c)  \myk^{(a)}_{-}\left(w-\frac{c}{2}\right) \right)\,. \nonumber
\end{tcolorbox}	
\noindent Comparing to the rational case \eqref{eq:QuadraticFields}, there are two structural differences:
\begin{enumerate}
\item There is a central charge $c$, and we also define $C=e^{\beta c}$, where $\beta$ is a parameter that can be rescaled away but has been introduced to facilitate the comparison to other formulations of the algebras.
\item There are now two sets of $\psi^{(a)}(z)$ fields: $\psi^{(a)}_{\pm}(z)$; and from the first two equations of \eqref{eq:QuadraticFieldsTE}, they commute only when the central charge $c$ vanishes.
\end{enumerate}
Now let us explain the ingredients in \eqref{eq:QuadraticFieldsTE}.
\begin{enumerate}
\item The bonding factor \eqref{eq:BondingFactorDef} is generalized to
\begin{equation}\label{eq-def-balanced}
\myphi^{a\Leftarrow b}(z,w)\equiv \left(Z W\right)^{\frac{\mathfrak{t}}{2}\chi_{ab}}\varphi^{a\Leftarrow b}(z-w)\;,
\end{equation}
where the chirality $\chi_{ab}$ was defined in \eqref{eq:chiabDef} and 
\begin{equation}\label{eq-def-extra}
\mathfrak{t}\equiv\begin{cases}
\begin{aligned}
&0&\qquad &\textrm{rational}\\
&       1&\qquad &\textrm{trig./elliptic/GCT}\,.
\end{aligned}
\end{cases}
\end{equation}
Furthermore, $\varphi^{a\Leftarrow b}(z-w)$ is defined to encompass \eqref{eq:BondingFactorDef} and share its structure:
\begin{equation}\label{eq:BondingFactorDefTE}
\varphi^{a\Leftarrow b} (u) 
\equiv 
e^{\pi i t_{ab}}\frac{\prod_{I\in \{a\rightarrow b\}} \zeta(u+h_I)}{\prod_{J\in \{b\rightarrow a\}} \zeta(u-h_J)}\,,
\end{equation}
where $\zeta(u)$ is the inverse function of the logarithm $\ell_{F}(u)$ of the formal group law $F_E(x,y)$ of the (generalized) cohomology $E^{*}(-)$, see \cite[Sec.\ 6]{Galakhov:2021vbo}:
\begin{equation}
\zeta(u)=\ell^{-1}_{F}(u) \,.
\end{equation}
When the generalized cohomology $E^{*}(-)$ reduces to the ordinary cohomology/\\K-theory/elliptic cohomology, we reproduce the familiar result shown in the first three lines below:
\begin{equation}\label{eq-def-zeta}
\zeta(u)=\begin{cases}
\begin{aligned}
&u &\qquad &\textrm{rational (cohomology)}\\
&\textrm{Sin}_{\beta}{(u)} &\qquad &\textrm{trig. (K-theory)}\\
&\Theta_q(u) &\qquad &\textrm{elliptic (elliptic cohomology)}\\
&\ell^{-1}_{F}(u) &\qquad &\textrm{GCT (generalized cohomology)}\,,
\end{aligned}
\end{cases} 
\end{equation}
where we have defined	
\begin{equation}\label{notations}
\begin{aligned}
\textrm{Sin}_{\beta}(z)&\equiv 2\sinh{\frac{\beta z}{2}}= Z^{\frac{1}{2}}-Z^{-\frac{1}{2}} \,,\\
\Theta_q(z)&\equiv- \frac{\theta_q(z)}{Z^{1/2}}= (Z^{\frac{1}{2}}-Z^{-\frac{1}{2}})\prod_{n=1}^{\infty} \left(1-Z^{-1} q^n\right) \left(1-Zq^{n}\right) \,,
\end{aligned}
\end{equation}
with $Z\equiv e^{\beta z}$ and  $W\equiv e^{\beta w}$.
The three different factors in \eqref{eq-def-zeta} explain why the corresponding algebras are called rational, trigonometric, and elliptic types, respectively.
Finally, the $(ZW)$ factor in the definition \eqref{eq:BondingFactorDefTE} is to ensure the reciprocity condition 
\begin{equation}\label{eq:ReciprocityTE}
 e^{\pi i (s_{ab}+s_{ba})} \, \myphi^{a\Leftarrow b} (z,w)\,\myphi^{b\Leftarrow a} (w,z)=1\,,
\end{equation}
generalizing \eqref{eq:Reciprocity}.

\item The central term $c$ is
\begin{equation}\label{eq:cDef}
c\equiv\begin{cases}
\begin{aligned}
&0&\qquad &\textrm{rational}\\
&       c&\qquad &\textrm{trig./elliptic/GCT}\,.
\end{aligned}
\end{cases}
\end{equation}

\item The propagator is defined via
\begin{equation}\label{eq:propagatorDef}
\myp(z)\equiv\begin{cases}
\begin{aligned}
&\frac{1}{z} &\qquad &\textrm{rational}\\
&\delta(Z)\equiv\sum\lm_{n\in\IZ}Z^{n}&\qquad &\textrm{trig./elliptic/GCT} \,.
\end{aligned}
\end{cases}
\end{equation}

\item Finally, the fields $\psi^{(a)}_{\pm}(z)$, $e^{(a)}(z)$, and $f^{(a)}(z)$ have mode expansions that can be determined after their actions on the poset representations are fixed.
These mode expansions depend on the type of the algebra and can be computed similarly to \eqref{eq:ModeExpansions}, for more details see \cite[Sec.\ 2.2]{Galakhov:2021vbo}. 

\end{enumerate}

\subsection{Poset representation when \texorpdfstring{$c=0$}{c=0}}

The poset representations discussed in the main text are representations of these new quiver algebras only when $c=0$, since only then do all the $\psi^{(a)}_{\pm}(z)$ commute, and can thus be Cartan generators. 

When $c=0$, the action of the rational/trigonometric/elliptic/GCT quiver algebras on an arbitrary state $\Pi$ in the poset $\mathcal{P}(Q,W)$ is summarized as follows:
\begin{tcolorbox}[ams align]\label{eq:ActionTE}
\myk^{(a)}_{\pm}(u)|\Pi\rangle&= \left[\Psi^{(a)}_{\Pi}(u)\right]_\pm|\Pi\rangle \;,\\
e^{(a)}(z)|\Pi\rangle &=\sum_{p^{(a)} \,\in \,\textrm{Add}(\Pi)}  \myp\left(u-p^{(a)}\right)\, \nonumber\\
&\qquad\qquad\qquad 
e^{\frac{\pi i }{4} (t_{aa}+|a\rightarrow a|)}\left( \lim_{u\rightarrow p^{(a)}}\, \zeta(u-p^{(a)})\,\Psi^{(a)}_{\Pi}(u) \right)^{\frac{1}{2}}\,|\Pi+p^{(a)}\rangle \;, \nonumber\\
f^{(a)}(z)|\Pi\rangle &=\sum_{p^{(a)}\, \in\, \textrm{Rem}(\Pi)}\myp\left(u-p^{(a)}\right)\, \nonumber\\
&\qquad \qquad \qquad
 e^{\frac{\pi i }{4} (t_{aa}+|a\rightarrow a|)}\left( \lim_{u\rightarrow p^{(a)}}\, \zeta(u-p^{(a)})\,\Psi^{(a)}_{\Pi}(u) \right)^{\frac{1}{2}}\,|\Pi-p^{(a)}\rangle \;,\nonumber
\end{tcolorbox} 
\noindent with $\Psi^{(a)}_{\Pi}(u)$ given by
\begin{equation}
\Psi^{(a)}_{\Pi}(u)={}^{\sharp}\psi^{(a)}_0(u)\cdot \prod_{b\in Q_0} \prod_{\sqbox{$b$}\in\Pi} \myphi^{a\Leftarrow b} \left(u, p^{(b)}\right) \,,
\end{equation}
and $[\star]_{\pm}$ defined as in \cite[Appx.\ A]{Galakhov:2021vbo}.
The ground state contribution is given by the framing of the quiver
\begin{equation}\label{eq-GS-FQ}
   {}^{\sharp}\psi^{(a)}_0(u)=\myphi^{a\Leftarrow \infty} (u, 0) \,,
\end{equation}
similar to \eqref{eq:psi0Framing}.
The choice of the signs for the square root follows a prescription similar to the rational case, described in Sec.\ \ref{ssec:Bootstrap}.
More precisely, if we define
\begin{equation}\label{eq:sqrtTE}
\begin{aligned}
\Psi^{(a)}_{\Pi+p^{(b)}}(z) ^{\frac{1}{2}}&=  \myphi^{a\Leftarrow b} (z, p^{(b)})^{
\frac{1}{2}} \,  \Psi^{(a)}_{\Pi}(z)^{\frac{1}{2}}\,,\\
\myphi^{a\Leftarrow b}(z,w)^{\frac{1}{2}}&= \left(Z W\right)^{\frac{\mathfrak{t}}{4}\chi_{ab}}\varphi^{a\Leftarrow b}(z-w)^{\frac{1}{2}}\;,\\
\varphi^{a\Leftarrow b} (z)^{\frac{1}{2}} &=     e^{\frac{\pi i}{2} t_{ab}}\frac{\prod_{I\in \{a\rightarrow b\}} \zeta(u+h_I)^{\frac{1}{2}}}{\prod_{J\in \{b\rightarrow a\}} \zeta(u-h_J)^{\frac{1}{2}}}\,,\\
\left( \lim_{u\rightarrow p^{(a)}}\, \zeta(u-p^{(a)})\,\Psi^{(a)}_{\Pi}(u) \right)^{\frac{1}{2}}&=\lim_{u\rightarrow p^{(a)}}\zeta(u-p^{(a)})^{\frac{1}{2}} \Psi^{(a)}_{\Pi}(u)^{\frac{1}{2}}\,,\\
\varphi^{a\Leftarrow a}(0)^{\frac{1}{2}}&=e^{\frac{\pi i}{2}(t_{aa}+|a\rightarrow a|)}    \,,
\end{aligned}
\end{equation}
the action \eqref{eq:ActionTE} respects the algebraic relations \eqref{eq:QuadraticFieldsTE}, irrespective of the signs of the individual $\zeta(u\pm h_I)^{\frac{1}{2}}$.
Finally, to derive the algebraic relations \eqref{eq:QuadraticFieldsTE} from the action, we can first derive the version with $c=0$, then turn on $c$ and fix the relations by associativity. 
This computation is exactly the same as in \cite[Appx.\ B]{Galakhov:2021vbo}.

%\nocite{*}
\bibliographystyle{utphys} 
\bibliography{biblio}

\providecommand{\href}[2]{#2}\begingroup\raggedright\begin{thebibliography}{100}

\bibitem{Harvey:1996gc}
J.~A. Harvey and G.~W. Moore, ``{On the algebras of BPS states},''
  \href{http://dx.doi.org/10.1007/s002200050461}{{\em Commun. Math. Phys.}
  {\bfseries 197} (1998) 489--519},
  \href{http://arxiv.org/abs/hep-th/9609017}{{\ttfamily arXiv:hep-th/9609017}}.

\bibitem{Galakhov:2021xum}
D.~Galakhov, W.~Li, and M.~Yamazaki, ``{Shifted quiver Yangians and
  representations from BPS crystals},''
  \href{http://dx.doi.org/10.1007/JHEP08(2021)146}{{\em JHEP} {\bfseries 08}
  (2021) 146}, \href{http://arxiv.org/abs/2106.01230}{{\ttfamily
  arXiv:2106.01230 [hep-th]}}.

\bibitem{Alday:2009aq}
L.~F. Alday, D.~Gaiotto, and Y.~Tachikawa, ``{Liouville Correlation Functions
  from Four-dimensional Gauge Theories},''
  \href{http://dx.doi.org/10.1007/s11005-010-0369-5}{{\em Lett. Math. Phys.}
  {\bfseries 91} (2010) 167--197},
  \href{http://arxiv.org/abs/0906.3219}{{\ttfamily arXiv:0906.3219 [hep-th]}}.

\bibitem{Wyllard:2009hg}
N.~Wyllard, ``{A(N-1) conformal Toda field theory correlation functions from
  conformal N = 2 SU(N) quiver gauge theories},''
  \href{http://dx.doi.org/10.1088/1126-6708/2009/11/002}{{\em JHEP} {\bfseries
  11} (2009) 002}, \href{http://arxiv.org/abs/0907.2189}{{\ttfamily
  arXiv:0907.2189 [hep-th]}}.

\bibitem{Nekrasov:2009uh}
N.~A. Nekrasov and S.~L. Shatashvili, ``{Supersymmetric vacua and Bethe
  ansatz},'' \href{http://dx.doi.org/10.1016/j.nuclphysbps.2009.07.047}{{\em
  Nucl. Phys. B Proc. Suppl.} {\bfseries 192-193} (2009) 91--112},
  \href{http://arxiv.org/abs/0901.4744}{{\ttfamily arXiv:0901.4744 [hep-th]}}.

\bibitem{Nekrasov:2009ui}
N.~A. Nekrasov and S.~L. Shatashvili, ``{Quantum integrability and
  supersymmetric vacua},'' \href{http://dx.doi.org/10.1143/PTPS.177.105}{{\em
  Prog. Theor. Phys. Suppl.} {\bfseries 177} (2009) 105--119},
  \href{http://arxiv.org/abs/0901.4748}{{\ttfamily arXiv:0901.4748 [hep-th]}}.

\bibitem{Nekrasov:2009rc}
N.~A. Nekrasov and S.~L. Shatashvili,
  \href{http://dx.doi.org/10.1142/9789814304634_0015}{``{Quantization of
  Integrable Systems and Four Dimensional Gauge Theories},''} in {\em {16th
  International Congress on Mathematical Physics}}, pp.~265--289.
\newblock 8, 2009.
\newblock \href{http://arxiv.org/abs/0908.4052}{{\ttfamily arXiv:0908.4052
  [hep-th]}}.

\bibitem{Galakhov:2022uyu}
D.~Galakhov, W.~Li, and M.~Yamazaki, ``{Gauge/Bethe correspondence from quiver
  BPS algebras},'' \href{http://arxiv.org/abs/2206.13340}{{\ttfamily
  arXiv:2206.13340 [hep-th]}}.

\bibitem{Dedushenko:2021mds}
M.~Dedushenko and N.~Nekrasov, ``{Interfaces and Quantum Algebras, I: Stable
  Envelopes},'' \href{http://arxiv.org/abs/2109.10941}{{\ttfamily
  arXiv:2109.10941 [hep-th]}}.

\bibitem{Bullimore:2021rnr}
M.~Bullimore and D.~Zhang, ``{3d $\mathcal{N}=4$ Gauge Theories on an Elliptic
  Curve},'' \href{http://dx.doi.org/10.21468/SciPostPhys.13.1.005}{{\em SciPost
  Phys.} {\bfseries 13} no.~1, (2022) 005},
  \href{http://arxiv.org/abs/2109.10907}{{\ttfamily arXiv:2109.10907
  [hep-th]}}.

\bibitem{Gu:2022dac}
J.~Gu, Y.~Jiang, and M.~Sperling, ``{Rational $Q$-systems, Higgsing and Mirror
  Symmetry},'' \href{http://arxiv.org/abs/2208.10047}{{\ttfamily
  arXiv:2208.10047 [hep-th]}}.

\bibitem{Li:2020rij}
W.~Li and M.~Yamazaki, ``{Quiver Yangian from Crystal Melting},''
  \href{http://dx.doi.org/10.1007/JHEP11(2020)035}{{\em JHEP} {\bfseries 11}
  (2020) 035}, \href{http://arxiv.org/abs/2003.08909}{{\ttfamily
  arXiv:2003.08909 [hep-th]}}.

\bibitem{Galakhov:2021omc}
D.~Galakhov, ``{On Supersymmetric Interface Defects, Brane Parallel Transport,
  Order-Disorder Transition and Homological Mirror Symmetry},''
  \href{http://arxiv.org/abs/2105.07602}{{\ttfamily arXiv:2105.07602
  [hep-th]}}.

\bibitem{Ooguri:2008yb}
H.~Ooguri and M.~Yamazaki, ``{Crystal Melting and Toric Calabi-Yau
  Manifolds},'' \href{http://dx.doi.org/10.1007/s00220-009-0836-y}{{\em Commun.
  Math. Phys.} {\bfseries 292} (2009) 179--199},
\href{http://arxiv.org/abs/0811.2801}{{\ttfamily arXiv:0811.2801 [hep-th]}}.
%%CITATION = ARXIV:0811.2801;%%.

\bibitem{Kucharski:2017poe}
P.~Kucharski, M.~Reineke, M.~Stosic, and P.~Su\l{}kowski, ``{BPS states, knots
  and quivers},'' \href{http://dx.doi.org/10.1103/PhysRevD.96.121902}{{\em
  Phys. Rev. D} {\bfseries 96} no.~12, (2017) 121902},
  \href{http://arxiv.org/abs/1707.02991}{{\ttfamily arXiv:1707.02991
  [hep-th]}}.

\bibitem{Kucharski:2017ogk}
P.~Kucharski, M.~Reineke, M.~Stosic, and P.~Su\l{}kowski, ``{Knots-quivers
  correspondence},'' \href{http://dx.doi.org/10.4310/ATMP.2019.v23.n7.a4}{{\em
  Adv. Theor. Math. Phys.} {\bfseries 23} no.~7, (2019) 1849--1902},
  \href{http://arxiv.org/abs/1707.04017}{{\ttfamily arXiv:1707.04017
  [hep-th]}}.

\bibitem{Cecotti:2011rv}
S.~Cecotti and C.~Vafa, ``{Classification of complete N=2 supersymmetric
  theories in 4 dimensions},'' \href{http://arxiv.org/abs/1103.5832}{{\ttfamily
  arXiv:1103.5832 [hep-th]}}.

\bibitem{Alim:2011ae}
M.~Alim, S.~Cecotti, C.~Cordova, S.~Espahbodi, A.~Rastogi, and C.~Vafa, ``{BPS
  Quivers and Spectra of Complete N=2 Quantum Field Theories},''
  \href{http://dx.doi.org/10.1007/s00220-013-1789-8}{{\em Commun. Math. Phys.}
  {\bfseries 323} (2013) 1185--1227},
  \href{http://arxiv.org/abs/1109.4941}{{\ttfamily arXiv:1109.4941 [hep-th]}}.

\bibitem{Alim:2011kw}
M.~Alim, S.~Cecotti, C.~Cordova, S.~Espahbodi, A.~Rastogi, and C.~Vafa,
  ``{$\mathcal{N} = 2$ quantum field theories and their BPS quivers},''
  \href{http://dx.doi.org/10.4310/ATMP.2014.v18.n1.a2}{{\em Adv. Theor. Math.
  Phys.} {\bfseries 18} no.~1, (2014) 27--127},
  \href{http://arxiv.org/abs/1112.3984}{{\ttfamily arXiv:1112.3984 [hep-th]}}.

\bibitem{Creutzig:2018pts}
T.~Creutzig and Y.~Hikida, ``{Rectangular W-algebras, extended higher spin
  gravity and dual coset CFTs},''
  \href{http://dx.doi.org/10.1007/JHEP02(2019)147}{{\em JHEP} {\bfseries 02}
  (2019) 147}, \href{http://arxiv.org/abs/1812.07149}{{\ttfamily
  arXiv:1812.07149 [hep-th]}}.

\bibitem{Eberhardt:2019xmf}
L.~Eberhardt and T.~Proch\'azka, ``{The matrix-extended $W_{1+\infty}$
  algebra},'' \href{http://dx.doi.org/10.1007/JHEP12(2019)175}{{\em JHEP}
  {\bfseries 12} (2019) 175}, \href{http://arxiv.org/abs/1910.00041}{{\ttfamily
  arXiv:1910.00041 [hep-th]}}.

\bibitem{Creutzig:2019qos}
T.~Creutzig and Y.~Hikida, ``{Rectangular W algebras and superalgebras and
  their representations},''
  \href{http://dx.doi.org/10.1103/PhysRevD.100.086008}{{\em Phys. Rev. D}
  {\bfseries 100} no.~8, (2019) 086008},
  \href{http://arxiv.org/abs/1906.05868}{{\ttfamily arXiv:1906.05868
  [hep-th]}}.

\bibitem{Rapcak:2019wzw}
M.~Rap\v{c}\'ak, ``{On extensions of $
  \mathfrak{gl}\widehat{\left(\left.m\right|n\right)} $ Kac-Moody algebras and
  Calabi-Yau singularities},''
  \href{http://dx.doi.org/10.1007/JHEP01(2020)042}{{\em JHEP} {\bfseries 01}
  (2020) 042}, \href{http://arxiv.org/abs/1910.00031}{{\ttfamily
  arXiv:1910.00031 [hep-th]}}.

\bibitem{MR1324698}
V.~Ginzburg, M.~Kapranov, and E.~Vasserot, ``Langlands reciprocity for
  algebraic surfaces,'' \href{http://dx.doi.org/10.4310/MRL.1995.v2.n2.a4}{{\em
  Math. Res. Lett.} {\bfseries 2} no.~2, (1995) 147--160}.

\bibitem{Ding:1996mq}
J.-t. Ding and K.~Iohara, ``{Generalization and deformation of Drinfeld quantum
  affine algebras},'' \href{http://dx.doi.org/10.1023/A:1007341410987}{{\em
  Lett. Math. Phys.} {\bfseries 41} (1997) 181--193}.

\bibitem{Miki2007}
K.~Miki, ``A $(q,\gamma)$ analog of the $\mathcal{W}_{1+\infty}$ algebra,''
  \href{http://dx.doi.org/10.1063/1.2823979}{{\em Journal of Mathematical
  Physics} {\bfseries 48} (12, 2007) 123520--123520}.

\bibitem{MR2793271}
B.~Feigin, E.~Feigin, M.~Jimbo, T.~Miwa, and E.~Mukhin, ``Quantum continuous
  {$\mathfrak{gl}_\infty$}: semiinfinite construction of representations,''
  \href{http://dx.doi.org/10.1215/21562261-1214375}{{\em Kyoto J. Math.}
  {\bfseries 51} no.~2, (2011) 337--364}.

\bibitem{Feigin:2013fga}
B.~Feigin, M.~Jimbo, T.~Miwa, and E.~Mukhin, ``{Branching rules for quantum
  toroidal gl$_n$},'' \href{http://dx.doi.org/10.1016/j.aim.2016.03.019}{{\em
  Adv. Math.} {\bfseries 300} (2016) 229--274},
\href{http://arxiv.org/abs/1309.2147}{{\ttfamily arXiv:1309.2147 [math.QA]}}.
%%CITATION = ARXIV:1309.2147;%%.

\bibitem{MR2566895}
B.~Feigin, K.~Hashizume, A.~Hoshino, J.~Shiraishi, and S.~Yanagida, ``A
  commutative algebra on degenerate {$\mathbb{CP}^1$} and {M}acdonald
  polynomials,'' \href{http://dx.doi.org/10.1063/1.3192773}{{\em J. Math.
  Phys.} {\bfseries 50} no.~9, (2009) 095215, 42}.

\bibitem{Bezerra:2019dmp}
L.~Bezerra and E.~Mukhin, ``{Quantum toroidal algebra associated with
  $\mathfrak{gl}_{m|n}$},''
\href{http://arxiv.org/abs/1904.07297}{{\ttfamily arXiv:1904.07297 [math.QA]}}.
%%CITATION = ARXIV:1904.07297;%%.

\bibitem{MR3262444}
Y.~Saito, ``Elliptic {D}ing-{I}ohara algebra and the free field realization of
  the elliptic {M}acdonald operator,''
  \href{http://dx.doi.org/10.4171/PRIMS/139}{{\em Publ. Res. Inst. Math. Sci.}
  {\bfseries 50} no.~3, (2014) 411--455}.

\bibitem{Nieri:2015dts}
F.~Nieri, ``{An elliptic Virasoro symmetry in 6d},''
  \href{http://dx.doi.org/10.1007/s11005-017-0986-3}{{\em Lett. Math. Phys.}
  {\bfseries 107} no.~11, (2017) 2147--2187},
  \href{http://arxiv.org/abs/1511.00574}{{\ttfamily arXiv:1511.00574
  [hep-th]}}.

\bibitem{Galakhov:2021vbo}
D.~Galakhov, W.~Li, and M.~Yamazaki, ``{Toroidal and elliptic quiver BPS
  algebras and beyond},'' \href{http://dx.doi.org/10.1007/JHEP02(2022)024}{{\em
  JHEP} {\bfseries 02} (2022) 024},
  \href{http://arxiv.org/abs/2108.10286}{{\ttfamily arXiv:2108.10286
  [hep-th]}}.

\bibitem{fomin2007clusteralgebrastriangulatedsurfaces}
S.~Fomin, M.~Shapiro, and D.~Thurston, ``Cluster algebras and triangulated
  surfaces. part i: Cluster complexes,'' 2007.
\newblock \url{https://arxiv.org/abs/math/0608367}.

\bibitem{LabardiniFragoso2008}
D.~Labardini-Fragoso, ``Quivers with potentials associated to triangulated
  surfaces,'' {\em Proceedings of the London Mathematical Society} {\bfseries
  98} no.~3, (Nov., 2008) 797–839.

\bibitem{labardinifragoso2009quiverspotentialsassociatedtriangulated}
D.~Labardini-Fragoso, ``Quivers with potentials associated to triangulated
  surfaces, part ii: Arc representations,'' 2009.
\newblock \url{https://arxiv.org/abs/0909.4100}.

\bibitem{Gaiotto:2024fso}
D.~Gaiotto, N.~Grygoryev, and W.~Li, ``{Categories of Line Defects and
  Cohomological Hall Algebras},''
  \href{http://arxiv.org/abs/2406.07134}{{\ttfamily arXiv:2406.07134
  [hep-th]}}.

\bibitem{Galakhov:2020vyb}
D.~Galakhov and M.~Yamazaki, ``{Quiver Yangian and Supersymmetric Quantum
  Mechanics},'' \href{http://arxiv.org/abs/2008.07006}{{\ttfamily
  arXiv:2008.07006 [hep-th]}}.

\bibitem{Ekholm:2018eee}
T.~Ekholm, P.~Kucharski, and P.~Longhi, ``{Physics and geometry of
  knots-quivers correspondence},''
  \href{http://dx.doi.org/10.1007/s00220-020-03840-y}{{\em Commun. Math. Phys.}
  {\bfseries 379} no.~2, (2020) 361--415},
  \href{http://arxiv.org/abs/1811.03110}{{\ttfamily arXiv:1811.03110
  [hep-th]}}.

\bibitem{Ekholm:2019lmb}
T.~Ekholm, P.~Kucharski, and P.~Longhi, ``{Multi-cover skeins, quivers, and 3d
  $\mathcal{N}=2$ dualities},''
  \href{http://dx.doi.org/10.1007/JHEP02(2020)018}{{\em JHEP} {\bfseries 02}
  (2020) 018}, \href{http://arxiv.org/abs/1910.06193}{{\ttfamily
  arXiv:1910.06193 [hep-th]}}.

\bibitem{Szendroi}
B.~Szendr{\H{o}}i, ``Non-commutative {D}onaldson-{T}homas invariants and the
  conifold,'' {\em Geom. Topol.} {\bfseries 12} no.~2, (2008) 1171--1202,
  \href{http://arxiv.org/abs/0705.3419}{{\ttfamily arXiv:0705.3419 [math.AG]}}.

\bibitem{Mozgovoy:2008fd}
S.~Mozgovoy and M.~Reineke, ``{On the noncommutative Donaldson-Thomas
  invariants arising from brane tilings},''
  \href{http://arxiv.org/abs/0809.0117}{{\ttfamily arXiv:0809.0117 [math.AG]}}.

\bibitem{Ginzburg:2006fu}
V.~Ginzburg, ``{Calabi-Yau algebras},''
  \href{http://arxiv.org/abs/math/0612139}{{\ttfamily arXiv:math/0612139}}.

\bibitem{bocklandt2008graded}
R.~Bocklandt, ``Graded calabi yau algebras of dimension 3,'' {\em Journal of
  pure and applied algebra} {\bfseries 212} no.~1, (2008) 14--32.

\bibitem{Kirillov}
A.~Kirillov~Jr., {\em {Quiver Representations and Quiver Varieties}}, vol.~174
  of {\em Graduate Studies in Mathematics}.
\newblock AMS, Providence, USA, 2016.

\bibitem{Gaberdiel:2018nbs}
M.~R. Gaberdiel, W.~Li, and C.~Peng, ``{Twin-plane-partitions and
  $\mathcal{N}=2$ affine Yangian},''
  \href{http://dx.doi.org/10.1007/JHEP11(2018)192}{{\em JHEP} {\bfseries 11}
  (2018) 192}, \href{http://arxiv.org/abs/1807.11304}{{\ttfamily
  arXiv:1807.11304 [hep-th]}}.

\bibitem{Rapcak:2018nsl}
M.~Rapcak, Y.~Soibelman, Y.~Yang, and G.~Zhao, ``{Cohomological Hall algebras,
  vertex algebras and instantons},''
  \href{http://dx.doi.org/10.1007/s00220-019-03575-5}{{\em Commun. Math. Phys.}
  {\bfseries 376} no.~3, (2019) 1803--1873},
  \href{http://arxiv.org/abs/1810.10402}{{\ttfamily arXiv:1810.10402
  [math.QA]}}.

\bibitem{Rapcak:2020ueh}
M.~Rapcak, Y.~Soibelman, Y.~Yang, and G.~Zhao, ``{Cohomological Hall algebras
  and perverse coherent sheaves on toric Calabi\textendash{}Yau $3$-folds},''
  \href{http://dx.doi.org/10.4310/CNTP.2023.v17.n4.a2}{{\em Commun. Num. Theor.
  Phys.} {\bfseries 17} no.~4, (2023) 847--939},
  \href{http://arxiv.org/abs/2007.13365}{{\ttfamily arXiv:2007.13365
  [math.QA]}}.

\bibitem{Mozgovoy:2020has}
S.~Mozgovoy and B.~Pioline, ``{Attractor invariants, brane tilings and
  crystals},'' \href{http://arxiv.org/abs/2012.14358}{{\ttfamily
  arXiv:2012.14358 [hep-th]}}.

\bibitem{Gaiotto:2009hg}
D.~Gaiotto, G.~W. Moore, and A.~Neitzke, ``{Wall-crossing, Hitchin systems, and
  the WKB approximation},''
  \href{http://dx.doi.org/10.1016/j.aim.2012.09.027}{{\em Adv. Math.}
  {\bfseries 234} (2013) 239--403},
  \href{http://arxiv.org/abs/0907.3987}{{\ttfamily arXiv:0907.3987 [hep-th]}}.

\bibitem{Denef:2002ru}
F.~Denef, ``{Quantum quivers and Hall / hole halos},''
  \href{http://dx.doi.org/10.1088/1126-6708/2002/10/023}{{\em JHEP} {\bfseries
  10} (2002) 023}, \href{http://arxiv.org/abs/hep-th/0206072}{{\ttfamily
  arXiv:hep-th/0206072}}.

\bibitem{Douglas:1996sw}
M.~R. Douglas and G.~W. Moore, ``{D-branes, quivers, and ALE instantons},''
  \href{http://arxiv.org/abs/hep-th/9603167}{{\ttfamily arXiv:hep-th/9603167}}.

\bibitem{Diaconescu:1997br}
D.-E. Diaconescu, M.~R. Douglas, and J.~Gomis, ``{Fractional branes and wrapped
  branes},'' \href{http://dx.doi.org/10.1088/1126-6708/1998/02/013}{{\em JHEP}
  {\bfseries 02} (1998) 013},
  \href{http://arxiv.org/abs/hep-th/9712230}{{\ttfamily arXiv:hep-th/9712230}}.

\bibitem{Diaconescu:1999dt}
D.-E. Diaconescu and J.~Gomis, ``{Fractional branes and boundary states in
  orbifold theories},''
  \href{http://dx.doi.org/10.1088/1126-6708/2000/10/001}{{\em JHEP} {\bfseries
  10} (2000) 001}, \href{http://arxiv.org/abs/hep-th/9906242}{{\ttfamily
  arXiv:hep-th/9906242}}.

\bibitem{Douglas:2000ah}
M.~R. Douglas, B.~Fiol, and C.~Romelsberger, ``{Stability and BPS branes},''
  \href{http://dx.doi.org/10.1088/1126-6708/2005/09/006}{{\em JHEP} {\bfseries
  09} (2005) 006}, \href{http://arxiv.org/abs/hep-th/0002037}{{\ttfamily
  arXiv:hep-th/0002037}}.

\bibitem{Douglas:2000qw}
M.~R. Douglas, B.~Fiol, and C.~Romelsberger, ``{The Spectrum of BPS branes on a
  noncompact Calabi-Yau},''
  \href{http://dx.doi.org/10.1088/1126-6708/2005/09/057}{{\em JHEP} {\bfseries
  09} (2005) 057}, \href{http://arxiv.org/abs/hep-th/0003263}{{\ttfamily
  arXiv:hep-th/0003263}}.

\bibitem{Hanany:2005ve}
A.~Hanany and K.~D. Kennaway, ``{Dimer models and toric diagrams},''
\href{http://arxiv.org/abs/hep-th/0503149}{{\ttfamily arXiv:hep-th/0503149
  [hep-th]}}.
%%CITATION = HEP-TH/0503149;%%.

\bibitem{Franco:2005rj}
S.~Franco, A.~Hanany, K.~D. Kennaway, D.~Vegh, and B.~Wecht, ``{Brane dimers
  and quiver gauge theories},''
  \href{http://dx.doi.org/10.1088/1126-6708/2006/01/096}{{\em JHEP} {\bfseries
  01} (2006) 096}, \href{http://arxiv.org/abs/hep-th/0504110}{{\ttfamily
  arXiv:hep-th/0504110}}.

\bibitem{Franco:2005sm}
S.~Franco, A.~Hanany, D.~Martelli, J.~Sparks, D.~Vegh, and B.~Wecht, ``{Gauge
  theories from toric geometry and brane tilings},''
  \href{http://dx.doi.org/10.1088/1126-6708/2006/01/128}{{\em JHEP} {\bfseries
  01} (2006) 128},
\href{http://arxiv.org/abs/hep-th/0505211}{{\ttfamily arXiv:hep-th/0505211
  [hep-th]}}.
%%CITATION = HEP-TH/0505211;%%.

\bibitem{Herzog:2003zc}
C.~P. Herzog, ``{Exceptional collections and del Pezzo gauge theories},''
  \href{http://dx.doi.org/10.1088/1126-6708/2004/04/069}{{\em JHEP} {\bfseries
  04} (2004) 069}, \href{http://arxiv.org/abs/hep-th/0310262}{{\ttfamily
  arXiv:hep-th/0310262}}.

\bibitem{Herzog:2004qw}
C.~P. Herzog, ``{Seiberg duality is an exceptional mutation},''
  \href{http://dx.doi.org/10.1088/1126-6708/2004/08/064}{{\em JHEP} {\bfseries
  08} (2004) 064}, \href{http://arxiv.org/abs/hep-th/0405118}{{\ttfamily
  arXiv:hep-th/0405118}}.

\bibitem{Aspinwall:2004bs}
P.~S. Aspinwall and S.~H. Katz, ``{Computation of superpotentials for
  D-branes},'' \href{http://dx.doi.org/10.1007/s00220-006-1527-6}{{\em Commun.
  Math. Phys.} {\bfseries 264} (2006) 227--253},
  \href{http://arxiv.org/abs/hep-th/0412209}{{\ttfamily arXiv:hep-th/0412209}}.

\bibitem{Hanany:2006nm}
A.~Hanany, C.~P. Herzog, and D.~Vegh, ``{Brane tilings and exceptional
  collections},'' \href{http://dx.doi.org/10.1088/1126-6708/2006/07/001}{{\em
  JHEP} {\bfseries 07} (2006) 001},
  \href{http://arxiv.org/abs/hep-th/0602041}{{\ttfamily arXiv:hep-th/0602041}}.

\bibitem{Gaiotto:2009we}
D.~Gaiotto, ``{N=2 dualities},''
  \href{http://dx.doi.org/10.1007/JHEP08(2012)034}{{\em JHEP} {\bfseries 08}
  (2012) 034}, \href{http://arxiv.org/abs/0904.2715}{{\ttfamily arXiv:0904.2715
  [hep-th]}}.

\bibitem{Cecotti:2012gh}
S.~Cecotti and M.~Del~Zotto, ``{4d N=2 Gauge Theories and Quivers: the
  Non-Simply Laced Case},''
  \href{http://dx.doi.org/10.1007/JHEP10(2012)190}{{\em JHEP} {\bfseries 10}
  (2012) 190}, \href{http://arxiv.org/abs/1207.7205}{{\ttfamily arXiv:1207.7205
  [hep-th]}}.

\bibitem{Katz:1996fh}
S.~H. Katz, A.~Klemm, and C.~Vafa, ``{Geometric engineering of quantum field
  theories},'' \href{http://dx.doi.org/10.1016/S0550-3213(97)00282-4}{{\em
  Nucl. Phys. B} {\bfseries 497} (1997) 173--195},
  \href{http://arxiv.org/abs/hep-th/9609239}{{\ttfamily arXiv:hep-th/9609239}}.

\bibitem{Katz:1996th}
S.~H. Katz and C.~Vafa, ``{Geometric engineering of N=1 quantum field
  theories},'' \href{http://dx.doi.org/10.1016/S0550-3213(97)00283-6}{{\em
  Nucl. Phys. B} {\bfseries 497} (1997) 196--204},
  \href{http://arxiv.org/abs/hep-th/9611090}{{\ttfamily arXiv:hep-th/9611090}}.

\bibitem{Cecotti:2010fi}
S.~Cecotti, A.~Neitzke, and C.~Vafa, ``{R-Twisting and 4d/2d
  Correspondences},'' \href{http://arxiv.org/abs/1006.3435}{{\ttfamily
  arXiv:1006.3435 [hep-th]}}.

\bibitem{Denef:2007vg}
F.~Denef and G.~W. Moore, ``{Split states, entropy enigmas, holes and halos},''
  \href{http://dx.doi.org/10.1007/JHEP11(2011)129}{{\em JHEP} {\bfseries 11}
  (2011) 129}, \href{http://arxiv.org/abs/hep-th/0702146}{{\ttfamily
  arXiv:hep-th/0702146}}.

\bibitem{Kontsevich:2008fj}
M.~Kontsevich and Y.~Soibelman, ``{Stability structures, motivic
  Donaldson-Thomas invariants and cluster transformations},''
  \href{http://arxiv.org/abs/0811.2435}{{\ttfamily arXiv:0811.2435 [math.AG]}}.

\bibitem{Gaiotto:2010okc}
D.~Gaiotto, G.~W. Moore, and A.~Neitzke, ``{Four-dimensional wall-crossing via
  three-dimensional field theory},''
  \href{http://dx.doi.org/10.1007/s00220-010-1071-2}{{\em Commun. Math. Phys.}
  {\bfseries 299} (2010) 163--224},
  \href{http://arxiv.org/abs/0807.4723}{{\ttfamily arXiv:0807.4723 [hep-th]}}.

\bibitem{Dimofte:2009bv}
T.~Dimofte and S.~Gukov, ``{Refined, Motivic, and Quantum},''
  \href{http://dx.doi.org/10.1007/s11005-009-0357-9}{{\em Lett. Math. Phys.}
  {\bfseries 91} (2010) 1}, \href{http://arxiv.org/abs/0904.1420}{{\ttfamily
  arXiv:0904.1420 [hep-th]}}.

\bibitem{Descombes:2021snc}
P.~Descombes, ``{Cohomological DT invariants from localization},''
  \href{http://arxiv.org/abs/2106.02518}{{\ttfamily arXiv:2106.02518
  [math.AG]}}.

\bibitem{Cirafici:2021yda}
M.~Cirafici, ``{On the M2\textendash{}Brane Index on Noncommutative Crepant
  Resolutions},'' \href{http://dx.doi.org/10.1007/s11005-022-01579-2}{{\em
  Lett. Math. Phys.} {\bfseries 112} no.~5, (2022) 88},
  \href{http://arxiv.org/abs/2111.01102}{{\ttfamily arXiv:2111.01102
  [hep-th]}}.

\bibitem{Kontsevich:2010px}
M.~Kontsevich and Y.~Soibelman, ``{Cohomological Hall algebra, exponential
  Hodge structures and motivic Donaldson-Thomas invariants},''
  \href{http://dx.doi.org/10.4310/CNTP.2011.v5.n2.a1}{{\em Commun. Num. Theor.
  Phys.} {\bfseries 5} (2011) 231--352},
  \href{http://arxiv.org/abs/1006.2706}{{\ttfamily arXiv:1006.2706 [math.AG]}}.

\bibitem{Morrison:2011rz}
A.~Morrison, S.~Mozgovoy, K.~Nagao, and B.~Szendroi, ``{Motivic
  Donaldson-Thomas invariants of the conifold and the refined topological
  vertex},'' \href{http://arxiv.org/abs/1107.5017}{{\ttfamily arXiv:1107.5017
  [math.AG]}}.

\bibitem{Cordova:2013cea}
C.~Cordova and D.~L. Jafferis, ``{Complex Chern-Simons from M5-branes on the
  Squashed Three-Sphere},''
  \href{http://dx.doi.org/10.1007/JHEP11(2017)119}{{\em JHEP} {\bfseries 11}
  (2017) 119}, \href{http://arxiv.org/abs/1305.2891}{{\ttfamily arXiv:1305.2891
  [hep-th]}}.

\bibitem{Noshita:2021ldl}
G.~Noshita and A.~Watanabe, ``{A note on quiver quantum toroidal algebra},''
  \href{http://dx.doi.org/10.1007/JHEP05(2022)011}{{\em JHEP} {\bfseries 05}
  (2022) 011}, \href{http://arxiv.org/abs/2108.07104}{{\ttfamily
  arXiv:2108.07104 [hep-th]}}.

\bibitem{Iqbal:2007ii}
A.~Iqbal, C.~Kozcaz, and C.~Vafa, ``{The Refined topological vertex},''
  \href{http://dx.doi.org/10.1088/1126-6708/2009/10/069}{{\em JHEP} {\bfseries
  10} (2009) 069}, \href{http://arxiv.org/abs/hep-th/0701156}{{\ttfamily
  arXiv:hep-th/0701156}}.

\bibitem{Tsymbaliuk:2014fvq}
A.~Tsymbaliuk, ``{The affine Yangian of $\mathfrak{gl}_1$ revisited},''
  \href{http://dx.doi.org/10.1016/j.aim.2016.08.041}{{\em Adv. Math.}
  {\bfseries 304} (2017) 583--645},
\href{http://arxiv.org/abs/1404.5240}{{\ttfamily arXiv:1404.5240 [math.RT]}}.
%%CITATION = ARXIV:1404.5240;%%.

\bibitem{Prochazka:2015deb}
T.~Proch\'azka, ``{$ \mathcal{W} $ -symmetry, topological vertex and affine
  Yangian},'' \href{http://dx.doi.org/10.1007/JHEP10(2016)077}{{\em JHEP}
  {\bfseries 10} (2016) 077}, \href{http://arxiv.org/abs/1512.07178}{{\ttfamily
  arXiv:1512.07178 [hep-th]}}.

\bibitem{Datta:2016cmw}
S.~Datta, M.~R. Gaberdiel, W.~Li, and C.~Peng, ``{Twisted sectors from plane
  partitions},'' \href{http://dx.doi.org/10.1007/JHEP09(2016)138}{{\em JHEP}
  {\bfseries 09} (2016) 138}, \href{http://arxiv.org/abs/1606.07070}{{\ttfamily
  arXiv:1606.07070 [hep-th]}}.

\bibitem{Gaberdiel:2017dbk}
M.~R. Gaberdiel, R.~Gopakumar, W.~Li, and C.~Peng, ``{Higher Spins and Yangian
  Symmetries},'' \href{http://dx.doi.org/10.1007/JHEP04(2017)152}{{\em JHEP}
  {\bfseries 04} (2017) 152}, \href{http://arxiv.org/abs/1702.05100}{{\ttfamily
  arXiv:1702.05100 [hep-th]}}.

\bibitem{Young2007}
B.~Young, ``Computing a pyramid partition generating function with dimer
  shuffling,'' 2007.
\newblock \url{https://arxiv.org/abs/0709.3079}.

\bibitem{bryan2010generating}
J.~Bryan and B.~Young, ``Generating functions for colored 3d young diagrams and
  the donaldson-thomas invariants of orbifolds,'' {\em Duke Mathematical
  Journal} {\bfseries 152} no.~1, (2010) 115--153.

\bibitem{Lee:2016dbm}
S.-J. Lee and P.~Yi, ``{Witten Index for Noncompact Dynamics},''
  \href{http://dx.doi.org/10.1007/JHEP06(2016)089}{{\em JHEP} {\bfseries 06}
  (2016) 089}, \href{http://arxiv.org/abs/1602.03530}{{\ttfamily
  arXiv:1602.03530 [hep-th]}}.

\bibitem{Duan:2020qjy}
Z.~Duan, D.~Ghim, and P.~Yi, ``{5D BPS Quivers and KK Towers},''
  \href{http://dx.doi.org/10.1007/JHEP02(2021)119}{{\em JHEP} {\bfseries 02}
  (2021) 119}, \href{http://arxiv.org/abs/2011.04661}{{\ttfamily
  arXiv:2011.04661 [hep-th]}}.

\bibitem{Bridgeland:2001xf}
T.~Bridgeland, A.~King, and M.~Reid, ``{The McKay correspondence as an
  equivalence of derived categories},''
  \href{http://dx.doi.org/10.1090/S0894-0347-01-00368-X}{{\em J. Am. Math.
  Soc.} {\bfseries 14} (2001) 535--554}.

\bibitem{McKay:1980}
J.~McKay, ``{Graphs, Singularities, and Finite Groups},''
  \href{http://dx.doi.org/10.1090/pspum/037}{{\em Proc. Symp. Pure Math.}
  {\bfseries 37} (1980) 183}.

\bibitem{Ueda:2020yon}
M.~Ueda, ``{Affine super Yangians and rectangular W-superalgebras},''
  \href{http://dx.doi.org/10.1063/5.0076638}{{\em J. Math. Phys.} {\bfseries
  63} no.~5, (2022) 051701}, \href{http://arxiv.org/abs/2002.03479}{{\ttfamily
  arXiv:2002.03479 [math.RT]}}.

\bibitem{Bao:2022jhy}
J.~Bao, ``{Quiver Yangians and $\mathcal{W}$-Algebras for Generalized
  Conifolds},'' \href{http://arxiv.org/abs/2208.13395}{{\ttfamily
  arXiv:2208.13395 [hep-th]}}.

\bibitem{Freyd1985}
P.~Freyd, D.~Yetter, J.~Hoste, W.~B.~R. Lickorish, K.~Millett, and A.~Ocneanu,
  ``A new polynomial invariant of knots and links,'' {\em Bulletin of the
  American Mathematical Society} {\bfseries 12} no.~2, (1985) 239--246.

\bibitem{przytycki1988invariants}
J.~Przytycki and P.~Traczyk, ``Invariants of links of conway type,'' {\em Kobe
  journal of mathematics} {\bfseries 4} no.~2, (1988) 115--139.

\bibitem{Ooguri:1999bv}
H.~Ooguri and C.~Vafa, ``{Knot invariants and topological strings},''
  \href{http://dx.doi.org/10.1016/S0550-3213(00)00118-8}{{\em Nucl. Phys. B}
  {\bfseries 577} (2000) 419--438},
  \href{http://arxiv.org/abs/hep-th/9912123}{{\ttfamily arXiv:hep-th/9912123}}.

\bibitem{Labastida:2000zp}
J.~M.~F. Labastida and M.~Marino, ``{Polynomial invariants for torus knots and
  topological strings},'' \href{http://dx.doi.org/10.1007/s002200100374}{{\em
  Commun. Math. Phys.} {\bfseries 217} (2001) 423--449},
  \href{http://arxiv.org/abs/hep-th/0004196}{{\ttfamily arXiv:hep-th/0004196}}.

\bibitem{Jankowski:2022qdp}
J.~Jankowski, P.~Kucharski, H.~Larragu\'\i{}vel, D.~Noshchenko, and
  P.~Su\l{}kowski, ``{Quiver Diagonalization and Open BPS States},''
  \href{http://dx.doi.org/10.1007/s00220-023-04753-2}{{\em Commun. Math. Phys.}
  {\bfseries 402} no.~2, (2023) 1551--1584},
  \href{http://arxiv.org/abs/2212.04379}{{\ttfamily arXiv:2212.04379
  [hep-th]}}.

\bibitem{Meinhardt2014}
S.~Meinhardt and M.~Reineke, ``Donaldson-thomas invariants versus intersection
  cohomology of quiver moduli,'' {\em Journal f{\"u}r die reine und angewandte
  Mathematik (Crelles Journal)} {\bfseries 2019} no.~754, (2019) 143--178,
  \href{http://arxiv.org/abs/1411.4062}{{\ttfamily arXiv:1411.4062 [math.RT]}}.

\bibitem{franzen2018semistable}
H.~Franzen and M.~Reineke, ``Semistable chow--hall algebras of quivers and
  quantized donaldson--thomas invariants,'' {\em Algebra \& Number Theory}
  {\bfseries 12} no.~5, (2018) 1001--1025,
  \href{http://arxiv.org/abs/1512.03748}{{\ttfamily arXiv:1512.03748
  [math.RT]}}.

\bibitem{efimov_2012}
A.~I. Efimov, ``Cohomological hall algebra of a symmetric quiver,''
  \href{http://dx.doi.org/10.1112/S0010437X12000152}{{\em Compositio
  Mathematica} {\bfseries 148} no.~4, (2012) 1133–1146},
  \href{http://arxiv.org/abs/1103.2736}{{\ttfamily arXiv:1103.2736 [math.RT]}}.

\bibitem{Jankowski:2021flt}
J.~Jankowski, P.~Kucharski, H.~Larragu\'\i{}vel, D.~Noshchenko, and
  P.~Su\l{}kowski, ``{Permutohedra for knots and quivers},''
  \href{http://dx.doi.org/10.1103/PhysRevD.104.086017}{{\em Phys. Rev. D}
  {\bfseries 104} no.~8, (2021) 086017},
  \href{http://arxiv.org/abs/2105.11806}{{\ttfamily arXiv:2105.11806
  [hep-th]}}.

\bibitem{Stosic:2017wno}
M.~Stosic and P.~Wedrich, ``{Rational links and DT invariants of quivers},''
  \href{http://dx.doi.org/10.1093/imrn/rny289}{{\em Int. Math. Res. Not.}
  {\bfseries 2021} no.~6, (2021) 4169--4210},
  \href{http://arxiv.org/abs/1711.03333}{{\ttfamily arXiv:1711.03333
  [math.QA]}}.

\bibitem{Stosic:2020xwn}
M.~Stosic and P.~Wedrich, ``{Tangle addition and the knots-quivers
  correspondence},'' \href{http://arxiv.org/abs/2004.10837}{{\ttfamily
  arXiv:2004.10837 [math.QA]}}.

\bibitem{Panfil:2018sis}
M.~Panfil, M.~Sto\v{s}i\'c, and P.~Su\l{}kowski, ``{Donaldson-Thomas
  invariants, torus knots, and lattice paths},''
  \href{http://dx.doi.org/10.1103/PhysRevD.98.026022}{{\em Phys. Rev. D}
  {\bfseries 98} no.~2, (2018) 026022},
  \href{http://arxiv.org/abs/1802.04573}{{\ttfamily arXiv:1802.04573
  [hep-th]}}.

\bibitem{Panfil:2018faz}
M.~Panfil and P.~Su\l{}kowski, ``{Topological strings, strips and quivers},''
  \href{http://dx.doi.org/10.1007/JHEP01(2019)124}{{\em JHEP} {\bfseries 01}
  (2019) 124}, \href{http://arxiv.org/abs/1811.03556}{{\ttfamily
  arXiv:1811.03556 [hep-th]}}.

\bibitem{Kimura:2020qns}
T.~Kimura, M.~Panfil, Y.~Sugimoto, and P.~Su\l{}kowski, ``{Branes, quivers and
  wave-functions},''
  \href{http://dx.doi.org/10.21468/SciPostPhys.10.2.051}{{\em SciPost Phys.}
  {\bfseries 10} no.~2, (2021) 051},
  \href{http://arxiv.org/abs/2011.06783}{{\ttfamily arXiv:2011.06783
  [hep-th]}}.

\bibitem{Kucharski:2020rsp}
P.~Kucharski, ``{Quivers for 3-manifolds: the correspondence, BPS states, and
  3d $ \mathcal{N} $ = 2 theories},''
  \href{http://dx.doi.org/10.1007/JHEP09(2020)075}{{\em JHEP} {\bfseries 09}
  (2020) 075}, \href{http://arxiv.org/abs/2005.13394}{{\ttfamily
  arXiv:2005.13394 [hep-th]}}.

\bibitem{Ekholm:2021irc}
T.~Ekholm, A.~Gruen, S.~Gukov, P.~Kucharski, S.~Park, M.~Sto\v{s}i\'c, and
  P.~Su\l{}kowski, ``{Branches, quivers, and ideals for knot complements},''
  \href{http://dx.doi.org/10.1016/j.geomphys.2022.104520}{{\em J. Geom. Phys.}
  {\bfseries 177} (2022) 104520},
  \href{http://arxiv.org/abs/2110.13768}{{\ttfamily arXiv:2110.13768
  [hep-th]}}.

\bibitem{Ekholm:2021gyu}
T.~Ekholm, P.~Kucharski, and P.~Longhi, ``{Knot homologies and generalized
  quiver partition functions},''
  \href{http://arxiv.org/abs/2108.12645}{{\ttfamily arXiv:2108.12645
  [hep-th]}}.

\bibitem{Witten:1988hf}
E.~Witten, ``{Quantum Field Theory and the Jones Polynomial},''
  \href{http://dx.doi.org/10.1007/BF01217730}{{\em Commun. Math. Phys.}
  {\bfseries 121} (1989) 351--399}.

\bibitem{Witten:1992fb}
E.~Witten, ``{Chern-Simons gauge theory as a string theory},'' {\em Prog.
  Math.} {\bfseries 133} (1995) 637--678,
  \href{http://arxiv.org/abs/hep-th/9207094}{{\ttfamily arXiv:hep-th/9207094}}.

\bibitem{Dimofte:2010tz}
T.~Dimofte, S.~Gukov, and L.~Hollands, ``{Vortex Counting and Lagrangian
  3-manifolds},'' \href{http://dx.doi.org/10.1007/s11005-011-0531-8}{{\em Lett.
  Math. Phys.} {\bfseries 98} (2011) 225--287},
  \href{http://arxiv.org/abs/1006.0977}{{\ttfamily arXiv:1006.0977 [hep-th]}}.

\bibitem{Terashima:2011qi}
Y.~Terashima and M.~Yamazaki, ``{SL(2,R) Chern-Simons, Liouville, and Gauge
  Theory on Duality Walls},''
  \href{http://dx.doi.org/10.1007/JHEP08(2011)135}{{\em JHEP} {\bfseries 08}
  (2011) 135}, \href{http://arxiv.org/abs/1103.5748}{{\ttfamily arXiv:1103.5748
  [hep-th]}}.

\bibitem{Dimofte:2011ju}
T.~Dimofte, D.~Gaiotto, and S.~Gukov, ``{Gauge Theories Labelled by
  Three-Manifolds},'' \href{http://dx.doi.org/10.1007/s00220-013-1863-2}{{\em
  Commun. Math. Phys.} {\bfseries 325} (2014) 367--419},
  \href{http://arxiv.org/abs/1108.4389}{{\ttfamily arXiv:1108.4389 [hep-th]}}.

\bibitem{Yagi:2013fda}
J.~Yagi, ``{3d TQFT from 6d SCFT},''
  \href{http://dx.doi.org/10.1007/JHEP08(2013)017}{{\em JHEP} {\bfseries 08}
  (2013) 017}, \href{http://arxiv.org/abs/1305.0291}{{\ttfamily arXiv:1305.0291
  [hep-th]}}.

\bibitem{Lee:2013ida}
S.~Lee and M.~Yamazaki, ``{3d Chern-Simons Theory from M5-branes},''
  \href{http://dx.doi.org/10.1007/JHEP12(2013)035}{{\em JHEP} {\bfseries 12}
  (2013) 035}, \href{http://arxiv.org/abs/1305.2429}{{\ttfamily arXiv:1305.2429
  [hep-th]}}.

\bibitem{Fuji:2012nx}
H.~Fuji, S.~Gukov, and P.~Sulkowski, ``{Super-A-polynomial for knots and BPS
  states},'' \href{http://dx.doi.org/10.1016/j.nuclphysb.2012.10.005}{{\em
  Nucl. Phys. B} {\bfseries 867} (2013) 506--546},
  \href{http://arxiv.org/abs/1205.1515}{{\ttfamily arXiv:1205.1515 [hep-th]}}.

\bibitem{Hwang:2017kmk}
C.~Hwang, P.~Yi, and Y.~Yoshida, ``{Fundamental Vortices, Wall-Crossing, and
  Particle-Vortex Duality},''
  \href{http://dx.doi.org/10.1007/JHEP05(2017)099}{{\em JHEP} {\bfseries 05}
  (2017) 099}, \href{http://arxiv.org/abs/1703.00213}{{\ttfamily
  arXiv:1703.00213 [hep-th]}}.

\bibitem{Reineke2011DegenerateCH}
M.~Reineke, ``{Degenerate cohomological Hall algebra and quantized
  Donaldson-Thomas invariants for $m$-loop quivers},'' {\em Documenta
  Mathematica} (2011) , \href{http://arxiv.org/abs/1102.3978}{{\ttfamily
  arXiv:1102.3978 [math.RT]}}.

\bibitem{Mozgovoy:2021iwz}
S.~Mozgovoy, ``{Operadic approach to wall-crossing},''
  \href{http://dx.doi.org/10.1016/j.jalgebra.2021.12.032}{{\em J. Algebra}
  {\bfseries 596} (2022) 53--88},
  \href{http://arxiv.org/abs/2101.07636}{{\ttfamily arXiv:2101.07636
  [math.AG]}}.

\bibitem{Arguz:2021zpx}
H.~Arg\"uz and P.~Bousseau, ``{The flow tree formula for
  Donaldson\textendash{}Thomas invariants of quivers with potentials},''
  \href{http://dx.doi.org/10.1112/S0010437X22007801}{{\em Compos. Math.}
  {\bfseries 158} no.~12, (2022) 2206--2249},
  \href{http://arxiv.org/abs/2102.11200}{{\ttfamily arXiv:2102.11200
  [math.RT]}}.

\bibitem{Gaberdiel:2017hcn}
M.~R. Gaberdiel, W.~Li, C.~Peng, and H.~Zhang, ``{The supersymmetric affine
  Yangian},'' \href{http://dx.doi.org/10.1007/JHEP05(2018)200}{{\em JHEP}
  {\bfseries 05} (2018) 200}, \href{http://arxiv.org/abs/1711.07449}{{\ttfamily
  arXiv:1711.07449 [hep-th]}}.

\bibitem{Ekholm:2024ceb}
T.~Ekholm, P.~Longhi, and L.~Nakamura, ``{The worldsheet skein D-module and
  basic curves on Lagrangian fillings of the Hopf link conormal},''
  \href{http://arxiv.org/abs/2407.09836}{{\ttfamily arXiv:2407.09836
  [math.SG]}}.

\bibitem{Feng:2000mi}
B.~Feng, A.~Hanany, and Y.-H. He, ``{D-brane gauge theories from toric
  singularities and toric duality},''
  \href{http://dx.doi.org/10.1016/S0550-3213(00)00699-4}{{\em Nucl. Phys. B}
  {\bfseries 595} (2001) 165--200},
  \href{http://arxiv.org/abs/hep-th/0003085}{{\ttfamily arXiv:hep-th/0003085}}.

\end{thebibliography}\endgroup

\end{document}